\documentclass[11pt, a4paper, oneside]{Thesis} 

\graphicspath{{Pictures/}} 

\usepackage[square,sort,compress,numbers]{natbib}
\usepackage{cite}
\usepackage{amsmath,latexsym} 
\usepackage{float}
\usepackage{hyperref}
\hypersetup{colorlinks,citecolor=red}
\hypersetup{colorlinks=true,urlcolor=black}
\title{\ttitle} 
\DeclareUnicodeCharacter{2212}{-}
\DeclareUnicodeCharacter{2212}{+}
\DeclareMathOperator{\sgn}{sgn}
\begin{document}

\setstretch{1.3} 

\fancyhead{} 
\rhead{\thepage} 
\lhead{} 

%

\thesistitle{Bouncing Scenario and Cosmic Dynamics in Modified Theories of Gravity}
\documenttype{Thesis}
\supervisor{Prof. Bivudutta Mishra}
\supervisorposition{Professor}
\supervisorinstitute{BITS-Pilani, Hyderabad Campus}
\examiner{}
\degree{Ph.D. Research Scholar}
\coursecode{DOCTOR OF PHILOSOPHY}
\coursename{Thesis}
\authors{\textbf{Agrawal Amarkumar Shyamsunder}}
\IDNumber{2018PHXF0463H}
\addresses{}
\subject{}
\university{\texorpdfstring{\href{http://www.bits-pilani.ac.in/} 
                {Birla Institute of Technology and Science, Pilani}} 
                {Birla Institute of Technology and Science, Pilani}}
\UNIVERSITY{\texorpdfstring{\href{http://www.bits-pilani.ac.in/} 
                {BIRLA INSTITUTE OF TECHNOLOGY AND SCIENCE, PILANI}} 
                {BIRLA INSTITUTE OF TECHNOLOGY AND SCIENCE, PILANI}}


\department{\texorpdfstring{\href{http://www.bits-pilani.ac.in/pilani/Mathematics/Mathematics} 
                {Mathematics}} 
                {Mathematics}}
\DEPARTMENT{\texorpdfstring{\href{http://www.bits-pilani.ac.in/pilani/Mathematics/Mathematics} 
                {Mathematics}} 
                {Mathematics}}
\group{\texorpdfstring{\href{Research Group Web Site URL Here (include http://)}
                {Research Group Name}} 
                {Research Group Name}}
\GROUP{\texorpdfstring{\href{Research Group Web Site URL Here (include http://)}
                {RESEARCH GROUP NAME (IN BLOCK CAPITALS)}}
                {RESEARCH GROUP NAME (IN BLOCK CAPITALS)}}
\faculty{\texorpdfstring{\href{Faculty Web Site URL Here (include http://)}
                {Faculty Name}}
                {Faculty Name}}
\FACULTY{\texorpdfstring{\href{Faculty Web Site URL Here (include http://)}
                {FACULTY NAME (IN BLOCK CAPITALS)}}
                {FACULTY NAME (IN BLOCK CAPITALS)}}

\maketitle

\clearpage
\frontmatter 
\Certificate
\setstretch{1.3} 
\Declaration
\pagestyle{empty} 
\pagenumbering{gobble}

\Dedicatory{\bf \begin{LARGE}
To 
\end{LARGE} 
\\
\vspace{3cm}
 My Family Members\\
 \vspace{1cm}
 }

\addtocontents{toc}{\vspace{2em}}

\begin{acknowledgements}

It gives me great joy to have the chance to offer my sincere gratitude to everyone whose inspirational influences have supported me as I pursue my doctorate.

My supervisor, \textbf{Prof. Bivudutta Mishra}, Professor of the Department of Mathematics, BITS-Pilani, Hyderabad Campus, is to be sincerely thanked. His tenacity and unwavering excitement motivate me. The chance to collaborate with him has been a wonderful privilege. Along with his superior mathematical knowledge and teaching abilities, his humanism has served as a wonderful model for me. This dissertation would not exist without his committed help and direction. 

I would also like to thank my DAC members, \textbf{Prof. Pradyumn Kumar Sahoo} and \textbf{Dr. Swastik Bhattacharya}, for their insightful questions, insightful comments, and constant feedback throughout this research journey.
 
I would like to extend my sincere gratitude to the Head of the Department, the DRC Convener, and the entire faculty members of the Department of Mathematics at BITS-Pilani Hyderabad Campus for their assistance, support, and encouragement in carrying out my research.

I am also grateful to the Associate Dean, AGRSD BITS-Pilani, Hyderabad Campus. 


I gratefully acknowledge BITS-Pilani, Hyderabad Campus, for providing me with the necessary facilities to carry out my research work.

I express my gratitude to the University Grants Commission (UGC) for its provision of support during the course of my research endeavors.

I would like to take this opportunity to thank, {\bf Dr. Sunil Kumar Tripathy}, Department of Physics, IGIT Saranga, Dhenkanal, Odisha and {\bf Dr. Saikat Chakraborty}, for providing me with the opportunity to collaborate with them. I have gained a great deal of knowledge from them, for which I am quite grateful.

I would like to express my gratitude to my research team {\bf Pratik, Sankarsan, Siddheshwar, Santosh, Lokesh, Shubham, Rahul}.

A special thanks to my friends {\bf Vinod, Shravani, Vipin, Kshma, Anjali, Tapsvini, Tushar, Tapas, Debananda, Sunita, Ankur, Pankaj, Shivangi, Amrutha, Sangeeta} and everyone who has helped me along the way.

Most importantly, I would like to thank my parents, brothers, family members, and friends for their love, care, and support for my personal life.

\end{acknowledgements}

\clearpage

\begin{abstract}       
When writing this thesis, a larger readership—including experts and newcomers—was taken into consideration. The main objective of this study is to investigate the phenomenon of the bouncing scenario of the universe. In the bouncing scenario, the universe undergoes contraction from $t\leq 0$, followed by a bounce at $t=0$, and then expands for $t\geq 0$. The most widely recognized cosmological framework is the standard cosmological model, sometimes referred to as the Big Bang Model. This is mainly because of its inherent properties and its consistent alignment with recent observational studies. However, the standard cosmological model faces some challenges concerning the physical conditions at the initial epochs. Some of these issues include the initial singularity problem, flatness problem, horizon problem etc. Some of these challenges could potentially be addressed by incorporating the inflationary scenario into the cosmological framework of the universe. However, the inflationary mechanism is not able to tackle the occurrence of the initial singularity.  The bouncing cosmology offers a probable solution to this initial singularity issue. In addition, it is capable of addressing some other issues that may arise during the early stages.

 Cosmological singularity occurring at an initial epoch is inherent within Einstein General Relativity. However,  this problem can be solved by introducing a material component with typical properties or by altering the gravitational forces in the standard field equations of classical physics. First, by altering the gravitational action the field equations, we construct the bouncing scenario. For example, we have geometrically modified gravity theories such as the $f(R)$ gravity, $f(R,T)$ gravity, $f(Q)$ gravity, and $f(Q,T)$ gravity to construct cosmological models witnessing bouncing scenarios. Here, $R$, $T$, and $Q$ stand for the Ricci scalar, the stress of the energy-momentum tensor, and the non-metricity tensor, respectively. An isotropic, homogeneous, flat background is used to build cosmological models within modified gravity theories. In order to meet the requirements for a successful bounce, the recommended functional forms of the scale factor are chosen. This method is used to get the solution of the gravitational field equations assuming that there is no initial singularity in the universe. A detailed analysis of the functional form of the scale factors is carried out, which includes multiple restrictions on the parameters. Through these studies, we could able to reconstruct the evolution history of the universe and investigate the dynamical aspects. Further, a model is reconstructed within the framework of  $f(Q)$ gravity that provides a bouncing solution to the universe. 

As a specific investigation, a compact phase space analysis is conducted on a scalar field theory with a Lagrangian that can be expressed as $F(X)-V(\phi)$. Particular attention has been given to a kinetic term of the form $F(X)=\beta X^{m}$ that has the exponential potential and power law potential of the scalar field. Examining the genericity of non-singular bounce in these models and the cosmic future of the bouncing cosmologies when they are generic are the main goals of this work. The analysis is conducted using a global dynamical system formulation that is well-suited for the investigation of non-singular bouncing cosmologies. It is demonstrated that within a specific range of parameters, the occurrence of a nonsingular bounce is generic. In order to answer the key issues concerning nonsingular bouncing solutions, our analysis highlights the significance of a global phase space analysis—a concept that may and should be used for such solutions even in other theories.  

\end{abstract}

\clearpage


\addtocontents{toc}{\vspace{1em}}
\tableofcontents 
\addtocontents{toc}{\vspace{1em}}
\lhead{\emph{List of Figures}}
\listoffigures 
\addtocontents{toc}{\vspace{1em}}
\lhead{\emph{List of Tables}}
\listoftables 
\addtocontents{toc}{\vspace{1em}}



\lhead{\emph{List of symbols}}
\listofsymbols{ll}{

\begin{tabular}{cp{0.6\textwidth}}
$H$ & : Hubble parameter\\
$q$ & : Deceleration parameter\\
$s$ & : Snap parameter\\
$\mathrm{j}$ & : Jerk parameter\\
$l$ & : Lerk parameter\\
$L_{m}$ & : Matter Lagrangian\\
$g_{ij}$ &: Lorentzian metric\\
 $g$ &: Determinant of $g_{ij}$  \\ 
$G_{ij}$ & : Einstein tensor\\
$\Lambda(T)$ & : Effective time variable cosmological constant\\
 $\Gamma^{k}_{ij}$  &: General affine connection (Christoffel symbols) \\ 
 $\{^{k} _{ij}\}$ &: Levi-Civita connection\\  
$\nabla_{i}$ &: Co-variant derivative \\ 
$R^{k}_{lij}$ &: Riemann tensor \\
$R_{ij}$ &: Ricci tensor \\
$R$ &: Ricci scalar \\
$S_{M}$ & : Matter action\\
$T_{ij}$ & : Energy-momentum tensor\\
$T$ & : Trace of the energy-momentum tensor \\
$C_{s}^{2}$ & : Adiabatic speed of sound\\
$\delta(t)$ & : Deviation of Hubble parameter\\
$\delta_{m}(t)$ & : Deviation of energy density\\
$\rho_{c}$ & : Critical density\\
$Q$ & : Non-metricity scalar\\
$L^{k}_{ij}$  &: Disformation tensor \\ 
$P^{k}_{ij}$  &: Super potential \\
$n_{s}$ & : Spectral index\\
$r$ & : Tensor to scalar ratio\\ 
\end{tabular}
}
\addtocontents{toc}{\vspace{1em}}
 
\addtocontents{toc}{\vspace{1em}}

\newpage
\begin{abbreviation}

\textbf{GR}: \textbf{G}eneral--\textbf{R}elativity \\
\textbf{SCM}: \textbf{S}tandard--\textbf{C}osmological--\textbf{M}odel \\
\textbf{$\Lambda$CDM}: \textbf{L}ambda--\textbf{C}old--\textbf{D}ark--\textbf{M}atter\\
\textbf{CMB}: \textbf{C}osmic--\textbf{M}icrowave--\textbf{B}ackground \\
\textbf{FLRW}: \textbf{F}riedmann--\textbf{L}ema\^itre--\textbf{R}obertson--\textbf{W}alker \\
\textbf{EoS}: \textbf{E}quation--of--\textbf{S}tate \\
\textbf{GUP}: \textbf{G}rand--\textbf{U}nified--\textbf{T}heory \\
\textbf{WEC}: \textbf{W}eak--\textbf{E}nergy--\textbf{C}ondition \\
\textbf{NEC}: \textbf{N}ull--\textbf{E}nergy--\textbf{C}ondition \\
\textbf{DEC}: \textbf{D}ominant--\textbf{E}nergy--\textbf{C}ondition \\
\textbf{SEC}: \textbf{S}trong--\textbf{E}nergy--\textbf{C}ondition \\
\textbf{LQC}: \textbf{L}oop--\textbf{Q}uantum--\textbf{C}osmology \\
\textbf{BKL}: \textbf{B}elinski--\textbf{K}halatnikov--\textbf{L}ifshitz \\
\textbf{DHOST}: \textbf{D}egenerate--\textbf{H}igher--\textbf{O}rder--\textbf{S}calar--\textbf{T}ensor 
\end{abbreviation}

%
%


\clearpage 

\lhead{\emph{Glossary}} 

 


\mainmatter 

\pagestyle{fancy} 



\chapter{Introduction} 
\label{Chapter1}

\lhead{Chapter 1. \emph{Introduction}} 


\newpage 

This thesis titled \textbf{bouncing scenario and cosmic dynamics in modified theories of gravity}, aims to resolve the initial singularity problem. The standard cosmological model is the most promising one because it gives almost all the information about the universe. However, the standard cosmological model possesses some early universe problems. The inflationary scenario could solve some of these but could not solve the initial singularity issue. The bouncing scenario is a possible solution for the initial singularity problems and other early universe problems.  

\section{Introduction}\label{ch1FDE}
The most effective theory of gravity ever presented is currently thought to be general relativity (GR). An unmatched level of accuracy in observational confirmation has been achieved for its astounding predictions regarding the gravitational redshift and light refraction by the Sun. However, recent cosmological observations \cite{Riess-1998-116, Perlmutter-1999, Akrami-2020} suggest our universe entered an accelerated expansion phase. These results suggest that traditional GR might not be able to properly explain gravitational events on galactic and cosmological scales, even though it has many successes with the solar system test. Since normal GR cannot explain dark matter and dark energy concerns, it may not be the final theory of gravitational force. In addition, GR is an imperfect physical model because Einstein's standard theory predicts the existence of space-time singularities in the Big Bang and inside black holes. It is likely necessary to extend GR consistently into the quantum domain in order to resolve the singularity problem.

Recently, numerous classical approaches have been proposed to explain cosmological observational results. There is still no adequate theory of gravity, though. The Ricci scalar $R$ may be replaced with any arbitrary function $f$ of the Ricci scalar $R$ in the gravitational action as one way to develop a new gravitational theory by expanding Einstein's gravity \cite{Buchdahl-1970,Barrow-1983}. This line of inquiry leads to $f(R)$ gravity \cite{Sotiriou-2010,De-Felice-2010-13}. The dark matter puzzle can also be geometrically solved within this framework. Assuming that geometry and matter have a non-minimal link is a second strategy for extending the Einstein-Hilbert action. This line of inquiry results in two separate categories of gravitational theories, denoted by the letters $f(R, L_m)$ gravity \cite{Harko-2010-70} and $f(R, T)$ gravity \cite{Harko-2011-84}, where $L_m$ and $T$ stand for matter Lagrangian and trace of the energy-momentum tensor, respectively.     

Weyl made the first attempt to develop a more universal geometry than the Riemannian one, which is a prime illustration of how mathematics and physics can successfully interact (see Ref. \cite{Weyl-1918}). Achieving a geometrical union of electromagnetic and gravitation was the main objective of Weyl's studies. The metric-compatible Levi-Civita connection, which enables length comparison, is the essential idea of Riemann geometry. Weyl introduced a connection that does not contain any information to determine the length of a vector in the parallel transport and replaced the metric field with the class of all conformally equivalent metrics. In order to learn the length of the vector. Weyl added a second connection known as the length connection that lacks information on the direction of a vector during parallel transport. The conformal factor is fixed or measured just by the length connection. In Weyl's theory, the covariant divergence of the metric tensor is non-zero, and this fact can be mathematically represented in terms of a brand-new geometric concept termed non-metricity.  

Another significant mathematical advancement with practical implications was made by Weitzenb-\"{o}ck, who introduced the Weitzenb\"{o}ck spaces \cite{Weitzenbock-1923}. These spaces have since been recognized as an important development in mathematics with important physical applications. A manifold that satisfies the conditions $\nabla_{k}g_{ij}=0$, $T^{k}_{ij}\neq 0$, and $R^{k}_{ijl}=0$ is known as a Weitzenb\"{o}ck manifold. Here, $g_{ij}$, $T^{k}_{ij}$, and $R^{k}_{ijl}$ represent the metric tensor, torsion tensor, and curvature tensor of the manifold, respectively. In the case where $T^{k}_{ij}=0$, the Weitzenb\"{o}ck manifold is transformed into an Euclidean manifold. The values of the torsion tensor exhibit complexity across various regions of the Weitzenb\"{o}ck manifold. The teleparallel approach to gravity replaces the spacetime metric $g_{ij}$ with a series of tetrad vectors $e^{i}_{j}$. Instead of curvature, the tetrad fields' torsion can properly characterize the gravitational effect. Thus, one can obtain the ``teleparallel equivalent of general relativity" (TEGR) or $f(T)$ gravity theory \cite{Pellegrini-1963,Hayashi-1979,Cai-2016-79}.

Based on the aforementioned presentation, it has been established that GR can be expressed through two comparable geometrical representations, namely the curvature representation (wherein the non-metricity and torsion are zero) and the teleparallel representation (wherein the non-metricity and curvature are zero). A third equivalent representation is available wherein the fundamental geometric variable that characterizes the properties of the gravitational interaction is denoted by the non-metricity $Q$ of the metric. The theory of symmetric teleparallel gravity has been expanded upon to create the $f(Q)$ gravity theory, also referred to as coincident GR \cite{Jiemenez-2018}, which is commonly recognized as nonmetric gravity.

A novel class of theories where the non-metricity $Q$ is non-minimally related to the matter Lagrangian was introduced within the framework of the metric-affine formalism as an extension of symmetric teleparallel gravity \cite{Harko-2018-98}. The Lagrangian $L=f_{1}(Q)+f_{2}(Q)L_m$, where $f_{1}$ and $f_{2}$ are generic functions of $Q$ and $L_m$ is a matter Lagrangian, was taken into consideration. The energy-momentum tensor is conserved as a result of this non-minimal coupling, which also causes an additional force to emerge in the geodesic equation of motion. Another extension of $f(Q)$ gravity, $f(Q, T)$ gravity \cite{Xu-2019-79}, is based on the non-minimal relationship between the non-metricity $Q$ and the trace $T$ of the matter-energy-momentum tensor. 

There are various types of geometries that can be classified based on their connection features.
 \begin{table}[H]
     \centering
     \begin{tabular}{c c}\hline
  Type of geometry & Connection properties \\ \hline
  Riemann  & $Q_{ijk}=T^{k}_{ij}=0$ \\ \hline
  Symmetric teleparallel & $R^{l}_{ijk}=T^{k}_{ij}=0$ \\ \hline
  Weitzenb\"ock  & $R^{l}_{ijk}=Q_{ijk}=0$  \\ \hline
  Teleparallel  & $R^{l}_{ijk}=0$ \\ \hline
  Torsion free  & $T^{k}_{ij}=0$ \\ \hline
     \end{tabular}
     \caption{There are subclasses of metric-affine geometry based on the connection properties.}
     \label{tab:my_label3}
 \end{table}

The initial singularity is another important issue that GR has encountered, among other issues, during the early universe. Friedmann \cite{Friedmann-1922} claimed that the occurrence of initial singularity was during the beginning of the evolution of the universe. It is believed that the singularity issue occurred before the inflation \cite{Brout-1978, Starobinsky-1980-91}. One possible solution might be that the universe does not attain singularity during the contraction, but expands after experiencing a bounce. This concept is known as the big bounce. Recent discoveries \cite{Perlmutter-1998-517, Tegmark-2004-69, Abazajian-2004, Spergel-2003-148, Hinshaw-2013-208, Parkinson-2012}, have revealed that our universe is undergoing a late time accelerated expansion phase, which is explained by dark energy, time-independent vacuum energy, according to the Lambda cold dark matter ($\Lambda$CDM) model. The cosmological constant $\Lambda$ \cite{Weinberg-1989-61}, scalar fields (including quintessence, phantom, quintom, tachyon, and others) \cite{KAMENSHCHIK-2001, CALDWELL-2002-545, Sadeghi-2009-48} are possibilities for describing dark energy scenarios. Modified gravity theory has advantages over other models since it avoids expensive numerical computations and is consistent with current data for a late-phase accelerating universe and dark energy.

The widely accepted cosmological model, commonly referred to as the Big Bang cosmology or the standard cosmological model (SCM), suggests that the universe originated from a massive explosion. This model encompasses all the information currently available about the universe as a whole. This allows for the tracking of the cosmological evolution of the universe, which, over the course of the past century, has helped to reinforce the theoretical foundations of cosmology with increasingly accurate cosmological data \cite{Riess-1998-116, Perlmutter-1999, Akrami-2020}. At the classical level, the standard cosmological model has achieved a number of its goals, the most notable of which are (i) the discovery of the expansion of the universe, (ii) the discovery of the black body nature of the cosmic microwave background (CMB), and (iii) the development of a framework for the investigation of the process of the construction of cosmic structures. Despite the fact that the mainstream cosmological models have the best fit with the cosmological evidence, these models nevertheless have a number of flaws and limitations. There are a few of them that have cosmic relics, and some of them are monopoles \cite{Kolb:1990vq}, gravitons \cite{Khlopov-1984-138, Ellis-1984-145}, and baryon asymmetry \cite{Dolgov:1991wv}. 

It is noteworthy to mention that Albert Einstein demonstrated a keen awareness of the issue of singularities within the framework of GR \cite{Hawking:1979ig}. Following the contributions of Friedmann, a prominent Russian physicist who initially derived the equation in 1922. As a meteorologist, Friedmann began his scientific career. Subsequently, the individual acquired knowledge in the field of GR through self-study and employed Einstein's field equations to elucidate the dynamics of a spatially homogeneous and isotropic universe, specifically its expansion or contraction over time. Friedmann presented his initial findings, and suggested that space might be expanding or contracting \cite{Ryden-2016}. In the early 1930s, Einstein made multiple attempts to regularize certain solutions to his theory, including the renowned Einstein-Rosen bridge. The theory of GR is founded upon a fundamental distinction between the gravitational field and matter. The validity of the equations for extremely high densities cannot be assumed, thus leaving open the possibility that a singularity may not exist in a unified theory \cite{Lorentz-1952}. In an effort to address several issues with the standard cosmological model, the oscillating universe has been investigated in various circumstances. The first instance of this universe was described in Lema\^itre's seminal paper \cite{Lemaitre:1933gd}, which stated the solutions in which the universe successively expands and contracts. This was the first time that such a universe had been described. However, Lema\^itre did not produce an explicit solution for the cyclic universe.

It is evident that the scientific literature has consistently featured a non-singular universe. Although the concept of a cosmological bounce has a long history, the initial explicit solutions for a bouncing geometry were derived by Novello and Salim \cite{Novello-1979}, as well as Melnikov and Orlov \cite{Melnikov-1979}, during the late 1970s. It is a valid inquiry to question why these proposed solutions failed to attract the attention of the community at that time. At the onset of the 1980s, it became evident that the SCM was experiencing a state of crisis. The problem developed from the issues mentioned above. However, during that period, the singularity theorems were widely regarded as the ultimate authority on the presence of a singularity in cosmological models that are considered reasonable. The finding of the acceleration of the universe at the end of the 1990s revived the notion of a nonsingular universe by bringing back to the fore the possibility that $\rho+ 3p$ may be negative, which is precisely one of the criteria required for a cosmic bounce in GR. The standard cosmological model faces several challenges, including the initial singularity problem as well as other issues that are outlined below.
 
\section{Friedmann equation}\label{Friedmann:GR}
The Einstein-Hilbert action \cite{Harko-2018b} can be written as
\begin{equation}\label{eqn:HE_action}
    S_{GR}=\frac{1}{2\kappa}\int R\sqrt{-g}d^{4}x,
\end{equation}
$\kappa=\frac{8\pi G}{c^4}$, $G$ be the Newton's gravitational constant, considering $c=\hslash=8\pi G=1$, where $G$ is defined as $1/M_{P}^{2}$ and $M_{P}$ represents the Planck mass, specifically $1.2\times 10^{19}$ GeV and $\hslash$ denotes the reduced Planck constant. $g$ is determinant of the metric tensor $g_{ij}$, the quadratic differential form $ds^{2}=g_{ij}dx^{i}dx^{j}$ in which $g_{ij}$ act as gravitational potentials. The upper and lower indices in this analysis are limited to the values 0, 1, 2, and 3 unless otherwise specified.

In general, the homogeneous and isotropic Friedmann-Lema\^itre-Robertson-Walker (FLRW) spacetime, particularly, the spatially flat FLRW metric is given by
\begin{equation}\label{spacetime:GR}
    ds^{2}=-dt^{2}+a^{2}(t)\left[dx^{2}+dy^{2}+dz^{2}\right].
\end{equation}
$(-,+,+,+)$ metric signatures is considered throughout work. $a(t)$ is the scale factor of the FLRW space-time. The scale factor describes how a homogeneous, isotropic universe expands or contracts with time. The Hubble parameter can measure the rate of expansion or contraction. The Hubble parameter $H(t)=\dot{a}(t)/a(t)$, where a dot over a variable signifies a time derivative of that quantity, can be used to represent the Einstein equation in the context of cosmology. 

Energy-momentum tensor for perfect fluid is defined as 
\begin{equation}\label{EMT_GR}
T_{ij}=(\rho+p)u_{i}u_{j}+p g_{ij},    
\end{equation}
where $u^{i}$ is the four-velocity of the fluid, normalised as $u^{i}u_{i}=-1$. In the above equation, $\rho$ and $p$ represent the energy density and the pressure, respectively.

The Einstein field equation can be written as
\begin{equation}\label{EFE_GR}
    R_{ij}-\frac{1}{2}R g_{ij}=T_{ij},
\end{equation}
where $R_{ij}$ is the (\textit{Ricci tensor}), the Ricci tensor takes the form
\begin{equation}
    R_{ij}=\frac{\partial \Gamma ^{k}_{ij}}{\partial x^{k}}-\frac{\partial \Gamma^{k}_{ik}}{\partial x^{j}}+\Gamma^{\lambda}_{ij}\Gamma^{k}_{\lambda k}-\Gamma^{\lambda}_{ki}\Gamma^{k}_{j\lambda}.  
\end{equation}
The Christoffel symbol in the above equation is
\begin{equation}
\Gamma^{k}_{ij}=\frac{1}{2}g^{kl}\left(\frac{\partial g_{lj}}{\partial x^{i}}+\frac{\partial g_{li}}{\partial x^{j}}-\frac{\partial g_{ij}}{\partial x^{l}}\right),    
\end{equation}
where $g^{ij}$ is the conjugate tensor of $g_{ij}$.

Contracting the Ricci tensor gives the $R$ (\textit{Ricci scalar}), defined as
\begin{equation}
R=R^{i}_{i}=g^{ij}R_{ij}.    
\end{equation}
From equations \eqref{spacetime:GR}, \eqref{EMT_GR}, and \eqref{EFE_GR} the Friedmann equation takes the form 
\begin{equation}\label{Friedmann_GR1}
    3H^{2}=\rho.
\end{equation}
The universe consists of nonrelativistic matter, radiation, and dark energy. Furthermore, it is possible that it contains even more exotic components. Fortunately for the sake of simplicity, the energy density and pressure of many components of the universe are additive. Assume that there are $N$ components in the universe and that the energy density of the $i^{th}$ component is $\rho_{i}$. The total energy density, denoted as $\rho$, can be expressed as the sum of the energy densities of the various components \cite{Ryden-2016},   
\begin{eqnarray}
    \rho=\sum_{i=1}^{N} \rho_{i}. 
\end{eqnarray}
There exists a relationship between pressure and energy density,
\begin{equation}
    p_{i}=\omega_{i} \rho_{i}.
\end{equation}
The parameter $\omega_{i}$ denotes the equation of state (EoS) of the specific component. 
The total pressure $p$ is the sum of the pressures of the different components:
\begin{eqnarray}
    p=\sum_{i=1}^{N} \omega_{i}\rho_{i}.
\end{eqnarray} 
Further studies of the Friedmann equation have provided valuable insights into both the present and early universe. In the following section, we will discuss some of the problems that have been detected during the early stages of the evolution of the universe, based on a detailed analysis. It is helpful to discuss energy density using the dimensionless density parameter while talking about the curvature of the universe,
\begin{equation}
    \Omega_{\text{total}}(t)=\frac{\rho}{\rho_{c}}=\frac{\sum_{i=1}^{N} \rho_{i}}{\rho_{c}}.
\end{equation}
The critical density of matter, denoted as $\rho_{c}=3H^{2}$. In this context, the value of $8\pi G$ has been assumed to be equal to 1 as stated above.

\section{The fluid equation}
We looked at the Friedmann equation \eqref{Friedmann_GR1} in Section \eqref{Friedmann:GR}. While the Friedmann equation is a valuable tool, it is insufficient to explain the time evolution of the scale factor $a(t)$ on its own. For the purpose of solving for $a$ and $\rho$ as a function of time, we require an additional equation involving $a(t)$ and $\rho$. We examine another application of the idea of energy conservation in Newtonian physics, which is the first law of thermodynamics:
\begin{equation}
    dQ=dE+pdV,
\end{equation}
where the heat flow into or out of a volume is denoted by $dQ$. The changes in integral energy are denoted by $dE$ and volume by $dV$. For every volume, $dQ=0$ if the universe is perfectly homogenous; in other words, there is no bulk heat flow. The energy conservation equation can be written as
\begin{equation}\label{FE_1}
    \dot{E}+p\dot{V}=0
\end{equation}
For the sake of concreteness, let us suppose a sphere of comoving radius $r_{s}$ that expands along with the universal expansion, such that $R_{s}(t)=a(t)r_{s}$ is its proper radius. The sphere has a volume,
\begin{equation}
 V(t)=\frac{4}{3}\pi r_{s}^{3}a^{3}(t)   
\end{equation}
So the rate of change of the volume of the sphere is
\begin{equation}\label{FE_2}
    \dot{V}=\frac{4}{3}r_{s}^{3}(3a^{2}\dot{a})=V\left(3\frac{\dot{a}}{a}\right).
\end{equation}
The internal energy of the sphere is 
\begin{equation}
    E(t)=V(t)\rho(t),
\end{equation}
the rate of change of the sphere's internal energy takes the form
\begin{equation}\label{FE_3}
    \dot{E}=V\dot{\rho}+\dot{V}\rho=V\left(\dot{\rho}+3\frac{\dot{a}}{a}\rho\right)
\end{equation}
Equations \eqref{FE_1}, \eqref{FE_2}, and \eqref{FE_3} are combined to get the following form for the first law of thermodynamics in an expanding (or contracting) universe:
\begin{equation}
    V\left(\dot{\rho}+3\frac{\dot{a}}{a}\rho+3\frac{\dot{a}}{a}p \right)=0 \implies \dot{\rho}+3\frac{\dot{a}}{a}(\rho+p)=0
\end{equation}
this equation is called the \textit{fluid equation}, which describes the expansion of the universe.

The fluid equation must hold separately for each component, as long as there is no interaction between them because the energy densities and pressures add in this manner \cite{Ryden-2016}. 
\begin{eqnarray}
    \dot{\rho}_{i}+3H(\rho_{i}+p_{i})=0.
\end{eqnarray}    
If this condition holds true, then the component that has the EoS parameter $\omega_{i}$ follows the equation
\begin{eqnarray}
\dot{\rho}_{i}+3H(1+\omega_{i})\rho_{i}=0\label{continuity_equation}. 
\end{eqnarray}

Equation \eqref{continuity_equation} can be written as
\begin{eqnarray}
    \frac{d\rho_{i}}{\rho_{i}}=-3(1+\omega_{i})\frac{da}{a}.
\end{eqnarray}
If we assume that $\omega_{i}$ is constant, then
\begin{eqnarray}
    \rho_{i}(a)=\rho_{i,0}a^{-3(1+\omega_{i})}.
\end{eqnarray}
It is concluded that the energy density $\rho_{m}$ corresponding to nonrelativistic matter with $\omega_{m}=0$ and the energy density of radiation, $\rho_{r}$ with $\omega_{r}=1/3$, may be expressed as
\begin{subequations}
    \begin{eqnarray}
        \rho_{m}(a)&=&\frac{\rho_{m,0}}{a^{3}},\\
        \rho_{r}(a)&=&\frac{\rho_{r,0}}{a^{4}}.
    \end{eqnarray}
\end{subequations}

\section{Cosmographic parameters}
Cosmologists are interested in determining the scale factor $a(t)$ that characterizes the expansion of the universe. The scale factor for a model universe whose contents are precisely known can be obtained using the Friedmann equation. However, determining the function $a(t)$ for the universe poses significant challenges. The scale factor is not directly observable; it can only be deduced indirectly from our imperfect and incomplete observations of the universe \cite{Ryden-2016}. 

The scale factor appears to be the single degree of freedom governing the universe, according to the cosmological principle. We can expand the scale factor $a(t)$ around $a_{0}$ (i.e., $a_{0}=a(t_{0})$) in a Taylor series \cite{Dunsby-2016} as follows:
\begin{eqnarray}\label{Taylor_series}
a(t)=a_0+\sum_{n=1}^{\infty} \left.\frac{1}{n!}\frac{d^na}{dt^n}\right\vert_{(t=t_0)} (t-t_0)^n,\label{eq:32}
\end{eqnarray}
In the given context, $t_{0}$ represents the current cosmic time, whereas $n$ is an integer that can take on values of 1, 2, 3, and so on. The coefficients of expansion are commonly known as the cosmographic coefficients. The cosmographic coefficients, due to their inclusion of several orders of derivatives of the scale factor, possess the potential to offer an enhanced geometric depiction of the model. At each given time $t$, it is possible to establish certain geometric parameters as follows:
\begin{eqnarray}\label{cosmographic_prameter}
H = \frac{\dot{a}}{a}, \qquad q = -\frac{a\ddot{a}}{\dot{a}^2}, \qquad \mathrm{j} = \frac{\dot{\ddot{a}}}{aH^3}, \qquad
s = \frac{a^{(4)}}{aH^4}.
\end{eqnarray}
where the fourth-order derivatives of the scale factor are denoted by $a^{(4)}$. The Hubble parameter and the deceleration parameter are respectively denoted by the symbols $H$ and $q$. The parameters $\mathrm{j}$ and $s$ represent the jerk and snap parameters, respectively.

Take note of the decision to define $q$. When $q$ is positive, $\ddot{a}(t)<0$, which indicates that the relative velocity of any two points is decreasing, indicates that the expansion of the universe is decelerating. A negative value of the variable $q$ is associated with the condition $\ddot{a}(t)>0$, indicating that the relative velocity between any two points exhibits a positive trend over time. The reason behind the sign selection of $q$ and choosing it as the deceleration parameter came from its initial definition in the mid-1950s when the available data supported a universe dominated by matter with $\ddot{a}(t)<0$. However, the deceleration parameter $q$ can have either sign if there is a sufficiently large cosmological constant in the universe. A positive $\mathrm{j}$, on the other hand, suggests that $q$ changes sign as the universe expands, and this is true for all the other parameters as well. The conventional practice is to assign the designations \textit{jerk} and \textit{snap} to the letters $\mathrm{j}$ and $s$ correspondingly \cite{Aviles-2012-86}.
\section{Early universe problem}
The flatness problem, the horizon problem, and the magnetic monopoles problem are some of the early universe problems. The statement \textit{``the universe is nearly flat today and was even flatter in the past"} sums up the issue of flatness. The horizon problem can be summed up as follows: \textit{``the universe is nearly isotropic and homogeneous today, and was even more so in the past".} The monopole problem can be summed up by the statement, \textit{``the universe does not appear to contain any magnetic monopoles".} In the following sections, we will briefly discuss the early universe problems \cite{Ryden-2016}.

\begin{itemize}
    \item {\bf{The flatness problem}}\\
    The Friedmann equation can be written as
\begin{equation}
    |\Omega_{\text{total}}-1|\equiv 0.
\end{equation}
Since the data currently available shows that $\Omega_{\text{total}}$ is very close to 1, it follows that if $\Omega_{\text{total}}$ initially approached 1, it must have been incredibly close to one \cite{Brandenberger-2017-47}. The phenomenon referred to is commonly known as the flatness problem. In the context of a cyclic universe, the parameter $\Omega_{\text{total}}$ begins to deviate from unity solely when the variable $a$ reaches its maximum value. Given that the maximum value increases proportionally to the number of cycles, it is plausible that in a cyclic universe of considerable age, a significant amount of time may be required for the total density parameter, $\Omega_{\text{total}}$ to deviate significantly from unity.
    \item{\bf{The horizon problem}}\\
In the context of the SCM, it is observed that light signals have the ability to travel a limited distance between the initial singularity and a specific time point, denoted as $t$. This is dependent upon the condition that the rate of change of energy density is faster than the inverse square of the scale factor, represented as $a^{-2}$. In order to get the universe to its high level of homogeneity, microphysics would not have enough time. According to the cyclic model, the age of the universe is calculated by adding the durations of all prior cycles. If correlations can successfully cross the bounce, this would solve the horizon problem.
    \item{\bf{The monopole problem}}\\
Combining the Big Bang scenario with the particle physics concept of a grand unified theory (GUT) leads to the monopole problem, the seeming lack of magnetic monopoles in the universe. A GUT is a particle physics field theory that unifies electromagnetic, weak nuclear, and strong nuclear forces. According to the GUT, the GUT phase transition creates magnetic monopole-like topological defects. The GUT phase transition energy density of radiation is ten orders of magnitude higher than the magnetic monopole energy density. Thus, magnetic monopoles would not have prevented a radiation-dominated universe at the GUT phase transition. Since magnetic monopoles are so massive, they would rapidly become non-relativistic with matter energy density. However, observation shows that magnetic monopoles do not rule the universe today \cite{Ryden-2016}.
\end{itemize}
\section{The Riemann curvature tensor and its properties}
The quantification of curvature can be achieved through the use of the Riemann tensor, which is generated from the connection. The purely geometric quantity $R^{k}_{~lij}$ (\textit{Riemann curvature tensor}) \cite{Carroll-2019} can be written as
\begin{equation}\label{Riemann_curvature}
R^{k}_{~lij}=\frac{\partial \Gamma ^{k}_{lj}}{\partial x^{i}}-\frac{\partial \Gamma^{k}_{li}}{\partial x^{j}}+\Gamma^{\lambda}_{lj}\Gamma^{k}_{\lambda i}-\Gamma^{\lambda}_{il}\Gamma^{k}_{j\lambda}.    
\end{equation}
The Riemann curvature tensor, as defined in equation \eqref{Riemann_curvature}, exhibits the significant characteristic of antisymmetry in relation to its final two indices, denoted as $i$ and $j$ respectively,
\begin{equation}
    R^{k}_{~lij}=-R^{k}_{~lji}.
\end{equation}
One way to generate a purely covariant curvature tensor is by lowering the upper index of the curvature tensor, which is given by
\begin{equation}
    R_{\lambda lij}=g_{k\lambda}R^{k}_{~lij}.
\end{equation}
This tensor also possesses a number of key symmetric features, including
\begin{equation}
 R_{\lambda lij}=-R_{l\lambda ij}=-R_{\lambda lji}, ~~~~~~~~ R_{\lambda lij}=R_{l \lambda ji}.   
\end{equation}
Therefore, the obtained results indicate that the covariant Riemann curvature tensor $R_{\lambda lij}$ exhibits intriguing symmetry properties. Specifically, it demonstrates antisymmetry in relation to both the first pair of indices $\lambda l$ and the second pair of indices $ij$. Furthermore, the symmetry of the system is observed when the interchange of the two pairs of indices $\lambda l$ and $ij$ is performed.
\begin{itemize}
    \item{\bf{Bianchi Identity}}\\
The geodesic coordinate system, which is a local coordinate system in Riemannian geometry, has the property that the first derivatives of all the metric tensor $g_{ij}$ components cancel out at a particular point $P$, being identifiably equal to zero. As a result, in this coordinate frame, all Christoffel symbols have a value of zero. The use of locally geodesic coordinate systems offers extremely effective methods for identifying and demonstrating tensor identities because, if one can demonstrate that a tensor is zero in a geodesic coordinate system, then the given tensor must also be zero in any other coordinate system because of the mathematical structure of the rules governing its transformations. It is discovered that the Riemann curvature tensor in such a system has a simple form by employing this characteristic of the local geodesic coordinate system,
\begin{equation}
    \left. R^{k}_{lij}\right\vert_{P}=\frac{\partial \Gamma^{k}_{lj}}{\partial x^{i}}-\frac{\partial \Gamma^{k}_{li}}{\partial x^{j}}.
\end{equation}
The indices $i$, $j$, and $\lambda$ are cyclically transposed in the aforementioned definition, and the resulting expressions are subsequently summed. Therefore, following a straightforward computation and replacing the standard derivatives with the covariant derivatives, the significant \textit{Bianchi identity}, expressed in a general frame of reference as follows,
\begin{equation}\label{Bianchi_identity}
\nabla_{\lambda} R^{k}_{lij}+\nabla_{j} R^{k}_{l\lambda i}+\nabla_{i} R^{k}_{li\lambda}=0.   
\end{equation} 
\end{itemize}
\section{Energy conditions}\label{GR_EC}
It can be useful to consider Einstein's equations without mentioning the particular theory of matter on $T_{ij}$ is based. 
It is observed that every metric $g_{ij}$ is the solution of Einstein's field equation for some associated energy-momentum tensor $T_{ij}$. The Bianchi identity \eqref{Bianchi_identity} will automatically preserve it. The main issue is whether Einstein's equation can be solved in the face of a \textit{realistic} source of energy and momentum, whatever that may be. Consideration of a certain source, such as a scalar field, dust, or electromagnetic field, is one technique. However, there are times when we want to comprehend the aspects of Einstein's equations that apply to a different source. In this case, it is practical to set energy constraints that restrict the arbitrary nature of $T_{ij}$ \cite{Raychaudhuri-1955, Carroll-2019}.   

Energy conditions refer to limitations imposed on the energy-momentum tensor that are invariant under changes in coordinates. Hence, it is necessary to form scalars by contracting $T_{ij}$ with either arbitrary time-like vectors $t^{i}$ or null vector $\xi^{i}$ \cite{Curiel2017}. There are four main energy conditions studied in the literature: weak energy condition (WEC), null energy condition (NEC), dominant energy condition (DEC), and strong energy condition (SEC). These energy conditions, with their physical and geometrical properties, can be studied as follows:
\begin{itemize}
    \item An illustration of the WEC can be seen in the inequality $T_{ij}t^{i}t^{j}\geq 0$, which holds true for any timelike vectors $t^{i}$. In order to facilitate physical understanding, the perfect fluid is commonly defined as outlined in equation \eqref{EMT_GR}. The pressure is isotropic when it satisfies the condition $T_{ij}t^{i}t^{j}\geq 0$ for all timelike vectors $t^{i}$, given that both $T_{ij}u^{i}u^{j}\geq 0$ and $T_{ij}\xi^{i}\xi^{j}\geq 0$ hold true for some null vector $\xi^{i}$. It gives the following relation
    \begin{equation}
    T_{ij}u^{i}u^{j}=\rho, ~~~ T_{ij}\xi^{i}\xi^{j}=(\rho+p)\left(u_{i}\xi^{i}\right)^{2}.    
    \end{equation}
    The WEC implies that the $\rho\geq 0$ and the $\rho + p\geq 0$. It means that the energy density must not be negative and that the pressure must not be excessively high in relation to the energy density. Naturally, there is no necessity to confine our analysis solely to perfect fluids; rather, we employ them as a means to acquire a deeper understanding of the constraints imposed by the energy conditions. 

    \item The NEC is a fundamental principle in physics, which asserts that the contraction of the stress-energy tensor $T_{ij}$ with any null vector $\xi^{i}$ should always yield a non-negative value, i.e., $T_{ij}\xi^{i}\xi^{j}\geq 0$. Alternatively, this condition can be expressed as the requirement $\rho+p\geq0$. This particular instance pertains to the WEC, wherein the presence of a time-like vector is substituted with a null vector. The energy density can exhibit a negative value, provided that there exists a corresponding positive pressure to balance it.

     \item The SEC is defined as follows: for any time-like vectors $t^{i}$, the inequality $T_{ij}t^{i}t^{j}-\frac{1}{2}T^{k}_{~k}t^{l}t_{l}\geq 0$ holds. Alternatively, this condition can be expressed as $\rho+p \geq 0$ and $\rho +3p\geq 0$. It is noteworthy to mention here that the SEC does not imply the WEC. Along with excluding extremely high negative pressures, it implies the NEC. Furthermore, the validation of the SEC implies that gravitational forces are attractive in nature. The energy conditions and their physical and geometrical significance have been discussed in the following table.

    \item The DEC involves the WEC and states that $T^{ij}t_{i}$ is a nonspacelike vector, meaning that $T_{ij}t^{i}t^{j}\geq 0$. In the case of a perfect fluid, the conditions mentioned are equivalent to the straightforward criterion that $\rho\geq |p|$. This ensures that the energy density is nonnegative and at least as large as the magnitude of the pressure.

\end{itemize}
\begin{table}[H]
    \centering
    \begin{tabular}{|c|c|c|c|} \hline
    Energy conditions & Physical form & Geometrical form & Perfect fluid \\ \hline
     \begin{tabular}{@{}c@{}} WEC\end{tabular} & $T_{ij}t^{i}t^{j}\geq 0$ & $G_{ij}t^{i}t^{j}\geq 0$ & \begin{tabular}{@{}c@{}}$\rho\geq 0$,\\ $\rho +p\geq 0$ \end{tabular}\\ \hline
      \begin{tabular}{@{}c@{}}NEC\end{tabular} & $T_{ij}\xi^{i}\xi^{j}\geq 0$ & $R_{ij}\xi^{i}\xi^{j}\geq 0$ & $\rho +p\geq 0$ \\ \hline
      \begin{tabular}{@{}c@{}}SEC \end{tabular}& $\left(T_{ij}-\frac{T}{2}g_{ij}\right)t^{i}t^{j}\geq 0$ & $R_{ij}t^{i}t^{j}\geq 0$ & \begin{tabular}{@{}c@{}} $\rho +p\geq 0$,\\ $\rho +3p\geq 0$ \end{tabular} \\ \hline
      \begin{tabular}{@{}c@{}}DEC \end{tabular} & \begin{tabular}{@{}c@{}}$T_{ij}t^{i}t^{j}\geq 0$ and \\ $T_{ij}t^{i}$ is not space like \end{tabular} & \begin{tabular}{@{}c@{}}$G_{ij}t^{i}t^{j}\geq 0$ and \\ $G_{ij}t^{i}$ is not space like \end{tabular} & $\rho \geq |p|$ \\ \hline
    \end{tabular}
    \caption{Energy conditions and their physical and geometrical significance.}
    \label{tab:Energy_con}
\end{table}
{\centering
WEC $\implies$ NEC, ~~~ SEC $\implies$ NEC, ~~~ and ~~~ DEC $\implies$ WEC. \\}

{\bf{Note}}: If the NEC is violated, it would result in the violation of all other energy conditions defined above.

In light of the fact that we are operating within the bouncing scenario of the universe, it has been proposed that the NEC should be violated at least during the epoch of the bounce. The violation of NEC means that all other energy conditions have also been violated. The SEC is a crucial energy condition that provides insights into the accelerating phenomena of the universe. Therefore, the subsequent debate will exclusively focus on the NEC and SEC, omitting the DEC from consideration. 

\section{Bouncing cosmology: an alternative to SCM}

The SCM, also known as the Big Bang model, is based on Einstein's general theory of relativity and implies that our universe is spatially homogeneous and isotropic on extremely large scales. This concept, according to modern cosmology, cannot explain the formation of structures in our universe unless the initial conditions at the moment of the Big Bang were highly fine-tuned. The inflationary scenario \cite{Guth-1981-23} is the current paradigm of early universe cosmology. Inflation resolves a number of issues within standard Big Bang cosmology. Despite its phenomenological effectiveness, the concept of inflation faces a number of conceptual obstacles. The singularity problem is a crucial obstacle to inflation. If inflation is realized by the dynamics of scalar matter fields coupled with Einstein's gravity, the Hawking-Penrose singularity theorem \cite{Penrose-1989} can be extended to demonstrate that an inflationary universe is geodesically incomplete. Thus, a singularity must exist prior to the onset of inflation.  Consequently, the inflationary scenario cannot account for the entire history of the early universe. For this reason, a successful theory of the very early universe must address at least some central issues of Big Bang cosmology. Inflationary cosmology is one of the two extant theories of the early universe, with the other being bounce cosmology, in which the theoretical contradictions of the Big Bang description of our universe are addressed. If one wants to avoid the Big Bang singularity near $t=0$ then one can model a universe where the scale factor $a(t)$ never becomes zero at $t=0$. In these models, the universe contracts during the time $-\infty<t\leq 0$ and expands during $t\geq 0$. The phenomenon of bouncing models gained significant attention in the late 1990s and early 21st century. This was primarily due to certain models, derived from string theory, which indicated that a bouncing geometry could potentially address the issues resolved by inflation. A bouncing cosmological scenario naturally avoids this singularity problem, but at the cost of introducing new physics to achieve the bounce. In the context of bouncing cosmology, we will discuss these issues and how they are solved.

\begin{itemize}
\item For the case of non-singular bounce, the bouncing scenario behaves as a contracting nature formulated by the scale factor which decreases with time, i.e., $\dot{a}<0$, which means the Hubble parameter is negative in the contracting phase, i.e., $H =\dot{a}/a<0$.
\item For bouncing epoch, the contracting nature of scale factor to a non-zero finite critical size is obtained as a result of which the Hubble parameter vanishes at bounce making $H_{t=0}=0$.
\item The nature of the scale factor increases with time in a positive acceleration, so as the Hubble parameter becomes positive after the bounce, i.e., $H=\dot{a}/a>0$.
\item In the situation of a near-bouncing epoch, the Hubble parameter transitions from negative to positive, i.e., $\dot{H}>0$  which is suitable for the ghost (phantom) behavior of the model in GR.
\end{itemize}

The possibility that bouncing cosmology can be derived as a cosmological solution of loop quantum cosmology (LQC) \cite{Ashtekar-2006-74} is another intriguing topic that can be discussed in relation to this branch of cosmology. The matter bounce scenario has drawn a lot of attention in the non-singular bouncing models because of its potential implications. This is due to the fact that the evolution of the universe, even in late times, can be compared to an epoch when the universe becomes matter-dominated. It is hypothesized that the universe went through a period of contraction and is now capable of expanding without coming into contact with any kind of initial singularity, as described by the matter bounce scenario. In other words, the universe goes through an initial matter-dominated contraction phase, which is then followed by a non-singular bounce and, finally, a causal genesis for fluctuation. The cosmological models substitute the cosmic singularity of the Big Bang with a scenario known as the big bounce, which describes the smooth transition from the phase of contraction to the phase of expansion \cite{Brandenberger-2011, Elizalde-2015}. However, one noteworthy finding is that even in a universe with no singularities, the presence of non-singular bounce might cause a violation of the constraint that there is no net energy flow. In point of fact, it is possible to violate the NEC under generalized Galileon theories that provide support for the possibility of non-singular cosmology \cite{Kobayashi-2016}. Another problem with the bouncing models is that they frequently give the impression of being unstable. This is one of the problems. On the other hand, some people believe that stable bouncing cosmologies can be articulated in a way that goes beyond the Horndeski theory and the effective field theory \cite{Creminelli-2016, Kolevatov-2017, Cai-2017}. In the framework of modified theories of gravity, the possibility of the occurrence of a big bounce scenario as a replacement for the Big Bang singularity would be an interesting topic to discuss.

\section{Cosmological bounce criteria}
Einstein's field equations \eqref{EFE_GR} can be defined for the flat FLRW space-time with perfect fluid as follows:
\begin{subequations}
\begin{eqnarray}
H^{2}&=&\frac{1}{3} \rho, \label{Friedmann_GR}\\
\dot{H}&=&-\frac{1}{2}(p+\rho).\label{Raychoudhury_GR}
\end{eqnarray}    
\end{subequations}
The acceleration equation takes the form
\begin{equation}
    \frac{\ddot{a}(t)}{a(t)}=-\frac{1}{6}(\rho +3p).
\end{equation}
From this equation, we can conclude that acceleration expansion occurs for $\rho +3p<0$ \cite{Bari-2018-50}.

The avoidance of singularity can be achieved through the implementation of a bouncing universe, wherein the scale factor remains finite as $t \rightarrow 0$, by means of the violation of the NEC in the context of GR. As previously established, the scale factor and Hubble parameter must meet specific requirements in order to generate the bounce scenario, which can be expressed as
\begin{equation}\label{BC}
    a(t=0)\neq 0, ~~~\dot{a}(t=0)=0, ~~~ \left.H\right|_{t=0}=0, ~~~ \left.\dot{H}\right|_{t=0}>0.
\end{equation}
From \eqref{Raychoudhury_GR} and \eqref{BC}, in the FLRW space-time during bounce epoch
\begin{equation}
\left[p+\rho\right]_{t=0}<0.    
\end{equation}
According to the aforementioned requirement, the NEC must be violated at the bounce point. The violation of the NEC, from a cosmic perspective, necessitates exotic matter, as was implied in \cite{Peter-2001}, and also demonstrates how to imagine something that violates the NEC in the early universe \cite{Rubakov-2014}. To account for a cosmological bounce, one might even modify the gravitational theory. 

 From equation \eqref{Raychoudhury_GR} one can write
 \begin{equation}
 \dot{H}=-\frac{1}{2}\rho (1+\omega).    
 \end{equation}
The equation presented above can be interpreted as indicating that for values of $\omega$ less than $-1$, this condition holds in the vicinity of the bounce epoch. The universe must enter the hot Big Bang era following the bounce; otherwise, a universe filled with matter with an EoS $\omega<-1$ will reach the big rip singularity, which is what happens to the phantom dark energy \cite{Feng-2005}. In order for this to take place, the EoS of the matter must shift from $\omega<-1$ to $\omega>-1$.
\section{Realizations of a bounce phase}
The Hawking-Penrose theorems \cite{Hawking-1970} states that within the framework of GR, in a homogeneous and isotropic model, the presence of an initial cosmological singularity is unavoidable. This is contingent upon the matter that interacts minimally with gravity adhering to the NEC. There are several approaches available to achieve a cosmological model characterized by bouncing. It is possible to incorporate the principles of Einstein's theory of gravity while introducing matter that violates the NEC \cite{Peter-2002}. The obstacle in this route is one must avoid instabilities such as ghosts \cite{Cline-2004} and gradient instabilities. On the other hand, the incorporation of quantum theory into the framework of gravity necessitates the inclusion of additional terms in the effective gravitational action that surpasses the Einstein-Hilbert term. Therefore, it may be more advantageous to endeavor to achieve bouncing cosmologies within the framework of modified theories of gravity. An additional method of categorizing bouncing models pertains to their singularity status within the framework of an effective field theory, such as the pre-big-bang or initial Ekpyrotic scenarios. Attempts in this direction will be briefly reviewed in the following sections.

\section{Bouncing cosmologies from modified gravity}
Terms in the gravitational action other than the standard Einstein-Hilbert term are predicted in any approach to quantum gravity. These terms could be non-local and higher derivatives of the effective gravitational action. We assume a standard matter theory, represented by an energy-momentum tensor $T_{ij}$. In this case, the effective equations of motion with leading quantum correction can be expressed as follows \cite{Brandenberger-2017-47}:
\begin{equation}\label{eq:D}
    \mathcal{D}_{ij}(g_{kl})= T_{ij},
\end{equation}
where $\mathcal{D}_{ij}$ is an operator containing the quantum gravitational higher derivative and/or non-local terms, as well as the Einstein tensor $G_{kl}$ as a leading term. The leading term can be taken out and written
\begin{equation}
    \mathcal{D}_{ij}(g_{kl})=G_{ij}+\Tilde{\mathcal{D}}_{ij}(g_{kl}).
\end{equation}
Equivalently, one can write \eqref{eq:D} as
\begin{equation}
    G_{ij}= T^{\text{eff}}_{ij},
\end{equation}
 with
 \begin{equation}
T^{\text{eff}}_{ij}=T_{ij}-\Tilde{\mathcal{D}}_{ij}(g_{kl}).     
 \end{equation}
Now, it is simple to see that if the gravitational contribution $\Tilde{\mathcal{D}}_{ij}(g_{kl})$ is such that the effective energy-momentum tensor violates the NEC, then a bounce can be obtained even with a matter energy-momentum tensor that obeys the NEC. 

\section{Bouncing cosmologies from modified matter}
Another approach to achieving a non-singular bouncing model with modified matter involves the use of a field with a kinetic energy term of the opposite sign, commonly referred to as a ghost field. In order to achieve the phenomenon of bounce, it is necessary for the scalar field to exhibit a violation of the NEC, which may potentially lead to the emergence of instabilities. Several studies have been conducted in an attempt to mitigate these instabilities through the utilization of ghost condensate scalar fields \cite{Buchbinder-2007-76, Arkani-Hamed-2004}. Additionally, it is necessary to ensure that during the contraction phase, the energy density of the ghost field increases in magnitude compared to that of regular matter. As an illustration, regular matter can be characterized as a perfect fluid \cite{Peter-2002} or as a massive scalar field $\phi$ with an energy density that decreases with the scale factor $a$ as $a^{-3}$. On the other hand, ghost matter can be described by a free scalar field $\psi$ with a kinetic term of opposite sign, where the dominant contribution to its energy density arises from the $\dot{\psi}^{2}$ term. Consequently, the energy density of ghost matter scales as $a^{-6}$ \cite{Cai-2008-2008}. 

The action equation \cite{Brandenberger-2017-47} can be defined as
\begin{equation}
    S=\int \left(\frac{1}{2}\partial^{i}\phi \partial_{i}\phi -V(\phi)-\frac{1}{2}\partial^{i}\psi \partial_{i}\psi\right)\sqrt{-g}d^{4}x.
\end{equation}
The kinetic component is represented by the symbol $X$, which can be expressed as $X=-\dfrac{1}{2}\partial^{i}\phi \partial_{i}\phi$. The potential is denoted as $V(\phi)$, it may represented as $V(\phi)=\dfrac{1}{2}m^{2}\phi^{2}$. When the energy densities of $\phi$ and $\psi$ are equivalent, it has been observed that the non-singular bounce occurs.

The \textit{ghost condensate} method can produce a better nonsingular bounce \cite{Arkani-Hamed-2004,Abramo-2007}. The Higgs mechanism in the potential sector is analogous to ghost condensation in the kinetic sector. In the context of the Higgs field denoted as $\phi$, the theoretical framework exhibits the presence of a tachyon when the expansion is performed around $\phi=0$. However, this tachyonic behavior ceases to manifest when the expansion is conducted in the vicinity of the actual minimum of the potential. 

One further issue that the model experiences is gradient instability. The resolution of this issue can be achieved by substituting the Lagrangian of the ghost condensate with that of a Galileon Lagrangian \cite{Nicolis-2009-79}, and by taking into account Galileon bounces \cite{Qiu-2011-2011,Easson-2011,Cai-2012-08}. There are several Galileon terms that can be taken into consideration. Some early designs were successful in eliminating gradient instability at the bounce point, however at the expense of such instability arising at a later stage \cite{Ijjas-2016-117,Libanov-2016-037,Kobayashi-2016-94}. However, more recently, models free of any ghost and gradient instabilities were reported in \cite{Libanov-2016-037,Ijjas-2017-289}. Criteria were discussed under which some instability was generic. Chapter 7 discusses the generic solution for modified scalar field models to provide a comprehensive overview.

\section{Literature review}
The literature includes several bounce models, such as symmetric bounce, super bounce, oscillatory cosmology, and matter bounce. Researchers are particularly interested in the matter bounce theory among these bounce cosmologies \cite{Caruana-2020}. In order to include the dark energy era and to address the late-time cosmic acceleration issue in the matter bounce scenario, the LQC approach may be adhered to \cite{Bojowald-2009-26, Ashtekar-2011-28, Cai-2011, Cailleteau-2012, Quintin-2014}. To note, $f(R)$ gravity theory \cite{Sotiriou-2010, De-Felice-2010-13} is an excellent alternative to the standard gravity model to study the dark energy cosmological models. In the $f(R)$ modified gravity framework, Odintsov and Oikonomou \cite{Odintsov-2017} have investigated a bouncing cosmology with a Type IV singularity at the bouncing point. Odintsov et al. \cite{Odintsov-2020-37} have proposed a cosmological model that merges a non-singular bounce to a matter-dominated epoch and space-time-dominated to a late-time accelerating epoch; i.e., the model is similar to a generalized matter bounce model, which is also compatible with the late dark energy dominant phase of the cosmic evolution. In their study, Odintsov et al. \cite{Odintsov-2021} looked into a Chern-Simons corrected $f(R)$ gravity theory of a non-singular bounce to a dark energy epoch. The Chern-Simons coupling function is thought to behave in a power law way with the Ricci scalar. The modified Friedmann equation in LQC has been transferred to the Palatini $f(R)$ theory \cite{Olmo-2009} by Olmo and Singh whereas Olmo \cite{Olmo-2011}, has discovered the necessary $f(R)$ function that must be taken into account to create a bouncing cosmology of this type of LQC. In the generalized $f(R)$ theory, Nojiri et al. \cite{Nojiri-2019} have studied non-singular bounce cosmology in the context of the Lagrange multiplier. Starobinsky \cite{Starobinsky-2007} suggested a $f(R)$ gravity model that is in line with cosmological conditions and accords with laboratory experiments and observations of the solar system. However, Cognola et al. \cite{Cognola-2008} developed and investigated the exponential gravity model. This model accurately captures the natural inflation of the early universe and the accelerated expansion of the present universe.

Using a perfect fluid as only matter content, Shabani and Ziaie \cite{Shabani-2018} investigated the classical bouncing solutions in the context of $f(R,T)$ gravity in a flat FLRW background. Singh et al. \cite{Singh-2018} conducted a study on the bounce scenario of the universe, focusing on the application of certain Hubble parameters inside the $f(R,T)$ gravity theory. This study looks into a bouncing scenario of a flat, homogeneous, and isotropic universe using the reconstruction method for the power-law parametrization of the Hubble parameter in a modified gravity theory with higher-order curvature and the trace of the energy-momentum tensor components. The cosmological initial singularity can be avoided by proving that bouncing criteria are met \cite{Singh-2023}. Additional research on modified gravity theory has been conducted \cite{Amoros-2013-87, Haro-2014, Logbo-2019, Elizalde-2020-954}.

In the following section, some modified theories that are used to study the bouncing scenario of the universe have been presented. 

\section{Gravitational theories, geometries, and connections}
\subsection{\texorpdfstring{$F(R)$}{} gravity field equation}
The $R$ in the Einstein Hilbert action \eqref{eqn:HE_action} can be replaced by a functional form of Ricci scalar $R$, we can get the action for $F(R)$ gravity \cite{Sotiriou-2010,De-Felice-2010-13} defined as,
\begin{equation}\label{eq:action_f(R)}
S_{F(R)}=\int\sqrt{-g}\frac{F(R)}{2\kappa}d^{4}x,    
\end{equation}
varying action  \eqref{eq:action_f(R)} with respect to $g_{ij}$, the $F(R)$ gravity field equations can be obtained as,

\begin{equation}\label{eq:field_f(R)}
F_{R}R_{ij}-\frac{1}{2}F g_{ij}-\nabla_{i}\nabla_{j}F_{R}+g_{ij}\square F_{R}=0.
\end{equation}
For the flat FLRW space, the last two terms in the above equation can be obtained as 
\begin{eqnarray}\label{eq:f(R)_term}
    \left(g_{ij}\square-\nabla_{i}\nabla_{j}\right)F_{R}&=&-F_{RR}\bigg\{\frac{\partial^{2}R}{\partial x^{i}\partial x^{j}}-\Gamma^{t}_{ij}\frac{\partial R}{dt}-g_{ij}\left[\left(\frac{\partial g^{tt}}{\partial t}+g^{tt}\frac{\partial \ln(\sqrt{-g})}{\partial t}\right)\frac{\partial R}{\partial t}+g^{tt}\frac{\partial^{2}R}{\partial t^{2}}\right]\bigg\} \nonumber \\
    &&-F_{RRR}\bigg\{\frac{\partial R}{\partial x^{i}}\frac{\partial R}{\partial x^{j}}-g_{ij}\left[g^{tt}\left(\frac{\partial R}{\partial t}\right)^{2}\right]\bigg\}.
\end{eqnarray}
Here, $F_{R}=\frac{dF}{dR}$, $\nabla_{i}$ represents the covariant derivative, and $\square\equiv g^{ij}\nabla_{i}\nabla_{j}$ is the d'Alembert operator. 

\subsection{\texorpdfstring{$f(R,T)$}{} gravity field equation}
The $f(R,T)$ theory \cite{Harko-2011-84} is a modification of the general theory of relativity in which a matter Lagrangian $L_{m}$ can be described by the combination of $R$ and $T$. Where $R$ and $T$ be respectively the Ricci scalar and trace of the energy-momentum tensor, the total gravitational action of $f(R,T)$ gravity becomes
\begin{equation}\label{eq:act:f(R,T)}
S_{f(R,T)}=\int{\left[\frac{f(R,T)}{16\pi}+L_{m} \right]}\sqrt{-g}d^{4}x,
\end{equation}
where $L_{m}$ be the matter Lagrangian and the stress-energy tensor of matter defined as,
\begin{equation}\label{eq:EMT:f(R,T)}
T_{ij}=-\frac{2}{\sqrt{-g}}\frac{\delta(\sqrt{-g}L_{m})}{\delta g^{ij}}.
\end{equation}
The non-minimal matter geometry coupling is considered as, $f(R,T)=f(R)+f(T)$. Varying action \eqref{eq:act:f(R,T)} with respect to the metric tensor $g_{ij}$, the field equations of $f(R,T)$ gravity with non-minimal matter coupling can be obtained as \cite{Harko-2011-84},
\begin{equation}
f_{R}(R)R_{ij}-\frac{1}{2}f(R)g_{ij}-(\nabla_{i}\nabla_{j}-g_{ij})f_{R}(R)=8\pi T_{ij}+f_{T}(T)T_{ij}+\left[f_{T}(T)p+\frac{1}{2}f(T)\right]g_{ij}. \label{eq:field:f(R,T)}
\end{equation}

In equation \eqref{eq:field:f(R,T)}, the notations are $f_{R}(R)=\partial f(R)/\partial R$ and $f_{T}(T)=\partial f(T)/\partial T$.

\subsection{\texorpdfstring{$f(Q)$}{} gravity field equation}
The metric tensor $g_{ij}$ is the generalization of gravitational potential and the affine connection $\Gamma^{k}_{~ ij}$ describes the parallel transport and covariant derivatives \cite{Jimenez-2020-101}. Some assumptions on the affine connection specify the metric affine geometry. In differential geometry, the metric affine connection can be expressed in three independent components, 

\begin{equation}\label{eq:affine_connection}
\Gamma^{k}_{ij}=\{^{k} _{ij}\}+K^{k}_{~ij}+L^{k}_{~ij},   
\end{equation}
where the three terms on the right-hand side denote the Levi-Civita Connection, Contortion, and the disformation tensor respectively, can be expressed as,
\begin{subequations}
\begin{eqnarray}
\{^{k}_{ij}\}&\equiv&\frac{1}{2}g^{kl}\left(\partial_{i}g_{lj}+\partial_{j}g_{li}-\partial_{l}g_{ij}\right), \label{Levi-Civita_connection} \\
K^{k}_{~ij}&\equiv& \frac{1}{2}T^{k}_{~ij}+T_{(i ~ ~ j)}^{ ~ ~ k}; ~~~~ T^{k}_{~ ij}\equiv 2\Gamma^{k}_{[ij]}, \label{Contoetion_tensor} \\
L^{k}_{~ ij}&\equiv&\frac{1}{2}Q^{k}_{~ ij}-Q_{(ij)}^{\ \ \  k} .\label{eq:disformation_tensor}
\end{eqnarray}    
\end{subequations}

The nonmetricity conjugate is, 
\begin{equation}\label{eq:superpotential}
P^{k}_{~ ij}=-\frac{1}{2}L^{k}_{~  ij}+\frac{1}{4}\left(Q^{k}-\tilde{Q}^{k} \right)g_{ij}-\frac{1}{4}\delta^{k}_{(i}Q_{j)}, 
\end{equation} 
where $Q_{k}=g^{ij}Q_{kij}$ and $\tilde{Q}_{k}=g^{ij}Q_{ikj}$ with $Q_{kij}$ be the nonmetricity tensor. The nonmetricity scalar $Q$ can be expressed as,
\begin{equation}\label{eq:Q1}
Q=-Q_{kij}P^{kij}.
\end{equation}

The action of $f(Q)$ \cite{Jiemenez-2018, Jarv-2018-97} gravity is, 
\begin{equation}\label{eq:act_f(Q)}
S_{f(Q)}=\int d^{4}x\sqrt{-g}\left(-\frac{1}{2\kappa}f(Q)+L_{m}\right),
\end{equation}
where $g$ is the determinant of the metric $g_{ij}$; $L_{m}$ be the matter Lagrangian and $Q_{kij}=\nabla_{k}g_{ij}$. 

In the geometrical framework, the flat and torsion-free connection has been considered. The connection can be parameterized with a set of functions $\xi^{l}$ as
\begin{equation}
    \Gamma^{k}_{~ij}=\frac{\partial x^{k}}{\partial \xi^{l}}\partial_{i}\partial_{j}\xi^{l}.
\end{equation}
As a result, it is always feasible to make a coordinate choice that causes the connection to disappear. This coordinate is known as a coincident gauge, and they are specified here as, $\mathring{\Gamma}^{k}_{~ij}=0$. Thus, in the coincident gauge, $\mathring{Q}_{kij}=\partial_{k}g_{ij}$, where the over ring notation refers to the coincident gauge. While in the arbitrary gauge, $Q_{kij}=\partial_{k}g_{ij}-2\Gamma^{l}_{~k(i} g_{_{j)l}}$. 

So, the field equations of $f(Q)$ gravity can be expressed as,
\begin{equation}\label{eq:field_f(Q)}
\frac{2}{\sqrt{-g}}\nabla_{k}\left(\sqrt{-g}f_{Q}P^{k}_{~ij} \right)+\frac{1}{2}g_{ij}f+f_{Q}\left(P_{ikl}Q_{j}^{~kl}-2Q_{kli}P^{kl}_{~ ~ ~ j} \right)=T_{ij},
\end{equation}
where the subscript $Q$ in the function $f\equiv f(Q)$ is the partial derivative with respect to the nonmetricity scalar.

\subsection{\texorpdfstring{$f(Q,T)$}{} gravity field equation}
The action of $f(Q,T)$ gravity \cite{Xu-2019-79} is given as,

\begin{equation} \label{act:f(Q,T)}
S_{f(Q,T)}=\int\left(\dfrac{1}{16\pi}f(Q,T)+L_{m}\right)d^{4}x\sqrt{-g},
\end{equation}

The non-metricity $Q$ can be defined as,
\begin{equation} \label{eq:Q}
Q\equiv -g^{ij}( L^k_{~li}L^l_{~j k}-L^k_{~lk}L^l_{~ij}) , 
\end{equation}

where $L^k_{~ij}\equiv -\frac{1}{2}g^{kl}(\bigtriangledown_{j}g_{il}+\bigtriangledown_{i}g_{lj}-\bigtriangledown_{l}g_{ij})$. The energy-momentum tensor and the superpotential terms are respectively represented in the equations \eqref{eq:EMT:f(R,T)} and \eqref{eq:superpotential}. By varying the gravitational action \eqref{act:f(Q,T)}, the field equation of $f(Q,T)$ gravity can be obtained as, 
\begin{equation}\label{field_eq_f(Q,T)}
-\frac{2}{\sqrt{-g}}\bigtriangledown_{k}(f_{Q}\sqrt{-g}P^{k}_ {ij})-\frac{1}{2}fg_{ij}+f_{T}(T_{ij}+\Theta_{ij})-f_{Q}(P_{i kl} Q^{\;\;\; kl}_{j}-2Q^{kl}_{\;\;\;i} P_{klj})=8 \pi T_{ij},
\end{equation}
where $f_Q=\frac{\partial f(Q,T)}{\partial Q}$, $\Theta_{ij}\equiv g^{kl}\frac{\delta T_{kl}}{\delta g^{ij}}$.

This chapter provides an examination of bounce cosmology, with a particular focus on Einstein's general theory of relativity. The violation of the NEC in Einstein's GR has been observed, indicating the requirement for exotic matter. It is noticed that when the GR is modified there will be no need for any exotic matter. Hence, the possibility of modifying GR by altering either the matter or the geometric elements of the theory, as well as the development of modified gravity and its history, has been covered.

In chapters 2 and 3, the bounce scenario of the universe in curvature-based gravities like $f(R)$ and $f(R,T)$ has been discussed. Chapter 2 investigates a bouncing cosmological model of the universe within the context of an extended theory of gravity (also known as $f(R)$ gravity). We will spend a little bit of time discussing the geometrical parameters that are involved in this gravity. The scalar perturbation technique has been used to analyze the stability of the models under consideration. In chapter 3, the bouncing cosmological solution will be studied in $f(R,T)$, the Ricci scalar coupled with the energy-momentum tensor, and the effects of the bouncing scenario will be observed. We demonstrate the dynamic stability of the model by using a calculation called a linear homogeneous perturbation.

Chapters 4 and 5 are devoted to studying the bounce scenario of the universe in extended symmetric teleparallel gravity. The nature of the bounce scenario in $f(Q)$ gravity will be covered in chapter 4. According to LQC, the energy density parameter and the Hubble square term have been found to be related. The Hubble square term can be used to reconstruct the $f(R)$ model, resulting in a bounce model. Compared to $f(R)$ gravity, completing the task in $f(Q)$ gravity will be more straightforward since $R$ and $Q$ are related to the Hubble square term. The stability of the developed $f(Q)$ model has been investigated using the scalar perturbation method. The evolution of the geometrical parameters in $f(Q, T)$ gravity will be discussed in chapter 5, and the stability of the bouncing model will also be examined. 

Chapter 6 will present the generic solution for the bouncing scenario of the universe by employing a dynamical system approach. Incorporating an additional kinetic and potential component, the stability of the critical points has been examined.
\section{Conclusion}
The SCM encounters certain challenges during the early stages of the universe. Many of these difficulties can be addressed by incorporating an inflationary scenario. However, the question of the initial singularity may still require further resolution. Therefore, the concept of bouncing cosmology presents an alternative to the inflationary scenario, offering a potential resolution to the problem of the initial singularity as well as other challenges encountered during the early stages of the universe. The bounce scenario of the universe has been examined in the context of Einstein's GR. The restriction on the energy-momentum tensor of the fluid has been determined after studying the energy conditions.  It has been noticed that the NEC must be violated for a viable bounce scenario at the bounce epoch. The violation of the NEC implies that exotic matter may be present in the universe. It has been observed that the instability issue can be resolved by modifying GR, either in the matter part or in the geometry part.  

This chapter has discussed various modifications that could be made to Einstein's general theory of relativity. The modeling of matter or geometry aspects of Einstein's GR has been found to be effective in solving the early universe and late-time acceleration issues. The modification in the geometry part leads to different gravity, $f(R)$ gravity, $f(R, T)$ gravity. It has been observed that in the context of modified gravity, the matter energy-momentum tensor can satisfy the NEC. This means that under certain conditions, it is possible to achieve a bounce. The bounce occurs when the gravitational contribution from the modified term is such that the effective energy-momentum tensor violates the NEC. One alternative method for creating a non-singular bouncing model with modified matter is by incorporating a field that possesses a kinetic energy term of the opposite sign, often known as a ghost field. To achieve the phenomenon of bounce, the scalar field must violate the NEC, which could potentially result in the emergence of instabilities. The ghost condensate method, on the other hand, has the potential to generate a more effective nonsingular bounce. This model has the potential to encounter gradient instability. To resolve this issue, we can substitute the Lagrangian of the ghost condensate with a Galileon term.
\chapter{Bouncing Cosmology in Curvature-Based Extended Gravity} 

\label{Chapter2} 

\lhead{Chapter 2. \emph{Bouncing Cosmology in Curvature-Based Extended Gravity}} 

\vspace{10 cm}
* The work, in this chapter, is covered by the following publication: \\

\textbf{A.S. Agrawal} et al., ``Bouncing cosmological models in a functional form of $F(R)$ gravity", \textit{Gravitation and Cosmology}, \textbf{29}, 293-303 (2023).

\clearpage

\section{Introduction}\label{ch4intro}
In this chapter, the bounce cosmology in the $F(R)$ gravity will be discussed. The $F(R)$ theory of gravity is well-known and attracts a lot of interest from researchers because of its characteristics. It is applied to address the problem of late-time acceleration. Numerous studies have been conducted in $F(R)$ gravity, as was indicated in the preceding chapter. The Starobinsky and exponential models have been examined in $F(R)$ gravity, with the scale factor assuming that the universe did not begin with a singularity. The perturbation approach has been used to assess the stability of the model. The geometrical and physical significance of the model has been studied for completeness.  

\section{Field equations of \texorpdfstring{$F(R)$}{} gravity}
For the flat FLRW space-time \eqref{spacetime:GR} using equation \eqref{eq:f(R)_term}, the temporal and spatial components of the equation \eqref{eq:field_f(R)} becomes
\begin{subequations}
\begin{eqnarray}
&&0=-\frac{F}{2}+3\left(H^{2}+\dot{H}\right)F_{R}-18\left(4H^{2}\dot{H}+H\ddot{H}\right)F_{RR}, \label{eq:friedmann_f(R)}\\
&&0=\frac{F}{2}-3\left(H^{2}+\dot{H}\right)F_{R}+6\left(8H^{2}\dot{H}+4\dot{H}^{2}+6H\ddot{H}+\dot{\ddot{H}}\right)F_{RR}+36\left(4H\dot{H}+\ddot{H} \right)^{2}F_{RRR}.\nonumber \\ \label{eq:Raychaudhury_f(R)}
\end{eqnarray}    
\end{subequations}
Where $F_{RR}=\frac{d^{2}F}{dR^{2}}, ~F_{RRR}=\frac{d^{3}F}{dR^{3}}$. When the above equations are compared to the standard Friedmann equations, it is clear that $F(R)$ gravity contributes to the energy-momentum tensor, with its effective energy density $\rho_{\text{eff}}$ and effective pressure $p_{\text{eff}}$ \cite{Odintsov-2020-959} given by
\begin{subequations}\label{eq:_p_rho_friedmann_f(R)2}
\begin{eqnarray}\label{eq:friedmann_f(R)2}
\rho_{\text{eff}}&=&-\frac{f}{2}+3\left(H^{2}+\dot{H}\right)f_{R}-18\left(4H^{2}\dot{H}+H\ddot{H}\right)f_{RR},\\
p_{\text{eff}}&=&\frac{f}{2}-3\left(H^{2}+\dot{H}\right)f_{R}+6\left(8H^{2}\dot{H}+4\dot{H}^{2}+6H\ddot{H}+\dot{\ddot{H}}\right)f_{RR}+36\left(4H\dot{H}+\ddot{H} \right)^{2}f_{RRR}.\nonumber \\
\label{eq:Raychaudhury_f(R)2}
\end{eqnarray}    
\end{subequations}
Equations \eqref{eq:_p_rho_friedmann_f(R)2} can be reformulated using the Hubble parameter, as well as the derivatives of the functional form of $F(R)$ with respect to $R$. In these equations, the Ricci scalar, represented as $R=6\left(\frac{\ddot{a}}{a}+\frac{\dot{a}^{2}}{a^{2}} \right)$. Therefore, in order to obtain the energy density and pressure of the matter field and to further investigate the dynamics of the universe, a Hubble function is required. In order to investigate the phenomenon of late-time acceleration, it is necessary to analyze the behavior of the effective EoS parameter. This parameter can be derived by analysis,
\begin{equation}\label{eq:EoS_f(R)}
\omega_{\text{eff}}=\frac{p_{\text{eff}}}{\rho_{\text{eff}}} =-1+\frac{12\left(2H^{2}\dot{H}+4\dot{H}^{2}+3H\ddot{H}+\dot{\ddot{H}}\right)f_{RR}+72\left(4H\dot{H}+\ddot{H} \right)^{2}f_{RRR}}{f-6\left(H^{2}+\dot{H}\right)f_{R}+36\left(4H^{2}\dot{H}+H\ddot{H}\right)f_{RR}}.
\end{equation}

The function $f(R)$ is used to denote the deviation of $F(R)$ gravity from Einstein's gravity, where $F(R)$ is defined as $R+f(R)$. The nature of $F(R)$ determines the effective energy-momentum tensor, as predicted. Thus, by taking into account the bouncing scale factor and some of the functional forms of $F(R)$, the bouncing scenario and the late-time cosmic acceleration problem of the universe has been examined in the following sections.

\section{The bouncing model and the analysis}\label{sect-iii}

During the bouncing event, the value of $H$ transitions from a negative value to a positive value at the crosspoint $H=0$, while $\dot{H}$ remains greater than zero. This transition occurs at the epoch of the bounce. The necessary properties of bouncing cosmological models can be represented as follows:

\begin{itemize}
\item During the bouncing epoch, the scale factor $a$ undergoes a contraction to a non-zero finite size, while the Hubble parameter $H$ approaches zero and the deceleration parameter $q=-1-\frac{\dot{H}}{H^2}$ becomes singular.  

\item For the scale factor, the bounce is followed by an increase in slope. The Hubble parameter is negative during the period of matter contraction. During the process of matter expansion, it turns positive. 
\end{itemize}

To give the bouncing cosmological model the above properties, the bouncing scale factor $a(t)=\left(\frac{\alpha}{\chi}+t^{2}\right)^{\frac{1}{2\chi}}$ \cite{Abdussattar-2011} was taken into account. Here, $\alpha$ and $\chi$ are both positive constants. From this, the Hubble parameter $H=\frac{t}{\alpha +\chi t^{2}}$.

\begin{figure}[H]
	\begin{center}
		\includegraphics[width=7.5cm]{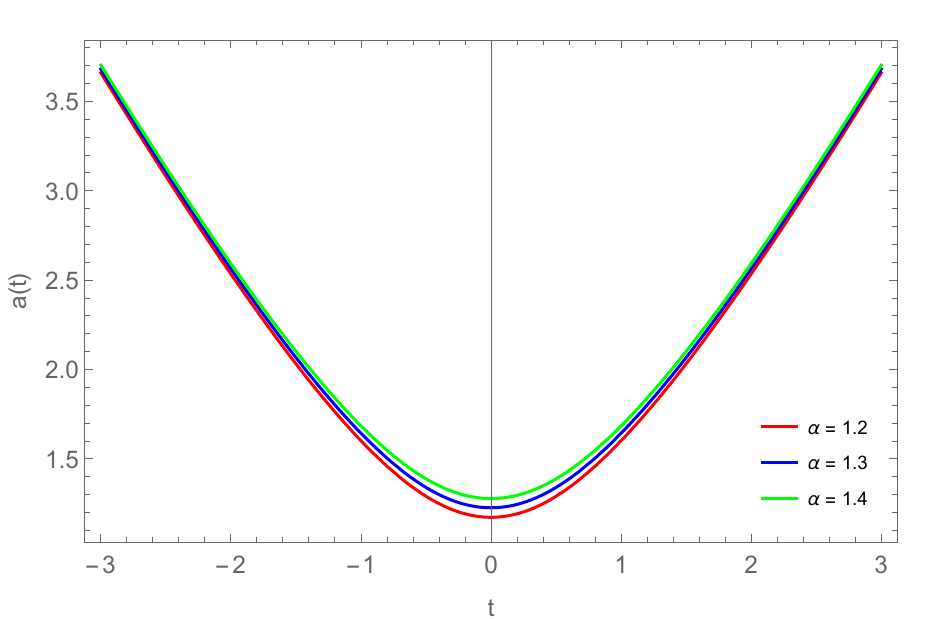}
		\includegraphics[width=7.5cm]{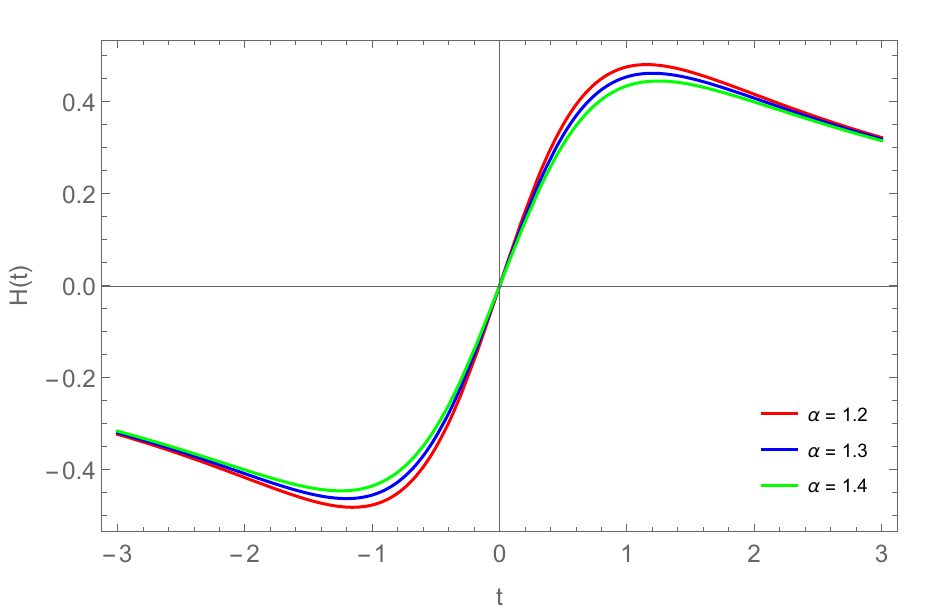}
	\end{center}
	\caption[Variation of the scale factor and the Hubble parameter with cosmic time.]{Variation of the scale factor (left panel) and Hubble parameter (right panel) with cosmic time.}\label{ch2_Fig:1}
\end{figure}

The development of the scale factor and Hubble parameter for cosmic time has been depicted for varying the $\alpha$ parameter with $\chi=0.9$. The graphical behavior of the scale factor and the Hubble parameter is shown in Figure \ref{ch2_Fig:1}. It can be observed that all of these quantities exhibit behavior consistent with the properties outlined in bouncing cosmology. The scale factor exhibits a decreasing trend from a higher value during the early stages, gradually reaching a small yet finite value at the point of bounce. Subsequently, it experiences an increase in value following the bounce. The Hubble parameter initiates from a higher negative value, reaches the null point of $H=0$ at the bounce event, and subsequently experiences further increments.
Thus, in this case, both the concurrent dynamic behavior and the bouncing conditions mentioned above are obeyed by our bouncing model with an assumed scale factor. When examining bouncing scenarios within the framework of $F(R)$ gravity theory, it is beneficial to take into account the generation era of the perturbation modes. In many bouncing models, the Hubble parameter typically becomes zero at the epoch of the bounce. Consequently, the comoving Hubble radius, denoted as $r_h=1/aH$, diverges. The universe in late time is accelerating or decelerating, as evidenced by the asymptotic behavior of the comoving Hubble radius. In certain instances where particular scale factors are employed, the Hubble radius exhibits a continuous decrease on either side of the bounce, eventually tending toward zero. This kind of behavior points to an accelerating universe in the late universe. Consequently, in such occurrences, the Hubble horizon diminishes to zero as cosmic time reaches large magnitudes, but exclusively the Hubble horizon exhibits an infinite magnitude in proximity to the bouncing point. Nevertheless, in the case of alternative selections of the bouncing scale factors, it can be shown that the Hubble radius exhibits divergence at later periods, which suggests a deceleration of the universe. Unlike previous situations where the perturbation modes are formed around the bouncing era, in such scenarios they are created at very large negative cosmic times, corresponding to the low curvature regime of the contracting era. Because all primordial modes are contained in the horizon only at that point, the primordial perturbation modes relevant to the current era are created for cosmic times close to the bouncing point. As the horizon gets smaller, the modes emerge from it and become significant for current observations \cite{Odintsov-2020-959}. The Hubble radius tends to approach 0 asymptotically when all perturbation modes are contained within the horizon at that particular time. Consequently, the perturbations take place in close range to the bounce event. The compatibility between the Planck limitations \cite{Ade-2014,Ade-2016} and the $F(R)$ gravity theory has been shown. In their study, Odintsov et al. \cite{Odintsov-2020-37} found that the $F(R)$ gravity theory exhibits a viable bounce phenomenon. This bounce is only achievable when perturbations arise in close range to the bounce itself, as determined through the utilization of observational indices in a bottom-up approach. 

To assess the compatibility of the scale factor employed in this study with the generation of perturbation modes and to determine the feasibility of incorporating a bouncing scenario within the framework of $F(R)$ gravity, the cosmic Hubble radius as a time-dependent function is presented. This analysis was conducted for various values of the parameter $\alpha$, while keeping $\chi$ fixed at a specific value of 0.9. Based on the available options, it can be observed that the cosmic Hubble radius exhibits a symmetrical and monotonic decrease around the bouncing epoch. Furthermore, it approaches zero asymptotically in both the positive and negative time domains. This observation aligns with the findings of Odintsov et al. \cite{Odintsov-2020-959}. They have reported that the cosmic Hubble radius, with the selected scale factor $a_F(t_F)=\left(a_0t_F^2+1\right)^{n}$, exhibits a consistent decrease on both sides of the bounce when $n>1/2$. 
\begin{figure}
\centering
\includegraphics[width=0.5\textwidth]{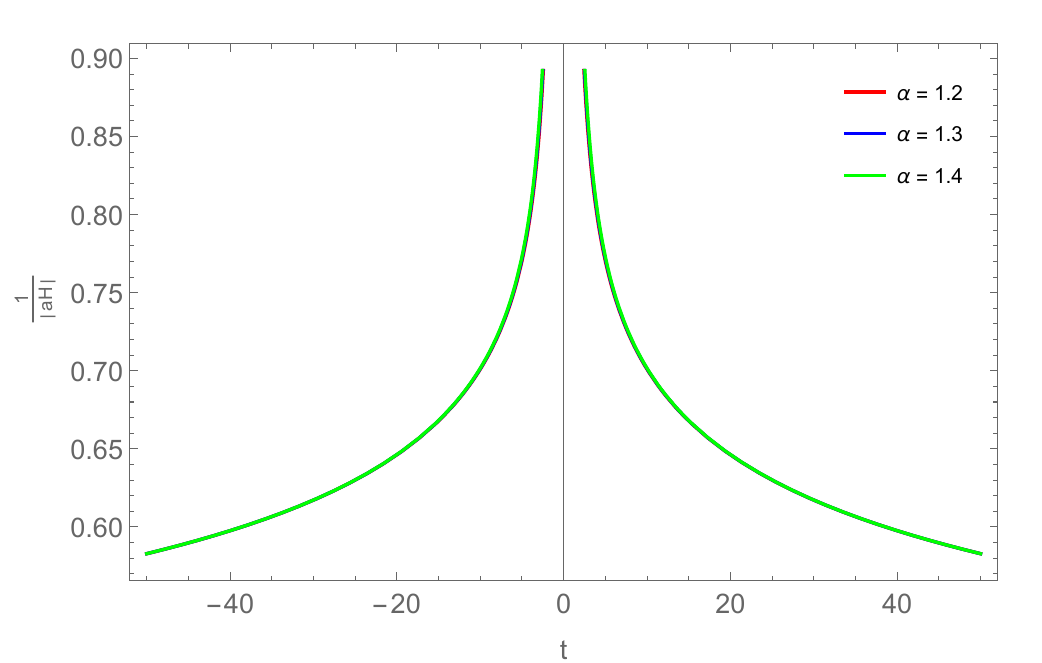}
\caption[Variation of the cosmic Hubble radius in cosmic time.]{Variation of the cosmic Hubble radius in cosmic time with varying $\alpha$ and $\chi = 0.9$.}
\label{ch2_Fig:2HR}
\end{figure}
Two distinct bouncing scenarios are obtained by considering the supplied symmetric bouncing scale factor along with two different functional forms of $F(R)$ in the following sections.
\section{Models}
In order to establish a functional form of $F(R)$, two widely recognized functional forms are utilized. The first one is the renowned $F(R)$ gravity model, which adheres to cosmological conditions and successfully passes solar system and laboratory tests. Starobinsky (model I) is the author who first proposed this model; (see \cite{Starobinsky-2007}). The exponential model (model II), as put forth by Cognola et al. \cite{Cognola-2008}, is the second functional form. This model effectively describes the inflation of the early universe and the current accelerated expansion of the universe in a coherent manner.
\begin{subequations}
    \begin{eqnarray}
        \rho+p&=&2 \dot{H} \left(F_{R}+12 F_{RR} \dot{H}-1\right),\\
    \end{eqnarray}
\end{subequations}
\subsection{Starobinsky model}
To construct the cosmological model, equations \eqref{eq:_p_rho_friedmann_f(R)2} are required to be solved so that the dynamical parameters can be obtained. To do so, a functional form for $F(R)$ is to be considered. The form of $F(R)$ \cite{Starobinsky-2007,Chen-2019} is considered as,
\begin{eqnarray}\label{eq:Starobinsky_model}
F(R)=R+\lambda R_{0}\left[\left(1+\frac{R^{2}}{R_{0}^{2}}\right)^{-n}-1\right],  \end{eqnarray}
Where $R_{0}$ represents the constant characteristic curvature, $\lambda$ and $n$ are additional constants. In order to align with the Starobinsky model, the value of the exponent $n=1$. By utilizing the functional form of $F(R)$ as given in equation \eqref{eq:Starobinsky_model}, equations \eqref{eq:_p_rho_friedmann_f(R)2}, and \eqref{eq:EoS_f(R)} can be simplified as follows:

\begin{subequations}\label{pp1}
\footnotesize
\begin{eqnarray}
\rho_{\text{eff}}&=& \frac{\lambda  R_{0} \left(72 H R_{0}^2 \left(R_{0}^2-3 R^2\right) \left(4 H \dot{H}+\ddot{H}\right)-12 R_{0}^2 R \left(\dot{H}+H^2\right) \left(R_{0}^2+R^2\right)+R^2 \left(R_{0}^2+R^2\right)^2\right)}{2 \left(R_{0}^2+R^2\right)^3}, \label{eq:rho_eff_model_1}\\
p_{\text{eff}}&=&\frac{4 \lambda  R_{0}^3 R \left(\dot{H}+3 H^2\right) \left(R_{0}^2+R^2\right)^{2}-24 \lambda  R_{0}^3 \left(R_{0}^2-3 R^2\right) \left(R_{0}^2+R^2\right) \left(4 \dot{H} \left(\dot{H}+2 H^2\right)+6 H \ddot{H}+\dot{\ddot{H}}\right)}{2 \left(R_{0}^2+R^2\right)^4}\nonumber \\ &&+\frac{1728 \lambda  R_{0}^3 R (R_{0}-R) (R_{0}+R) \left(4 H \dot{H}+\ddot{H}\right)^2-\lambda  R_{0} R^2 \left(R_{0}^2+R^2\right)^3}{2 \left(R_{0}^2+R^2\right)^4}, \label{eq:P_eff_model_1}\\
\omega_{\text{eff}}&=&\frac{4 R_{0}^2 R \left(\dot{H}+3 H^2\right) \left(R_{0}^2+R^2\right)^2+1728 R_{0}^2 R (R_{0}-R) (R_{0}+R) \left(4 H \dot{H}+\ddot{H}\right)^2-R^2 \left(R_{0}^2+R^2\right)^3}{\left(R_{0}^2+R^2\right) \left(72 H R_{0}^2 \left(R_{0}^2-3 R^2\right) \left(4 H \dot{H}+\ddot{H}\right)-12 R_{0}^2 R \left(\dot{H}+H^2\right) \left(R_{0}^2+R^2\right)+R^2 \left(R_{0}^2+R^2\right)^2\right)}\nonumber \\&&-\frac{24 R_{0}^2 \left(R_{0}^2-3 R^2\right) \left(R_{0}^2+R^2\right) \left(4 \dot{H} \left(\dot{H}+2 H^2\right)+6 H \ddot{H}+\dot{\ddot{H}}\right)}{\left(R_{0}^2+R^2\right) \left(72 H R_{0}^2 \left(R_{0}^2-3 R^2\right) \left(4 H \dot{H}+\ddot{H}\right)-12 R_{0}^2 R \left(\dot{H}+H^2\right) \left(R_{0}^2+R^2\right)+R^2 \left(R_{0}^2+R^2\right)^2\right)}. \nonumber \\ \label{eq:EoS_eff_model_1}
\end{eqnarray}    
\end{subequations}
In model I, the effective pressure continues to be negative as it gets closer and closer to the bouncing point, also known as the point when time equals zero in the evolution. Throughout the progression, the energy density stays in the positive area. The bounce at $t=0$ gets more noticeable with larger levels of $\alpha$. The energy density increases at first, exhibiting a sort of ditch that narrows close to and during the bounce before eventually declining. During the contracting phase, the effective EoS parameter decreases and crosses the phantom-divide line. During the bounce epoch, the effective EoS parameter exhibits phantom-like behavior, and it increases during the expanding phase of evolution. The phantom divide is crossed twice, once during the pre-bounce phase and once during the post-bounce phase, exhibiting symmetric behavior. It is clear that the current model mostly resides in the quintessence domain, displaying $\Lambda$CDM behavior in both positive and negative time scales as we move away from the bounce epoch and phantom-like behavior at the bounce epoch. The value of $\alpha$ determines the magnitude of the well that occurred here; the deeper the well, the higher the value of $\alpha$. The well is only visible near the bounce when $\alpha$ is tiny; otherwise, it usually stays in the quintessence phase.
\begin{figure}[H]
	\begin{center}
		\includegraphics[width=7.5cm]{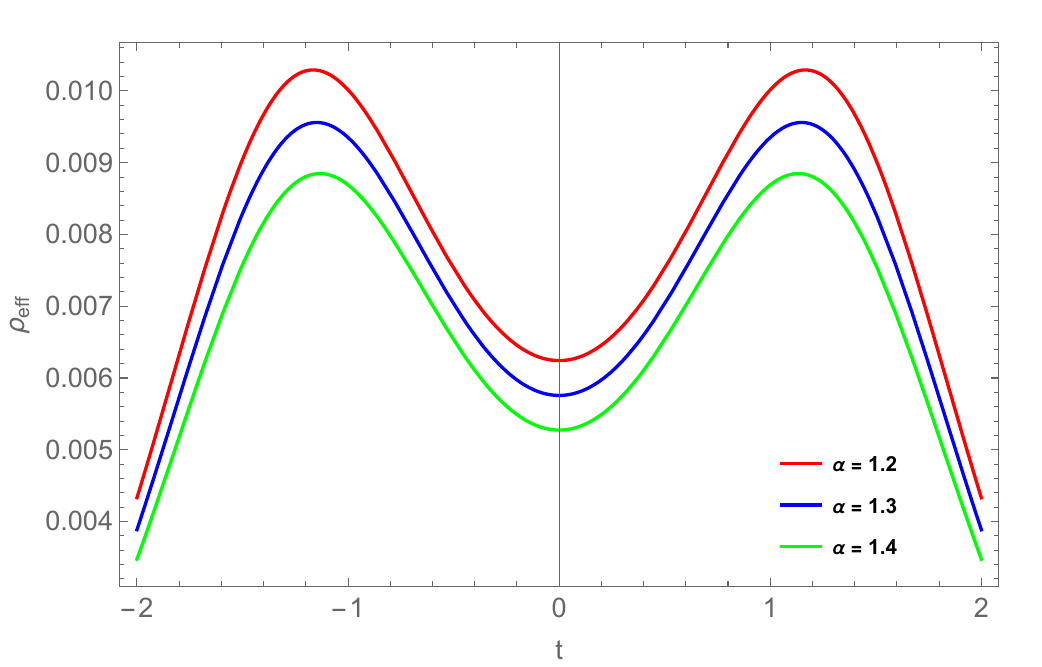}
		\includegraphics[width=7.5cm]{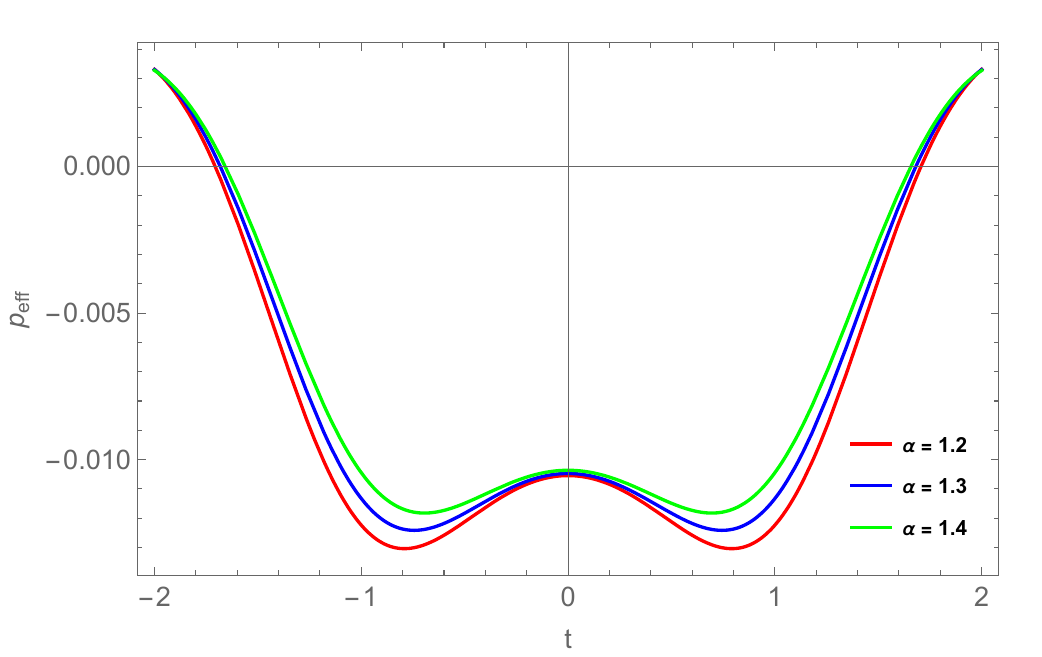}
		\includegraphics[width=7.5cm]{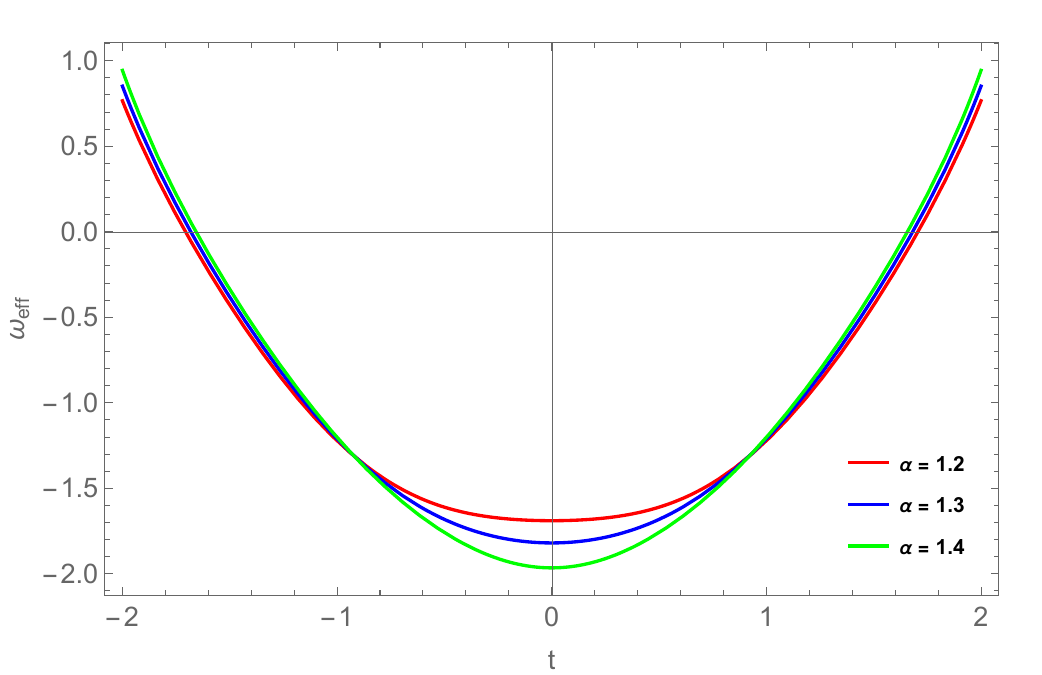}
	\end{center}
	\caption[Variation of physical parameters in cosmic time for model I.]{Variation of effective energy density (left panel), effective pressure (right panel), effective EoS parameter (lower panel) in cosmic time with varying $\alpha$ and $\chi =0.9$, $R_{0}=2$, $\lambda =0.01$, $n=1$ for model I.}
\label{ch2_Fig:3a}
\end{figure}
\subsection{Exponential model}
\begin{figure}[ht!]
	\begin{center}
		\includegraphics[width=7.5cm]{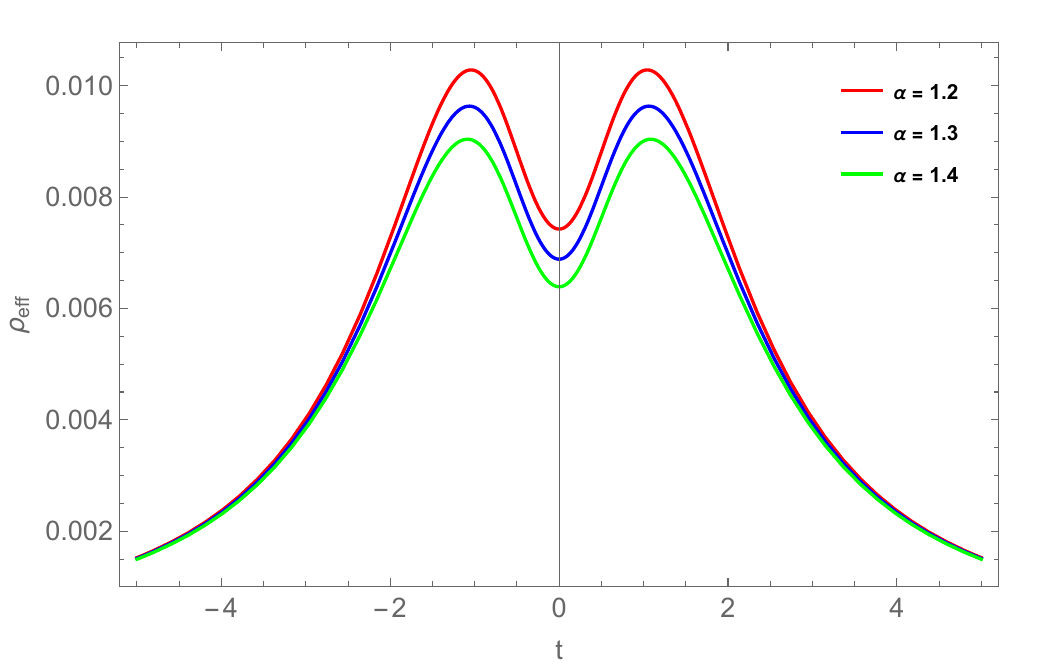}
		\includegraphics[width=7.5cm]{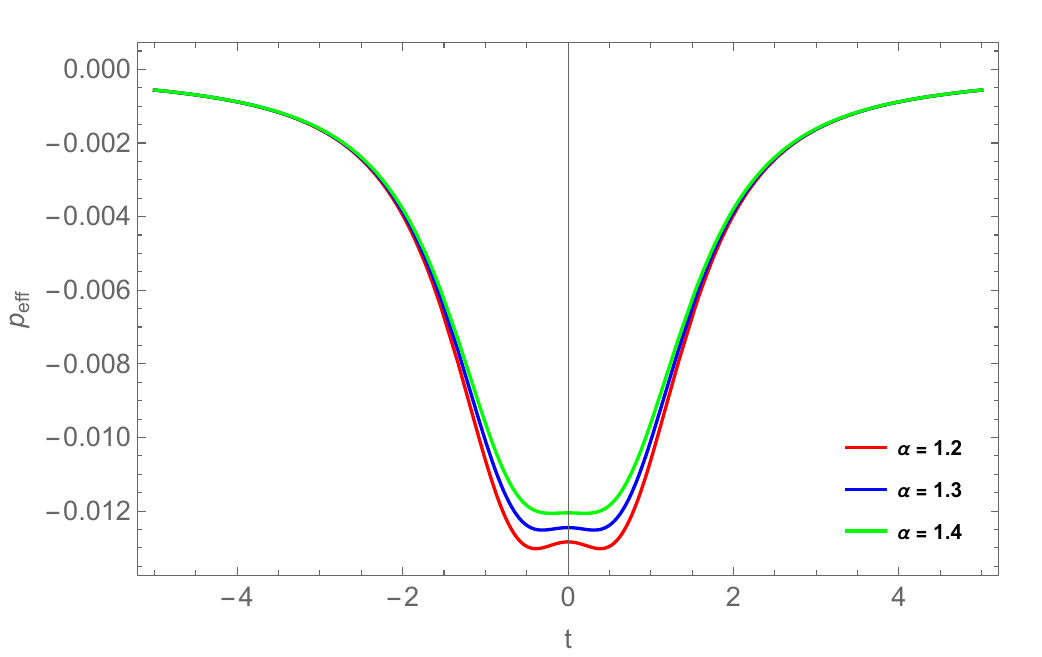}
		\includegraphics[width=7.5cm]{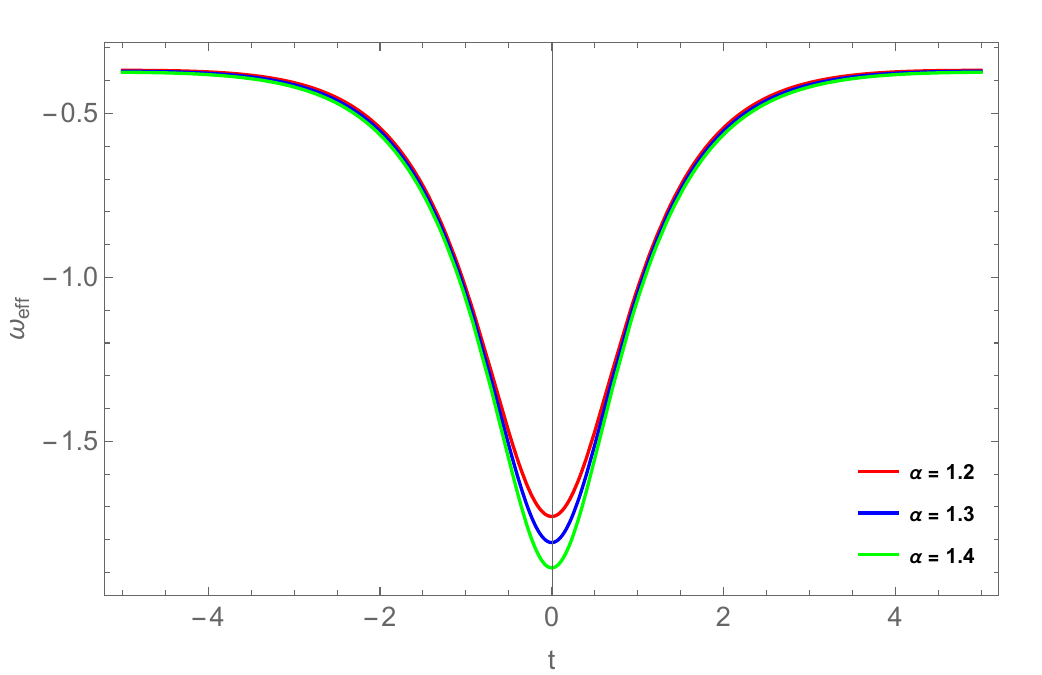}
	\end{center}
	\caption[Variation of physical parameters in cosmic time for model II.]{Variation of effective energy density (left panel), effective pressure (right panel), effective EoS parameter (lower panel) in cosmic time with varying $\alpha$, $\chi =0.9$, $R_{0}=2.5$, $\lambda=0.01$ for model II.}
\label{Cha2:Fig.4}
\end{figure}

The other form of the function $F(R)$ \cite{Cognola-2008} is considered as, 
\begin{eqnarray}\label{eq.13}
F(R)=R+R_{0}\lambda\left(e^{-\frac{R}{R_{0}}}-1\right),
\end{eqnarray}
where $R_0$ and $\lambda$ are constants. The same scale factor has been considered here as in model I. From equations \eqref{eq:_p_rho_friedmann_f(R)2}, the effective energy density, effective pressure, and effective EoS parameter for the exponential $F(R)$ form \eqref{eq.13} can be obtained as,
\begin{subequations}\label{pp2}
\footnotesize
\begin{eqnarray}
\rho_{\text{eff}}&=&\frac{\lambda  \left[e^{-\frac{R}{R_{0}}} \left(-6 \left(24 H^2+R_{0}\right) \dot{H}-R_{0} \left(6 H^2+R_{0}\right)-36 H \ddot{H}\right)+R_{0}^2\right]}{2 R_{0}},\label{eq:rho_eff_modelII}\\
p_{\text{eff}}&=&\frac{\lambda  e^{-\frac{R}{R_{0}}} \left[R_{0} \left(R_{0} \left(6 H^2-R_{0} e^{R/R_{0}}+R_{0}\right)+12 \dot{\ddot{H}}\right)+2 \dot{H} \left(R_{0} \left(48 H^2+R_{0}\right)-288 H \ddot{H}\right)\right]}{2 R_{0}^2}\nonumber 
\\ &&+\frac{\lambda  e^{-\frac{R}{R_{0}}}\left(48\dot{H}^2 \left(R_{0}-24 H^2\right)+72( H\ddot{H}R_{0}- \ddot{H}^2)\right)}{2 R_{0}^2},\label{eq:P_eff_modelII}\\
\omega_{\text{eff}}&=&-1-\frac{4 \left[\dot{H} \left(R_{0} \left(12 H^2+R_{0}\right)+144 H \ddot{H}\right)+12\dot{H}^2 \left(24 H^2-R_{0}\right)-3 \left(3 H R_{0} \ddot{H}+\dot{\ddot{H}} R_{0}-6\ddot{H}^2\right)\right]}{R_{0}^3 e^{R/R_{0}}-R_{0} \left(6 \left(24 H^2+R_{0}\right) \dot{H}+R_{0} \left(6 H^2+R_{0}\right)+36 H \ddot{H}\right)}.\nonumber 
\\ \label{eq:EoS_eff_modelII}
\end{eqnarray}    
\end{subequations}
As the universe evolves, we can see in Figure \ref{Cha2:Fig.4} that the behavior of the effective pressure is always negative. There is still a well at the bounce point for a lower $\alpha$, and the energy density is completely positive. Near the bounce epoch, the effective EoS parameter primarily stays in the phantom phase. Twice, both before and after the bounce, it traverses the phantom phase as needed by the bounce model. Similar to model I, as $\alpha$ increases, the effective EoS parameter becomes deeper.  When it advances away from the bounce period, it exhibits quintessential behavior after transitioning from phantom-like behavior during the bounce epoch. 
\section{Energy conditions}
In GR, Einstein's field equations address the causal metric and geodesic structure of space-time, so the energy-momentum tensor has to satisfy some conditions. We can define the energy conditions as the contractions of time-like or null vector fields with respect to Einstein tensor and the energy-momentum tensor from the matter side of Einstein's field equations \cite{Raychaudhuri-1955}. One can obtain four energy conditions which are defined in section \ref{GR_EC}. 
The extended theories of gravity are the straightforward extension of Einstein's GR, and so the $F(R)$ gravity. Any such extended theory should be confronted with the energy conditions. 

\subsection{Energy conditions for Starobinsky model}

The energy conditions of the bouncing $F(R)$ model can be obtained by using \eqref{pp1},
\begin{subequations}
\footnotesize
    \begin{eqnarray}
        \rho_{\text{eff}}+p_{\text{eff}}&=&-\frac{4 \lambda  R_{0}^3}{\left(R_{0}^2+R^2\right)^4} \bigg[3 \dot{\ddot{H}} \left(R_{0}^2+R^2\right) \left(R_{0}^2-3 R^2\right)+
        9 \ddot{H} \left(24 R \ddot{H} (R-R_{0}) (R_{0}+R)+H \left(R_{0}^2+R^2\right) \left(R_{0}^2-3 R^2\right)\right) \nonumber \\
        &&+\dot{H} \left(\left(R_{0}^2+R^2\right) \left(R \left(R_{0}^2+R^2\right)-12 H^2 \left(R_{0}^2-3 R^2\right)\right)+1728 H R \ddot{H} (R-R_{0}) (R_{0}+R)\right) \nonumber \\
        &&+12 \left(\dot{H}\right)^2 \left(-2 R_{0}^2 R \left(144 H^2+R\right)+3 R^3 \left(96 H^2-R\right)+R_{0}^4\right)\bigg],\\
        \rho_{\text{eff}}+3p_{\text{eff}}&=&\frac{\lambda  R_{0}}{2 \left(R_{0}^2+R^2\right)^4} \bigg[\left(R_{0}^2+R^2\right) \bigg(72 H R_{0}^2 \left(R_{0}^2-3 R^2\right) \left(\ddot{H}+4 H \dot{H}\right)-12 R_{0}^2 R \left(H^2+\dot{H}\right) \left(R_{0}^2+R^2\right)  \nonumber \\
        &&+R^2 \left(R_{0}^2+R^2\right)^2\bigg)+12 R_{0}^2 R \left(3 H^2+\dot{H}\right) \left(R_{0}^2+R^2\right)^2+5184 R_{0}^2 R (R_{0}-R) (R_{0}+R) \left(\ddot{H}+4 H \dot{H}\right)^2 \nonumber \\
        && -72 R_{0}^2 \left(R_{0}^2-3 R^2\right) \left(R_{0}^2+R^2\right) \left(\dot{\ddot{H}}+6 H \ddot{H}+4 \dot{H} \left(2 H^2+\dot{H}\right)\right)-3 R^2 \left(R_{0}^2+R^2\right)^3\bigg].
    \end{eqnarray}
\end{subequations}

\begin{figure}[H]
\centering
\minipage{0.45\textwidth}
\includegraphics[width=\textwidth]{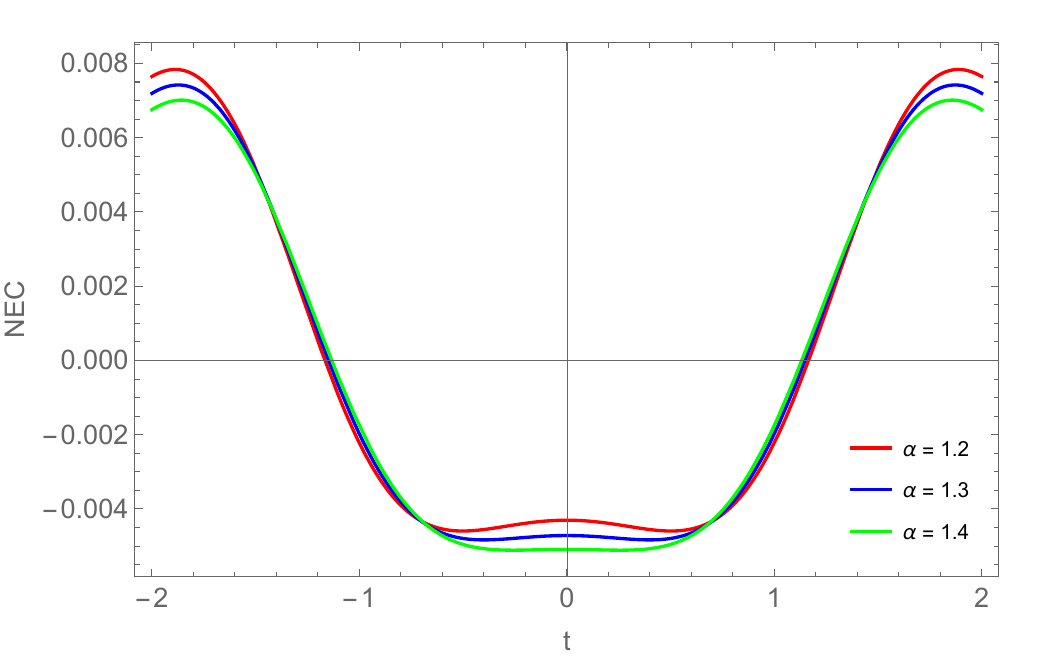}
\endminipage\hfill
\minipage{0.45\textwidth}
\includegraphics[width=\textwidth]{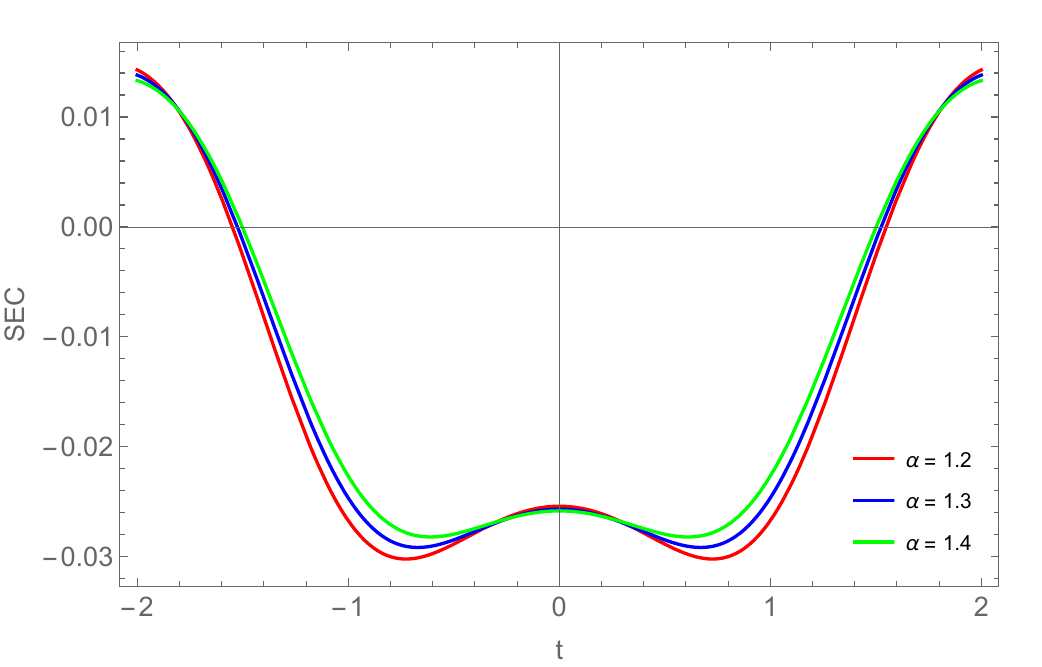}
\endminipage\hfill
\caption[Variation of the energy conditions in cosmic time for model I.]{Variation of null energy condition (left panel) and strong energy condition (right panel) in cosmic time with varying $\alpha$ with $\chi=0.9$, $R_{0}= 2$, $\lambda =0.01$ for model I.}
\label{Cha3;Fig.5}
\end{figure}

An energy condition representation with variable $\alpha$ can be seen in Figure \ref{Cha3;Fig.5}. The SEC and NEC are both violates at the bounce. In every energy condition, the symmetric behavior surrounding the bounce has been achieved. The NEC has a transitional behavior; for the most part, it stays in the positive domain in both the positive and negative time zones, but it also stays in the negative domain close to the bounce epoch, indicating a potential NEC violation. The bouncing model is realized by the violation of NEC during the bounce by Nojiri et al. \cite{Nojiri-2019}. Since the expanded theory of gravity likewise requires the violation of SEC, it is argued that the model under consideration also supports late-time cosmic acceleration. Furthermore, the energy conditions are consistent with the effective EoS parameter, as NEC and SEC violate for $\omega_{\text{eff}} \leq -1$ and $\omega_{\text{eff}} \leq -1/3$, respectively. Within the framework of contemporary cosmic dynamics, this allows us to further assert the validity of the model.

\subsection{Energy conditions for exponential model}
The energy conditions of model II can be obtained from \eqref{pp2},
\begin{subequations}
\footnotesize
    \begin{eqnarray}
        \rho_{\text{eff}}+p_{\text{eff}}&=&-\frac{2 \lambda  e^{-\frac{R}{R_{0}}}}{R_{0}^2} \bigg[-3 \left(\dot{\ddot{H}} R_{0}+3 H R_{0} \ddot{H}-6 \left(\ddot{H}\right)^2\right)+\dot{H} \left(R_{0} \left(12 H^2+R_{0}\right)+144 H \ddot{H}\right) \nonumber \\
        &&+12 \left(24 H^2-R_{0}\right) \left(\dot{H}\right)^2\bigg],\\
        \rho_{\text{eff}}+3p_{\text{eff}}&=&\frac{\lambda  e^{-\frac{R}{R_{0}}}}{R_{0}^2} \bigg[18 \left(\dot{\ddot{H}} R_{0}+5 H R_{0} \ddot{H}-6 \left(\ddot{H}\right)^2+4 H \dot{H} \left(H R_{0}-12 \ddot{H}\right)+4 \left(R_{0}-24 H^2\right) \left(\dot{H}\right)^2\right) \nonumber \\
        &&+R_{0}^2 \left(6 H^2-R_{0} e^{R/R_{0}}+R_{0}\right)\bigg].
    \end{eqnarray}
\end{subequations}
\begin{figure}[H]
\centering
\minipage{0.45\textwidth}
\includegraphics[width=\textwidth]{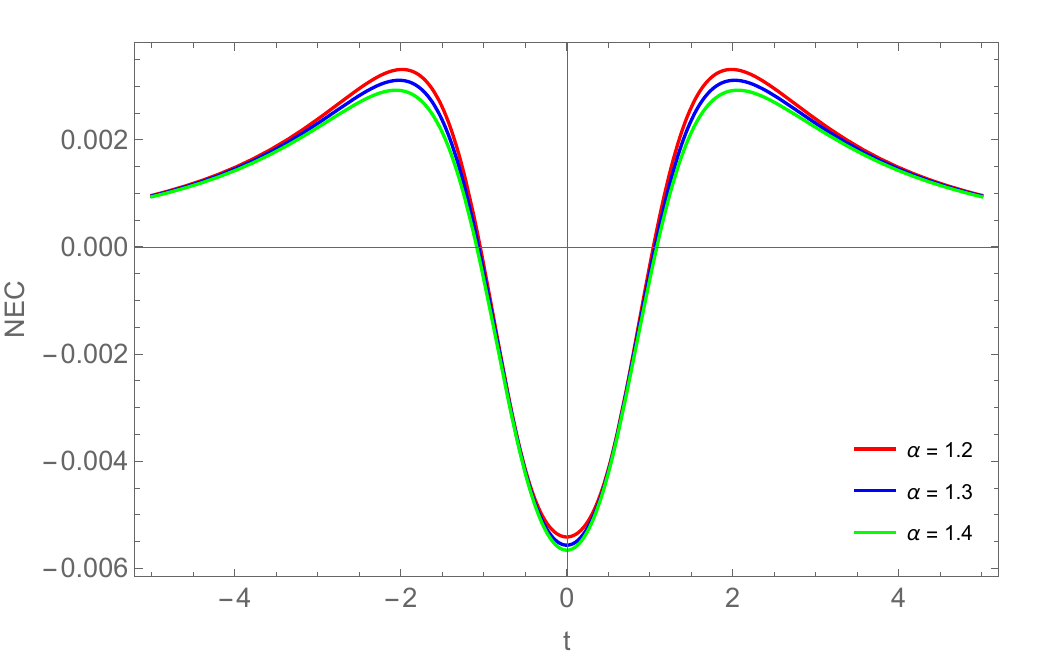}
\endminipage\hfill
\minipage{0.45\textwidth}
\includegraphics[width=\textwidth]{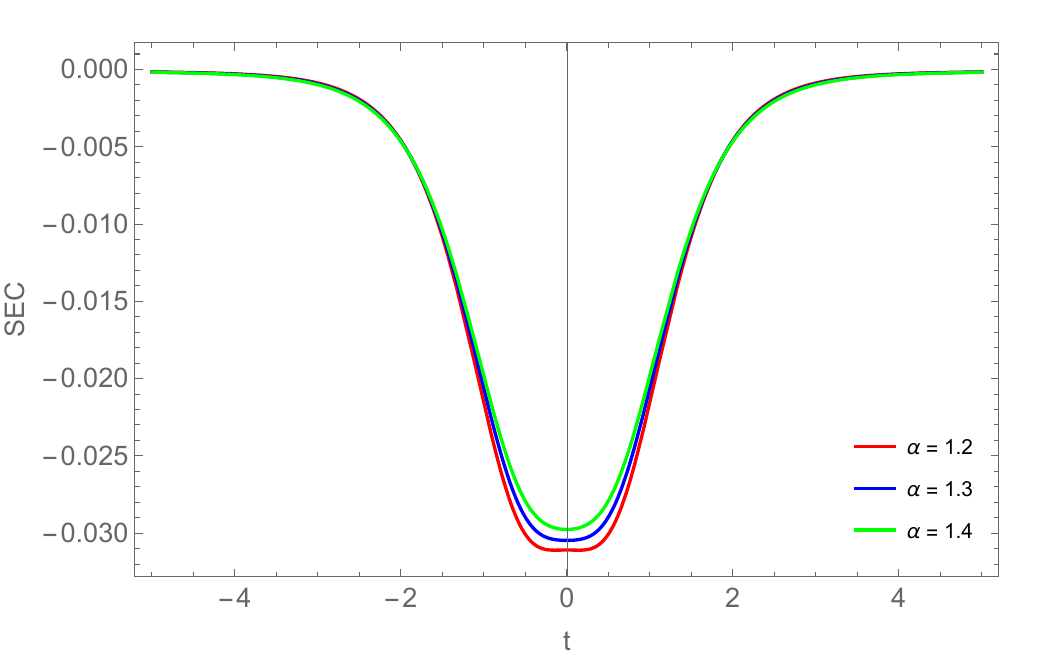}
\endminipage\hfill
\caption[Variation of the energy conditions in cosmic time for model II.]{Variation of null energy condition (left panel) and strong energy condition (right panel) in cosmic time with varying $\alpha$ with $\chi=0.9$, $R_{0}= 2.5$, $\lambda =0.01$ for model II.}
\label{Cha3;Fig.6}
\end{figure}
Figure \ref{Cha3;Fig.6} illustrates how the energy conditions of model II behave. At the bounce epoch, the model violates the NEC; however, as it moves away from the bouncing period, it fails to violate the positive and negative time zones. In their latest work, Nojiri et al. \cite{Nojiri-2019} examined a similar type of behavior. The SEC is completely violated; it starts off declining and then rises following the bounce. This energy condition result can validate the bouncing behavior of the model and its applicability in an extended gravity. It has been noted that the parametric value $\alpha$ influences the violation of energy conditions. The well is deeper at the bounce epoch for smaller values of $\alpha$, whereas the bounce appears flat for bigger values of $\alpha$.


\section{Stability analysis through scalar perturbation}
In theory, it is recommended that perturbations always be stated in terms of gauge invariant quantities. The momentum density $\delta T_{0i} = 0$ (where $T_{ij}$ is the effective energy-momentum tensor and the symbol $\delta$ signifies the associated perturbation) is what defines the comoving gauge in the current study. The scalar metric perturbation in this gauge \cite{Odintsov-2020-959,Odintsov-2020-37} is written as,
\begin{equation}
    \delta g_{ij}=a^{2}(t)(1-2\mathcal{R})\delta_{ij},
\end{equation}

where the scalar perturbation, also referred to as the comoving curvature perturbation and which is in fact a gauge invariant variable, is denoted by the symbol $\mathcal{R}(t, \Tilde{x})$. The second order action of $\mathcal{R}(t, \Tilde{x})$ is given by the additional metric perturbations $\delta g_{00}$ and $\delta g_{0i}$, which may be calculated in terms of $\mathcal{R}$ using perturbed gravitational equations
\begin{equation}\label{eq:deltaS}
\delta S_{\mathcal{R}}=\int dt d^{3}\Tilde{x} a(t)z^{2}(t)\left(\dot{\mathcal{R}}^{2}-\frac{1}{a^{2}(t)}(\partial_{i}\mathcal{R})^{2}\right),    
\end{equation}
$z(t)$ can be represented as follow
\begin{equation}\label{eq:z(t)}
z(t)=\frac{a}{\left(H+\frac{1}{2F_{R}}\frac{dF_{R}}{dt}\right)}\sqrt{\frac{3}{2F_{R}}\left(\frac{dF_{R}}{dt}\right)^{2}}.
\end{equation}
The sound speed of the scalar perturbation $(C_{s}^{2})$ equals unity, preventing superluminal modes and gradient instabilities. The unit sound speed for scalar perturbation is generic to $F(R)$ and scalar-tensor theories. A conformal transformation of the metric maps $F(R)$ and scalar-tensor theory to each other at the action level, therefore this sound speed equivalency is expected. Returning to the action \eqref{eq:deltaS}, the scalar perturbation has positive kinetic terms if $z^{2}(t) > 0$ holds, which is identical to the condition $F_{R}(R)>0$ as shown in \eqref{eq:z(t)}. It is demonstrated that the $F(R)$ models considered in earlier sections satisfy $F_{R}(R) > 0$ and stabilize the scalar perturbation. 

In terms of the Fourier transformed scalar perturbation variable  $\mathcal{R}_{k}(t)=\int d\Tilde{x}e^{-i\Tilde{k}\Tilde{x}}\mathcal{R}(r,\Tilde{x})$, the action equation \eqref{eq:deltaS} can be written as
\begin{equation}
\frac{1}{a(t)z^{2}(t)}\frac{d}{dt}\left(a(t)z^{2}(t)\dot{\mathcal{R}}\right)+\frac{k}{a^{2}(t)}\mathcal{R}_{k}(t)=0,   
\end{equation}
where $k$ is the wave number for the $k$-th perturbation mode. 
Taylor series around $t=0$ for the scale factor $a(t)=(\alpha/\chi +t^{2})^{1/2\chi}$ and keeping up-to quadratic order in cosmic time $(t)$ i.e., neglect the higher order of $t$ in the Taylor expansion of $a(t)$ as one can get the scale factor near the bounce point.
\begin{equation}
a_{b}(t)=\left(\frac{\alpha}{\chi}\right)^{\frac{1}{2\chi}}\left(1+\frac{t^{2}}{2\alpha}\right).    
\end{equation}
As mentioned earlier, near the bounce epoch one can easily get the equation for $F_{R}(R)$ as,
\begin{table}[H]
    \begin{tabular}{cc}
        \textbf{Model I}: & $\frac{dF}{dR}=1-\frac{2 \lambda  R_{0}^3 R}{\left(R_{0}^2+R^2\right)^2},$ \\
     \textbf{Model II}: & $\frac{dF}{dR} =1-\lambda  e^{-\frac{R}{R_{0}}}.$
    \end{tabular}
\end{table}

The model parameters $(R_{0},\lambda)$ are considered as $(2,0.01)$ and $(2.5,0.01)$ for the Starobinsky and exponential gravity models, respectively, and the scale factor parameters $\chi=0.9$ and $\alpha$ have three values, respectively, $1.2, 1.3,$ and $1.4$.  The behavior of $F_{R}$ with respect to cosmic time will be examined to assess the stability of the specified model for the given sets of model parameters. It is noted that the $F_{R} >0$ is near the bounce epoch for both models, implying that the models exhibit stable behavior \cite{Odintsov-2020-37}.
\begin{figure}[H]
\centering
\minipage{0.45\textwidth}
\includegraphics[width=\textwidth]{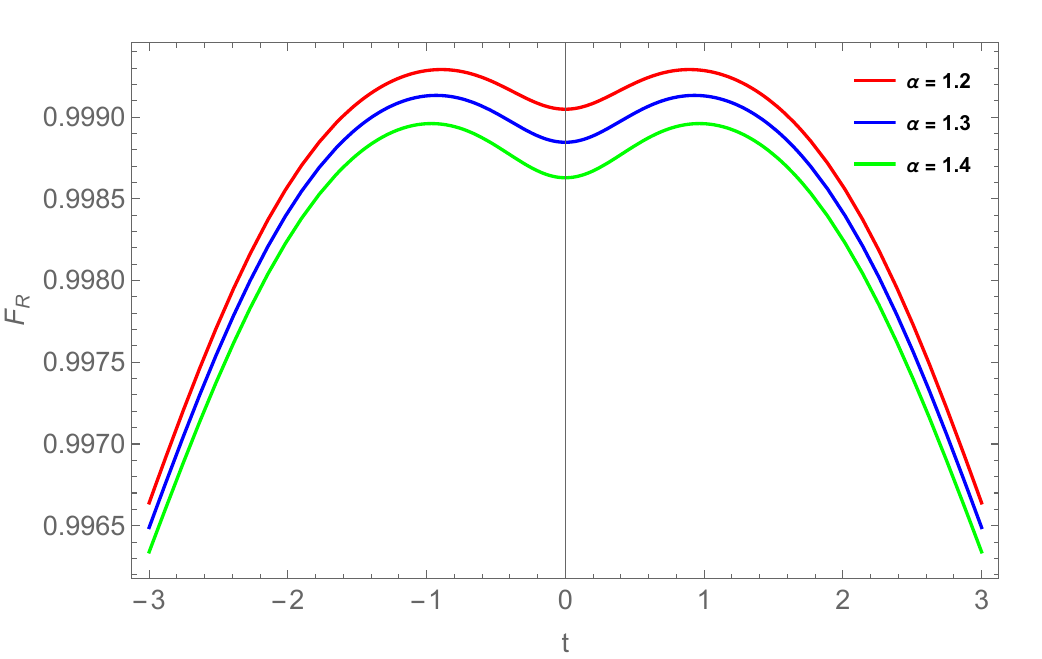}
\endminipage\hfill
\minipage{0.45\textwidth}
\includegraphics[width=\textwidth]{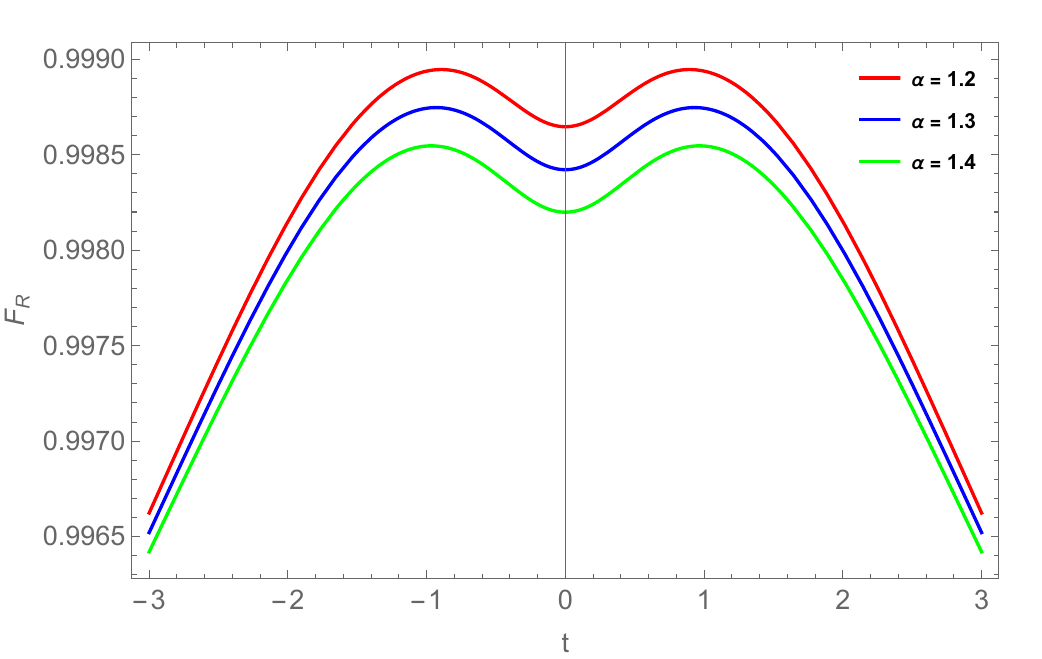}
\endminipage\hfill
\caption[Variation of $F_{R}$ for Starobinsky and exponential model in cosmic time.]{Variations of $F_{R}=\frac{dF}{dR}$ vs. $t$ for Starobinsky model (left panel) and exponential model (right panel) represented for different values of $\alpha$ with $\chi =0.9$.}
\label{ch2_Fig:8}
\end{figure}

\section{Conclusion}
Two cosmological models of the universe, namely the Starobinsky $(n=1)$ \cite{Starobinsky-2007} and the exponential gravity model \cite{Cognola-2008} have been presented in $F(R)$ theory of gravity. The $F(R)$ theory of gravity is derived from an action where the usual Ricci scalar is replaced by a minimally coupled function in $R$. Two well-recognized forms of $F(R)$ function have been considered with a bouncing scale factor. Both model I and model II show the bouncing behavior at $t=0$. Moreover, the values of the model parameters and the scale factor parameter are chosen so that the effective energy density of the models lies in the favorable profile. On the other hand, the effective pressure lies in the negative profile throughout the evolution of the universe. A bounce at the epoch $t=0$ has been observed based on the behavior of the geometrical parameters. The Hubble parameter reaches zero at the bouncing epoch, providing confirmation of the cosmological bounce. Furthermore, for the chosen bouncing scale factor, the value of parameter $\chi=0.9$ makes $n>1/2$ in $(a_{0}t_{F}^{2}+1)^{n}$ with $\alpha=1.2, 1.3$ and $1.4$ value for which the Hubble radius diverges at the bounce point and falls monotonically on both sides of the bounce before asymptotically shrinking to zero, indicating an accelerating late-time universe \cite{Odintsov-2020-959}. Also, such a behavior of the scale factor is required for the compatibility of the given $F(R)$ theory with Planck constraints and generates the required perturbation modes near the bounce. The effective EoS parameter curve crosses two times the phantom-divide line in both models, to support the bouncing behavior. Further, the accelerated expansion of the models was validated through the behavior of effective EoS and deceleration parameters. It can be mentioned here that, the presence of a finite non-zero value of $R_{0}$ throughout the bouncing epoch removes the singularity in the effective EoS parameter.\\

Further, the behavior of the effective EoS parameter is also determined by the scale parameter of the scale factor. The violation of NEC and SEC in both models is shown. These violations are inevitable in the context of the modified theory of gravity and the bouncing scale factor. To note, the phantom phase might develop in the model with a positive Hubble parameter slope due to the violation of null energy requirements. Moreover, the stability of the model has been detected from the behavior of the $F_{R}$ with cosmic time, both the models show stable behavior throughout the evolution. In conclusion, these two models may give some more insight into resolving the initial singularity issue of the early universe.\\


\chapter{Bouncing Cosmological Model in the Framework of \texorpdfstring{$f(R,T)$}{} Gravity} 

\label{Chapter3} 

\lhead{Chapter 3. \emph{Bouncing Cosmological Model in the Framework of \texorpdfstring{$f(R,T)$}{} Gravity}} 

\vspace{10 cm}
* The work, in this chapter, is covered by the following publication: \\

\textbf{A.S. Agrawal} et al., ``Bouncing cosmology in extended gravity and its reconstruction as dark energy model", \textit{Fortschritte der Physik}, \textbf{70}, 202100065 (2022).

\clearpage
       
\section{Introduction}
One of the alternative theories that offer a different perspective and increased efficiency is $f(R,T)$ gravity, which incorporates the Ricci scalar curvature $R$ and the trace of the stress-energy momentum tensor $T$. Based on the theoretical framework, the presence of an arbitrary coupling constant between matter and geometry is proposed as the underlying factor contributing to a source term. This source term is responsible for generating the matter-stress-energy tensor associated with the metric. The selection of a distinct Lagrangian $L_{m}$ would result in a specific set of field equations. In the context of $f(R,T)$ gravity, a cosmological model is developed inside a flat, homogeneous, and isotropic background, adopting a particular formulation of the scale factor. The suggested functional form of the scale factor is selected in a manner that satisfies the conditions for a successful bounce. This approach is applied to determine a solution for the gravitational field equations, under the assumption that the universe does not possess an initial singularity. The functional form of the scale factor, which is thoroughly examined, incorporates numerous limitations on the parameters. Furthermore, investigate the physical and geometrical implications of the model in light of the imposed limitations. Moreover, illustrates the bounce scenario, which is implemented in the model, using specific values of the model parameters. Consequently, it is observed that all the requisite criteria are met for a prosperous bounce model.

\section{\texorpdfstring{$f(R,T)$}{} gravity field equations in FLRW metric} 

The expression $f(R,T)=R+2f(T)$ has been examined in \cite{Harko-2011-84}. A significant number of cosmological models with $f(R)=R$ and $f(T)=\beta_{1} T+ \Lambda_{0}$, where the coupling constant is $\beta_{1}$, are reported in the literature. In this step, the time-independent cosmological constant $\Lambda_{0}$ in $f(T)$, which results in $f(R,T)=R+2\beta_{1} T+2\Lambda_{0}$.

For the flat FLRW space-time \eqref{spacetime:GR}, the field equation \eqref{eq:field:f(R,T)} can now be expressed as,
\begin{equation}\label{eq:field2}
G_{ij}=(8\pi+2\beta_{1})T_{ij}+\Lambda(T)g_{ij},
\end{equation}
where the effective time variable cosmological constant is denoted by $\Lambda(T)=(2p+T)\beta_{1}+\Lambda_{0}$. It should be noted that, with the discovery of supernovae, the cosmological constant $\Lambda$ has gained significance in the investigation of accelerating cosmological models; previously, $\Lambda$ was taken to be zero. However, according to the current extended gravity theory, it changes as the universe expands and manifests as a cosmic time function. Interestingly, with a vanishing $\beta_{1}$, $\Lambda$ simplifies to a pure constant $\Lambda_{0}$. It is now possible to reduce the field equations \eqref{eq:field2} to,

\begin{equation}\label{eq:field3}
G_{ij}=(8\pi+2\beta_{1})T_{ij}+[(2p+T)\beta_{1}+\Lambda_{0}]g_{ij}.
\end{equation}

A non-dissipative perfect fluid distribution throughout the universe is considered, where the matter content energy-momentum tensor is mentioned in equation \eqref{EMT_GR}.

Consequently, the field equations of $f(R,T)$ gravity  \eqref{eq:field2} can be obtained as,
\begin{subequations}
\begin{eqnarray}
2\dot{H}+3H^{2}&=& - \eta p+\beta_{1} \rho +\Lambda_{0},   \label{eq:friedmann:f(R,T)}  \\
3H^{2}&=&\eta \rho- \beta_{1} p+\Lambda_{0},   \label{eq:Raychaudhary:f(R,T)}
\end{eqnarray}    
\end{subequations}
where $\eta=8\pi +3\beta_{1}$ is the equation being used. After performing some algebraic manipulations among the equations \eqref{eq:friedmann:f(R,T)} and \eqref{eq:Raychaudhary:f(R,T)}, one can extract the matter-energy density $\rho$, the matter pressure $p$ and the EoS parameter $\omega=\frac{p}{\rho}$ in terms of the Hubble parameter as,
\begin{subequations}
\begin{eqnarray}
\rho&=& \frac{1}{(\eta^2-\beta_{1}^2)}\left[-2\beta_{1} \dot{H}+3(\eta-\beta_{1})H^2-(\eta-\beta_{1})\Lambda_0 \right], \label{eq.rho1:f(R,T)}\\
p&=& -\frac{1}{(\eta^2-\beta_{1}^2)}\left[2\eta\dot{H}+3(\eta-\beta_{1})H^2-(\eta-\beta_{1})\Lambda_0\right], \label{eq.p1:f(R,T)} \\  
\omega&=&-1+\left[\frac{2(\eta+\beta_{1}) \dot{H}}{2\beta_{1} \dot{H}-3(\eta-\beta_{1})H^2+(\eta-\beta_{1})\Lambda_0}\right].\label{eq.omega1:f(R,T)}
\end{eqnarray}    
\end{subequations}

Expressing the parameters in Hubble terms allows for the investigation of the bouncing behavior of the model. This bouncing behavior is characterized by a scale factor $a(t)$. The universe bounces in a classical way after contracting to a tiny, finite size. In order to ignore the effect of quantum gravity, the energy density should be observed to be lower than the Planck scale. This particular change can take place when the NEC is violated for a finite amount of time, during which the bouncing epoch may also be present. Thus, the model with the bouncing scale factor has been discussed in the last chapter has been considered.  

\subsection{Dynamical parameters}
In the current study, energy density, pressure, and the EoS parameter can be derived for the bouncing scale factor discussed in the last chapter $a(t)=\left(\frac{\alpha}{\chi}+t^{2}\right)^{\frac{1}{2\chi}}$ is under consideration,
\begin{subequations}
\begin{eqnarray}
\rho&=& \frac{1}{(\eta^2-\beta_{1}^2)}\left[\frac{-2\beta_{1}(\alpha-\chi t^2)+3(\eta-\beta_{1})t^2}{(\alpha+\chi t^2)^2}\right]-\frac{\Lambda_0}{(\eta+\beta_{1})}, \label{eq.rho2:f(R,T)} \\
p&=&-\frac{1}{(\eta^2-\beta_{1}^2)}\left[\frac{2\eta(\alpha-\chi t^2)+3(\eta-\beta_{1})t^2}{(\alpha+\chi t^2)^2}\right]+\frac{\Lambda_0}{(\eta+\beta_{1})}, \label{eq.p2:f(R,T)} \\
\omega&=& -1+\left[\frac{2(\eta+\beta_{1})(\alpha-\chi t^2)}{2\beta_{1}(\alpha-\chi t^2)-3(\eta-\beta_{1})t^2+(\eta-\beta_{1})(\alpha+\chi t^2)^2 \Lambda_0}\right]. \label{eq.omega2:f(R,T)}
\end{eqnarray}    
\end{subequations}

\begin{figure}[ht!]
	\begin{center}
		\includegraphics[width=7.5cm]{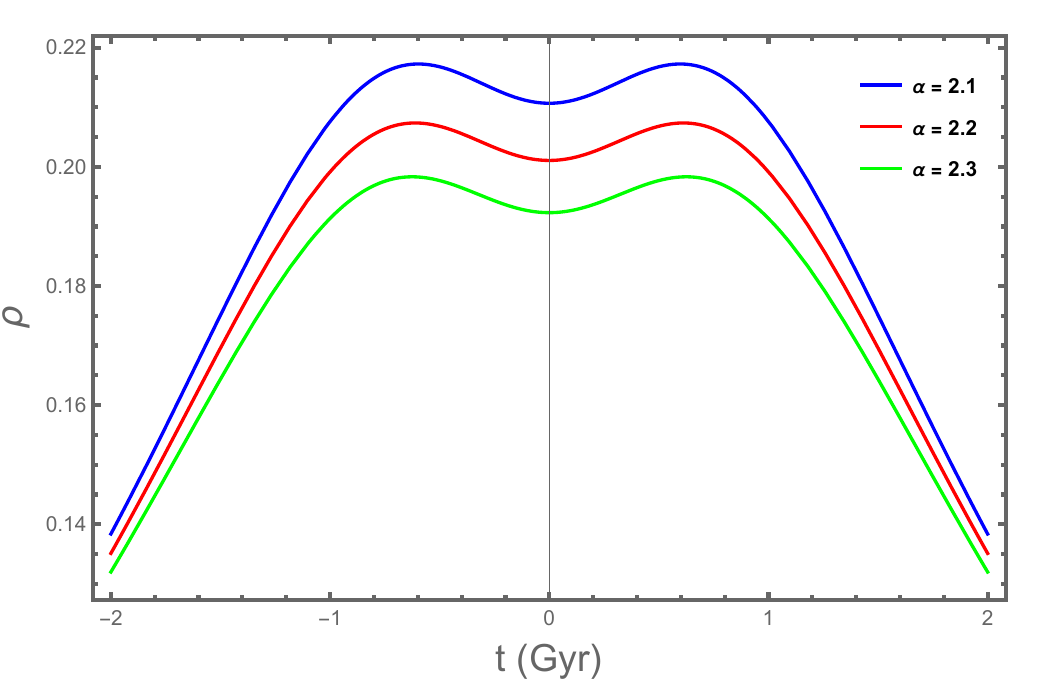}
		\includegraphics[width=7.5cm]{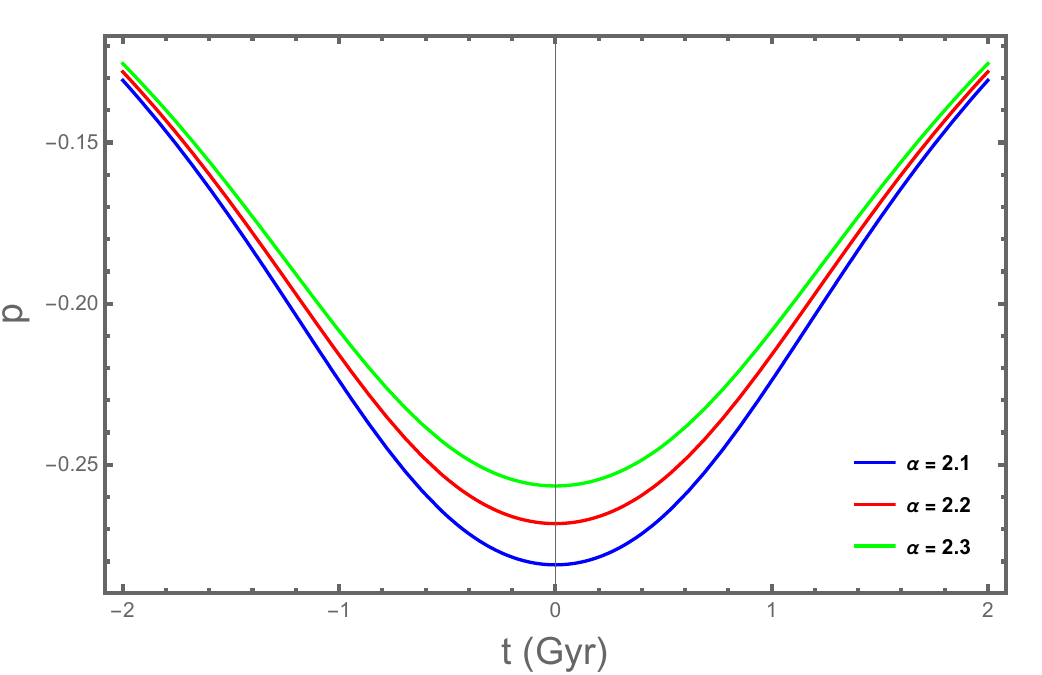}
            \includegraphics[width=7.5cm]{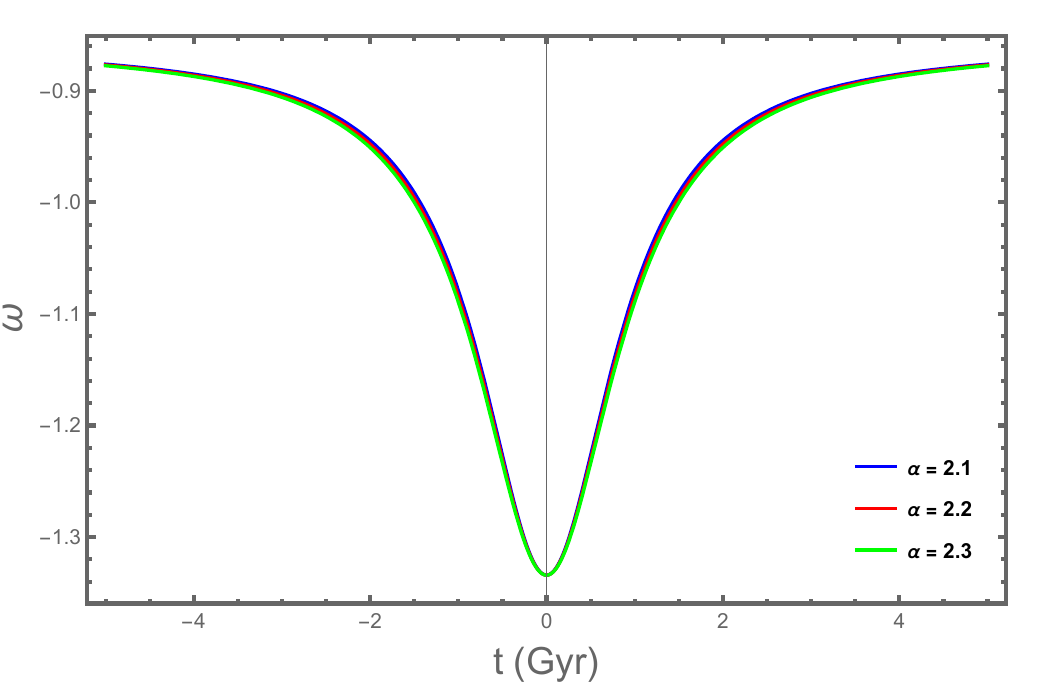} 
	\end{center}
	\caption[Variation of the dynamical parameters with cosmic time.]{Variation of the energy density (upper left panel), pressure (upper right panel), and EoS parameter (lower panel) with cosmic time.}\label{ch3_Fig:2}
 \end{figure}

The graphical representation of the evolutionary behavior of the energy density, matter pressure, and EoS parameter are given in Figure \ref{ch3_Fig:2}. In order to guarantee that the energy density stays positive during the bounce and in both the positive and negative time zones, the constraints on the parameters have been placed. Moreover, a bump is visible close to the bouncing epoch. During the contracting phase, there is a significant increase in energy density as the time approaches the bounce event. Subsequently, the energy density reaches its maximum value before declining at the moment of the bounce. After the bounce, it grows for a short time before decreasing as cosmic time progresses. The energy density curve exhibits a decreasing slope as the parameter $\alpha$ increases. Furthermore, it should be noted that during the bounce, the energy density experiences a decrease, particularly for larger values of $\alpha$. Perhaps this energy density behavior is a necessary component to set off a bounce followed by an expanding phase. For the three representative values of the model parameters $\alpha$, the matter pressure exhibits a well-shaped bounce at $t=0$ and stays in the negative profile all the way through. When the value of the parameter $\alpha$ is increased, there is a corresponding reduction in the depth of the well. It is found that the EoS parameter is well-shaped close to the bounce. There is no observed significant impact on the EoS parameter of the parameter $\alpha$ on the depth of the well. In this study, it was observed that for all selected values of $\alpha$, the curves of the EoS parameter exhibit a significant convergence near the point of bounce. However, in the tail region, both in the negative and positive time domain, the curve of the EoS parameter exhibits a slight divergence for various values of $\alpha$. The EoS parameter evolves from the $\omega<-1$ on both sides of the bouncing epoch. As it moves away from the bouncing epoch, the model crosses the $\omega=-1$ line and stays in the $\omega>-1$. A similar pattern of behavior has been observed in the evolution of the EoS parameter in both the negative and positive time zones. 

\subsection{Energy conditions}
Here, equations \eqref{eq.rho2:f(R,T)} and \eqref{eq.p2:f(R,T)} are used to present the NEC and SEC formulations, in the context of $f(R,T)$ gravity,
\begin{subequations}
\begin{eqnarray}\label{eq:EC:f(R,T)}
\rho+p &=& -\frac{2}{(\eta -\beta_{1})} \left[\frac{\alpha-\chi t^2}{(\alpha+\chi t^2)^2}\right], \label{eq:NEC:f(R,T)} \\
\rho+3p &=& \frac{1}{(\eta^2-\beta_{1}^2)}\left[\frac{(-2\beta_{1}-6\eta)(\alpha-\chi ^2t)-6(\eta-\beta_{1})t^2}{(\alpha+ \chi t^2)^2}\right]+\frac{2\Lambda_0}{(\eta+\beta_{1})}.\label{eq:SEC:f(R,T)}
\end{eqnarray}    
\end{subequations}

\begin{figure}[H]
	\begin{center}
		\includegraphics[width=7.5cm]{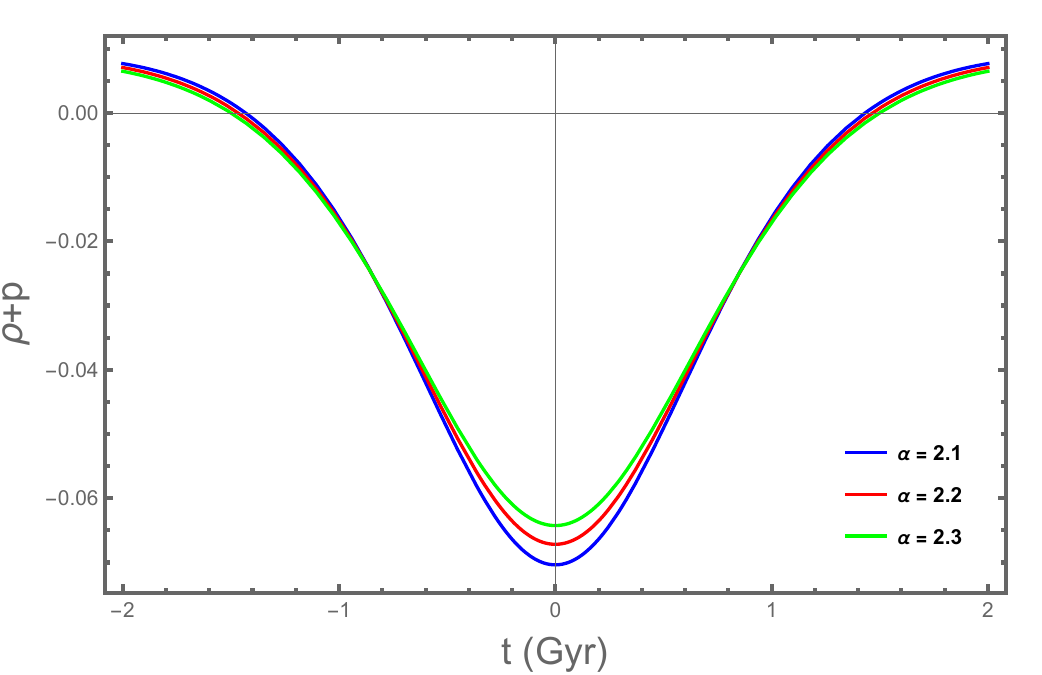}
		\includegraphics[width=7.5cm]{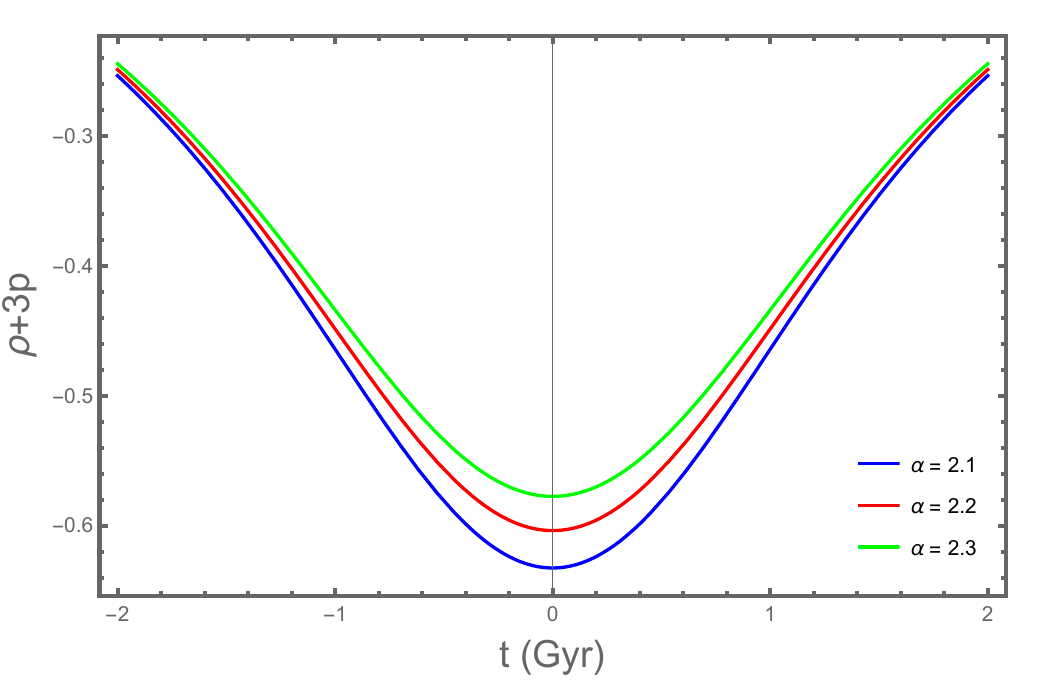}
	\end{center}
	\caption[Variation of the energy conditions with cosmic time.]{Variation of the null energy condition (left panel) and strong energy condition (right panel) with cosmic time.}\label{ch3_Fig:3}
\end{figure}
Figure \ref{ch3_Fig:3} displays the graphical behavior of energy conditions. The bouncing phenomenon necessitates the violation of the NEC during the bouncing phase, as clearly depicted in the left panel of Figure \ref{ch3_Fig:3}. NEC violations result in SEC violations, however, they are still not as significant in the analysis. In order to reconstruct the extended gravity model, the bouncing scale factor is employed inside a simplified extended gravity framework. Consequently, by utilizing this reconstructed model, one can observe the bouncing situation without the need to invoke any dissipative fluid within the matter field. To demonstrate that bouncing scenarios might be conceivable in the extended gravity theory, one can look at the contribution of a perfect fluid in a geometrically modified gravity theory. Furthermore, it has been demonstrated that the model satisfies the bounce conditions, while also ensuring the necessary violation of NEC.


\subsection{Cosmography}
Cosmography is the mathematical framework of the cosmological parameters to describe the universe. Visser \cite{Visser-2004} has extended the idea of cosmography which was originally mentioned by Weinberg \cite{Weinberg-1972}. It depends on the Copernican principle leads to the FLRW metric. We have seen the nature of the Hubble parameter, here the other cosmographic parameters can be represented as, 
\begin{subequations}\label{cosmographic_parameters_model_I_t}
\begin{eqnarray}
q&=&-1-\frac{\alpha}{t^{2}}+\chi \label{eq:q:f(R,T)} \\
\mathrm{j}&=&\frac{(2 \chi -1) \left[t^2 (\chi -1)-3 \alpha \right]}{t^2}, \label{eq:j:f(R,T)} \\
s&=&-\frac{(2 \chi -1) \left[3 \alpha ^2+t^4 (\chi -1) (3 \chi -1)+6 \alpha  t^2 (1-3 \chi )\right]}{t^4}. \label{eq:s:f(R,T)}
\end{eqnarray}    
\end{subequations}
\begin{figure}[H]
	\begin{center}
		\includegraphics[width=7.5cm]{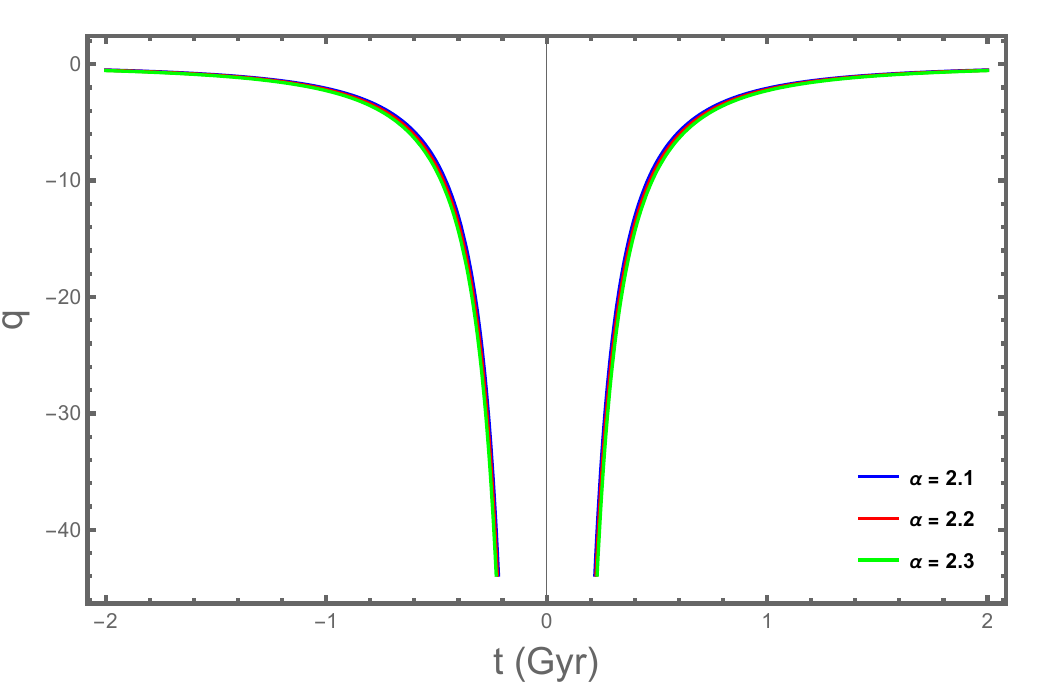}
		\includegraphics[width=7.5cm]{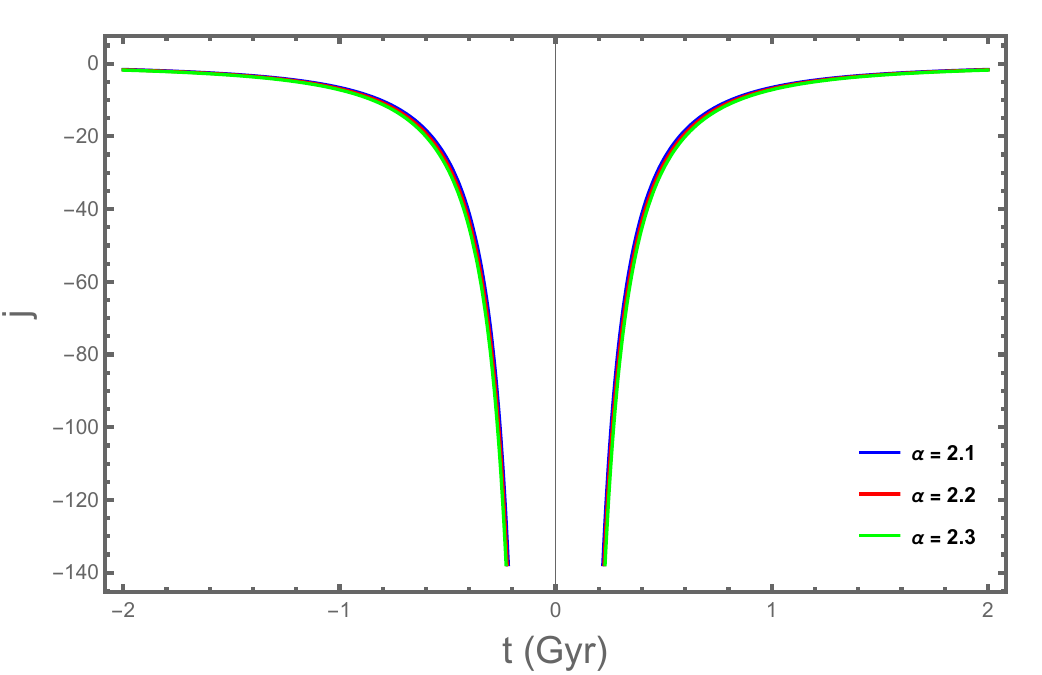}
		\includegraphics[width=7.5cm]{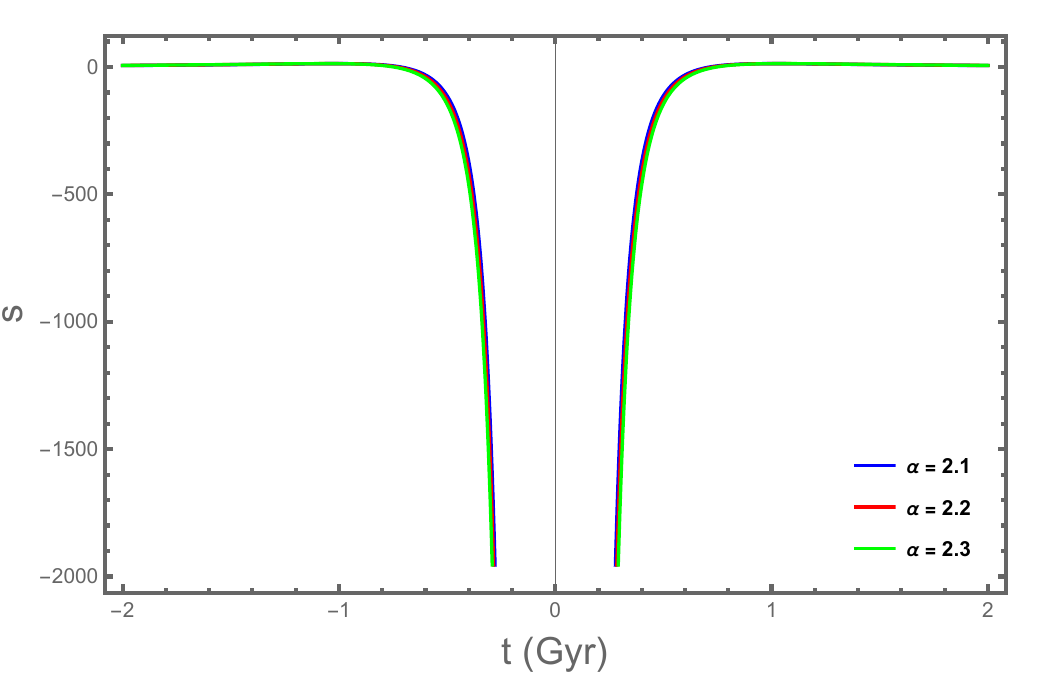}
	\end{center}
	\caption[Variation of the cosmographic parameters with cosmic time.]{Variation of the cosmographic parameters with cosmic time. deceleration parameter (upper left panel), jerk parameter (upper right panel), snap parameter (lower panel).} \label{ch3_Fig:4}
\end{figure}
Figure \ref{ch3_Fig:4} displays the graphical behavior of various cosmographic parameters. The evolution of the EoS parameter from the phantom region and its overall stay in the quintessence zone as we move away from the bouncing epoch have already been seen. Figure \ref{ch3_Fig:4} (top left) illustrates how the deceleration parameter changes with time, starting at a negative value of $-1$ and ending at an asymptotic value of $-1$ at a late time. It also changes during the pre-bounce and post-bounce eras. Singularity occurs in the deceleration parameter at the bouncing epoch. With the fundamental parameters behaving in a way that makes sense, we can now go on to the scale factor that will be used to frame the cosmological model and analyze it. The jerk parameter, which is determined by the third derivative of the scale factor, exhibits a unique behavior during the bouncing phase. Analogous behavior has also been observed for the snap parameter, which encompasses the fourth derivative of the scale factor.

\subsection{Stability of the model}
Furthermore, it is of interest to examine the stability of the aforementioned model in the context of homogeneous and isotropic linear perturbations. A pressureless dust FLRW background has been analyzed, for which $H(t)=H_{b}(t)$ could be a possible general solution. In accordance with the findings of Dombriz et al. \cite{Dombriz-2012-29}, the perturbations in the Hubble parameter and the energy density have been examined around the arbitrary solutions $H_{b}(t)$.
\begin{eqnarray}
H(t) &=& H_b\left(1+\delta (t)\right),\\
\rho (t) &=& \rho_b\left(1+\delta_m (t)\right).
\end{eqnarray}
The symbols $\delta_m(t)$ and $\delta(t)$ represent the deviations from the background energy density and the Hubble parameter, respectively. In the current framework of extended gravity theory, the considered model for this formalism is $f(R,T)=R+2\beta_{1} T+2\Lambda_0$. This function can be expanded in terms of $R_b$ and $T_b$ using a power series,
\begin{equation}
f(R,T)=f_b+(R-R_b)+2\beta_{1}(T-T_b)+\mathcal{O}^2,
\end{equation}
where the term $\mathcal{O}^2$ includes all the higher powers of $R$ and $T$. 

Using the perturbative approach in the equivalent FLRW equation, one can obtain
\begin{equation}
6H_b^2\delta(t)=\eta \rho_b\delta_m(t).\label{eq:22}
\end{equation}
It demonstrates how geometry and matter perturbations are related algebraically. This brings us to the 
\begin{equation}
\rho_b=\frac{3H_b^2-\Lambda_0}{\eta+\beta_{1}}.
\end{equation}
From the conservation equation, one may obtain 
\begin{equation}
\dot{\delta}_m(t)+3H_b(t)\delta(t)=0.\label{eq:24}
\end{equation}

From equations \eqref{eq:22} and \eqref{eq:24}, the evolution equation for the perturbation in the Hubble parameter may be obtained as
\begin{equation}
\dot{\delta}(t)+\frac{\eta\rho_b}{2H_b}\delta(t)=0.\label{eq:25}
\end{equation}
For a vanishingly small value of $\Lambda_0$, the evolution equation \eqref{eq:25} reduces to
\begin{equation}
\dot{\delta}(t)+\frac{3\eta H_b}{2\left(\eta+\beta_{1}\right)}\delta(t)=0.\label{eq:26}
\end{equation}

For a bouncing scenario as prescribed in section \ref{sect-iii}, equation \eqref{eq:26} may be integrated to obtain the geometrical perturbation as
\begin{equation}
\delta(t)=C_k\left(\alpha+\chi t^2\right)^{-\frac{3\eta}{4\chi\left(\eta+\beta_{1}\right)}},\label{eq:27}
\end{equation}
where $C_k$ is an integration constant.

The matter perturbation may be obtained in a straightforward manner from equation \eqref{eq:22} as
\begin{equation}
\delta_m(t)=\frac{2C_k\left(\eta+\beta_{1} \right)}{\eta}\left(\alpha+\chi t^2\right)^{-\frac{3\eta}{4\chi\left(\eta+\beta_{1}\right)}}.\label{eq:28}
\end{equation}

\begin{figure}[ht!]
	\begin{center}
		\includegraphics[width=7.5cm]{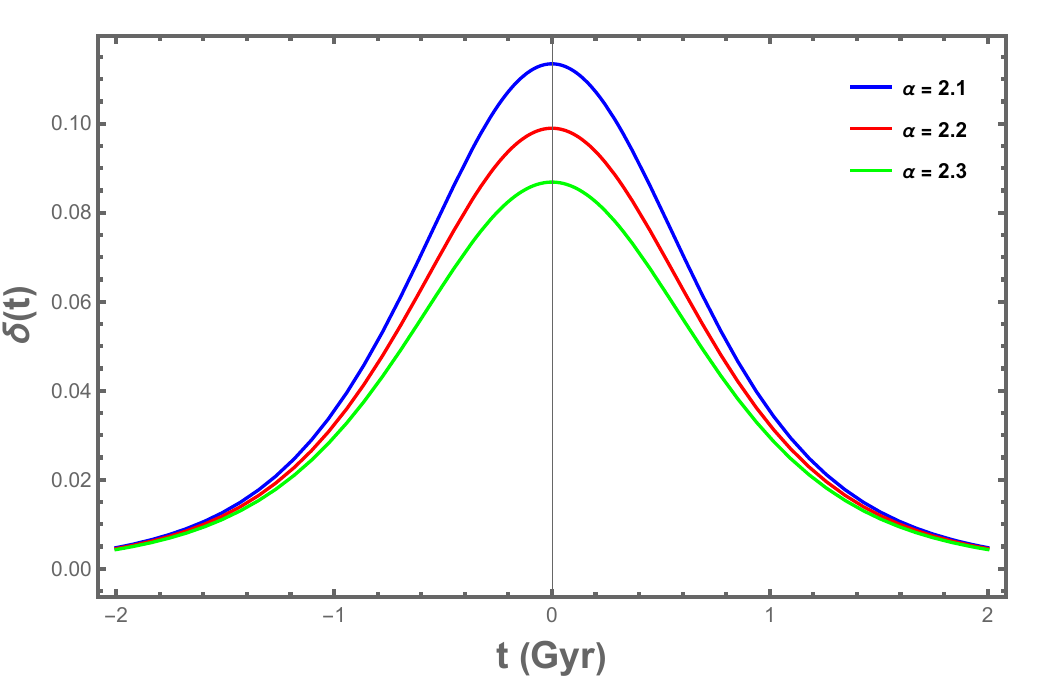}
		\includegraphics[width=7.5cm]{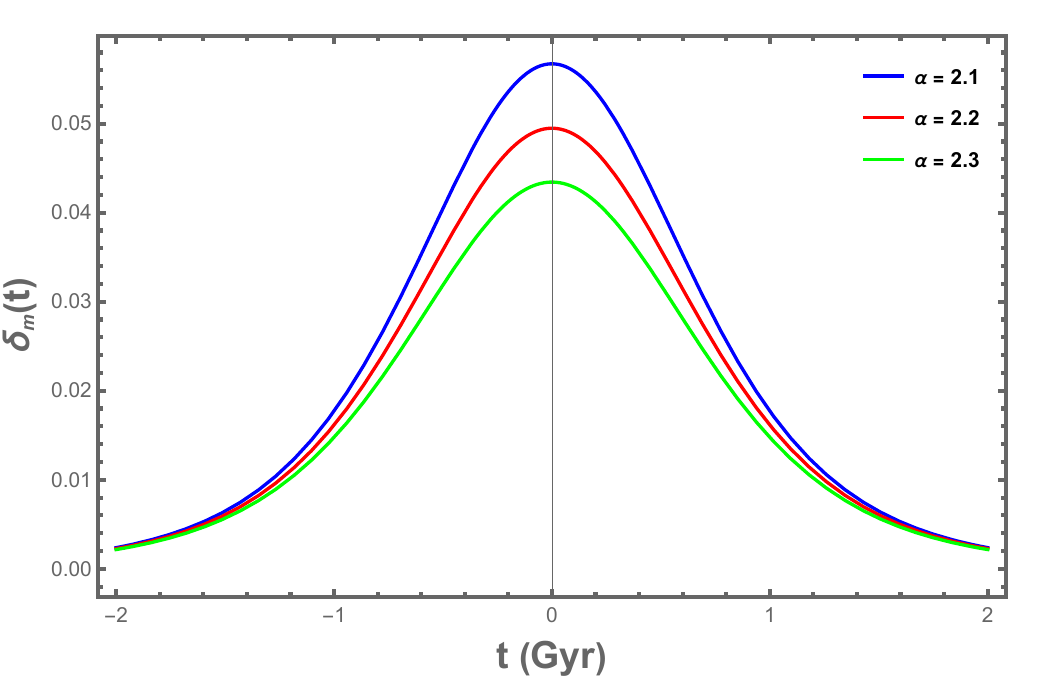}
	\end{center}
	\caption[Variation of the deviations with cosmic time.]{Evolution of the deviation in the Hubble parameter (left panel) and energy density (right panel) with cosmic time. }\label{ch3_Fig:6}
\end{figure}
Based on the preceding discussion, it is concluded that geometric and matter perturbations $\delta(t)$ and $\delta_m(t)$ behave as $a^{-\frac{\eta}{2(\eta+\beta_{1})}}$. In this study, the selected parameters are $\eta$ and $\eta+\beta_{1}$ to be positive values. As a result, the linear perturbations in the Hubble parameter and the energy density decrease symmetrically as time progresses away from the bouncing epoch. In Figure \ref{ch3_Fig:6}, the behavior of these linear homogeneous perturbations for three different values of the model parameters has been illustrated. The stability of the model under a bouncing scenario is clearly represented by the behavior of the perturbations as they gradually decrease over time.
\section{Conclusion}
In this chapter, the bouncing cosmological model of the universe in $f(R,T)$ gravity, where the matter field is represented by a perfect fluid, has been explored. The energy density and pressure behaviors, entirely within the positive and negative regions, support the bouncing behavior. During the bounce, the value of $\omega$ is less than $-1$. However, as it evolves, it transitions through the $\omega=-1$ line and eventually enters the $\omega>-1$. The behavior of the deceleration parameter at the bounce exhibits singularity, similar to other cosmographic parameters like the jerk parameter and snap parameter. An exciting feature of the bouncing model is that the jerk and snap parameters reach a singularity at their negative profile. Another criterion for the bouncing scenario is the violation of NEC. The violation of NEC within the range where the bounce occurs has been successfully identified. From the perturbation approach, the observed linear perturbations gradually decay over time, which contributes to the stability of the presented model. 

Finally, a non-singular bouncing scenario within an extended gravity theory is stable in the pre-and post-bouncing epochs. A bouncing cosmology often consists of a contracting phase, a bounce, and a hot expansion phase. The contracting phase was thought to be the way to thicken and smooth out the cosmic background and make almost scale-invariant density fluctuations larger than the Hubble radius. These fluctuations would be the seeds of structure in the universe after the bounce. The ekpyrotic contraction phase, which is characterized by slow contraction, can be viewed as a cosmological phase that smoothes out and maintains homogeneity, isotropy, and flatness throughout the universe.  However, the smoothness issue has not been taken into account in our model. Within the formalism in this work, it is still possible to think about a time of slow contraction to deal with these kinds of problems.

\chapter{Bouncing Cosmology in Non-Metricity-Based Extended Gravity} 

\label{Chapter4} 

\lhead{Chapter 4. \emph{Bouncing Cosmology in Non-Metricity-Based Extended Gravity}} 

\vspace{10 cm}
* The work, in this chapter, is covered by the following publication: \\

\textbf{A.S. Agrawal}, B. Mishra and P.K. Agrawal, ``Matter bounce scenario in extended symmetric teleparallel gravity", \textit{The European Physical Journal C}, \textbf{83}, 113 (2023).

 \clearpage
 
\section{Introduction}
$f(R)$ and $f(R,T)$ are examples of curvature-based gravitational theories that were covered in the preceding chapter. We shall talk about the bouncing scenario of the universe in non-metricity-based gravity. In fact, GR can be described in two geometrical representations: (i) curvature representation, where the torsion and non-metricity are zero; and (ii) the teleparallel representation, where the curvature and non-metricity vanish. Another possible representation is the symmetric teleparallel gravity, where the geometric variable can be represented by the non-metricity $Q$. The symmetric teleparallel gravity has been further developed into the $f(Q)$ gravity. We had a conversation about the bouncing solution of the universe in various modified gravity. The fact that bouncing cosmology can be obtained as a cosmic solution of LQC is another fascinating aspect of the topic \label{Chap6intro}\cite{Ashtekar-2006-74, Sami-2006-74, Copeland-2008-77, Corichi-2009-80, Bojowald-2009-26, Ashtekar-2011-28}. Within the non-singular bouncing models, there has been much discussion of the matter bounce scenario. This is due to the fact that the evolution of the universe, even in recent times, is comparable to an epoch dominated by matter. Moreover, the matter bounce scenario gives rise to a primordial power spectrum that exhibits nearly scale-invariant characteristics \cite{Odintsov-2020-959, Odintsov-2021}. Additionally, it results in a phase of expansion dominated by matter in its later stages \cite{Cai-2009-80, Quintin-2014-90, Haro-2015-47}. According to this hypothesis, primordial space-time perturbations are generated deep inside the comoving Hubble radius during a period in the contracting age with large negative time, from which the universe emerged. The comoving Hubble radius, denoted as $r_{h}$, exhibits a continuous increase over time and ultimately approaches an infinite value in the far future. As a result, the deceleration stage occurred during the late expansion phase. In the majority of bouncing models those are grounded in modified theories of gravity, the comoving Hubble radius exhibits an increase in magnitude as cosmic time progresses. It will be difficult to convey the presence of the dark energy epoch, but it will be possible to experience the decelerating age of the universe in the far future. It is important to highlight that the matter bounce scenario is subject to a significant defect. The Belinski-Khalatnikov-Lifshitz (BKL) instability arises \cite{Belinskii-1970-19}, wherein the anisotropic energy density of the space-time increases at a quicker rate than that of the bouncing agent during the contracting phase. Consequently, the background evolution exhibited instability. Additionally, the perturbation evolution revealed a substantial tensor-to-scalar ratio, indicating that the amplitudes of the scalar and tensor perturbations were comparable.

The extended symmetric teleparallel gravity, referred to as $f(Q)$ gravity, is a contemporary geometrically modified theory of gravity that has been formulated utilizing the non-metricity approach. In this context, the symbol $Q$ represents the non-metricity. On the other hand, the matter bounce scenario in this gravitational theory has not been thoroughly investigated. We will look at the matter bounce scenario in this chapter, which is based on LQC. Specifically, we will investigate the functional form of $f(Q)$ within the context of FLRW space-time.

The non-metricity term in the e-folding parameter has been expressed considering a dust-fluid-dominated background cosmology. In light of the invalidity of the slow roll criteria in the bouncing context, a conformal equivalence is employed between the function $f(Q)$ and the scalar-tensor model. This enabled us to utilize the bottom-up reconstruction technique in the context of the bouncing model. Phase space analysis has been used to examine the dynamics of the model in order to analyze the various states of the universe, yielding both stable and unstable nodes. Furthermore, an examination of the stability of the model has been conducted by considering the first-order scalar perturbation of the Hubble parameter and matter-energy density.

\section{\texorpdfstring{$f(Q)$}{} gravity field equations} \label{Formalism_f(Q)}
In this discussion, one can examine the properties of the homogeneous and isotropic FLRW space-time denoted as \eqref{spacetime}. When the lapse function is expressed in standard form, denoted as $N(t)=1$, the quantity $Q$ can be determined as $Q=6H^2$. Due to the application of the diffeomorphism in establishing the coincident gauge, the selection of any lapse function is prohibited. The energy-momentum tensor corresponds to the ideal fluid distribution as described by equation \eqref{EMT_GR}.

Now, the field equations of $f(Q)$ gravity can be obtained as, 
\begin{subequations}\label{eq:p_rho_f(Q)}
\begin{eqnarray} 
6f_{Q}H^{2}-\frac{1}{2}f&=&\rho,  \label{eq:Friedmann_f(Q)} \\ 
\left(12H^{2}f_{QQ}+f_{Q}\right)\dot{H}&=&-\frac{1}{2}\left(\rho +p\right). \label{eq:Raychoudhuri_f(Q)}
\end{eqnarray}    
\end{subequations}
Because the nonmetricity variable $Q$ is directly linked to the Hubble parameter, reconstructing $f(Q)$ is much easier than reconstructing $f(R)$, which is based on curvature. In this study, the objective is to investigate the matter bounce situation within the framework of $f(Q)$ gravity. 

\section{\texorpdfstring{$f(Q)$}{} gravity in matter bounce scenario} \label{III}
Obtaining a cosmological model with a bouncing scenario in geometrically modified theories of gravity is a significant challenge. Hence, in the majority of scenarios, the models of bouncing phenomena are reconstructed using principles derived from gravitational theory. In this context, we will proceed to reconstruct a model within the framework of nonmetricity-based gravitational theory. Essentially, the goal would be to rebuild a model whose Hubble squared parameter would be,  
\begin{eqnarray}\label{eq:Hubble_square}
H^{2}=\frac{ \rho_{m}(\rho_{c} - \rho_{m})}{3\rho_{c}}.
\end{eqnarray}
With the matter-energy density being equal to
\begin{eqnarray} \label{eq:energy_density_matter_1}
\rho_{m} =\frac{{\rho_{c}}}{\left(\frac{3}{4} {\rho_{c}} t^2+1\right)}.
\end{eqnarray}  
In the context of the matter bounce scenario under zero pressure conditions, the continuity equation and the energy density can be expressed as follows:
\begin{eqnarray} \label{eq:energy_density_matter_2}
\dot{\rho}_{m}=-3H\rho_{m} \hspace{1cm} \text{and}  \hspace{1cm} \rho_{m}=\rho_{m0}a^{-3}. 
\end{eqnarray}  

It should be noted that the aforementioned equation can be derived from the holonomy-corrected Friedmann equations within the framework of Loop Quantum Cosmology (LQC) for a universe dominated by matter \cite{Haro-2014}. The matter-energy density and critical energy density are denoted as $\rho_{m}$ and $\rho_{c}$, correspondingly. Moreover, the critical energy density,
\begin{equation} \label{eq:Critical_density}
\rho_{c}=(c^{2}\sqrt{3})/(32\pi^{2}\gamma^{3}G_{N}l_{p}^{2}), 
\end{equation}
$\gamma=0.2375$ and $l_{p}=\sqrt{\hbar G_{N}/c^{3}}$ denote the Barbero-Immirzi parameter and the Planck length, respectively. In the subsequent section, the utilization of Planck units, denoted as $c=\hbar=G_{N}=1$, will be employed. It may be seen from equation \eqref{eq:Hubble_square} that a bounce occurs when the matter-energy density approaches its critical value, $H^{2}=0$. 

By solving equations \eqref{eq:Hubble_square} and \eqref{eq:energy_density_matter_2}, while considering factor \eqref{eq:energy_density_matter_1}, the subsequent answers for the scale factor $a(t)$ and the Hubble parameter $H(t)$ inside the matter bounce scenario are obtained as,
\begin{eqnarray} \label{eq:rho_m_Hubble_SF}
H(t)=\frac{2\rho_{c}t}{3\rho_{c}t^{2}+4},  \hspace{1.5cm} a(t)=\left({\frac{3}{4}\rho_{c} t^{2}+1}\right)^{\frac{1}{3}}. 
\end{eqnarray}
Now for the considered matter energy density \eqref{eq:energy_density_matter_2}, the Hubble squared parameter becomes 
\begin{equation}\label{eq:Hubble_square2}
H^{2}=\frac{\rho_{c}}{3}\left(\frac{1}{a^{3}}-\frac{1}{a^{6}}\right).
\end{equation}
The relation between the e-folding parameter and the scale factor can be defined as, $e^{-N}=\frac{a_{0}}{a}$, $a_0$ be the present value of the scale factor. Applying this in equation \eqref{eq:Hubble_square2},
\begin{equation}\label{eq:Hubble_square3}
H^{2}=\frac{\rho_{c}}{3a_{0}^{3}}\left(e^{-3N}-\frac{e^{-6N}}{a_{0}^{3}}\right).
\end{equation}
The parameters $A$ and $b$ can be considered as
\begin{equation}\label{eq:parameters}
A=\frac{\rho_{c}}{3a_{0}^{3}}, \hspace{2cm} b=\frac{1}{a_{0}^{3}}.
\end{equation}
Now, from \eqref{eq:Hubble_square3},  one can easily write the nonmetricity scalar in the form of an e-folding parameter as,
\begin{equation}\label{eq:Q_N}
Q=6A\left[e^{-3N}-be^{-6N}\right].
\end{equation}
On solving, the value of an e-folding parameter can be obtained as,
\begin{equation}\label{eq:N_Q}
N=-\frac{1}{3}Log\left(\frac{3A+\sqrt{9A^{2}-6AbQ}}{6Ab}\right).
\end{equation}
In addition, the energy density \eqref{eq:Friedmann_f(Q)} is of form,
\begin{equation}
\rho=\sum_{i}\rho_{i0}a_{0}^{-3(1+\omega_{i})}e^{-3N(1+\omega_{i})}.\end{equation}\label{eq:rho_N}
By setting $S_{i}=\rho_{i0}a_{0}^{-3(1+\omega_{i})}$, the energy density becomes
\begin{equation}\label{eq:rho_Q}
\rho=\sum_{i}S_{i}\left(\frac{3A+\sqrt{9A^{2}-6AbQ}}{6Ab}\right)^{(1+\omega_{i})}.
\end{equation}
Substituting equation \eqref{eq:rho_Q} in \eqref{eq:Friedmann_f(Q)}, 
\begin{equation}\label{eq:field_equation2}
Qf_{Q}-\frac{1}{2}f-\sum_{i}S_{i}\left(\frac{3A+\sqrt{9A^{2}-6AbQ}}{6Ab}\right)^{(1+\omega_{i})}=0.
\end{equation}
The pressure term becomes zero when we assume that the universe is just filled with dust fluid, which suggests that the EoS parameter disappears. By utilizing equation \eqref{eq:parameters}, the matter-energy density at the initial condition $\omega_{i}=0$ can be readily determined,
\begin{equation}\label{eq:fieldeqn2}
Qf_{Q}-\frac{1}{2}f-\left(\frac{\rho_{c}+\sqrt{\rho_{c}(\rho_{c}-2Q)}}{2}\right)=0.
\end{equation}
On solving,
\begin{equation}\label{eq:f(Q)_model}
f(Q)=-\sqrt{\rho_{c} (\rho_{c}-2 Q)}-\sqrt{2\rho_{c} Q} \arcsin{\left(\frac{\sqrt{2} \sqrt{Q}}{\sqrt{\rho_{c}}}\right)}-\rho_{c}.
\end{equation}
The aforementioned expression of $f(Q)$ yields the evolutionary trajectory of the universe known as the matter bounce. The generation of primordial perturbation modes occurs during the deep contracting era, at a significant distance from the bounce. This is a time when all perturbation modes are contained within the horizon, mostly due to the expansion of the Hubble radius. Nevertheless, in this particular scenario, it can be observed that the cosmic Hubble radius exhibits a divergence as we go further from the bounce epoch \cite{Odintsov-2020-959}. 

\begin{figure}[H]
\centering
\minipage{0.45\textwidth}
\includegraphics[width=\textwidth]{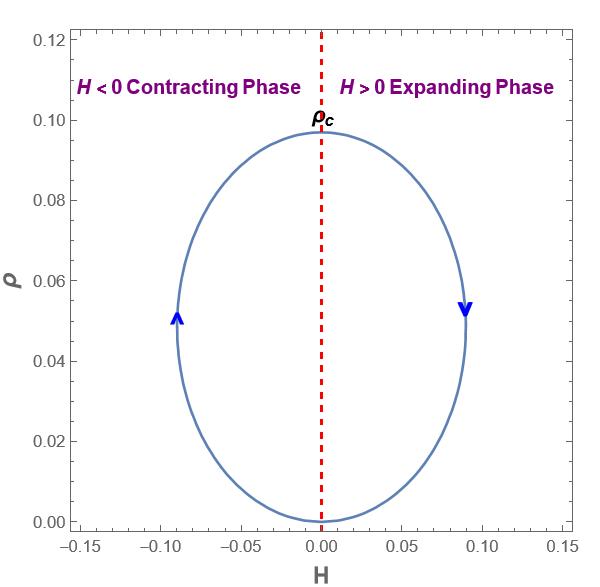}
\endminipage\hfill
\minipage{0.45\textwidth}
\includegraphics[width=\textwidth]{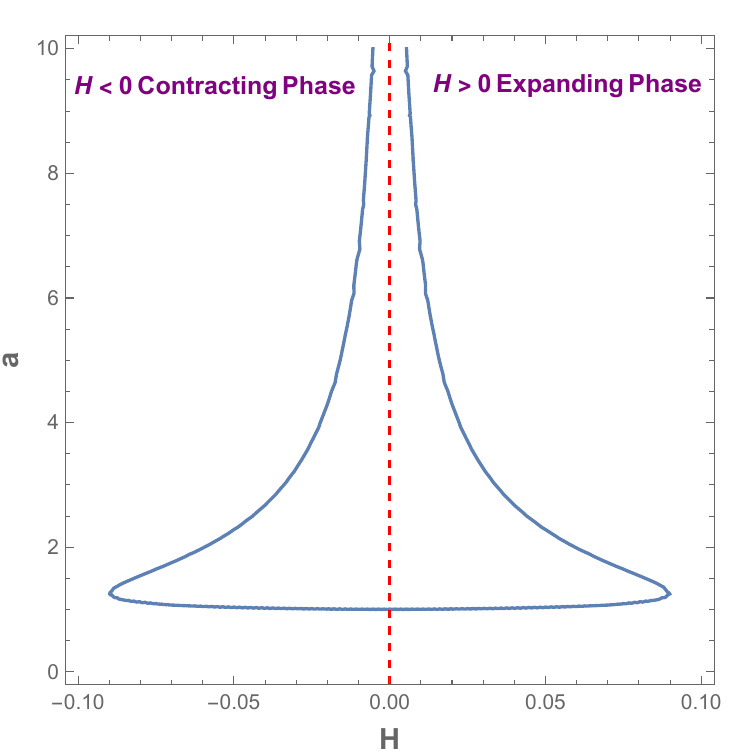}
\endminipage
\caption[Variation of energy density and scale factor in Hubble parameter.]{Evolutionary behavior of energy density (left panel) and scale factor (right panel) in Hubble parameter.}
\label{Ch6_FIG.2}
\end{figure}

The left panel of Figure \ref{Ch6_FIG.2} illustrates the evolutionary behavior of energy density in relation to the Hubble parameter. Similarly, the right panel of Figure \ref{Ch6_FIG.2} demonstrates the relationship between the scale factor and the Hubble parameter. The elliptic curve can be observed in the Figure \ref{Ch6_FIG.2} (left panel). The right portion of the figure depicts the expanding universe, where the Hubble parameter maintains a positive value. On the contrary, the left portion depicts the contracting phase of the universe. Furthermore, it has been observed that the Hubble parameter reaches a value of zero (shown by a red dotted line) at the smallest and greatest values of the energy density, specifically at $0$ and $\rho_{c}$, respectively \cite{Amoros-2013-87}. By examining equation \eqref{eq:energy_density_matter_2}, it is evident that the universe undergoes a clockwise trajectory along the elliptical path, transitioning from a contracting phase to an expanding phase. The fact that energy density increases during the contracting phase and then decreases during the expansion phase can explain this pattern. When the scale factor reaches its smallest value, the Hubble parameter becomes zero. The scale experiences a fall in value during the contraction phase of the universe, then bounces back to its lowest point. Afterward, during the expansion phase of the universe, the scale undergoes an increase in value.

\section{Conformal transformation}\label{IV}
The bottom-up reconstruction technique was inspired by \cite{Odintsov-2018-388}, which tested $F(R)$ gravity in a slow roll inflationary scenario. Later, when this technique was used in a bounce scenario in $F(R)$ gravity \cite{Odintsov-2020-959}, it was observed that due to the fact that the slow roll conditions do not hold true in general, there is no such general expression of the tensor-to-scalar ratio in a $F(R)$ bouncing model, unlike in the inflationary case where the slow roll conditions are true and the ratio has a general expression like $r=48\epsilon^{2}$ irrespective of the form of $F(R)$ \cite{Odintsov-2018-388}. More specifically, the slow roll parameters can be used to define observable quantities like the spectral index and tensor-to-scalar ratio in $F(R)$ inflationary scenarios, but not in bouncing scenarios. The bottom-up reconstruction technique proposed in \cite{Odintsov-2018-388} is difficult to implement in a present bouncing scenario since the slow roll requirements are not valid in a bounce model.

The slow roll conditions are the only reason the bottom-up strategy works in an inflationary scenario but not in a bouncing model. The conformal correspondence of a bounce model with an inflationary one when the slow roll conditions are valid may be the next attempt to apply the bottom-up reconstruction process in a bouncing model. The objective is to get a conformal transformation of the space-time metric that can transfer an $f(Q)$ theory to a scalar-tensor theory where the scalar field potential depends on $f(Q)$. The scale factor and proper time of one frame are related to the other due to a conformal connection.

Let us take the scalar field $\varphi$ as an independent variable and consider the action functional \cite{Jarv-2018-97},
\begin{equation}\label{eq:action_eqn_CT_f(Q)}
S=\frac{1}{2}\int d^{4}x\sqrt{-g}\left[Q-\mathcal{B}(\varphi)g^{ij}\partial_{i}\varphi \partial_{j}\varphi -2\mathcal{V}(\varphi)+2\lambda_{k}^{~jil}R^{k}_{~jil}+2\lambda_{k}^{~ij}T^{k}_{~ij}\right].
\end{equation}
The conditions under which the theory simplifies to the symmetric teleparallel equivalent to general relativity (STEGR) are denoted as $\mathcal{B}(\varphi)=\mathcal{V}(\varphi)=0$. The Lagrange multipliers are assumed to adhere to the antisymmetry of the interconnected geometric entities. Hence, the imposition of vanishing curvature $R^{k}_{~jil}=0$ and torsion $T^{k}_{~ij}=0$ is anticipated within the context of symmetric teleparallelism, as dictated by equations $\lambda_{k}^{~jil}=\lambda_{k}^{~j [il]}$ and $\lambda_{k}^{~ij}=\lambda_{k}^{~[ij]}$.
   
The aforementioned action can be represented as a scalar-tensor (ST) action through the application of a mapping that corresponds to the transformation of the metric 
\begin{equation}\label{eq:CT_f(Q)}
\tilde{g}_{ij}=e^{-\sqrt{2/3}\Omega(\varphi)}g_{ij}.
\end{equation} 
In the $f(Q)$ and scalar-tensor frames, the non-metricity scalar is denoted as $Q$ and $\tilde{Q},$ respectively. The relation between $Q$ and $\tilde{Q}$ can be written as,
\begin{equation}\label{eq:Q_CT_f(Q)}
Q=e^{-\sqrt{2/3}\Omega(\varphi)}\left[\tilde{Q}-\left(\frac{d\Omega}{d\varphi}\right)^{2}\tilde{g}^{ij} \partial_{i}\varphi \partial_{j}\varphi -e^{\sqrt{2/3}\Omega(\varphi)}\frac{d\Omega}{d\varphi}(Q^{i}-\tilde{Q}^{i})\partial_{i}\varphi \right].
\end{equation}
The action mentioned above \eqref{eq:action_eqn_CT_f(Q)} for the scalar-tensor frame takes the following form,
\begin{equation}\label{eq:act_CT-f(Q)2}
S=\frac{1}{2}\int d^{4}x\sqrt{-\tilde{g}}\left[\tilde{Q}-\mathcal{B}\tilde{g}^{ij}\partial_{i}\varphi \partial_{j}\varphi -2\mathcal{V}(\varphi)\right].
\end{equation}
As a consequence of $Q^{i}=Q^{ij}_{~~j}$ and $\tilde{Q}^{i}\equiv Q_{j}^{~ji}$, respectively. The initial action does not include the component $(Q^{i}-\tilde{Q}^{i})\partial_{i}\Omega$, but, the additional component corresponding to $\left(\frac{d\Omega}{d\varphi}\right)^{2}\tilde{g}^{ij} \partial_{i}\varphi \partial_{j}\varphi$ can be incorporated into the redefinition of the kinetic term, $\mathcal{B}(\varphi)=\left(\frac{d\Omega}{d\varphi}\right)^{2}$, of the scalar field.

In the context of perturbation theory, it is advantageous to express the metric in terms of conformal time $\eta$ rather than cosmic time $t$, where the relationship between the two is given by $dt = a d\eta$. The model space-time of $f(Q)$ is described by a FLRW metric, where $\eta$ represents the conformal time and $a(\eta)$ denotes the scale factor, 
\begin{equation}\label{eq:spacetime_STframe1}
ds^{2}=a^{2}(\eta)[-d\eta^{2}+\delta_{ij}dx^{i}dx^{j}].
\end{equation} 
Subsequently, the metric of the scalar-tensor model undergoes modifications,
\begin{eqnarray}
d\tilde{s}^{2}&=&e^{-\sqrt{2/3}\Omega(\varphi)}\left[ a^{2}(\eta)(-d\eta^{2} +\delta_{ij}dx^{i}dx^{j}) \right],\nonumber \\
&=& a_{s}^{2}(\eta)[-d\eta^{2}+\delta_{ij}dx^{i}dx^{j}].\label{eq:spacetime_STframe2}
\end{eqnarray}
In the context of the scalar-tensor model, the symbol $a_{s}(\eta)=e^{-\sqrt{1/6}\Omega(\varphi)}a(\eta)$ represents the scale factor. In both frames, the conformal time remains constant, whereas the cosmic time experiences a transformation according to formula $dt_{s}=e^{-\sqrt{1/6}\Omega(\varphi)}dt$ \cite{Odintsov-2017, Nandi-2020-809}. Notations will be used in this chapter: cosmic time, and scale factor in the $f(Q)$ frame and in the scalar-tensor frame are represented by the symbols ($t,a(t)$) and ($t_{s}, a_{s}(t_{s})$) respectively. In the $f(Q)$ frame, the Hubble parameter will be denoted as $H$, while in the scalar-tensor frame, it will be denoted as $H_{s}$. Additionally, in this chapter, the suffix $s$ will indicate that the parameter is used to describe the scalar-tensor frame.
 
One can get the field equations in the scalar-tensor frame,
\begin{subequations}
\begin{eqnarray}
3H_{s}^{2}&=&\frac{1}{2}\mathcal{B}\left(\frac{d\varphi}{dt_{s}}\right)^{2}+\mathcal{V}, \label{eq:Friedmann_ST}\\
2\frac{dH_{s}}{dt_{s}}+3H_{s}^{2}&=&-\frac{1}{2}\mathcal{B}\left(\frac{d\varphi}{dt_{s}}\right)^{2}+\mathcal{V}. \label{eq:Raychaudhary_ST}
\end{eqnarray}    
\end{subequations}
Using the scalar field equation,
\begin{equation}\label{eq:continuity}
\mathcal{B}\frac{d^{2}\varphi}{dt_{s}^{2}}+3H_{s}\mathcal{B}\frac{d\varphi}{dt_{s}}+\frac{1}{2}\frac{d\mathcal{B}}{dt_{s}}\frac{d\varphi}{dt_{s}}+\frac{d\mathcal{V}}{d\varphi}=0,
\end{equation}
where $\frac{d}{dt_{s}}=\frac{1}{a_{s}(\eta)}\frac{d}{d\eta}$.

From the preceding equations, one can readily obtain $2\frac{dH_{s}}{dt_{s}}=-\mathcal{B}\left(\frac{d\varphi}{dt_{s}}\right)^{2}$. In the scalar-tensor model, the slow roll conditions continue to be applicable due to the presence of an inflationary scenario. The slow roll condition occurs when certain characteristics, which are believed to be less than unity, are inputted during an inflationary period.
\begin{equation}\label{eq:slow_roll_parameter}
\epsilon_{1}=-\frac{1}{H_{s}^{2}}\frac{dH_{s}}{dt_{s}}, \hspace{1cm} \epsilon_{2}=\frac{1}{H_{s}}\frac{d^{2}\varphi/dt_{s}^{2}}{d\varphi/dt_{s}}, \hspace{1cm} \epsilon_{4}=\frac{1}{2\mathcal{B}H_{s}}\frac{d\mathcal{B}}{dt_{s}}.
\end{equation}  

There is another slow roll parameter defined as $\epsilon_{3}=\frac{1}{2H_{s}\mathcal{G}_{\tilde{Q}}}\frac{d\mathcal{G}_{\tilde{Q}}}{dt_{s}}, ~\mathcal{G}_{\tilde{Q}}=\partial \mathcal{G}/\partial \tilde{Q}$ in more general actions like $S=\frac{1}{2}\int d^{4}x\sqrt{-\tilde{g}}\left[\mathcal{G}(\tilde{Q},\varphi)-\mathcal{B}\tilde{g}^{ij}\partial_{i}\varphi \partial_{j}\varphi -2\mathcal{V}(\varphi)\right]$ (where $\mathcal{G}(\tilde{Q},\varphi)$ is any analytic function of $\tilde{Q}$ and $\varphi$), but in the current situation, i.e., for action eq. \eqref{eq:act_CT-f(Q)2} $\mathcal{G}(\tilde{Q},\varphi)=\tilde{Q}$, and thus the slow roll parameter $\epsilon_{3}$ vanishes \cite{Nojiri-2017-692, Hwang-2005-71}. With the condition $\epsilon_{i}\ll 1$, the spectral index ($n_{s}$) for curvature perturbation and the tensor-to-scalar ratio ($r$) of the scalar-tensor model is given by
\begin{subequations}
\begin{eqnarray}
n_{s}&=&1-4\epsilon_{1}-2\epsilon_{2}-2\epsilon_{4}, \label{eq:SI1} \\
r&=&\frac{8\mathcal{B}}{H_{s}^{2}}\left(\frac{d\varphi}{dt_{s}}\right)^{2}. \label{eq:TSR1}
\end{eqnarray}    
\end{subequations}

The gravitational equation $\frac{2dH_{s}}{dt_{s}}=-\mathcal{B}\left(\frac{d\varphi}{dt_{s}}\right)^{2}$ provides the following simplified form of the tensor-to-scalar ratio:
\begin{equation} \label{eq:TSR2}
r=-\frac{16}{H_{s}^{2}}\frac{dH_{s}}{dt_{s}}=16\epsilon_{1}.
\end{equation}
Additionally, the equations of motion can be approximated due to the slow roll conditions.
\begin{equation}\label{eq:Hubble_potential}
3H_{s}^{2}\simeq \mathcal{V}(\varphi),
\end{equation}
and 
\begin{equation}\label{eq:continuity2}
\frac{1}{2}\frac{d\mathcal{B}}{d\varphi}\left(\frac{d\varphi}{dt_{s}}\right)^{2}+3H_{s}\mathcal{B}\frac{d\varphi}{dt_{s}}+\frac{d\mathcal{V}}{d\varphi}=0.
\end{equation}   
After establishing the context, let us now examine a proposed tensor-to-scalar ratio expressed in terms of the e-folding number,
\begin{equation}\label{eq:TSR3}
r(N_{s})=16e^{\alpha_{1}(N_{s}-N_{f})}.
\end{equation} 
In the given context, $\alpha_{1}$ represents a model parameter that does not have any dimensions, while $N_{s}$ represents the e-folding parameter in the scalar-tensor frame. It is important to note that the e-folding number can be defined as either $N_{s}=\int_{t_{h}}^{t_{s}}{H_{s}dt_{s}}$ or $N_{s}=\int_{t_{s}}^{t_{\text{end}}}{H_{s}dt_{s}}$. In this definition, $t_{h}$ represents the onset point of inflation, while $t_{\text{end}}$ represents the endpoint. In the former situation, $dN_{s}/dt_{s}>0$, the e-folding parameter increases monotonically with the cosmic time $t_{s}$, whereas in the latter instance, $dN_{s}/dt_{s}<0$, the e-folding parameter drops monotonically with the cosmic time $t_{s}$. The most crucial component is to see if the $r(N_{s})$ decision results in observable conformity with the Planck restrictions. Using the relation $\frac{d}{dt_{s}}=H_{s}\frac{d}{dN_{s}}$, on comparing equations \eqref{eq:TSR2} and \eqref{eq:TSR3}. 
\begin{equation}\label{eq:Hubble_ST_N}
\frac{1}{H_{s}}\frac{dH_{s}}{dN_{s}}=-e^{\alpha_{1}(N_{s}-N_{f})}.  
\end{equation} 
The Hubble parameter, in the form of an e-folding parameter, can be written as 
\begin{equation}\label{eq:Hubble_ST_N2}
H_{s}(N_{s})=H_{s0}~Exp\left(-\frac{1}{\alpha_{1}}e^{\alpha_{1}(N_{s}-N_{f})}\right).
\end{equation}
It is noteworthy to mention here that the ansatz that is considered in equation \eqref{eq:TSR2} allows an inflationary scenario of the universe having an exit at $N_{s}=N_{f}$ ie., at $t_{s}=t_{\text{end}}$. On the other hand, near the beginning of the inflation, the Hubble parameter follows a quasi-de-sitter evolution.

Using the relation $\frac{dN_{s}}{dt_{s}}=H_{s}(N_{s})$ the conformal time can be defined as follow,
\begin{equation}\label{eq:conformal_time_N}
\eta(N_{s})=-\frac{Exp\left(\frac{1}{\alpha_{1}}e^{-\alpha_{1} N_{f}}\right)}{H_{s0}(1-e^{-\alpha_{1} N_{f}})}e^{-(1-e^{-\alpha_{1} N_{f}})N_{s}}.
\end{equation} 
The next step is to find the conformal factor $\Omega(\varphi)$ in such a way that the conformally transformed $f(Q)$ frame scale factor results in a non-singular bounce after the inflationary scenario in the scalar-tensor frame has been verified. The considered form is, 
\begin{equation}\label{eq.44}
\Omega(\varphi(N_{s}))=\sqrt{6}~ln\left[e^{-N_{s}}\left(\frac{3}{4}\rho_{c}\eta^{2}(N_{s})+1\right)^{\frac{1}{3}}\right].
\end{equation}  
It is simple to see that the conformally connected $f(Q)$ frame scale factor exhibits the following behavior because of the aforesaid form of $\Omega(\varphi)$.
\begin{equation}\label{eq.45}
a(\eta)=\left(\frac{3}{4}\rho_{c}\eta^{2}+1\right)^{\frac{1}{3}}.
\end{equation}
It is simple to demonstrate that the scale factor indicated above causes a non-singular bounce at $\eta=0$. Additionally, close to $\eta=0$, the $f(Q)$ frame scale factor can be approximated as $a(\eta)=1+\frac{1}{4}\rho_{c}\eta^{2}$, and as a result, the conformal time is connected to the $f(Q)$ cosmic time by $t=\int a(\eta)d\eta =\eta+\frac{\rho_{c}\eta^{3}}{12}\approx \eta$. Because of this, the scale factor in terms of cosmic time turns out to be $a(t)=\left(\frac{3}{4}\rho_{c}t^{2}+1\right)^{\frac{1}{3}}$ in the $f(Q)$ frame.
 
Using the relation $\mathcal{B}(\varphi)=\left(d\Omega/d\varphi\right)^{2}$ the spectral index can be defined as follows
\begin{equation}\label{eq:SI2}
n_{s}=1+\frac{4}{H_{s}^{2}}\frac{dH_{s}}{dt_{s}}+\frac{2\left(3\frac{dH_{s}}{dt_{s}}\mathcal{B}+3H_{s}\frac{d\mathcal{B}}{dt_{s}}+\frac{1}{2}\frac{d^{2}\mathcal{B}}{dt_{s}^{2}}\right)}{H_{s}\left(3H_{s}\mathcal{B}+\frac{1}{2}\frac{d\mathcal{B}}{dt_{s}}\right)}-\frac{1}{\mathcal{B}H_{s}}\frac{d\mathcal{B}}{dt_{s}}.
\end{equation} 
The scalar spectral index in terms of e-folding number is to be determined and for this reason, we need the following identities;
\begin{equation}\label{eq:SSI_N}
\frac{d}{dt_{s}}=H_{s}\frac{d}{dN_{s}}, \hspace{1cm} \frac{d^{2}}{dt_{s}^{2}}=H_{s}^{2}\frac{d^{2}}{dN_{s}^{2}}+H_{s}\frac{dH_{s}}{dN_{s}}\frac{d}{dN_{s}}.
\end{equation}
 To determine the value of spectral index the relation $\mathcal{B}(\varphi)=\left(d\Omega/d\varphi\right)^{2}$ can be used which provides the right-hand side of the equation \eqref{eq:SI2} as
 
\begin{eqnarray}
n_{s}&=&1-2e^{\alpha_{1} (N_{s}-N_{f})}-2\frac{d^{2}\Omega}{dN_{s}^{2}}\left(\frac{d\Omega}{dN_{s}}\right)^{-1}+\frac{2}{\left(\frac{d^{2}\Omega}{dN_{s}^{2}}+3-e^{\alpha_{1}(N_{s}-N_{f})\frac{d\Omega}{dN_{s}}}\right)}\nonumber \\
&&\times \bigg[-3\frac{d\Omega}{dN_{s}}e^{\alpha_{1}(N_{s}-N_{f})}+\left(\frac{d^{2}\Omega}{dN_{s}^{2}}-\frac{d\Omega}{dN_{s}}e^{\alpha_{1}(N_{s}-N_{f})}\right)\left(6-3e^{\alpha_{1}(N_{s}-N_{f})}+\frac{d^{2}\Omega}{dN_{s}^{2}}\right) \nonumber \\  
&&+\left(\frac{d^{3}\Omega}{dN_{s}^{3}}-\frac{d^{2}\Omega}{dN_{s}^{2}}e^{\alpha_{1}(N_{s}-N_{f})}-\alpha_{1}\frac{d\Omega}{dN_{s}}e^{\alpha_{1}(N_{s}-N_{f})}\right)\bigg].\label{eq:SI_STF2}
\end{eqnarray}
The integral can be performed for the limit $N_{s}\rightarrow{0}$ in equation \eqref{eq:conformal_time_N} i.e., near the horizon crossing time, which is sufficient in the current context because observable quantities such as spectral index and tensor to scalar ratio are eventually determined at the horizon crossing instance. As a result, the conformal factor in terms of the e-folding number takes the following form:
\begin{equation}\label{eq:CF1}
\Omega({N_{s}})=\sqrt{6}\left[-N_{s}+ln\left(\frac{3}{4}\rho_{c}\left(\frac{e^{\frac{1}{\alpha_{1}}e^{-\alpha_{1} N_{f}}-(1-e^{-\alpha_{1} N_{f}})N_{s}}}{H_{s0}(1-e^{-\alpha_{1} N_{f})}}\right)^{2}+1\right)^{\frac{1}{3}}\right].   
\end{equation}
The dimensionless parameter $\alpha_{1}$ determines the tensor-to-scalar ratio in eq. \eqref{eq:TSR2}, and $N_{f}-N(t_{h})=N_{T}$, where $N_{T}$ is the total e-folding of the inflationary epoch and $t_{h}$ is the horizon crossing instance. For $\alpha_{1}>0.092$, the tensor-to-scalar ratio is inside the Planck restrictions for $N_{T}=60$. So, for $\alpha_{1}=0.1$ and $N_{T}=60$, the spectral index derived in eq. \eqref{eq:SI_STF2} is consistent with Planck results. Now from equations  \eqref{eq:TSR2}, \eqref{eq:SI_STF2} and \eqref{eq:CF1} values for scalar spectral index and the tensor to scalar ratio in the scalar-tensor frame are $n_{s}=0.9649\pm 0.0042$ and $r<0.064$ respectively, from the Planck 2018 constraints \cite{Akrami-2020}.
\section{Phase space analysis} \label{V}
\begin{figure}[H]
\centering
\minipage{0.50\textwidth}
\includegraphics[width=\textwidth]{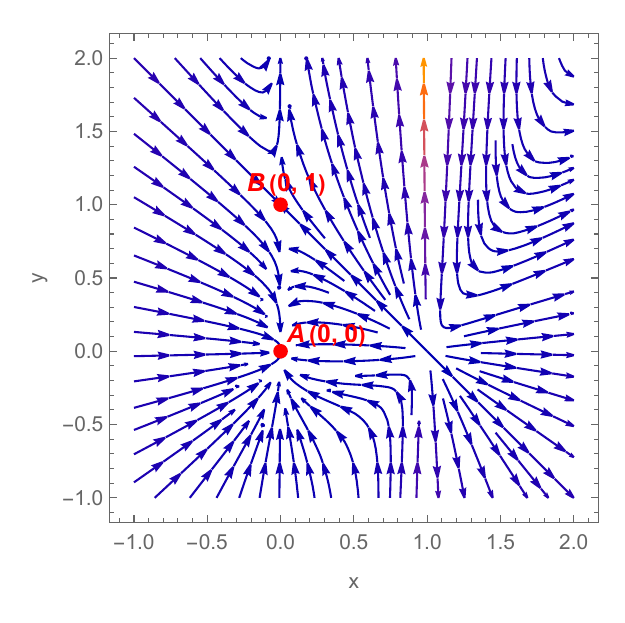}
\endminipage\hfill
\caption[Phase-space trajectories on the $x$-$y$ plane for $f(Q)$ gravity.]{Phase-space trajectories on the $x$-$y$ plane for $f(Q)$ gravity.}\label{Ch6_FIG.3}
\end{figure}

Phase space analysis is a study that involves representing all possible states of a system, with each state being assigned a unique point. This can also be described as the combination of all possible values in position space and momentum space. A phase space analysis of the system described by the function $f(Q)$, which is given by equation \eqref{eq:f(Q)_model} has been demonstrated. Here a general form of $f(Q)$ is presented as $Q+\psi(Q)$ \cite{Khyllep-2021-103, Narawade-2022-36} and accordingly equations \eqref{eq:p_rho_f(Q)} take the form,
\begin{subequations}\label{eq:p_rho_PSA_1}
\begin{eqnarray} 
3H^{2}&=&\rho+\frac{\psi}{2}-Q\psi_{Q}, \label{eq:Friedmann_PSA} \\  2\dot{H}+3H^{2}&=&-p-2\dot{H}(2Q\psi_{QQ}+\psi_{Q})+\left(\frac{\psi}{2}-Q\psi_{Q}\right).    \label{eq:Raychaudhury_PSA}
\end{eqnarray}    
\end{subequations}
When the universe comprises matter and radiation fluids, with the matter and radiation energy density represented respectively as  $\rho_{m}$ and $\rho_{r}$. One can obtain the following relation 
\begin{equation} \label{eq:total_density}
\rho=\rho_{m}+\rho_{r}, \hspace{2cm}  p=\frac{\rho_{r}}{3}.  
\end{equation}
Hence, the relation can written as
\begin{subequations}\label{eq:p_rho_PSA_2}
\begin{eqnarray} 
3H^{2}&=&\rho+\rho_{de},  \label{eq:Friedmann_PSA2} \\ 
2\dot{H}+3H^{2}&=&-p-p_{de}.    \label{eq:Raychaudhury_PSA2}
\end{eqnarray}    
\end{subequations}
Comparing equations \eqref{eq:p_rho_PSA_1} with \eqref{eq:p_rho_PSA_2}, the dark energy density and dark energy pressure contributions caused by the geometry can be separated as,
\begin{subequations}
\begin{eqnarray}
\rho_{de}&=&\frac{\psi}{2}-Q\psi_{Q}, \label{eq:DE_density} \\    
p_{de}+\rho_{de}&=&2\dot{H}(2Q\psi_{QQ}+\psi_{Q}). \label{eq:DE_pressure_density}   
\end{eqnarray}    
\end{subequations}
The density parameters for the matter-dominated, radiation-dominated, and dark energy phases are respectively denoted as $ \Omega_{m}=\frac{\rho_{m}}{3H^{2}}$, $\Omega_{r}=\frac{\rho_{r}}{3H^{2}}$ and $\Omega_{de}=\frac{\rho_{de}}{3H^{2}}$ with $\Omega_{m}+\Omega_{r}+\Omega_{de}=1$. Therefore, the effective EoS parameter can be expressed in the following form,
\begin{eqnarray}
\omega_{\text{eff}}&=&-1+\frac{\Omega_{m}+\frac{4}{3}\Omega_{r}}{2Q\psi_{QQ}+\psi_{Q}+1}.   \label{eq:omega_eff} 
\end{eqnarray}
In order to examine the dynamics of the model, the dimensionless variables, $x=\frac{\psi-2Q\psi_{Q}}{6H^{2}}$ and $ y=\frac{\rho_{r}}{3H^{2}}$ can be examined, which have been turned into an autonomous dynamical system. Moreover, the term \textit{prime} represents the derivative with respect to the number of e-folds of the universe, denoted as $N=\ln a$. Consequently, the equations of the model may be calculated by applying the chain rule.
\begin{equation}
\phi'=\frac{d\phi}{dN}=\frac{d\phi}{dt}\frac{dt}{da}\frac{da}{dN}=\frac{\dot{\phi}}{H}.     \label{eq:prime_defination}
\end{equation}
So, the autonomous dynamical system can be given as, 
\begin{subequations}
\begin{eqnarray}
x'&=&-2\frac{\dot{H}}{H^{2}}(\psi_{Q}+2Q\psi_{QQ}+x), \label{eq:x_prime}\\
y'&=&-2y\left(2+\frac{\dot{H}}{H^{2}}\right), \label{eq:y_prime}
\end{eqnarray}    
\end{subequations}
and with an algebraic manipulation, one can obtain the relation, $
 \frac{\dot{H}}{H^{2}}=-\frac{1}{2}\left(\frac{3-3x+y}{2Q\psi_{QQ}+\psi_{Q}+1}\right)$. If $f(Q)$ compared with \eqref{eq:f(Q)_model}, then $\psi(Q)$ can be represented as,  
\begin{equation}\label{eq:psi(Q)_model}
\psi(Q)=-\sqrt{\rho_{c} (\rho_{c}-2 Q)}-\sqrt{2\rho_{c} Q} \arcsin \left(\frac{\sqrt{2} \sqrt{Q}}{\sqrt{\rho_{c}}}\right)-\rho_{c}-Q,  \end{equation}
and 
\begin{equation}
2Q\psi_{QQ}+\psi_{Q}=-\frac{\rho_{c}}{\sqrt{\rho_{c} (\rho_{c}-2 Q)}}-1.   
\end{equation}
Now the dimensionless variables can be represented as,
\begin{subequations}
\begin{eqnarray}
x'&=&x\left[3(x-1)-y\right], \label{eq:x_prime2} \\
y'&=&-\frac{y [x (-3 x+y+4)+y-1]}{x-1}. \label{eq:y_prime2}
\end{eqnarray}    
\end{subequations}
The system of equations mentioned above possesses critical points in the coordinates (0,0) and (0,1). The stability of these critical points can be assessed by examining their respective eigenvalues. The eigenvalues $\{-3,-1\}$ were obtained for the critical point $(0,0)$, while $\{-4,1\}$ were obtained for the critical point $(0,1)$. Given that the eigenvalues at the critical point $(0,0)$ are both negative, it can be inferred that the system exhibits a stable node.  Conversely, at the point $(0,1)$, the eigenvalues exhibit a combination of positive and negative real components, so indicating instability at this location.  The EoS is denoted as equation \eqref{eq:omega_eff}, and the deceleration parameter can be derived in terms of the dynamical variables.
\begin{subequations}
\begin{eqnarray}
\omega_{\text{eff}}&=&-1+\frac{(x+1) (3 x-y-3)}{3 (x-1)}, \label{eq:omega_eff2} \\
q&=&-1+\frac{(x+1) (3 x-y-3)}{2 (x-1)}.    \label{eq:deceleration}  
\end{eqnarray}    
\end{subequations}
The details of the critical points and their behavior are given in the following phase portrait (Figure \ref{Ch6_FIG.3}) and the corresponding cosmology in Table \ref{TABLE_f(Q)}.

The dynamical system technique allows for the extraction of the cosmological properties of the model without the need to get the precise solution to the evolution equations. Furthermore, an analysis can also be conducted on the gravitational theory and cosmic evolution. The examination of the cosmic dynamics of the model involves the determination of critical points through the solution of the systems involving variables $x$ and $y$. Table \ref{TABLE_f(Q)} displays the composition of the system, which consists of two critical points, one of which exhibits stable behavior. The coordinates of the point $A(0,0)$ correspond to a universe that is dominated by matter. Additionally, the value of $\omega_{\text{eff}}=0$ signifies that the universe is in a phase of matter domination, which is also evident in the phase space portrait. The coordinates of point B, specifically (0,1), are associated with a phase characterized by radiation dominance. This is evident from the effective EoS parameter, denoted as $\omega_{\text{eff}}$, which has a value of 1/3, signifying the era of radiation dominance. Furthermore, it can be observed that the deceleration parameter exhibits positive values at both sites, indicating a decelerating trend. It is important to acknowledge that a significant body of research on bouncing cosmology has consistently indicated that the matter bounce scenario is insufficient in elucidating the late-time dark energy period. Specifically, this scenario fails to yield vital points that serve as indicators of the dark energy era.
\begin{table}[ht]
\caption{Critical points for the dynamical system of the model}
\begin{center}
\begin{tabular}{ |c|c|c|c|c|c|c|c|c } 
 \hline
 Point($x,y$)  & $\Omega_{m}$ & $\Omega_{r}$ &  $\Omega_{de}$ & $\omega_{\text{eff}}$ & Deceleration $(q)$ & Eigenvalues& Stability   \\ \hline
$A(0,0)$ & 1 & 0 & 0 &  0 &  1/2 & \{-3,-1\} & Stable node \\ \hline
$B(0,1)$ & 0 & 1 & 0 & 1/3 &  1 & \{-4,1\} & Unstable \\ \hline
\end{tabular}
\end{center}
\label{TABLE_f(Q)}
\end{table}

\section{Stability analysis with scalar perturbation} \label{VI}
\begin{figure}[H]
\centering
\minipage{0.40\textwidth}
\includegraphics[width=\textwidth]{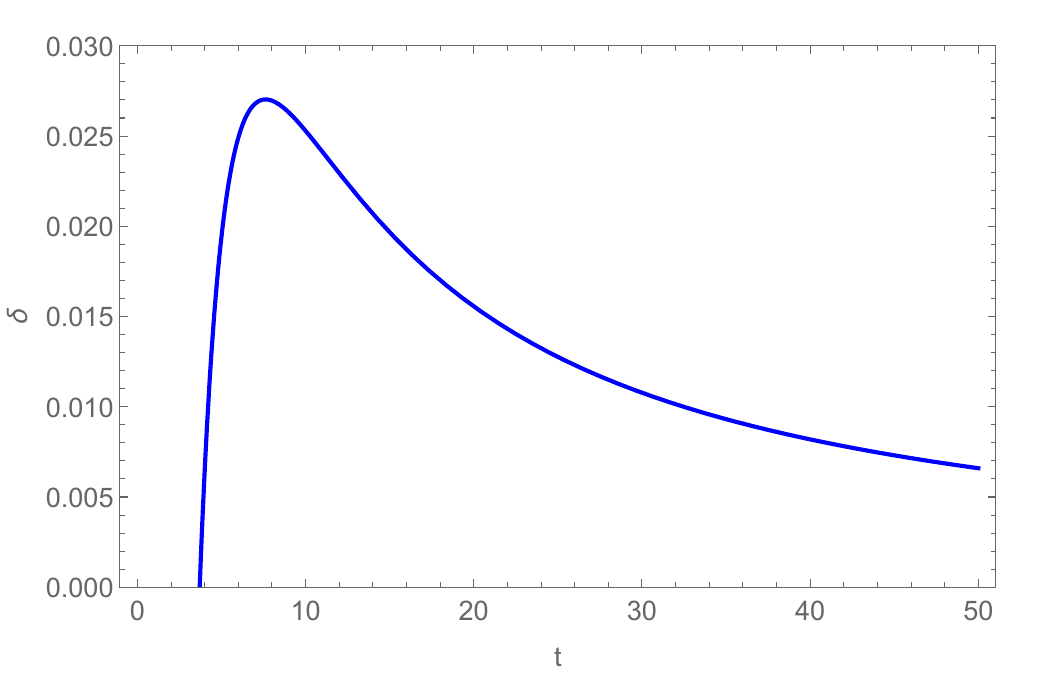}
\endminipage
\minipage{0.40\textwidth}
\includegraphics[width=\textwidth]{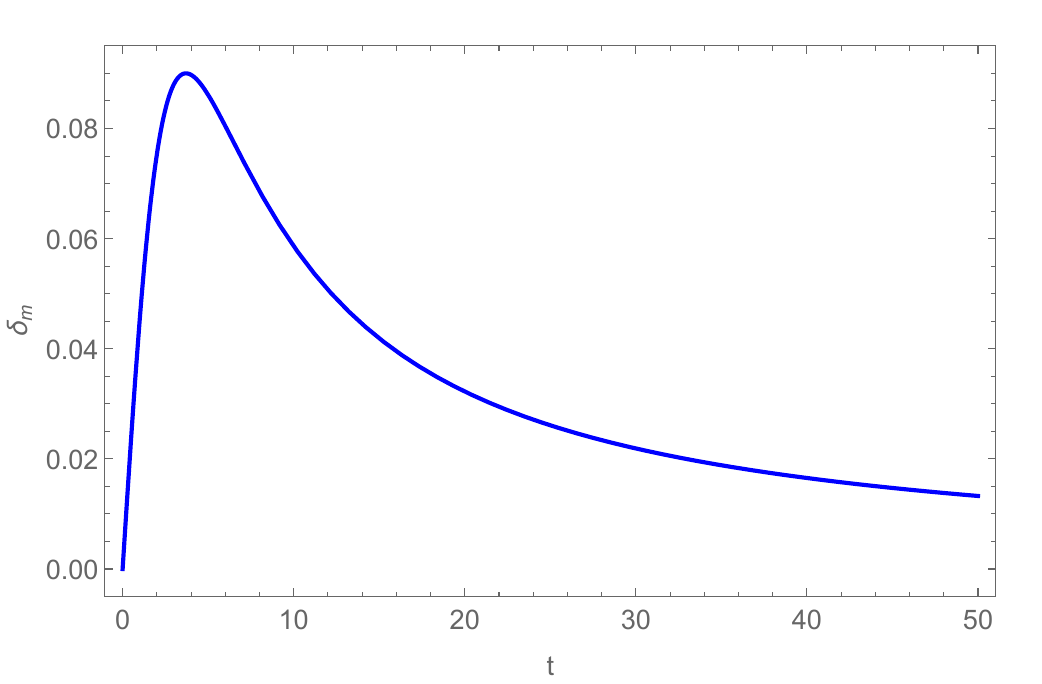}
\endminipage
\caption[Variation of deviations in cosmic time.]{Evolution of deviation of Hubble parameter $\delta$ and the deviation of energy density $\delta_{m}$ in cosmic time for $C_{1}=1$ and $C_{2}=-1/3$ (matter dominated case).}\label{Ch6_FIG.4}
\end{figure}

In this study, the objective is to conduct a scalar perturbation analysis to examine the stability characteristics of the reconstructed model within the framework of $f(Q)$ gravity. In this study, we will stick to the linear homogeneous and isotropic perturbation framework, focusing on the description of perturbations in energy density and the Hubble parameter. The initial perturbation in the FLRW backdrop, characterized by the perturbation geometry functions $\delta(t)$ and matter functions $\delta_m(t)$, can be mathematically represented as follows:
\begin{eqnarray}
H(t)\rightarrow {H_b(t)(1+\delta(t))}, \hspace{2cm} \rho(t)\rightarrow {\rho_b(t)(1+\delta_{m}(t))}. \label{eq:perturbation_Hubble_density}
\end{eqnarray}
Both the functions $\delta(t)$ and $\delta_{m}(t)$ can be interpreted as representing the isotropic deviations of the Hubble parameter and matter over-density, respectively. The calculation of the perturbation of the function $f(Q)$ and $f_Q(Q)$ can be determined as follows:
\begin{eqnarray}
\delta f(Q)=f_{Q}\delta Q, \hspace{2cm} \delta f_{Q}(Q)=f_{QQ}\delta Q. 
\end{eqnarray}
The symbol $\delta Q$ denotes the first-order perturbation of the variable $Q$. Now, if we ignore the higher power of $\delta(t)$, we can calculate the Hubble parameter as follows:
\begin{equation}\label{eq:perurbed_Hubble}
6H^{2}=6H_b^{2}(1+\delta(t))^{2}=6H_b^{2}(1+2\delta(t)),
\end{equation}
and subsequently equation \eqref{eq:Friedmann_f(Q)} can be reduced to
\begin{eqnarray}\label{eq:perurbed_density}
Q(2Qf_{QQ}+f_{Q})\delta =\rho \delta_{m}.
\end{eqnarray}
 The relationship between matter and geometric perturbation, as well as the perturbed Hubble parameter, can be derived from equation \eqref{eq:perturbation_Hubble_density}. In order to derive the analytical solution for the perturbation function, the perturbation continuity equation can be examined as,
\begin{eqnarray}
\dot{\delta}_{m}+3H(1+\omega)\delta =0, \label{eq:perurbed_density2}
\end{eqnarray}
and from equations \eqref{eq:perurbed_density}- \eqref{eq:perurbed_density2}, the following first-order differential equation can be obtained, 
\begin{equation} \label{eq:perurbed_density3}
\dot{\delta}_{m}+\frac{3H(1+\omega)\rho}{Q(2Qf_{QQ}+f_{Q})}\delta_{m}=0.   
\end{equation}
Further using the $tt$-component field equation and equation \eqref{eq:perurbed_density3}, the simplified relation can be obtained,
\begin{equation}\label{eq.73}
\dot{\delta}_{m}-\frac{\dot{H}}{H}\delta_{m}=0 ,   
\end{equation}
which provides $\delta_{m}=C_{1}H$, where $C_{1}$ is the integration constant. Subsequently from equation \eqref{eq:perurbed_density2},
\begin{equation}\label{eq:perturbed_Hubble2}
\delta=C_{2}\frac{\dot{H}}{H},    
\end{equation}
where, $C_{2}=-\frac{C_{1}}{3(1+\omega)}$. The evolution behavior of $\delta$ and $\delta_m$ are given in Figure \ref{Ch6_FIG.4}. 

In the scenario dominated by matter, the pressure term is observed to be zero, indicating that the EoS parameter also assumes a value of zero. The stability of the model has been assessed by examining the scalar perturbation of the Hubble parameter and energy density while considering a value of zero for the EoS parameter. It is evident that in the initial stages of both deviations, $\delta(t)$ and $\delta_{m}(t)$, there is an initial increase followed by a gradual decrease over time, eventually converging to zero in the later stages. Consequently, it can be asserted that while the model initially exhibits transient instability, it predominantly demonstrates stability when subjected to scalar perturbations.
\section{Conclusion}\label{VII}

In the context of the background level during the matter-dominated phase, a particular functional form of $f(Q)$ has been formulated, which exhibits the phenomenon of matter bounce. The determination of the e-folding value can also be derived by a logarithmic function with an expression that incorporates the nonmetricity scalar. In addition, the use of symmetric teleparallel gravity has led to the development of a bottom-up reconstruction technique that has successfully formulated a viable non-singular bounce model. The application of the bottom-up strategy is highly feasible within an inflationary framework, particularly when the observable quantities can be effectively characterized by the slow-roll parameter in general (due to slow-roll conditions). Since the slow roll criterion in the bouncing context is not true, a conformal equivalence is employed between the $f(Q)$ and the scalar-tensor model to apply the bottom-up reconstruction technique in the $f(Q)$ bouncing model. In the context of the scalar-tensor framework, the selection of the conformal factor is made in a manner that leads to the emergence of an inflationary epoch. In contrast, when a suitably analyzed conformal factor is taken into account, the $f(Q)$ frame scale factor exhibits a behavior described by the equation $a(\eta)=\left(3\rho_{c}\eta^{2}/4+1\right)^{1/3}$. This behavior suggests the occurrence of a non-singular bounce at $\eta=0$. The results of the dynamical stability analysis show that, as expected, the current model is unable to account for the dark energy era. The eigenvalues and their corresponding cosmology can be derived from the critical points. Two critical points have been identified, one representing a stable node and the other an unstable point. The positive deceleration parameters indicate that the universe underwent deceleration during its early stages. In order to assess the stability of the reconstructed model, the scalar perturbation approach has been employed. Based on the graphical analysis of the deviation functions $\delta(t)$ and $\delta_{m}(t)$ over cosmic time, it has been observed that both deviations tend to converge towards zero in the late-time regime. Based on our analysis, it can be inferred that the reconstructed bouncing model exhibits a certain degree of instability during the initial phase. However, it demonstrates stability throughout the majority of its evolution.

\chapter{Bouncing Cosmological Models in the Framework of \texorpdfstring{$f(Q,T)$}{} Gravity} 

\label{Chapter5} 

\lhead{Chapter 5. \emph{Bouncing Cosmological Models in the Framework of \texorpdfstring{$f(Q,T)$}{} Gravity}} 
\vspace{10 cm}
* The work, in this chapter, is covered by the following publication: \\

\textbf{A.S. Agrawal} et al., ``Matter Bounce Scenario and the Dynamical Aspects in $f(Q,T)$ Gravity", \textit{Physics of The Dark universe}, \textbf{33}, 100863 (2021).

\clearpage
  
\section{Introduction}
In the previous chapter, the bounce scenario is discussed in $f(Q)$ gravity. Again the $f(Q)$ gravity has been extended by framing the non-minimal coupling between non-metricity $Q$ and trace energy-momentum tensor of the matter $T$, to form $f(Q,T)$ gravity. The non-metricity tensor is a general description of the gravitational interaction. Of late, a lot of attention is being given to the $f(Q)$ and $f(Q,T)$ gravity because of the agreement on the late time cosmic acceleration and other issues of the early universe. In fact, bouncing cosmology can be studied by modifying both the gravitational actions and the matter field \cite{Boisseau-2015}. Since these extended theories of gravity have been successful in resolving the bouncing scenario, we are motivated here to study the bouncing scenario in another geometrically modified gravity, the $f(Q,T)$ gravity. 

The $f(Q,T)$ gravity theory remains applicable even when scalar fields are incorporated into the action, in place of ordinary matter. An additional potential application of the $f(Q,T)$ theory involves examining inflation within the context of scalar fields. This approach has the potential to offer a novel viewpoint on the geometrical, gravitational, and cosmological processes that were significant in the early dynamics of the universe. The primary aim of this chapter is to examine the role of $f(Q,T)$ gravity in offering cosmological models that effectively address the issue of late-time cosmic speed-up and explore its potential contribution to a bounce scenario. Our focus is on the mathematical simplification of extending symmetric teleparallel gravity while accounting for changes in geometry. Also, since other geometrically extended theories have been able to solve the problem of the initial singularity, it might be interesting to see if the combination of trace and non-metricity can do the same.

\section{A brief review of the \texorpdfstring{$f(Q,T)$}{} gravity and the cosmological taxonomy  }

In this discussion, we will now examine the cosmological implications of the $f(Q,T)$ theory. We will make the assumption that the universe may be characterized by a homogeneous, isotropic, and spatially flat FLRW space-time.
\begin{eqnarray}\label{spacetime}
ds^{2}=-N^{2}(t)dt^{2}+a^{2}(t)(dx^{2}+dy^{2}+dz^{2}),
\end{eqnarray}
where $N(t)$ is the lapse function. The dilation rate is defined as $\tilde{T}=\frac{\dot{N}(t)}{N(t)}$. By adopting the coincident gauge, in the covariant derivatives reduced to ordinary derivatives, the non-metricity for the flat FLRW space-time becomes, 
\begin{eqnarray}
 Q &=& 6\frac{H^2}{N^2}.
\end{eqnarray}
In the standard case, we have $N(t)=1$ and the non-metricity becomes, $Q=6H^{2}$. We consider a perfect fluid distribution of the universe and therefore, the energy-momentum tensor is written as, $T^{i}_{j}=\texttt{diag}(-\rho, p, p, p)$. 

Now, the field equations \eqref{field_eq_f(Q,T)} of $f(Q,T)$ gravity \cite{Xu-2019-79} in the standard case for the FLRW space-time can be derived as,
\begin{subequations}\label{Raychoudhari:f(Q,T)_1}
\begin{eqnarray}
\rho&=&\frac{1}{16\pi}\left[f-12F H^2-4\dot{\zeta}\kappa_{1}\right],\label{Raychoudhari:f(Q,T)}\\
p&=&-\frac{1}{16\pi}\left[f-12FH^2-4\dot{\zeta}\right],\label{friedmann:f(Q,T)}
\end{eqnarray}    
\end{subequations}
where $F=\frac{\partial f}{\partial Q}$ and $8\pi \kappa\equiv f_{T}=\frac{\partial f}{\partial T}$, $\kappa_{1}=\frac{\kappa}{1+\kappa}$ and $\zeta=FH$. Adding equations \eqref{Raychoudhari:f(Q,T)} and \eqref{friedmann:f(Q,T)}, the evolution equation for the Hubble function $H$ can be obtained as,
\begin{equation} \label{zetaprime}
\dot{\zeta}=\dot{F}H+F\dot{H}=4\pi(p+\rho)(1+\kappa).
\end{equation}
In comparison to the Friedman equations of Einstein's GR, $\rho_{\text{eff}}$ and $p_{\text{eff}}$ can be characterized as,
\begin{subequations}\label{eff_f(Q,T)}
\begin{eqnarray}
3H^2&=&\frac{1}{F}\left[\frac{f}{4}-4\pi[(1+\kappa)\rho+\kappa p] \right]=8\pi \rho_{\text{eff}}, \label{eff_rho_f(Q,T)}\\
2\dot{H}+3H^2&=&\frac{1}{F}\left[\frac{f}{4}-2\dot{F}H+4\pi[(1+\kappa)\rho +(2+\kappa)p]\right]=-8\pi p_{\text{eff}}.\label{eff_p_f(Q,T)}
\end{eqnarray}    
\end{subequations}
In the $f(Q,T)$ gravity theory, the matter-energy-momentum tensor exhibits non-conservation, wherein the non-conservation vector is dependent on $Q$, $T$, and the thermodynamical properties of the system \cite{Xu-2019-79}. Nonetheless, it has been noted that the conservation equation is satisfied by the effective thermodynamical quantities,
\begin{equation}
\dot{\rho}_{\text{eff}}+3H(\rho_{\text{eff}}+p_{\text{eff}})=0.    
\end{equation}
Considering cosmological applications, three forms for $f(Q,T)$ have been suggested, such as (i) $f(Q,T)=\lambda_1 Q+\lambda_2 T$, (ii) $f(Q,T)=\lambda_1 Q^{m}+\lambda_2 T$ , (iii) $f(Q,T)=-\lambda_1 Q-\lambda_2 T^2$ \cite{Xu-2019-79}. Here $\lambda_1$ and $\lambda_2$ are arbitrary constants. Based on the generality of the forms, we consider here, $f(Q,T)=\lambda_1 Q^{m}+\lambda_2 T$. One can easily get the first model for $m=1$. For this functional, we have $F=\lambda_1 mQ^{m-1}$, $\lambda_2=8\pi\kappa$, $\zeta=\lambda_1mQ^{m-1}H$. Also we have $\dot{F}=2(m-1)F\frac{\dot{H}}{H}$ and $\dot{\zeta}=F\dot{H}(2m-1)$. Now from equations \eqref{Raychoudhari:f(Q,T)_1}, we obtain the energy density and the pressure as, 
\begin{subequations}\label{eq:p_rho_f(Q,T)}
\begin{eqnarray}
\rho &=&   \frac{2\dot{\zeta}[3\kappa-(2+3\kappa)\kappa_{1}]+\lambda_1 (6H^2)^{m} (1-2m)}{4\pi[(2+\kappa)(2+3\kappa)-3\kappa^{2}]}, \label{eq:rho_f(Q,T)}\\
p&=& \frac{2\dot{\zeta}[2+\kappa -\kappa \kappa_{1}]-\lambda_1(6H^{2})^{m}(1-2m)}{4\pi [(2+\kappa)(2+3\kappa)-3\kappa^{2}]}.\label{eq:p_f(Q,T)}
\end{eqnarray}    
\end{subequations}
The EoS parameter $\omega=\frac{p}{\rho}$ can be obtained as,
\begin{equation}
\omega =\frac{p}{\rho } = -1+\frac{4\dot{\zeta}[(1+2\kappa )(1-\kappa_{1})]}{2\dot{\zeta}[3\kappa-(2+3\kappa)\kappa_{1}]+\lambda_1 (6H^2)^{m} (1-2m)}.   \label{eq:omega_f(Q,T)}
\end{equation}
The violation of the SEC is essential for the validity of the extended theory of gravity. Moreover, it has been observed that if the NEC is violated, then all the other pointwise energy conditions will be violated. In some bouncing models or models dominated by phantom fields, the NEC has to be violated. Now, we can express the general form of energy conditions in the context of $f(Q,T)=\lambda_1Q^m+\lambda_2T$ gravity as,
\begin{subequations}\label{EC_f(Q,T)}
\begin{eqnarray}
\rho+p&=& \frac{1}{4\pi}\left[(1-\kappa_1)\dot{\zeta}\right], \label{NEC_f(Q,T)}\\
\rho+3p&=&\frac{1}{16\pi}\left[-2f+24FH^2+4\dot{\zeta}(3-\kappa_1)\right], \nonumber \\
&=& \frac{1}{16 \pi(1+2\kappa)}\left[-2(1-2m)\lambda_1Q^m+2\dot{\zeta}(6+6\kappa-2\kappa_1-6\kappa\kappa_1)\right]. \label{SEC_f(Q,T)}
\end{eqnarray}    
\end{subequations}
In this study, we aim to examine various bouncing models and their implications for the violation of the NEC during cosmic evolution, particularly near the bounce epoch. The evolutionary behavior of the energy conditions will be examined in depth for the corresponding models. We aim to outline the essential prerequisites for the violation of the NEC within the context of $f(Q,T)$ gravity. The equation \eqref{NEC_f(Q,T)} guarantees that if there is a violation of the NEC, specifically when $\rho+p<0$, it will result in either $\kappa_{1}>1$ with $\dot{\zeta}>0$ or $\kappa_{1}<1$ with a  negative value for $\dot{\zeta}$. If the $\dot{\zeta}<0$, it follows that $\kappa=\frac{1}{8 \pi}\frac{\partial f}{\partial T} >-1$. Alternatively, if the second condition is satisfied, we can observe that $\dot{\zeta}$ is negative, resulting in $\dot{H}\frac{\partial f}{\partial Q}+\frac{\partial}{\partial t}\left(\frac{\partial f}{\partial Q}\right)H <0$. At the bouncing epoch, all of the symmetric bounce models meet the requirements $H=0$ and $\dot{H}>0$. Considering this, a violation of NEC with $\kappa_1>0$ necessitates $\frac{\partial f}{\partial Q}<0$, at least during the bouncing era. If $\kappa_{1}>0$ and the model parameter $\lambda_{1}<0$ are met, the necessary bouncing conditions can be met. Therefore, we will examine two different bouncing scale factors in order to analyze the dynamic behavior of the physical parameters. Specifically, we will set the value for $\lambda_{1}$ as $-0.5$.

\section{ Dynamical parameters of some cosmological models favoring bounce scenario}

In accordance with the formalism presented in the preceding section, the objective is to examine various bounce scenarios. These scenarios involve a contracting universe preceding a non-singular bouncing epoch, followed by an expansion dominated by matter. A proposed scenario has been put forth to prevent the occurrence of singularities in the Big Bang models. Given the aforementioned information, we will now proceed to examine two distinct bounce scenarios, the first of which is inspired by the principles of loop quantum gravity in the subsequent subsections. The quantum signature of the model is managed by adjusting specific model parameters. This tuning process aims to obtain viable bouncing models that can also offer a satisfactory explanation for the late-time cosmic speed-up phenomenon.

\subsection{The ansatz for scale factor: model I}
We consider the ansatz for the scale factor as,
\begin{equation}\label{SF1t_f(Q,T)}
a(t)=\left(\frac{\alpha}{\chi}+t^2 \right)^{\frac{1}{2\chi}}.
\end{equation} 
The scale factor parameters $\alpha$ and $\chi$ are selected appropriately to induce a bouncing behavior. The model exhibits a bounce at time $t_b=0$, where the scale factor $a(t_b)$ assumes a non-zero finite value given by $a(t_b)=\left(\frac{\alpha}{\chi}\right)^{\frac{1}{2\chi}}$. Prior to this period of oscillation, the scale factor undergoes symmetrical expansion on both sides of the bounce point, illustrating a phase of early contraction followed by a bounce and afterward expansion. During the transition from a contracting to an expanding stage, the scale factor undergoes fluctuations as a new era emerges, this stage transition leads to a non-singular bounce. The Hubble parameter can be expressed for this scale factor as.

\begin{equation}\label{HP1t_f(Q,T)}
H={t}(\alpha +t^2 \chi )^{-1}, 
\end{equation}
where $a_0$ is the scale factor at the present epoch. At the bounce point, the Hubble parameter shifts from $H(t)<0$ to $H(t)>0$ through $H=0$. For the bouncing scenario mentioned above, Figure \ref{Hubble_f(Q,T)} illustrates the evolution of the Hubble parameter. In order to plot the figure, we have used the parametric values of the constants as $\alpha=0.43$ and $\chi=1.001$. As the Hubble parameter starts at a negative value, passes through zero at $t=0$, and then exhibits positive behavior, it satisfies the requirements set forth for the bouncing cosmology.
\subsection{The ansatz for scale factor: model II}
As a second example, we consider the ansatz,
\begin{equation}\label{SF2t_f(Q,T)}
a(t)=\sqrt[3]{\frac{3 \rho_{c}t^2}{4}+1},
\end{equation}
 
where $\rho_c$ is a constant parameter for the scale factor $a(t)$. The model bounces at the epoch $t_b=0$ and the scale factor at bounce is $a (t_b)=1$. The Hubble parameter for the scale factor becomes, 
\begin{equation}\label{HP2t_f(Q,T)}
H=\frac{\rho_{c}t}{2}\left(\frac{3\rho_{c}t^2} {4}+1\right)^{-1}.
\end{equation}


 
The Hubble parameter is depicted in Figure \ref{Hubble_f(Q,T)} as a function of cosmic time. The parameter $\rho_c$ was selected to have a value of 0.75. The Hubble parameter fulfills the necessary criteria for the framework of bouncing cosmology. The Hubble parameter displays a negative trend during the initial phases of evolution, reaches a point of zero at time $t=0$, and subsequently demonstrates a positive trend during the later stages of evolution. To clarify, the value of $H$ is negative prior to the bounce event, becomes zero at the moment of bounce, and subsequently becomes positive during the post-bounce period.  Furthermore, it can be observed that the derivative of the Hubble parameter, denoted as $\dot{H}$, is greater than zero in the vicinity of the bounce epoch.  
\begin{figure}[H]
\centering
\minipage{0.50\textwidth}
\includegraphics[width=\textwidth]{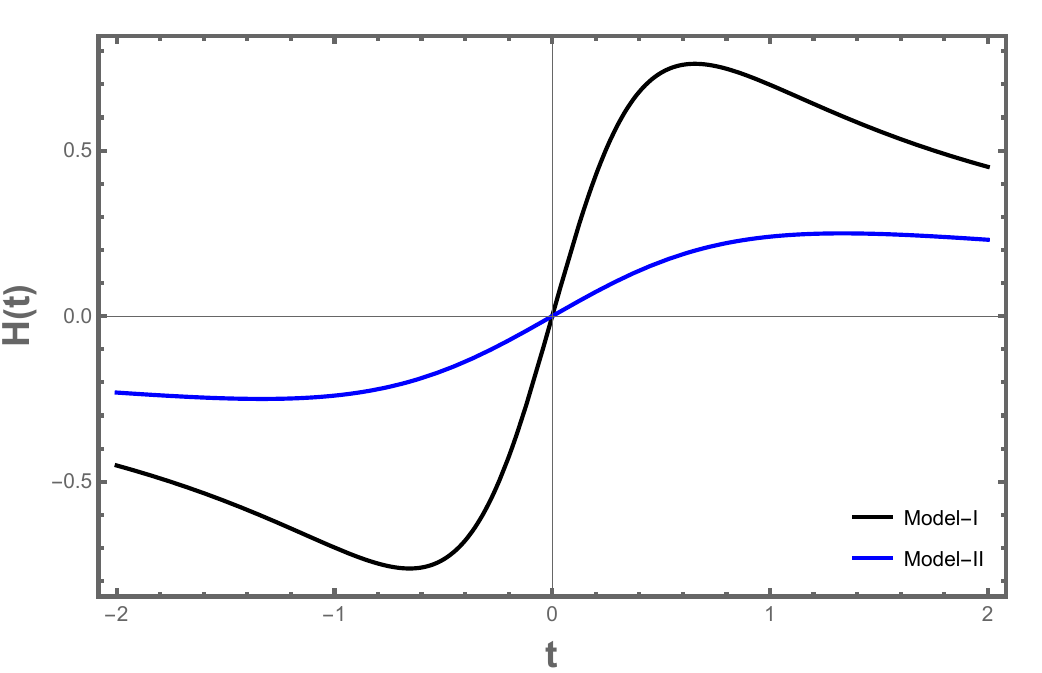}
\endminipage
\caption[Variation of the Hubble parameter in cosmic time for model I and model II.]{Plot for the variation of the Hubble parameter vs. cosmic time for model I with $\alpha=0.43$ and $\chi=1.001$ and for model II with $\rho_{c}=0.75$. ~[Equations  \eqref{HP1t_f(Q,T)} and \eqref{HP2t_f(Q,T)}].}
\label{Hubble_f(Q,T)}
\end{figure}
The existence of interacting dark energy places restrictions on the non-singular bounce model, which was the first model we examined in our study. This dark energy is characterized by a cosmological constant that varies with time \cite{Abdussattar-2011}. However, the second approach is derived from the principles of loop quantum cosmology. The main difference between these two models lies in their ability to change their dominance based on the value of the scale factor parameter. While the second model stays fixed in a matter-dominated state, the first model is flexible enough to change its dominance. Consider the case where $\chi<1$. In this case, the Hubble radius asymptotically approaches 0 and drops constantly on both sides of the bounce. This behavior corresponds to an accelerating late-time universe. Conversely, when $\chi>1$, the Hubble radius diverges asymptotically, indicating a decelerating universe in the later stages, similar results can be seen in \cite{Odintsov-2020-959}. In the study, it is worth considering the value $\chi>1$ as it can provide insights into a singular free universe exhibiting a late acceleration phenomenon. The behavior of the comoving Hubble radius we have discussed in the last chapter, if we consider the parameter value $\chi<1$, the cosmic Hubble radius decreases continuously on both sides of the bounce and eventually approaches zero asymptotically. The variation of the cosmic Hubble radius in Figure \ref{Fig.HR_f(Q,T)} for second ansatz. 

\begin{figure}[H]
\centering
\minipage{0.5\textwidth}
\includegraphics[width=\textwidth]{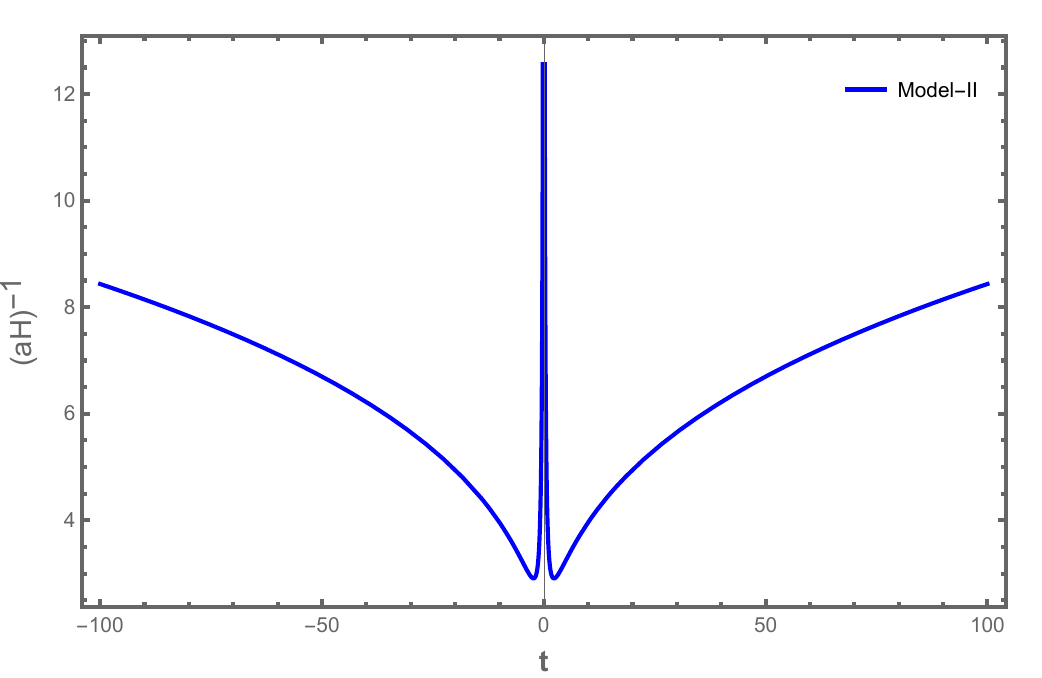}
\endminipage\hfill
\caption{Variation of comoving Hubble radius for model II with cosmic time.}
\label{Fig.HR_f(Q,T)}
\end{figure}
\subsection{Dynamical parameters for model I and its evolution}
Understanding the dynamical features of the universe requires an examination of the dynamical parameters of the model. If the model is bouncing or not, one may quickly tell by observing how the dynamical parameters behave. The dynamical parameters such as the pressure, energy density, and the EoS parameter for the modified symmetric teleparallel gravity theory $f(Q, T)$ can be derived in equations \eqref{eq:p_rho_f(Q,T)}-\eqref{eq:omega_f(Q,T)}, and the energy conditions are represented as \eqref{EC_f(Q,T)}, where we have used the shorthand notation $\tilde{\lambda}_2=\lambda_2+8\pi$. These equations can be found in the following sections for considered models.

The energy density, pressure, and EoS parameters for model I can be represented as,

\begin{subequations}\label{pp_f(Q,T)_model_I}
\begin{eqnarray}
\rho &=& -\frac{\lambda_{1} 2^{m-2} 3^{m-1} (2 m-1)}{(\lambda_{2}+4 \pi ) \tilde{\lambda}_{2} t^2} \left(-\alpha  \lambda_{2} m+\lambda_{2} t^2 (m \chi +3)+24 \pi  t^2\right)\left(\frac{t}{\alpha +t^2 \chi }\right)^{2m},\label{rho1_f(Q,T)}\\
p&=&\frac{\lambda_{1} 2^{m-2} 3^{m-1} (2 m-1)}{\lambda_{2}+4 \pi } \left(\frac{(3 \lambda_{2}+16 \pi ) m \left(\alpha -t^2 \chi \right)}{\tilde{\lambda}_2 t^2}+3\right)\left(\frac{t}{\alpha +t^2 \chi }\right)^{2m},\label{p1_f(Q,T)}\\
\omega &=&\frac{3 \lambda_{2} m t^2 \chi-3 \lambda_{2} \left(\alpha  m+t^2\right) +8 \pi  \left(t^2 (2 m \chi -3)-2 \alpha  m\right)}{\lambda_{2} t^2 (m \chi +3)-\alpha  \lambda_{2} m+24 \pi  t^2}\label{omega1_f(Q,T)}.
\end{eqnarray}    
\end{subequations}
\begin{figure}[H]
	\begin{center}
		\includegraphics[width=7.5cm]{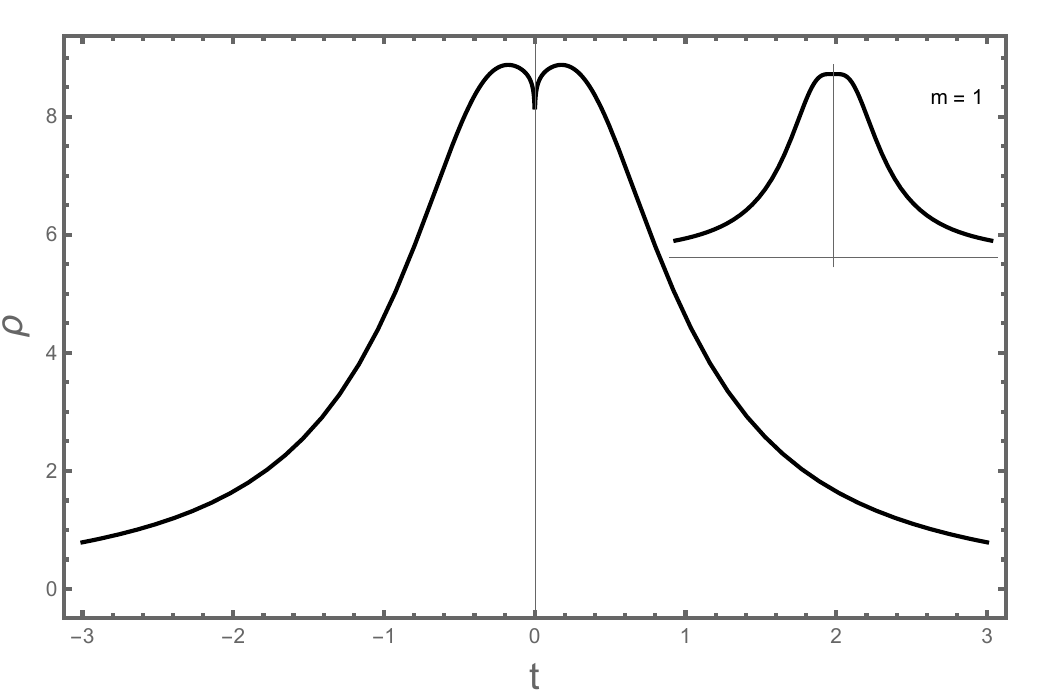}
		\includegraphics[width=7.5cm]{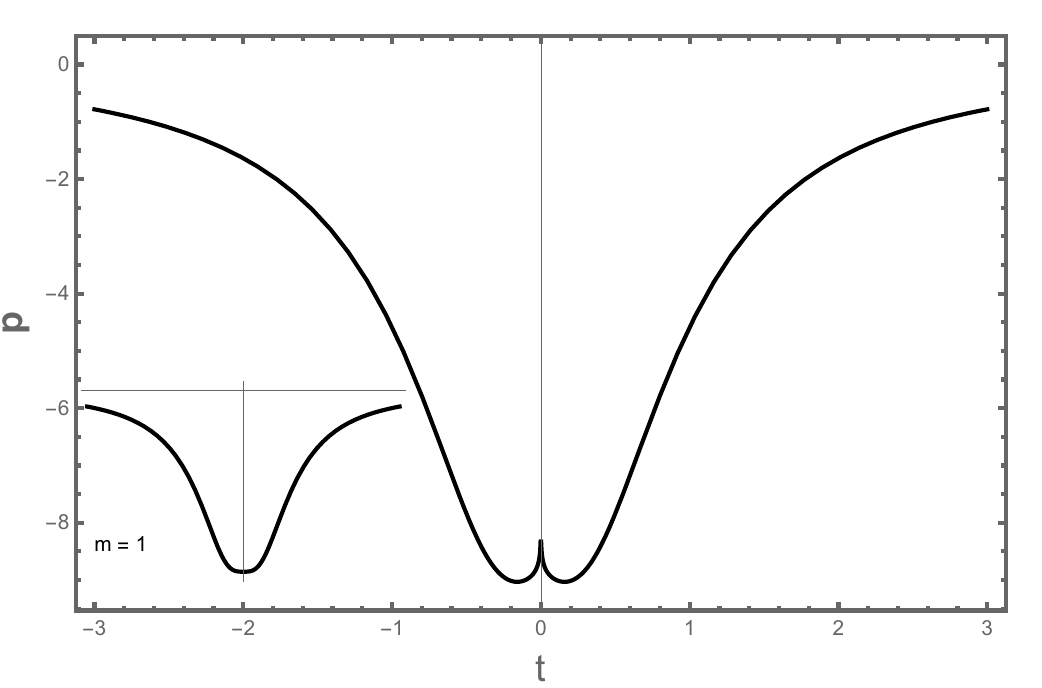}
		\includegraphics[width=7.5cm]{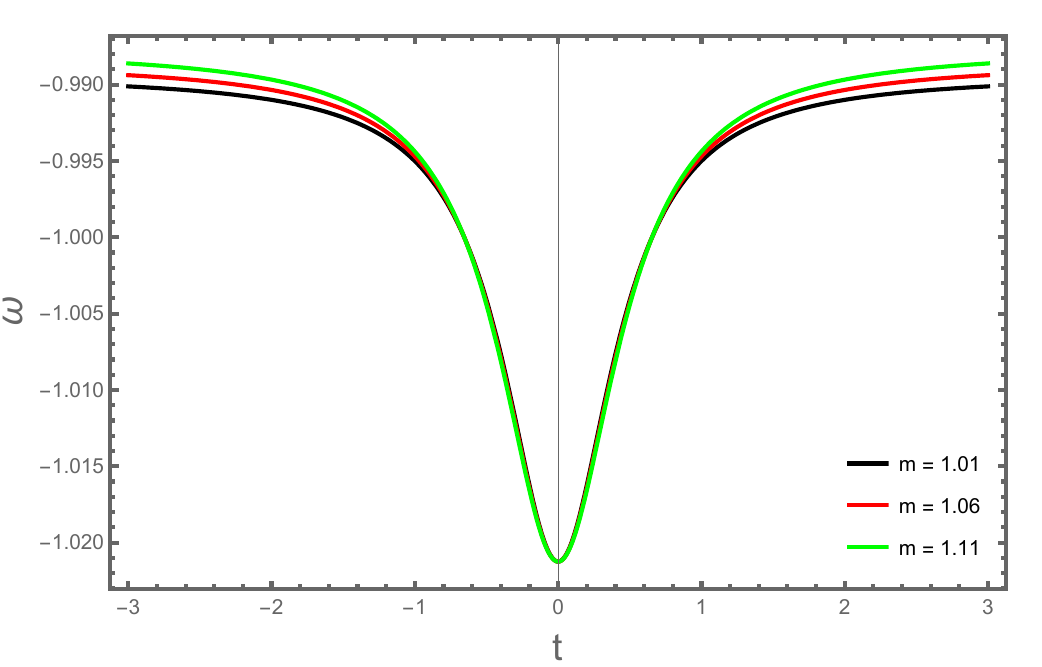}
	\end{center}
	\caption[Variation of physical parameters in cosmic time for model I.]{Variation of energy density (left panel), pressure (right panel), EoS parameter (lower panel) in cosmic time. We have used the parameter space $\alpha=0.43$, $\chi=1.001$, $\lambda_1=-0.5$, $\lambda_2=-12.5$ and $m=1.01$ for the model I, three different values have been considered to see the variation in EoS parameter.}
\label{Chap_5:fig_3}
\end{figure}
Figure \ref{Chap_5:fig_3} displays the pressure and energy density graphically as a function of cosmic time. To maintain a favorable energy density throughout cosmic history, encompassing the negative time zone, we have taken into account the values of $\lambda_1=-0.5$, $\lambda_2=-12.5$, and $m=1.01$. As we approach the bouncing point in the pre-bounce period, we notice an increase in the energy density for this set of parameters. After forming a ditch close to the bounce, it decreases in the post-bounce region. The ditch in the energy density near the bounce vanishes with a choice of the teleparallel gravity parameter $m=1$ (the behavior for $m=1$ is depicted in the embedded figure). The pressure graph appears to be a mirror image of the energy density curve from the bounce point. In the context of cosmic evolution, it is seen that the pressure exhibits a negative characteristic. Specifically, inside the pre-bounce zone, the pressure gradually diminishes from a small negative magnitude to a significantly larger negative magnitude at the point of bounce. Conversely, inside the post-bounce region, there is a transition from significantly negative numbers to a little negative value. At the bouncing epoch, a tiny bump forms in the pressure curve that vanishes for $m=1$. Considering the aforementioned, it is our contention that the selection of the model parameter $m$ significantly impacts the dynamic characteristics of the model.  The observation of $p<0$ and $\rho>0$ dynamics during the post-bounce phase provides evidence for an accelerating universe. 

We present the evolution of the EoS parameter as a function of cosmic time in Figure \ref{Chap_5:fig_3}. During the transition from the pre-bounce phase to the post-bounce phase, which occurs inside the bouncing epoch, it is noticed that the EoS parameter initially decreases. Subsequently, it crosses the phantom divide line and then increases once again after the formation of a well in the vicinity of the bounce. The depth of the well is contingent upon the selection of the parameter $m$. EoS parameter behavior in the right half indicates that it increases from a $\omega<-1$ to $\omega>-1$ during the post-bounce period. Since there is a greater rate of increment in $\omega$ for large values of $m$, the adjusted gravity parameter serves to separate the tail area of the EoS parameter. Notably, under the assumption of a scale factor parameter $\chi=0.99$, the comoving Hubble radius approaches to zero as we progress away from the bounce epoch. Conversely, for $\chi=1.001$, the comoving Hubble radius diverges as we move away from the bounce epoch. However, in both cases pressure term and energy density behavior remain consistent.

\subsection{Dynamical parameters for model II and its evolution}
The pressure, energy density, and EoS parameter for the current model are derived using equations \eqref{eq:p_rho_f(Q,T)} and \eqref{eq:omega_f(Q,T)} and the scaling factor in equation \eqref{SF2t_f(Q,T)} as,

\begin{subequations}\label{pp_f(Q,T)_model_II}
\begin{eqnarray}
\rho&=&-\frac{\lambda_{1} 24^{m-1} (2 m-1)\left(3 \rho_{c}  t^2 (\lambda_{2} (m+2)+16 \pi )-4 \lambda_{2} m\right)}{(\lambda_{2}+4 \pi ) \tilde{\lambda}_2 \rho_{c}  t^2}\left(\frac{\rho_{c} t}{3 \rho_{c}  t^2+4}\right)^{2m} ,\label{rho2_f(Q,T)}\\
p&=&-\frac{\lambda_{1} 24^{m-1} (2 m-1)\left(3 \rho_{c}  t^2 (\lambda_{2} (3 m-2)+16 \pi  (m-1))-4 (3 \lambda_{2}+16 \pi ) m\right)}{(\lambda_{2}+4 \pi ) \tilde{\lambda}_2 \rho_{c}  t^2}\left(\frac{\rho_{c} t}{3 \rho_{c}  t^2+4}\right)^{2m},\nonumber \\
\label{p2_f(Q,T)}\\
\omega&=&\frac{3 \rho_{c}  t^2 (\lambda_{2} (3 m-2)+16 \pi  (m-1))-4 (3 \lambda_{2}+16 \pi ) m}{3 \rho_{c}  t^2 (\lambda_{2} (m+2)+16 \pi )-4 \lambda_{2} m}.\label{omega2_f(Q,T)}     
\end{eqnarray}    
\end{subequations}

\begin{figure}[H]
	\begin{center}
		\includegraphics[width=7.5cm]{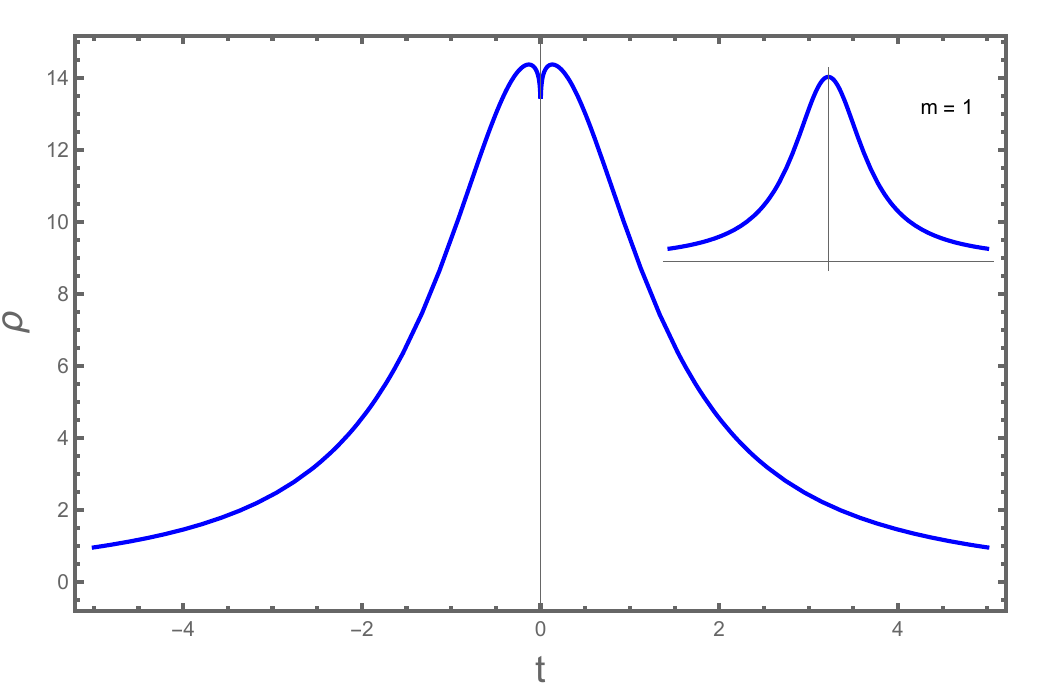}
		\includegraphics[width=7.5cm]{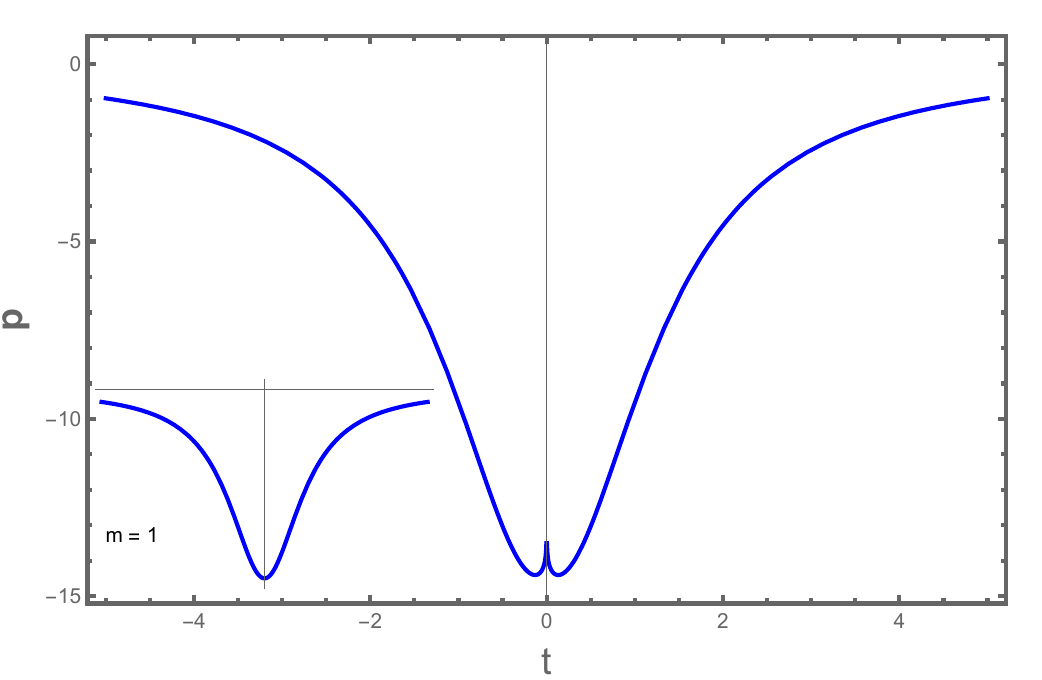}
		\includegraphics[width=7.5cm]{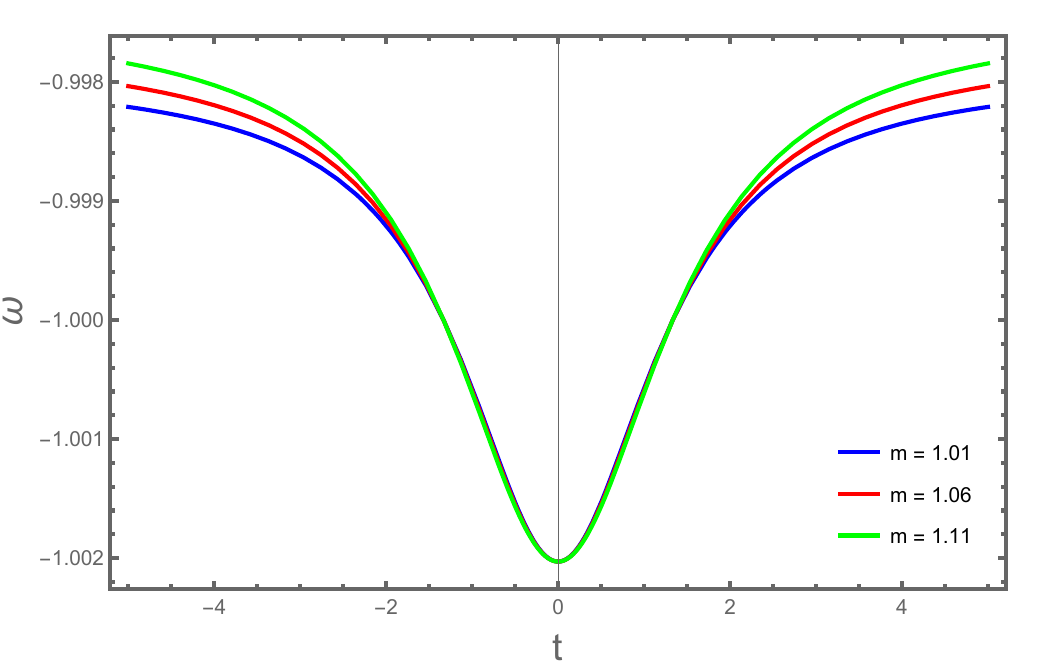}
	\end{center}
	\caption[Variation of physical parameters in cosmic time for model II.]{Variation of energy density (left panel), pressure (right panel), EoS parameter (lower panel) in cosmic time. We have used the parameter space $\rho_{c}=0.75$, $\lambda_1=-0.5$, $\lambda_2=-12.56$ and $m=1.01$ for model II, three different values have been considered to see the variation in EoS parameter.}
\label{Chap_5:fig_4}
\end{figure}

The graphical behavior of the pressure and energy density as a function of cosmic time is shown in Figure \ref{Chap_5:fig_4}. The bounce at $t=0$ is shown with a well-shaped curve in the energy density for the typical values of the model parameters. In order to guarantee that the energy density is positive during the bounce and in both the positive and negative time zones, the parameters have been set. The energy density exhibits behavior that is comparable to the earlier bouncing model covered in model I. From a modest positive value, it changes to form a ditch at bounce, and in the positive time zone, it falls to a lower positive value. The selection of the parameter $m$ influences both the maximum energy density and the creation of the ditch. We do not have such a trench in the energy density curve for $m=1$; instead, the energy density curve nearly follows a Gaussian pattern. Throughout the evolution of the universe, pressure has exhibited a consistent trend toward negative values. In the context of the negative time zone, as the cosmic time approaches the bouncing epoch, there is a notable acceleration in the rate at which the pressure decreases. However, after producing a bump close to the bounce, the pressure increases quickly in the post-bounce period. The emergence of the bump is contingent upon the selection of the variable $m$. As is customary, the bump in the pressure curve flattens out when $m=1$.

The graphical representation of the EoS parameter as a function of cosmic time is depicted in Figure \ref{Chap_5:fig_4}. During the course of cosmic evolution, the EoS parameter consistently maintains a value close to $\omega=-1$. Upon closer examination of the dynamical aspect of the EoS parameter, it can be observed that it undergoes an evolution from a value greater than $-1$ to a value less than $-1$ in the pre-bounce region. Subsequently, following the bounce, a transition occurs, and the EoS parameter climbs from a value less than $-1$ to a value greater than $-1$ in the post-bounce region. The behavior around the bounce point is not influenced by the value of model $m$. However, when considering a significant time period after the bounce epoch, it becomes evident that the EoS parameter becomes divided and exhibits many types of plots based on different values of $m$. The depth of the well in the EoS parameter is contingent upon the magnitude of the model parameter $m$. A higher value of $m$ corresponds to a greater depth, while a value of $m$ close to 1 indicates less depth in the well.

\section{Energy conditions for both models}

In Section-\ref{GR_EC}, we have mentioned the importance of the energy conditions in describing a model. Also, given the extension of the symmetric teleparallel gravity $f(Q,T)$ theory, we made a comprehensive discussion on what should be the additional conditions followed by a bouncing model. In fact, the prescribed condition to get the bouncing solution is the possible violation of the NEC at the bounce epoch. Here, we wish to present the energy conditions for the two models discussed in the present work.

The NEC and SEC for model-I may be expressed as,
\begin{subequations}
\begin{eqnarray}
\rho+p&=&\frac{\lambda_{1} 2^m 3^{m-1} m (2 m-1) \left(\alpha -t^2 \chi \right) }{\tilde{\lambda}_2 t^2}\left(\frac{t}{\alpha +t^2 \chi }\right)^{2m},\label{NEC1_f(Q,T)}\\
\rho+3p&=&-\frac{\lambda_{1} 6^{m-1} (2 m-1)\left[\lambda_{2} \left(t^2 (5 m \chi -3)-5 \alpha  m\right)-24 \pi  \left(\alpha  m+t^2 (1-m \chi )\right)\right] }{(\lambda_{2}+4 \pi ) \tilde{\lambda}_2 t^2}\nonumber \\
&&\times \left(\frac{t}{\alpha +t^2 \chi }\right)^{2m}.
\label{SEC1_f(Q,T)}
\end{eqnarray}
\end{subequations}


The NEC and SEC for model- II are expressed as,
\begin{subequations}
\begin{eqnarray}
\rho+p&=&-\frac{\lambda_{1} 2^{3 m-1} 3^{m-1} m (2 m-1) \left(3 \rho_{c}  t^2-4\right) }{\tilde{\lambda}_2 \rho_{c}  t^2}\left(\frac{\rho_{c} t}{3 \rho_{c}  t^2+4}\right)^{2m},\label{NEC2_f(Q,T)}\\
\rho+3p&=&-\frac{\lambda_{1} 2^{3 m-2} 3^{m-1} (2 m-1) \left(3 \rho_{c}  t^2 (\lambda_{2} (5 m-2)+8 \pi  (3 m-2))-4 (5 \lambda_{2}+24 \pi ) m\right)}{(\lambda_{2}+4 \pi ) \tilde{\lambda}_2 \rho_{c}  t^2} \nonumber \\ &&\times\left(\frac{\rho_{c} t}{3 \rho_{c}  t^2+4}\right)^{2m}.\label{SEC2_f(Q,T)}
\end{eqnarray}    
\end{subequations}
\begin{figure}[H]
\centering
\minipage{0.50\textwidth}
\includegraphics[width=\textwidth]{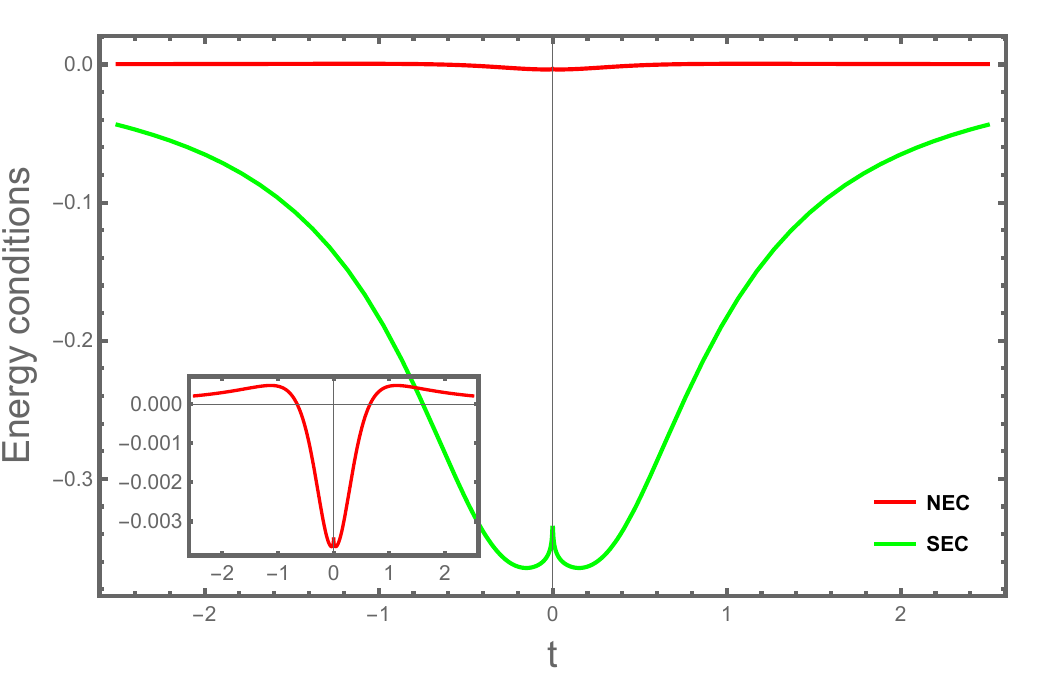}
\endminipage\hfill
\minipage{0.50\textwidth}
\includegraphics[width=\textwidth]{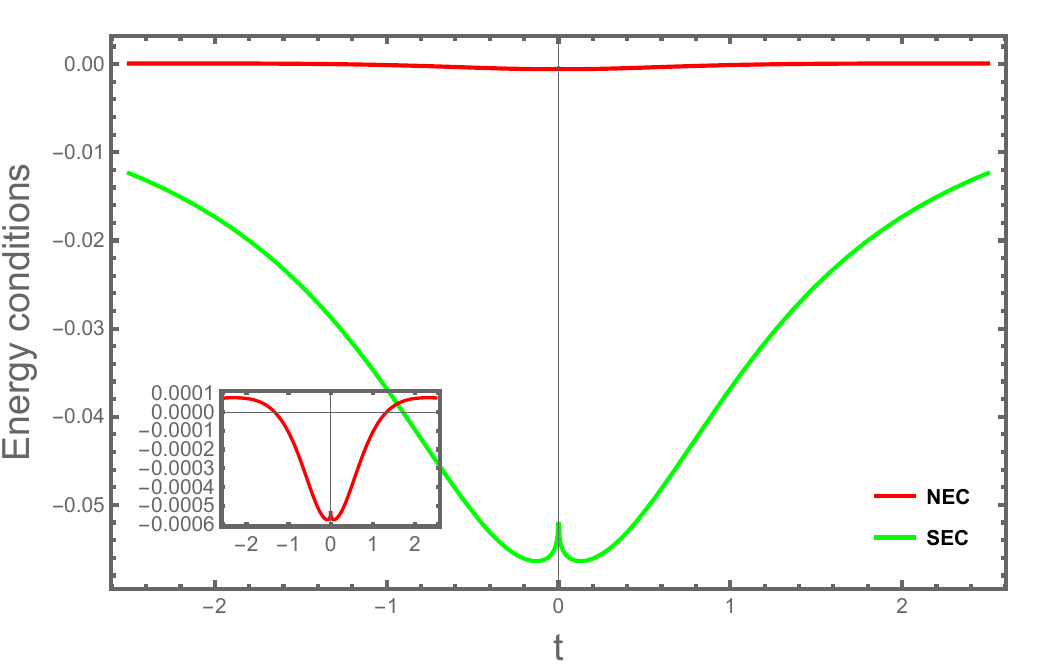} 
\endminipage
\caption[Variation of the energy conditions for model I and model II.]{Energy conditions: Plot of the energy conditions for model I (left panel), the model parameters are chosen as $\alpha=0.43$ and $\chi=1.001$. Evolution of energy conditions for model II (right panel) with the model parameter $\rho_{c}=0.75$. For both the models, we chose the $f(Q,T)$ parameters as $\lambda_1=-0.5$, $\lambda_2=-12.5$ and $m=1.01$.}
\label{Chap_5:fig_5}
\end{figure}
The energy condition violation is demonstrated in Figure \ref{Chap_5:fig_5} for both models. The left panel displays the temporal progression of the energy conditions in model I, whereas the right panel exhibits the corresponding evolution in model II. Furthermore, a notable occurrence arises when the SEC is violated in conjunction with the violation of the null energy requirement. Nevertheless, based on the graphical depiction, it has been noticed that the SEC is consistently violated throughout the entire evolutionary process. It is noteworthy to emphasize that the violation of the NEC in the early universe is associated with the non-singular bouncing solution. In the current study, similar observations have been made that provide support for the concept of a bouncing solution for the universe.

\section{Validation through cosmographic test}

\begin{figure}[H]
    \centering
    \minipage{0.45\textwidth}
    \includegraphics[width=\textwidth]{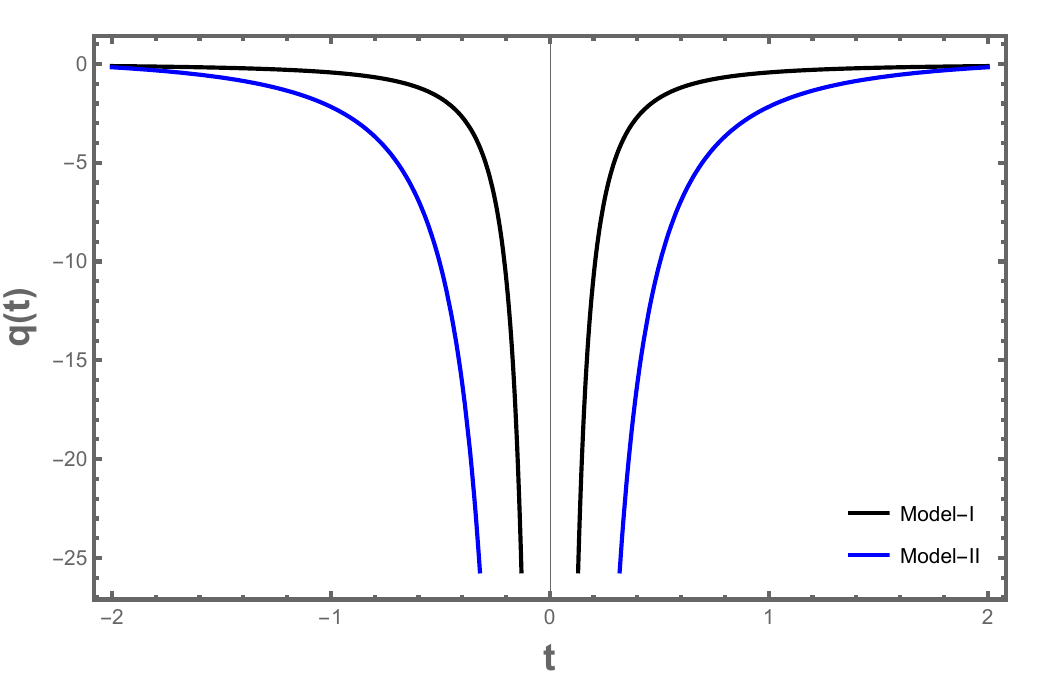}
    \endminipage\hfill
    \minipage{0.45\textwidth}
    \includegraphics[width=\textwidth]{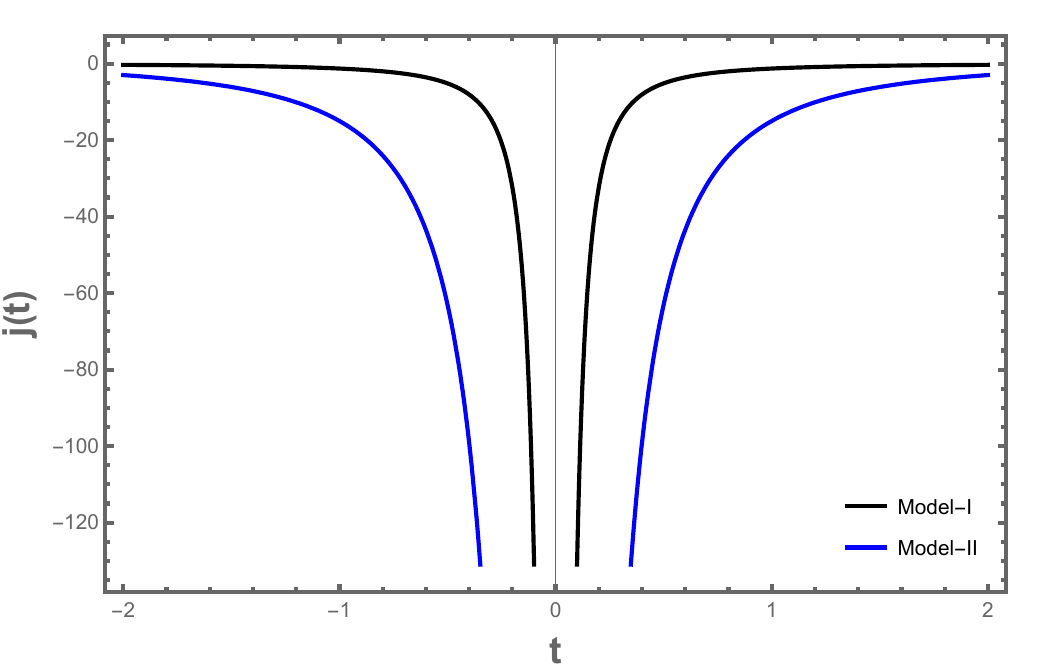}
    \endminipage\hfill
    \minipage{0.45\textwidth}
    \includegraphics[width=\textwidth]{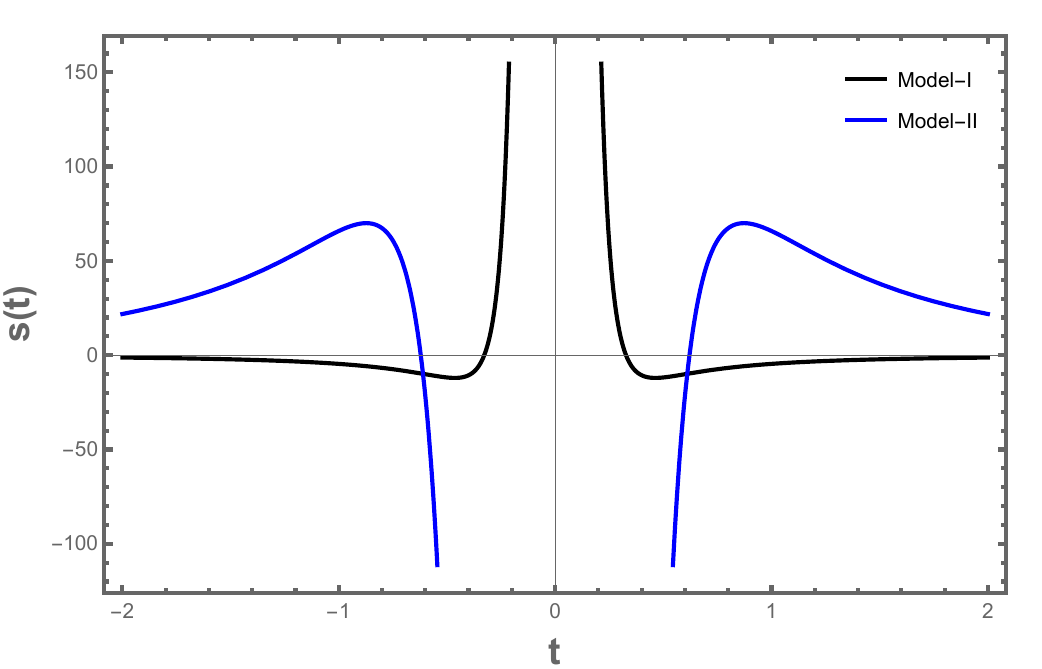}
    \endminipage\hfill
    \caption[Variation of the cosmographic parameters in cosmic time.]{Plot for the variation of the parameters: deceleration (upper left panel), jerk (upper right panel), and snap (lower panel) as functions of cosmic time for model I (black line) with $\alpha=0.43$, $\chi=1.001$ and model II (blue line).~~[Equations \eqref{SF1t_f(Q,T)} and \eqref{SF2t_f(Q,T)}].}
    \label{Chap_5:fig_6}
\end{figure}

The Hubble parameter has been previously examined in the corresponding bouncing models. In this section, we shall examine the cosmographic parameters that have yet to be addressed. The deceleration parameter $q$ serves as an indicator for discerning whether the universe is experiencing acceleration or deceleration. Put simply, a positive deceleration parameter signifies that the force of standard gravity is the dominant factor compared to other entities, whereas a negative value implies a repulsive influence that surpasses the standard gravitational pull. The symbol $\mathrm{j}$ serves as an indicator for the alteration in the dynamics of the universe, where a positive value denotes the presence of a transitional period during which the universe adapts its rate of expansion. It is important to acknowledge that the cosmographic parameters discussed are independent of the $f(Q,T)$ gravity theory and the specific values chosen for the parameters $m$, $\lambda_1$, and $\lambda_2$. The determination of cosmographic parameters is contingent upon the fixed parameters that are present in the scale factor. In this section, we will show the equations for the cosmographic parameters of the two models, as we have previously examined two distinct ansatzes for the scale factor that describes a bounce scenario. The cosmographic parameters for model I in cosmic time are defined in equation \eqref{cosmographic_parameters_model_I_t}.
The cosmographic parameters for model II in cosmic time are obtained as 
\begin{eqnarray} \label{CP2t_f(Q,T)}
q(t)&=&\frac{1}{2}-\frac{2}{\rho_{c}t^{2}}, \nonumber \\
\mathrm{j}(t)&=&1-\frac{12}{\rho_{c}t^{2}}, \nonumber \\
s(t)&=&\frac{12 \left(7 \rho_{c}  t^2-2\right)}{\rho_{c} ^2 t^4}-\frac{7}{2}.
\end{eqnarray}
The cosmographic characteristics of both models undergo changes across cosmic time, indicating their significance in relation to the developing nature of dark energy in the universe. Indeed, under the framework of the current extended symmetric teleparallel gravity theory, the nature of dark energy is exclusively determined to be of a geometric origin. Figure \ref{Chap_5:fig_6} presents graphical depictions of the deceleration, jerk, and snap parameters as a function of cosmic time. In both the negative and positive time zones, the deceleration parameter exhibits a negative value, featuring a singularity at the bounce point and a tendency to converge towards a value of $-0.1$ as it progresses further away from the epoch of the bounce. Both models exhibit identical behavior in terms of the jerk parameters. These parameters display a singularity at the bounce epoch and gradually approach zero as we travel further away from this epoch. In contrast, it is observed that the snap parameter exhibits contrasting behavior for both models in the vicinity of the bounce epoch. Specifically, for the first model, the singularity manifests in the positive range, but for the second model, the singularity is observed in the negative range. 
\section{Stability analysis}
In this section, we want to investigate the stability of the bouncing models within the framework of modified $f(Q, T)$ gravity, as previously introduced. The idea is that the universe is made up of a perfect fluid, for which the adiabatic speed of sound is defined as $C_{s}^{2}=dp/d\rho$. In order to maintain thermodynamic or mechanical stability within a system, it is necessary for the sound velocity, denoted as $C_{s}^{2}$, to remain positive. Consequently, stability is achieved when $C_{s}^{2}$ assumes a positive value. Additionally, it is imperative to maintain mechanical stability by ensuring that the value of $C_{s}^{2}$ does not exceed 1. Hence, the region defined by the inequality $0\leq C_{s}^{2}\leq 1$ yields stable solutions. Nevertheless, in the event of a violation of the NEC, it is conceivable that ghost fields may emerge, indicating possibly dangerous instabilities at either the classical or quantum scales. Moreover, it is not entirely possible to completely prevent the phenomenon of superluminality. The utilization of negative energy scalar fields, such as ghost condensates and conformal galileon, together with other ways, serves as the main driving force behind the implementation of bounces. However, it is important to acknowledge that these approaches may occasionally lead to instabilities, which necessitate appropriate measures to be taken. The relationship between the scale factor and redshift has been employed to ascertain the stability of the models, namely using the equation $1/(1+z)=a(t)$. In order to differentiate the equations with respect to redshift, equations \eqref{pp_f(Q,T)_model_I} and \eqref{pp_f(Q,T)_model_II} are utilized. The redshift parameter is employed to express the features of $C_{s}^{2}$.

For model I we get, 
\begin{equation} \label{eq:CSsqr1}
C_{s}^{2}=\frac{2\alpha^2\left((6\tilde{\lambda}_2-16\pi)m\chi-3\tilde{\lambda}_2\right)-\frac{\alpha\chi\left[(3\tilde{\lambda}_2-8\pi)(4m+1)\chi-9\tilde{\lambda}_2\right]}{(1+z)^{2\chi}}+\frac{\chi^{2}(\left(3\tilde{\lambda}_2-8\pi\right)m\chi-3\tilde{\lambda}_2)}{(1+z)^{4\chi}}}{2\alpha^2(2\lambda_2\chi+3\tilde{\lambda}_2)-\frac{\alpha\chi((4m+1)\lambda_2\chi+9\tilde{\lambda}_2)}{(1+z)^{2\chi}}+\frac{\chi^2(\lambda_2\chi m+3\tilde{\lambda}_2)}{(1+z)^{4\chi}}},
\end{equation}

and for model II, we obtain
\begin{equation}\label{eq:CSsqr2}
C_{s}^{2}=\frac{\lambda_2\left[\frac{3m}{(2z^{3}+6z^{2}+6z+1)^{-2}}-\frac{(z^{3}+3z^{2}+3z)}{(4z^{3}+12z^{2}+12z+5)^{-1}}-3\right]+16\pi\left[\frac{m}{(2z^{3}+6z^{2}+6z+1)^{-2}}-\frac{(z+1)^3}{(2z^{3}+6z^{2}+6z)^{-1}}-1\right]}{\lambda_2\left[m(2z^{3}+6z^{2}+6z+1)^2+\frac{(z^{3}+3z^{2}+3z)}{(4z^{3}+12z^{2}+12z+1)^{-1}}-1\right]+16\pi\frac{[2(z^{3}+3z^{2}+3z)+1]}{(z^{3}+3z^{2}+3z)^{-1}}}.
\end{equation}
\begin{figure}[H]
\centering
\minipage{0.50\textwidth}
\includegraphics[width=\textwidth]{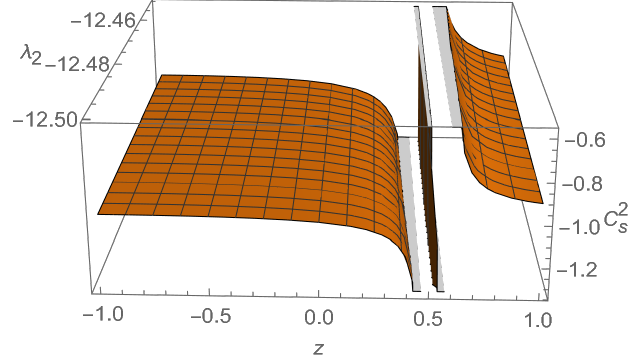}
\endminipage
\minipage{0.50\textwidth}
\includegraphics[width=\textwidth]{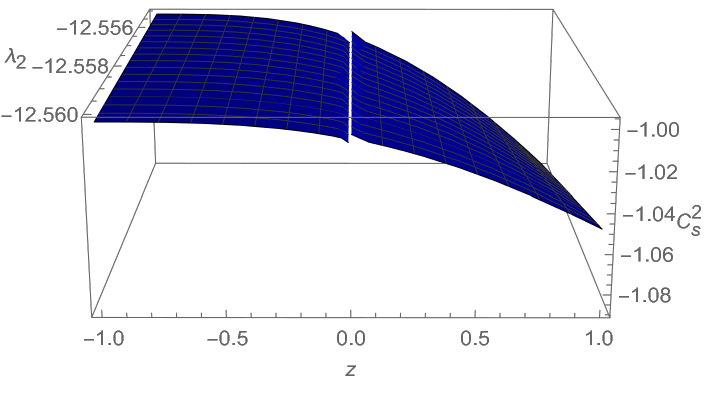} 
\endminipage
\caption[Variation of $C^{2}_{s}$ for model I and model II.]{Variation of $C^{2}_{s}$ with the parameters $\{\alpha=0.43$, $\chi=1.001$, $\lambda_1=-0.5$, $\lambda_2=-12.5$, $m=1.01\}$ (model I) and $\{\rho_{c}=0.75$, $\lambda_1=-0.5$, $\lambda_2=-12.56$, $m=1.01\}$ (model II).}
\label{Fig:Ch5:CSsqr1}
\end{figure}
The stability of the extended symmetric teleparallel gravity models is depicted in Figure \ref{Fig:Ch5:CSsqr1}. The stability criterion is not met in the cases of model I and II due to the persistent negative value of $C_{s}^{2}$. Given this perspective, it is possible that the models could exhibit certain instabilities. To mitigate instabilities, it is necessary to ensure that the value of $C_{s}^{2}$ is greater than zero. Consequently, it is possible to break the NEC without compromising stability.   
\section{Conclusion}
The current chapter has focused on the examination of the bouncing scenario of the universe within the framework of extended symmetric teleparallel gravity, specifically referred to as $f(Q,T)$ gravity.  In this study, we examine a general functional form denoted as $f(Q,T)=\lambda_1Q^m+\lambda_2T$, where $Q$ represents the non-metricity function. In the typical approach employed in the literature, the EoS parameter is commonly treated as either a constant or expressed in a parametrized form. This is done to effectively address the highly nonlinear equations of motion, ultimately leading to the determination of the scale factor. During this process, the dynamical behavior of the model is restricted by the selected model parameters, which impose limitations on the evolutionary behavior of the EoS parameter, as well as other dynamic qualities such as energy density and pressure. In this study, we examine a specific assumed ansatz for the scale factor and deduce the equations that describe the dynamic characteristics, including the energy density, pressure, and EoS parameters. The assessment of the role of modified gravity in the dynamical component of the model is evident in this phase. Two distinct ansatzes have been taken into account in order to describe the scale factors that characterize the non-singular bounce scenario at certain early epochs. This study examines the impact of extended symmetric teleparallel gravity on the dynamics of the model. By conducting a comprehensive study, it is evident that the bounce scenario can be accommodated within the framework of the $f(Q,T)$ gravity theory, while also exhibiting a favorable behavior for the EoS parameter. The dynamical parameters, namely energy density, pressure, and the EoS parameter exhibit a distinct pattern of a ditch/hump behavior in close proximity to the bounce event, indicating an evolutionary element. The manifestation of this behavior is contingent upon the selection of the $f(Q,T)$ parameters.

The models have been validated by calculating the cosmographic parameters and assessing the energy conditions, providing justification for their use. As is expected in any dark energy scenario, the violation of the SEC is found throughout cosmic development, both in the positive and negative temporal domains. For the scenario of bouncing to be realized, it is necessary for the NEC to be violated. While our models exhibit some deviations from the NEC, we address this issue by refining the selection of model parameters. Therefore, the occurrence of NEC violation in bouncing models results in both thermodynamic and mechanical instability inside the models. However, additional research is necessary to gain a comprehensive understanding of the stability and the problem of energy-momentum non-conservation inside this geometric theory.


\chapter{Global Phase Space Analysis for a Class of Single Scalar Field Bouncing Solutions} 

\label{Chapter6} 

\lhead{Chapter 6. \emph{Global Phase Space Analysis for a Class of Single Scalar Field Bouncing Solutions}} 

\vspace{10 cm}
* The work, in this chapter, is covered by the following publication: \\

\textbf{A.S. Agrawal} et al., ``Global phase space analysis for a class of single scalar field bouncing solutions in general relativity", \textit{European Physical Journal C}, \textbf{84}, 56 (2024).

\clearpage

\section{Introduction} 
    The focus of this chapter will be on a single scalar field model, characterized by a Lagrangian density expressed as $\mathcal{L}=F(X)-V(\phi)$. This model can generate nonsingular bouncing solutions in the presence of a phantom field \cite{Cai-2012-08, Cai:2013kja}. It has been established in the literature that in order to obtain nonsingular bouncing solutions within the framework of GR, it is necessary to permit the violation of the NEC and/or the SEC \cite{Novello:2008ra}. A spatially flat case is considered where an NEC violation is required. When examining the dynamics of inhomogeneous cosmic perturbations, it has been observed that scalar fields violating the NEC may exhibit ghost instability \cite{Carroll-2003, Garriga-2013, Sawicki-2013} and gradient instability \cite{Vikman-2005}. It is important to note that efforts to develop stable nonsingular bouncing models, despite the violation of the NEC, have led to the emergence of more complex and intricate models. These models incorporate ghost condensates \cite{Buchbinder-2007, Arkani-Hamed-2004} and Galileons \cite{Creminelli-2010-2010, Easson-2011, Qiu-2011-2011}, instead of the conventional scalar field. However, the primary objective of this study is not to analyze the dynamics of inhomogeneous perturbations but rather to investigate the stability of a bouncing solution with a minor perturbation in the initial condition.
 
    In this analysis, the generality of nonsingular bouncing solutions has been explored by considering a simplified form of the kinetic term, denoted as $F(X)$, and two specific examples for the potential, represented as $V(\phi)$. The canonical field, commonly known as the quintessence field with potential, is a fundamental scalar field \cite{Fang-2007}. However, the universe contains various intricate cosmological dynamics that cannot be fully explained by the canonical scalar field. The quintessence field model fails to provide an explanation for phenomena such as the crossing of the phantom divide line and the bouncing solution. This leads to a more comprehensive explanation of the non-canonical scalar field, which is a specific type of scalar field. The non-canonical scenario offers a solution to the coincidence problem without introducing any fine-tuning issues, making it advantageous compared to the canonical setup. Furthermore, non-canonical models exhibit a smaller tensor-to-scalar ratio compared to canonical models, leading to improved agreement with measurements of the CMB. The fascinating features of non-canonical scalar fields motivate us to study more about their cosmic dynamics \cite{Unnikrishnan-2012}.

	A dynamical system analysis of the models is performed and looks into the nonsingular bouncing solutions within the phase space in order to address the issue of the genericity of bouncing solutions. The dynamical systems technique has proven to be a valuable tool for comprehending the qualitative behaviors of cosmological models, even without the need to analytically solve the system of differential equations \cite{Ellis, Coley:2003mj}. By employing this technique, it is possible to reframe the dynamics of the universe as the phase flow within a phase space that has been appropriately defined. While this technique has been widely adopted for investigating inflationary and dark energy models \cite{Bahamonde:2017ize}, its application to nonsingular bouncing cosmologies is relatively uncommon. The primary reason is that the Hubble normalized dimensionless dynamical variables, which are commonly used for cosmological phase space analysis, exhibit divergence at a nonsingular bounce. On the other hand, the dynamical system formulation appears to be well-suited for addressing qualitative questions, such as determining the genericity of a cosmological solution. Suppose that one sets initial conditions at a random moment during the contracting phase of the evolution, numerically evolves the field equations, and obtains a bounce. The question is whether applying a random slight change in the initial condition would also result in a bounce. In order to study nonsingular bouncing solutions in the phase space picture, it is necessary to either establish a different set of dimensionless dynamical variables or confine the infinite phase space within a finite domain using a particular compactification method. In addition to addressing the question of genericity, the phase space picture also offers insights into the past and future asymptotics of a nonsingular bouncing cosmology. The future asymptotic is of great importance as it determines the ultimate states of the bouncing cosmologies. Do they eventually become asymptotically de-Sitter-like in the context of $\Lambda$CDM, or do they result in a big rip?  These are the main reasons why this work has been done.
 
	The cosmological phase space of models with the form $F(X)-V(\phi)$ has been studied in Ref. \cite{De-Santiago:2012ibi}. In this reference, the authors also explore bouncing solutions from a phase space perspective by introducing a different set of dynamical variables. The analysis presented in Ref. \cite{Panda:2015wya} is extended to the case of Bianchi-I. However, none of the aforementioned works conducts a comprehensive analysis of the phase space in a concise manner. In general, a system may have intriguing cosmological scenarios that involve fixed points hidden at infinity. To understand the overall dynamics of such a system, a comprehensive analysis is necessary, which involves compact analysis. In Ref. \cite{De-Santiago:2012ibi}, the authors were able to demonstrate phase trajectories that show a nonsingular bounce. However, it is important to note that this does not provide a definitive answer regarding the genericity and future asymptotics of the phenomenon. In this study, the same dynamical system formulation is used as in previous works (Refs. \cite{De-Santiago:2012ibi, Panda:2015wya}). However, the enhancement of these earlier studies by conducting a comprehensive phase space analysis, thereby addressing the aforementioned inquiries. In this analysis, the scenarios in which a bounce occurs generically for the functions $F(X)$ and $V(\phi)$ are explicitly identified. In addition, aims to determine whether these scenarios demonstrate asymptotically de-Sitter behavior or result in a big rip. It is important to understand that the lack of generic bouncing solutions does not mean that bouncing solutions cannot be achieved at all. It is still possible to achieve a bounce under certain specific initial conditions. However, in contrast to generic bounces, it cannot be guaranteed that a chosen initial condition that results in a bounce will maintain this property when subjected to a random slight perturbation. The cases in which the bounce is generic are especially intriguing when constructing models that involve bouncing.
	
	
	\section{Basic cosmological equations for \texorpdfstring{$\mathcal{L}=F(X)-V(\phi)$}{}}\label{sec:basics}
	
	The most general action of a minimally coupled scalar field theory
	is given by
	\begin{equation}\label{eq:action}
		S=\int d^4x \sqrt{-g}\left(\frac{M_{Pl}^2}{2} R+\mathcal{L}(\phi,X)\right)+S_m\,,
	\end{equation}
	where the kinetic component of the scalar field $\phi$ is denoted by $X$ (i.e., $X = -\frac{1}{2}\partial_{i} \phi \partial^{i} \phi$), the reduced Planck mass is $M_{Pl}$, and the final term $S_m$ is the action corresponds to the matter component assumed to be a perfect fluid.
	
	The Einstein field equations are obtained by taking the variation of \eqref{eq:action} with respect to the metric $g_{ij}$.
	\begin{equation}
		G_{ij}= T_{ij}^{(\phi)} + T_{ij}^{(m)}\,,
	\end{equation}
	the symbol $G_{ij}$ represents the Einstein tensor, while $T_{ij}^{(\phi)}$ denotes the energy-momentum tensor of the scalar field,
	\begin{equation}
		T_{ij}^{(\phi)}=\frac{\partial \mathcal{L}}{\partial X} \partial_{i} \phi \partial_{j} \phi-g_{ij} \mathcal{L}\,,
	\end{equation}
	and the matter energy-momentum tensor $T_{ij}^{(m)}$ is given by
	\begin{equation}\label{eq:EMT_mat}
		T_{ij}^{(m)}=(\rho_m+p_m) u_i u_j+p_m g_{ij}\,.
	\end{equation}
	The symbols $\rho_{m}$ and $p_m$ represent the energy density and pressure of the matter component, respectively. These quantities are associated with the four-velocity vector $u_i$. In this work, the FLRW cosmology is examined and is described by the metric \eqref{spacetime:GR}. Also, the focus is on a scalar field model whose generic form of the Lagrangian is given by 
	\begin{equation}
		\mathcal{L}(\phi,X) = F(X) - V(\phi)\,,
	\end{equation}
	where $V(\phi)$  is a scalar field potential and $F(X)$ is an arbitrary  function of $X$.
	
	The energy-momentum tensor \eqref{eq:EMT_mat} of  a perfect fluid under the FLRW cosmology is
	\begin{equation}
		T^{i (m)}_{j}= \text{diag}(-\rho_m, p_m, p_m, p_m). 
	\end{equation}
	The spatially flat FLRW space-time transforms the aforementioned Einstein field equations into the cosmological field equations shown below,
	\begin{subequations}
		\begin{eqnarray}
			&& H^{2} = \frac{1}{3M_{Pl}^{2}}\left[2XF_X - F + V + \rho_{m}\right], \\
			&& \dot{H} = -\frac{1}{2M_{Pl}^{2}}\left[2XF_X + (1+\omega_{m})\rho_{m}\right]\,, \label{eq:Raychaudhuri}
		\end{eqnarray}
	\end{subequations}
	where $\omega_{m}$ is the ideal fluid EoS parameter, defined as $p_m=\omega_{m} \rho_{m}$, and $H=\frac{\dot{a}}{a}$ is the Hubble parameter. A derivative with respect to $t$ is indicated by an over dot, while a derivative with respect to $X$ is indicated by the subscript $X$. According to the ordinary continuity equation, the fluid component scales in the absence of any energy exchange between the fluid and the field.
	\begin{equation}
		\dot{\rho}_m + 3H(1+\omega_m)\rho_m = 0 \quad \Rightarrow \quad \rho_{m}\propto a^{-3(1+\omega_{m})}\,,
	\end{equation}
	and the scalar field satisfies the generic Klein-Gordon equation
	\begin{equation}
		\frac{d}{dN}(2XF_X - F + V) + 6XF_X = 0,
	\end{equation}

	where $N=\ln a$. 
	
	In the following section, the corresponding autonomous system will be built for the cosmological equations mentioned above. Then, examine the bouncing scenarios through a comprehensive analysis of the global dynamical system.
	
	
	\section{Dynamical system formulation of \texorpdfstring{$\mathcal{L}=F(X)-V(\phi)$}{} models suitable for investigating nonsingular bounces}\label{sec:dsa}
	
	The dynamical system construction described in Ref.\cite{De-Santiago:2012ibi} was used to generate an independent system of equations suitable to explore nonsingular bouncing solutions from a phase space point of view. The dynamical variables can be defined as
 \begin{equation}\label{dynvar_def}
		\begin{aligned}
			& x = \frac{\sqrt{3}M_{Pl}H}{\sqrt{|\rho_k|}}, \quad 
			y = \sqrt{\frac{|V|}{|\rho_k|}}\sgn(V), \quad 
			\Omega_{m} = \frac{\rho_{m}}{\vert \rho_{k}\vert},\\
			& \sigma = -\frac{M_{Pl}V_{\phi}}{V}\sqrt{\frac{2X}{3|\rho_k|}}\sgn(\dot{\phi}) = -\frac{M_{Pl}}{\sqrt{3|\rho_k|}}\frac{dlog{V}}{dt}.
		\end{aligned}
	\end{equation}
	 The kinetic part of the energy density is denoted by
	\begin{equation}
		\rho_{k} = 2XF_{X} - F.
	\end{equation}
	The kinetic part of the pressure is
	\begin{equation}
		p_k = F.
	\end{equation}
	Motivated by this, one can define 
	\begin{equation}
		\omega_{k} \equiv \frac{p_k}{\rho_k} = \frac{F}{2XF_{X}-F},
	\end{equation}
	The expression can be understood as the EoS parameter corresponding to the kinetic component of the Lagrangian. The EoS for the scalar field can be derived by expressing it in terms of a new variable as,  
	\begin{equation}
		\omega_{\phi}=\frac{p_{\phi}}{\rho_{\phi}}=\frac{\omega_{k}x^{2}-y^{2}}{x^{2}+y^{2}}\,.
	\end{equation} 
	Next, two auxiliary variables are defined as
	\begin{equation}\label{auxvar_def}
		\Xi=\frac{X F_{XX}}{F_{X}}, \qquad \Gamma=\frac{V V_{\phi \phi}}{V_{\phi}^{2}},
	\end{equation}
	 which will be required to write the dynamical system. Lastly, the phase space-time variable is defined as \cite{De-Santiago:2012ibi}
	\begin{equation}
		d\tau = \sqrt{\frac{|\rho_{k}|}{3M_{Pl}^{2}}}dt.
	\end{equation}
	
	With respect to the dynamical variables and auxiliary variables defined in equation \eqref{dynvar_def} and equation \eqref{auxvar_def}, the Friedmann constraint and the dynamical equations become
	\begin{subequations}\label{dynsys_0}
 \footnotesize
		\begin{eqnarray}
			&& x^2 - y|y| - \Omega_{m} = 1\times \sgn(\rho_{k}),\label{constr_a}\\
			&& \frac{dx}{d\tau} = \frac{3}{2}x\left[(\omega_{k}+1)x-\sigma y|y|\sgn(\rho_{k})\right] - \frac{3}{2}\left[(\omega_{k}-\omega_{m})\sgn(\rho_{k})+(1+\omega_{m})(x^{2}-y|y|)\right],\\   
			&& \frac{dy}{d\tau} = \frac{3}{2}y\left[-\sigma+(\omega_{k}+1)x-\sigma y|y|\sgn(\rho_{k})\right],\\
			&& \frac{d\sigma}{d\tau} = -3\sigma^{2}(\Gamma-1) + \frac{3\sigma[2\Xi(\omega_{k}+1)+\omega_{k}-1]}{2(4\Xi+1)(\omega_{k}+1)}\left((\omega_{k}+1)x-\sigma y^{2}\right)\,.
		\end{eqnarray}
	\end{subequations}
	Because of the definition of the dynamical variable $y$, $y|y|$ can also be written as $y^{2}\sgn(V)$. Therefore, the dynamical system \eqref{dynsys_0} can also be written as
	\begin{subequations}\label{dynsys}
 \footnotesize
		\begin{eqnarray}
			&& x^2 - y^{2}\sgn(V) - \Omega_{m} = 1\times \sgn(\rho_{k}),\label{constr}\\
			&& \frac{dx}{d\tau} = \frac{3}{2}x\left[(\omega_{k}+1)x-\sigma y^{2}\sgn(V)\sgn(\rho_{k})\right] - \frac{3}{2}\left[(\omega_{k}-\omega_{m})\sgn(\rho_{k})+(1+\omega_{m})(x^{2}-y^{2}\sgn(V))\right]\label{eq:ds_x_gen}\\
			&& \frac{dy}{d\tau} = \frac{3}{2}y\left[-\sigma+(\omega_{k}+1)x-\sigma y^{2}\sgn(V)\sgn(\rho_{k})\right],\label{eq:ds_y_gen}\\
			&& \frac{d\sigma}{d\tau} = -3\sigma^{2}(\Gamma-1) + \frac{3\sigma[2\Xi(\omega_{k}+1)+\omega_{k}-1]}{2(2\Xi+1)(\omega_{k}+1)}\left((\omega_{k}+1)x-\sigma y^{2}\right)\,.\label{eq:ds_s_gen}
		\end{eqnarray}
	\end{subequations}
	
	Since $\Omega_m\geq0$, the phase space of the system \eqref{dynsys} is given by
	\begin{equation}
		\lbrace (x,y,\sigma) \in \mathbb{R}^{3}: x^2-y^2\sgn(V)-\sgn(\rho_k)\geq 0\rbrace\,.
	\end{equation}
	It is important to express the quantity $\frac{\dot{H}}{H^2}$ in terms of the dynamical variables as this will help us to find the cosmological evolution corresponding to a fixed point
	\begin{equation}\label{H_eq}
		\frac{\dot{H}}{H^{2}}=-\frac{3}{2x^{2}}\left[(1+\omega_{m})(x^{2}-y\vert y\vert)+(\omega_{k}-\omega_{m})\sgn(\rho_{k})\right]\,.
	\end{equation}
	
	Additionally, one can introduce the effective EoS $\omega_{\rm eff}$ given by
	\begin{equation}\label{eq:weff}
		\omega_{\rm eff}=-1-\frac{2}{3}\frac{\dot{H}}{H^2}=-1+\frac{1}{x^{2}}\left[(1+\omega_{m})(x^{2}-y\vert y\vert)+(\omega_{k}-\omega_{m})\sgn(\rho_{k})\right]\,.
	\end{equation} 
	After considering some specific examples of $F(X)-V(\phi)$ models in the following section, we will look for nonsingular bouncing solutions in the phase space.
	
\subsection{Specific case I: \texorpdfstring{\quad $F(X)=\beta X$}{}}
In this section, the scalar field Lagrangians is considered of the form
\begin{equation}
\mathcal{L}(\phi,X) = \beta X - V(\phi)\,,  \label{langragian}  
\end{equation}
Since $\rho_{\phi}+p_{\phi}=2XF_X=2\beta X$ and $X\geq0$, it reduces to a canonical scalar field when $\beta>0$ and to a phantom scalar field when $\beta <0$. One can notice from equation \eqref{eq:Raychaudhuri} that for $\dot{H}>0$ near the bounce one must necessarily require $\beta<0$, i.e. a phantom scalar field.
 
\subsubsection{Power law potential \texorpdfstring{$V(\phi)=V_0 \phi^n$}{}}\label{subsec:modelI}

 As a first example the specific case given by $F(X)=\beta X,\, V(\phi)=V_0 \phi^n$ has been considered, where $\beta, V_0$  are constants with suitable dimension and $n$ is a dimensionless constant. For this choice, we have
 
	\begin{equation}
		\Xi=0,\qquad 
	  \Gamma=1-\frac{1}{n}, \qquad \omega_{k}=1, \qquad \rho_{k}=\beta X.
	\end{equation}

 \paragraph{Finite fixed point analysis}:
 
	In this case, the system \eqref{dynsys} reduces to
	 \begin{subequations}\label{dynsys_pow}
  \footnotesize
 	\begin{eqnarray}
	 		&& \frac{dx}{d\tau} = \frac{3}{2}x\left[2x - \sigma y^2 \sgn(V)\sgn(\beta)\right] - \frac{3}{2}\left[(1-\omega_{m})\sgn(\beta) + (1+\omega_{m})(x^{2}-y^2 \sgn(V))\right],
            \label{eq:ds_x_pow}\\
 		&& \frac{dy}{d\tau} = \frac{3}{2}y\left[-\sigma + 2x - \sigma y^2 \sgn(V)\sgn(\beta)\right],\label{eq:ds_y_pow}\\
			&& \frac{d\sigma}{d\tau} = \frac{3}{n}\sigma^{2}\,.\label{eq:ds_s_pow}
		\end{eqnarray}
	 \end{subequations}
       The dynamical system \eqref{dynsys_pow} exhibits symmetry when reflected against the $y=0$ plane, or when $y\to-y$ is transformed. This suggests that the $y>0$ region of the phase space is all that needs to be considered, since the phase picture in the $y<0$ region will merely be a reflection of that against $y=0$. From a physical perspective, this indicates that we can focus just on $\sgn(V)=1$ and that the qualitative behavior of the model is independent of the signature of the potential. Furthermore, dynamics occurring in the positive branch of the potential can never cross into the negative branch of the potential, and vice versa, because the $y=0$ line functions as an invariant submanifold. Furthermore, the requirement $\Omega_{m}\geq 0$ is necessary for the model to be physically viable.

       The phase space of the system \eqref{dynsys_pow} is therefore constrained within the region given by
	\begin{equation}
		\lbrace (x,y,\sigma) \in \mathbb{R}^3: y \geq 0, x^2-y^2-\sgn(\beta)\geq 0\rbrace\,.
	\end{equation}
	
	
	\begin{table}[H]
		\begin{center}
			\begin{tabular}{ |c|c|c|c| } 
				\hline
				Point & Co-ordinate $(x,y,\sigma)$ &  Existence &  Physical viability $(\Omega_{m}\geq0)$\\ \hline
				$A_{1+}$ & $(1,0,0)$ & $\beta>0$ & Always \\ \hline
				$A_{1-}$ & $(-1,0,0)$ & $\beta>0$ & Always \\ \hline
			\end{tabular}
		\end{center}
		\caption[Existence and physical viability conditions for finite fixed points for a scalar field with kinetic term $F(X)=\beta X$ and potential $V(\phi)=V_0 \phi^n$.]{Existence and physical viability conditions for finite fixed points for a scalar field with kinetic term $F(X)=\beta X$ and potential $V(\phi)=V_0 \phi^n$, calculated from the system \eqref{dynsys_pow}.}
		\label{tab:finite_fixed_pts_pow}
	\end{table}
	
	
	\begin{table}[H]
		\begin{center}
			\begin{tabular}{ |c|c|c|c| } 
				\hline
				Point   & Co-ordinates $(x,y,\sigma)$ & Stability & Cosmology \\ \hline
				$A_{1+}$ & $(1,0,0)$ &  \begin{tabular}{@{}c@{}}unstable\end{tabular} & $a(t)=(t-t_{*})^{\frac{1}{3}}, ~t\geq t_{*}$ \\ \hline
				$A_{1-}$ & $(-1,0,0)$ & \begin{tabular}{@{}c@{}} saddle (NH) \end{tabular} & $a(t)=(t_{*}-t)^{\frac{1}{3}}, ~t\leq t_{*}$ \\ \hline
			\end{tabular}
		\end{center}
		\caption[Stability condition of physically viable fixed points given in Table \ref{tab:finite_fixed_pts_pow} along with their cosmological behavior.]{Stability condition of physically viable fixed points given in Table \ref{tab:finite_fixed_pts_pow} along with their cosmological behavior. Here and throughout the study, NH stands for nonhyperbolic.}
		\label{tab:stab_finite_fixed_pts_pow}
	\end{table}	
        \begin{figure}[H]
		\centering
		\includegraphics[width=7cm, height=7.2cm]{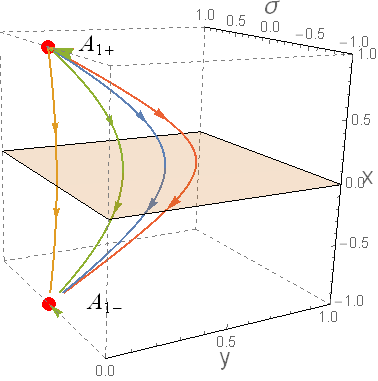}
		\caption[Phase portrait for $F(X)=\beta X,\,V(\phi)=V_0 \phi^n$ with $\omega_{m}=0$, $n=4$ and $\beta<0$.]{Phase portrait of system \eqref{dynsys_pow} for $\omega_{m}=0$, $n=4$ and $\beta>0$.}\label{fig:finte_pow}
	\end{figure}
 Two finite fixed points, $A_{1+}$ and $A_{1-}$, are present in the system \eqref{dynsys_pow} (see Table \ref{tab:finite_fixed_pts_pow}). Both points are only present when $\beta>0$, which is the canonical scalar field. Point $A_{1+}$ corresponds to a decelerated expansion of the universe, whereas point $A_{1-}$ corresponds to a decelerated contraction of the universe, as can be seen from the value of $x$-coordinate and the nature of scale factor $a(t)$ (see Table \ref{tab:stab_finite_fixed_pts_pow}). Here, it is observed that the point $A_{1-}$ is non-hyperbolic, and since it has an empty unstable subspace close to a point, the center manifold theory has been incorporated for additional investigation.  After conducting the study, it was discovered that $A_{1-}$ is a saddle point (see Appendix \ref{app:cmt} for details). Additionally, the trajectories move from a saddle point $A_{1-}$ to an unstable point $A_{1+}$. As a result, no bouncing solution can be extracted from the finite analysis; instead, recollapsing solutions are shown (see Figure  \ref{fig:finte_pow}).
	
	\paragraph{\bf{Fixed points at infinity}}:
	
	To get a global picture of the phase space, the following compact dynamical variables have been introduced
	\begin{equation}\label{3D_dynvar_comp_def}
		\Bar{x}=\frac{x}{\sqrt{1+x^{2}}}\,, \qquad \Bar{y}=\frac{y}{\sqrt{1+y^{2}}} \,, \qquad  \Bar{\sigma}=\frac{\sigma}{\sqrt{1+\sigma^{2}}}\,.
	\end{equation}
	The evolution equation \eqref{dynsys} can be converted to the following system of equations
	\begin{subequations}\label{dynsys_3d}
 \footnotesize
		\begin{eqnarray}
			\frac{d\Bar{x}}{d\tau}&=&\frac{3(1-\Bar{x}^{2})^{\frac{3}{2}}}{2}\left[(\omega_{k}-\omega_{m})\left(\frac{\Bar{x}^{2}}{1-\Bar{x}^{2}}-\sgn(\beta)\right)+\left(1+\omega_{m}-\frac{\Bar{\sigma}\Bar{x}\sgn(\beta)}{\sqrt{1-\Bar{\sigma}^2}\sqrt{1-\Bar{x}^{2}}}\right)\frac{\Bar{y}^{2}\sgn(V)}{1-\Bar{y}^{2}}\right],\\
			\frac{d\Bar{y}}{d\tau}&=&\frac{3}{2}\Bar{y}(1-\Bar{y}^{2})\left[-\frac{\Bar{\sigma}}{\sqrt{1-\Bar{\sigma}^2}} +\frac{(\omega_{k}+1)\Bar{x}}{\sqrt{1-\Bar{x}^{2}}}-\frac{\Bar{\sigma}\Bar{y}^{2}\sgn(V)}{\sqrt{1-\Bar{\sigma}^2}(1-\Bar{y}^{2})}\sgn(\beta)\right] \,,\\
			\frac{d\Bar{\sigma}}{d\tau}&=&\frac{3}{n} \Bar{\sigma}^2\sqrt{1-\Bar{\sigma}^2}\,. 
		\end{eqnarray}
	\end{subequations}
	The variables $\Bar{x}, \Bar{y}$, and $\Bar{\sigma}$ have values between $-1$ and 1, as you can see. While the dynamical system in equations \eqref{dynsys_3d} contains a pole of order $\frac{1}{2}$ at $\Bar{x}^2=1$ and $\Bar{\sigma}^2=1$, as well as a pole of order $1$ at $\Bar{y}^2=1$, it is not regular at the borders of the compact phase space $\Bar{x}^2=1,\,\Bar{y}^2=1$ and $\Bar{\sigma}^2=1$. According to a prescription from Ref.\cite{Bouhmadi-Lopez:2016dzw}, this can be regularized. Here, the phase space-time variable is proposed to be redefined as
	\begin{equation}\label{time_redef_pow}
		d\tau \rightarrow d\Bar{\tau} = \frac{d\tau}{(1-\Bar{x}^{2})^{\frac{1}{2}}(1-\Bar{y}^{2})(1-\Bar{\sigma}^2)^\frac{1}{2}}.
	\end{equation}
	With respect to this redefined time variable, the dynamical system \eqref{dynsys_3d} can be rewritten as
	\begin{subequations}\label{dynsys_3d_new}
 \footnotesize
		\begin{eqnarray}
			\frac{d\Bar{x}}{d\Bar{\tau}} &=& \frac{3}{2}(\omega_{k}-\omega_{m})\left(\Bar{x}^{2}-(1-\Bar{x}^{2})\sgn(\beta)\right)(1-\Bar{x}^{2})(1-\Bar{y}^{2}) \sqrt{1-\Bar{\sigma}^2}\nonumber\\
            && + \frac{3}{2}\left((1+\omega_{m})(1-\Bar{x}^{2}) \sqrt{1-\Bar{\sigma}^2}-\Bar{\sigma}\Bar{x}\sgn(\beta)\sqrt{1-\Bar{x}^{2}}\right)(1-\Bar{x}^{2})\Bar{y}^{2},\label{dynsys_3d_new_x}\\
			\frac{d\Bar{y}}{d\Bar{\tau}}& =& \frac{3}{2}\Bar{y}(1-\Bar{y}^{2})\left[(-\Bar{\sigma}\sqrt{1-\Bar{x}^{2}} +(\omega_{k}+1)\Bar{x}\sqrt{1-\Bar{\sigma}^2})(1-\Bar{y}^{2})-\Bar{\sigma}\Bar{y}^{2}\sqrt{1-\Bar{x}^{2}}\sgn(\beta)\right]\,,\label{dynsys_3d_new_y}\\
			\frac{d\Bar{\sigma}}{d\Bar{\tau}}&=&\frac{3}{n}\Bar{\sigma}^2(1-\Bar{\sigma}^2) \sqrt{1-\Bar{x}^2}(1-\Bar{y}^2)\,.\label{dynsys_3d_new_s}
		\end{eqnarray}
	\end{subequations}
	
	\begin{table}[H]
		\begin{center}
                \resizebox{\textwidth}{!}{
			\begin{tabular}{|*{4}{c|}} 
				\hline
				Point & Co-ordinate $(\Bar{x},\Bar{y},\Bar{\sigma})$ &  Existence &  \begin{tabular}{@{}c@{}}Physical viability\\ $(\Omega_{m}\geq0)$\end{tabular}\\ \hline
				$B_{1+}$ & $(1,1,\Bar{\sigma})$ & Always & Always \\ \hline
			$B_{1-}$ & $(-1,1,\Bar{\sigma})$ & Always &          Always\\ \hline
				$B_{2+}$ & $(1,0,\Bar{\sigma})$ & Always & Always \\ \hline
				\parbox[c][0.6cm]{0.5cm}{$B_{2-}$} & $(-1,0,\Bar{\sigma})$ & Always & Always \\ \hline
				\parbox[c][1cm]{0.5cm}{$B_{3+}$} & $\left(\Bar{x},0,1\right)$ & Always & \begin{tabular}{@{}c@{}}$\frac{1}{2}\leq \Bar{x}^2 \leq 1$ \& $\Bar{x}\neq 0$\\ if $\sgn(\beta)=1$\\$ \Bar{x}^2 \leq 1$\\ if $\sgn(\beta)=-1$ \end{tabular}  \\ \hline
				\parbox[c][1cm]{0.5cm}{$B_{3-}$} & $\left(\Bar{x},0,-1\right)$ & Always &\begin{tabular}{@{}c@{}}$\frac{1}{2}\leq \Bar{x}^2 \leq 1$ \& $\Bar{x}\neq 0$\\  if $\sgn(\beta)=1$\\$ \Bar{x}^2 \leq 1$\\ if $\sgn(\beta)=-1$ \end{tabular} \\ \hline
				$B_{4+}$ & $\left(\frac{(1+\omega_{m})\sqrt{1-\Bar{\sigma}^2}}{\sqrt{(1+\omega_{{m}})^2-\Bar{\sigma}^2\omega_{m}(\omega_{m}+2)}},1,\Bar{\sigma}\right)$ & \begin{tabular}{@{}c@{}}$\left(\left(\Bar{\sigma}^2\leq \frac{(1+\omega_{m})^2}{\omega_{m}(\omega_{m}+2)}\right)\wedge  \left((\beta<0)\wedge\left[ (\Bar{\sigma}\leq 0) \vee (\Bar{\sigma}=1)\right]\right)\right)\vee$ \\ $\left(\left(\Bar{\sigma}^2\leq \frac{(1+\omega_{m})^2}{\omega_{m}(\omega_{m}+2)}\right) \wedge\left((\beta>0)\wedge \left[(\Bar{\sigma}\geq 0)\vee(\Bar{\sigma}=-1)\right]\right)\right)$\end{tabular}& $\Bar{\sigma}=0$ \\ \hline
                $B_{4-}$ & $\left(-\frac{(1+\omega_{m})\sqrt{1-\Bar{\sigma}^2}}{\sqrt{(1+\omega_{{m}})^2-\Bar{\sigma}^2\omega_{m}(\omega_{m}+2)}},1,\Bar{\sigma}\right)$ & \begin{tabular}{@{}c@{}}$\left(\left(\Bar{\sigma}^2\leq \frac{(1+\omega_{m})^2}{\omega_{m}(\omega_{m}+2)}\right)\wedge \left((\beta<0)\wedge \left[(\Bar{\sigma}\geq 0)\vee(\Bar{\sigma}=-1)\right]\right)\right)\vee$\\ 
                $\left(\left(\Bar{\sigma}^2\leq \frac{(1+\omega_{m})^2}{\omega_{m}(\omega_{m}+2)}\right)\wedge\left((\beta>0)\wedge \left[(\Bar{\sigma}\leq 0)\vee (\Bar{\sigma}=1)\right]\right)\right)$\end{tabular} & $\Bar{\sigma}=0$ \\ \hline
				$B_{5+}$ & $\left(0,\frac{1}{\sqrt{2}},1\right)$ & $\beta <0$ & Always \\ \hline
				$B_{5-}$ &$\left(0,\frac{1}{\sqrt{2}},-1\right)$ & $\beta <0$ & Always \\ \hline				
			\end{tabular}}
		\end{center}
		\caption[Existence and physical viability condition for fixed points at infinity for a scalar field with kinetic term $F(X)=\beta X$ and potential $V(\phi)=V_0 \phi^n$.]{Existence and physical viability condition for fixed points at infinity for a scalar field with kinetic term $F(X)=\beta X$ and potential $V(\phi)=V_0 \phi^n$ calculated from the system \eqref{dynsys_3d_new}. }   
		\label{tab:infinite_fixed_pts_pow}
	\end{table}
 
	
	In terms of the compact variables, the constraint equation \eqref{constr} can be rewritten for this model as
	\begin{equation}\label{constr_comp}
		\frac{\Bar{x}^{2}}{1-\Bar{x}^{2}} - \frac{\Bar{y}^{2}}{1-\Bar{y}^{2}} - \sgn(\beta) = \Omega_{m} \,.   
	\end{equation}
	Therefore, for a canonical scalar field where $\sgn(\beta)=1$, the physical viability condition requires 
	\begin{equation} \label{physviab_can_1}
		\frac{\Bar{x}^{2}}{1-\Bar{x}^{2}}-\frac{\Bar{y}^{2}}{1-\Bar{y}^{2}}-1=\Omega_{m}\geq 0 \,.  
	\end{equation}
	It can be checked that the necessary and sufficient conditions for physical viability of canonical scalar field are given respectively as follows
	\begin{eqnarray}\label{physviab_can}
		\Bar{y}^{2}\leq 2-\frac{1}{\Bar{x}^{2}}, \hspace{1cm} \Bar{x}^{2}\geq \frac{1}{2}.
	\end{eqnarray}
	For a phantom scalar field where $\sgn(\beta)=-1$, the physical viability condition requires
	\begin{equation}
		\frac{\Bar{x}^{2}}{1-\Bar{x}^{2}}-\frac{\Bar{y}^{2}}{1-\Bar{y}^{2}}+1=\Omega_{m}\geq 0\,.
	\end{equation}
	The necessary \emph{and} sufficient condition for the above to be satisfied is
	\begin{eqnarray}\label{physviab_noncan}
		\Bar{y}^{2}\leq \frac{1}{2-\Bar{x}^{2}}.
	\end{eqnarray}
	The physically viable region of the entire 3-dimensional compact phase space for canonical and phantom scalar fields is specified by the limitations in equations \eqref{physviab_can} and \eqref{physviab_noncan}, respectively. It is consistent with the observation that a phantom scalar field is necessary to achieve a nonsingular bounce that the line $\Bar{x}=0$ is physical only for the situation of a phantom scalar field ($\beta<0$).
	\begin{table}[H]
		\begin{center}
			\begin{tabular}{|*{4}{c|}}
				\hline
				Point   & Co-ordinates $(\Bar{x},\Bar{y},\Bar{\sigma})$ & Stability & Cosmology \\ \hline
				\parbox[c][1.2cm]{0.5cm}{$B_{1+}$} & $(1,1,\Bar{\sigma})$ & \begin{tabular}{@{}c@{}}stable for $\Bar{\sigma}=0$ or \\ $~\{\Bar{\sigma} \neq 0, \sgn(\beta)\neq \sgn(\Bar{\sigma})\}$ \\ \& saddle for \\ $\{\Bar{\sigma} \neq 0, \sgn(\beta)= \sgn(\Bar{\sigma})\}$\end{tabular}& De Sitter\\ \hline
				\parbox[c][1.2cm]{0.5cm}{$B_{1-}$} &  $(-1,1,\Bar{\sigma})$& \begin{tabular}{@{}c@{}} unstable for $\Bar{\sigma}=0$ or \\ $~\{\Bar{\sigma}\neq 0, \sgn(\beta)= \sgn(\Bar{\sigma})\}$  \\ \& saddle for \\ $\{\Bar{\sigma} \neq 0, \sgn(\beta)\neq \sgn(\Bar{\sigma})\}$\end{tabular} & De Sitter \\ \hline
				\parbox[c][0.8cm]{0.5cm}{$B_{2+}$}   & $(1,0,\Bar{\sigma})$ & saddle always & $a(t)=(t-t_{*})^{\frac{2}{3(1+\omega_{m})}}, ~t\geq t_{*}$\\ \hline
				\parbox[c][0.8cm]{0.5cm}{$B_{2-}$}   & $(-1,0,\Bar{\sigma})$ & saddle always & $a(t)=(t_{*}-t)^{\frac{2}{3(1+\omega_{m})}}, ~t\leq t_{*}$\\ \hline
				$B_{3+}$ & $\left(\Bar{x},0,1\right)$ &  \begin{tabular}{@{}c@{}}stable if $\sgn(n)>0$\\saddle otherwise \end{tabular}  & depending on $\Bar{x}$ and $\beta$\\ \hline
				$B_{3-}$ & $\left(\Bar{x},0,-1\right)$  &  \begin{tabular}{@{}c@{}}unstable if $\sgn(n)>0$\\saddle otherwise \end{tabular}  &  depending on $\Bar{x}$ and $\beta$ \\ \hline
				$B_{5+}$ & $\left(0,\frac{1}{\sqrt{2}},1\right)$ &  \begin{tabular}{@{}c@{}}unstable if $\sgn(n)<0$\\saddle otherwise \end{tabular}  &$a(t)=$constant\\ \hline
				$B_{5-}$ & $\left(0,\frac{1}{\sqrt{2}},-1\right)$ & \begin{tabular}{@{}c@{}}stable if $\sgn(n)<0$\\saddle otherwise \end{tabular} & $a(t)=$constant\\ \hline
			\end{tabular}
		\end{center}
		\caption[Stability condition of physically viable fixed points given in Table \ref{tab:infinite_fixed_pts_pow} along with their cosmological behavior.]{Stability condition of physically viable fixed points given in Table \ref{tab:infinite_fixed_pts_pow} along with their cosmological behavior.}
		\label{tab:stab_infinite_fixed_pts_pow}
	\end{table}

	A total of six invariant submanifolds, $\bar{x}=\pm1,\,\Bar{y}=0,1$ and $\bar{\sigma}=\pm1$, are presented by the system \eqref{dynsys_3d_new}. Calculations for their stability are found in appendix \ref{app:stab_inv_sub}. Table \ref{tab:infinite_fixed_pts_pow} lists the fixed points for system \eqref{dynsys_3d_new}, and Table \ref{tab:stab_infinite_fixed_pts_pow} lists the stability conditions for these fixed points. While it may seem from Table \ref{tab:infinite_fixed_pts_pow} that there are five distinct pairs of fixed points at the infinity of the phase space, the pair $B_{4\pm}$ is only physically feasible when $\Bar{\sigma}=0$; in that case, they already lie on the $B_{1\pm}$ fixed point line. Consequently, at infinity, there are just four distinct pairings of fixed points. While $B_{2\pm}$ corresponds to cosmic phases dominated by the hydrodynamic matter component, $B_{1\pm}$ are de-Sitter solutions. Since $B_{2+}$ is a saddle and so always denotes an intermediate phase of evolution, it can be seen as the matter-dominated epoch for $\omega_m=0$. Different cosmological situations can be observed at fixed places on the line $B_{3\pm}$. For example, they can characterize a static world for $\Bar{x}=0,\,\beta<0$ and a matter-dominated universe for $\Bar{x}=1,\,\beta>0$. The fixed points $B_{5\pm}$, in conclusion, are consistent with a static universe.
 \begin{figure}[H]
		\centering
		\minipage{0.40\textwidth}
		\includegraphics[width=\textwidth]{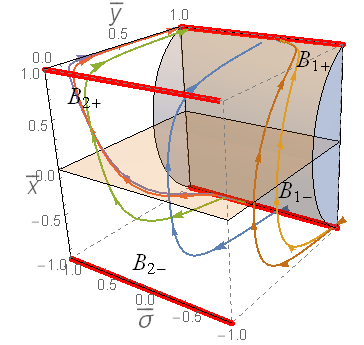}
		\endminipage
		\minipage{0.3\textwidth}
		\includegraphics[width=\textwidth]{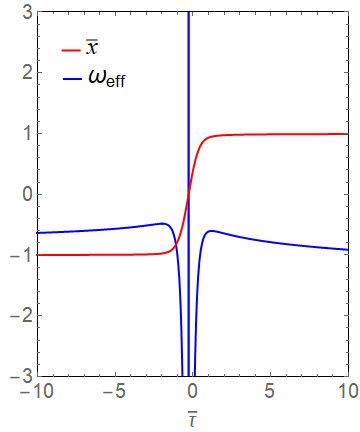}
		\endminipage\\
		\minipage{0.40\textwidth}
		\includegraphics[width=\textwidth]{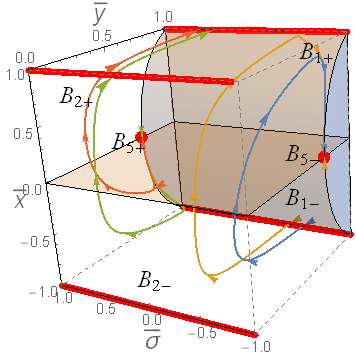}
  \endminipage
		\minipage{0.3\textwidth}
		\includegraphics[width=\textwidth]{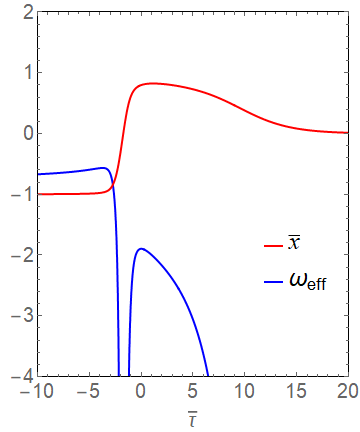}
		\endminipage
		\caption[Phase portrait for $F(X)=\beta X, V(\phi)=V_{0}\phi^{n}$ with $\omega_{m}=0$, $\beta<0$ moreover, $n=4$ in the upper left panel and $n=-4$ in the lower panel. A plot of variables $\Bar{x}$ and effective EoS $\omega_{\rm eff}$ is shown in the upper and lower right panel.]{Phase portrait of the system \eqref{dynsys_3d_new} with the shaded region representing the non-physical region of the phase space (upper and lower left panels). Here $\omega_{m}=0$, $\beta<0$ with $n=4$ in the upper left panel and $n=-4$ in the lower panel. A plot of variables $\Bar{x}$ and effective EoS $\omega_{\rm eff}$ is shown in the upper and lower right panel for the blue trajectories of the upper and lower left panel, which represents two characteristic types of nonsingular bouncing solutions, namely, asymptotically de-Sitter and asymptotically static in future.}\label{fig:infinte_pow}
	\end{figure}
	
	The phase portrait of the system \eqref{dynsys_3d_new} in the compact phase space corresponding to physically feasible nonsingular bouncing solutions is shown in the left panel of Figure \ref{fig:infinte_pow}. Additionally, as predicted, there is a violation of the NEC during bounce (see Figure \ref{fig:infinte_pow}, right panel). As predicted, the graphs also clearly demonstrate that the matter-dominated stationary spots $B_{2\pm}$ are saddles or intermediate epochs of development. 
 
   The phase trajectories in the upper left panel are nonsingular bouncing solutions connecting the contracting and expanding de-Sitter phases $B_{1-}$ and $B_{1+}$, respectively, and correspond to $F(X)=\beta X\,(\beta<0),\,V(\phi)=V_{0}\phi^4$. However, since $B_{1-}$ and $B_{1+}$ are not global repellers and attractors, one cannot claim that this represents the general behavior of phase trajectories. There are additional attractor $B_{3+}$ and repeller $B_{3-}$ in addition to $B_{1-}$ and $B_{1+}$. As a result, depending on how the trajectories evolve, heteroclinic trajectories connecting $B_{3-}$ to $B_{1+}$, $B_{1-}$ to $B_{3+}$, and $B_{3-}$ to $B_{3+}$ may or may not correlate to a nonsingular bouncing cosmology.

    $F(X)=\beta X\,(\beta<0),\,V(\phi)=V_{0}\phi^{-4}$ corresponds to the phase trajectories in the lower left panel. These trajectories show nonsingular bouncing solutions connecting two de-Sitter phases ($B_{1-} \rightarrow B_{1+}$) or a de-Sitter phase with a static universe ($B_{1-} \rightarrow B_{5-}$ or $B_{5+} \rightarrow B_{1+}$). Again, though, one cannot claim that this kind of behavior is universal. When $n=-4<0$, the fixed points $B_{3\pm}$ are in a saddle shape, but $B_{5\pm}$ turns into a pair that is attractor and repellent. Therefore, heteroclinic trajectories between $B_{5-}$ and $B_{5+}$ may exist; again, the evolution of the trajectories may or may not match a nonsingular bouncing cosmology.
    
    \subsubsection{Exponential potential \texorpdfstring{$V(\phi)=V_{0}e^{-\lambda \phi/M_{Pl}}$}{}: }\label{subsec:modelII}
    
    As a second example the specific case given by $F(X)=\beta X,\,V(\phi)=V_{0}e^{-\lambda \phi/M_{Pl}}$ has been considered, where $\beta, V_0$ are constants with suitable dimension, $\lambda$ is a dimensionless constant. For this choice, we have
        \begin{equation}
	\Xi=0, \qquad \Gamma=1, \qquad \omega_{k}=1, \qquad \rho_{k}=\beta X, \qquad \sigma = \frac{\sqrt{{2}/{3}}\lambda}{\sqrt{|\beta|}}.
	\end{equation}

 \paragraph{Finite fixed point analysis}:
	
	In this case, since $\sigma$ is a constant, the system \eqref{dynsys} reduces to a 2-dimensional dynamical system
	\begin{subequations}\label{dynsys_exp}
 \footnotesize
		\begin{eqnarray}
			&& \frac{dx}{d\tau} = \frac{3}{2}x\left[2x-\sigma y^{2}\sgn(V_{0})\sgn(\beta)\right] - \frac{3}{2}\left[(1-\omega_{m})\sgn(\beta)+(1+\omega_{m})(x^{2}-y^{2}\sgn(V_{0}))\right], \\
			&& \frac{dy}{d\tau} = \frac{3}{2}y\left[2x-\sigma\left(1+ y^{2}\sgn(V_{0})\sgn(\beta)\right)\right]\,.
		\end{eqnarray}
	\end{subequations}
 The dynamical system remains unaffected by the parameter $\lambda$. The dynamical system \eqref{dynsys_exp} is symmetric under reflection against the $y=0$ line, that is, under the transformation $y\to-y$, as can be seen, just as in the power law example. This suggests that the $y>0$ region of the phase space is all that needs to be considered, since the phase picture in the $y<0$ region will merely be a reflection of that against $y=0$. From a physical perspective, this indicates that we may focus on $\sgn(V_0)=1$ and that the qualitative behavior of the model is independent of the signature of the potential. This makes it extremely clear how the $y=0$ invariant submanifold should be interpreted in this situation. As previously, the requirement $\Omega_{m}\geq 0$ is necessary for the model to be physically viable.

	The phase space of the system \eqref{dynsys_pow} is therefore constrained within the region given by
	\begin{eqnarray}
		&&\lbrace (x,y) \in \mathbb{R}^2: y \geq 0, x^2-y^2-\sgn(\beta)\geq 0\rbrace\,.  
	\end{eqnarray}
	
	\begin{table}[H]
		\begin{center}
			\begin{tabular}{|*{4}{c|}} 
				\hline
				Point & Co-ordinate $(x,y)$ &  Existence &  Physical viability $(\Omega_{m}\geq0)$\\ \hline
				\parbox[c][0.6cm]{0.5cm}{$A_{1+}$} & $\left(1,0\right)$ & $\beta >0$ & Always \\ \hline
				\parbox[c][0.6cm]{0.5cm}{$A_{1-}$} & $\left(-1,0\right)$ & $\beta >0$ & Always \\ \hline
				\parbox[c][1cm]{0.5cm}{$A_{2+}$} & $\left(\frac{2}{|\sigma|},\frac{\sqrt{4+\sigma^{2}}}{|\sigma|}\right)$ & $\beta<0$ and $\sigma<0$ & Always \\ \hline
				\parbox[c][1cm]{0.5cm}{$A_{2-}$} & $\left(-\frac{2}{|\sigma|},\frac{\sqrt{4+\sigma^{2}}}{|\sigma|}\right)$ & $\beta<0$ and $\sigma>0$ & Always \\ \hline
				\parbox[c][1cm]{0.5cm}{$A_{3+}$} & $\left(\frac{2}{|\sigma|},\frac{\sqrt{4-\sigma^{2}}}{|\sigma|}\right)$ & $\beta>0$ and $0<\sigma<2$ & Always \\ \hline
				\parbox[c][1cm]{0.5cm}{$A_{3-}$} & $\left(-\frac{2}{|\sigma|},\frac{\sqrt{4-\sigma^{2}}}{|\sigma|}\right)$ & $\beta>0$ and $-2<\sigma<0$ & Always  \\ \hline
				\parbox[c][1cm]{0.5cm}{$A_{4}$} & $\left(\frac{\sigma}{(1+\omega_{m})},\sqrt{\frac{1-\omega_{m}}{1+\omega_{m}}}\right)$ & $\beta>0$ & $|\sigma |\geq\sqrt{2(1+\omega_{m})}$ \\ \hline
			\end{tabular}
		\end{center}
		\caption[Existence and physical viability conditions for finite fixed points for a scalar field with kinetic term $F(X)=\beta X$ and potential $V(\phi)=V_0 e^{-\lambda\phi/M_{Pl}}$.]{Existence and physical viability conditions for finite fixed points for a scalar field with kinetic term $F(X)=\beta X$ and potential $V(\phi)=V_0 e^{-\lambda\phi/M_{Pl}}$, calculated from the system \eqref{dynsys_exp}.}
		\label{tab:finite_fixed_pts_exp}
	\end{table}
\begin{table}[H]
		\begin{center}
			\begin{tabular}{|*{4}{c|}}
				\hline
				Point   & Co-ordinates $(x,y)$ & Stability & Cosmology \\ \hline
				\parbox[c][0.6cm]{0.5cm}{$A_{1+}$} & $\left(1,0\right)$ & unstable always & $a(t)=(t-t_{*})^{\frac{1}{3}}, ~t\geq t_{*}$\\ \hline
				\parbox[c][0.6cm]{0.5cm}{$A_{1-}$} & $\left(-1,0\right)$ & stable always & $a(t)=(t_{*}-t)^{\frac{1}{3}}, ~t\leq t_{*}$\\ \hline
				\parbox[c][1cm]{0.5cm}{$A_{2+}$} & $\left(\frac{2}{|\sigma|},\frac{\sqrt{4+\sigma^{2}}}{|\sigma|}\right)$ & N.H. always & $a(t)=\frac{1}{(t_{*}-t)^{\frac{4}{3\sigma^{2}}}}, ~t<t_{*}$ \\ \hline
				\parbox[c][1cm]{0.5cm}{$A_{2-}$} & $\left(-\frac{2}{|\sigma|},\frac{\sqrt{4+\sigma^{2}}}{|\sigma|}\right)$ & N.H. always & $a(t)=\frac{1}{(t-t_{*})^{\frac{4}{3\sigma^{2}}}}, ~t>t_{*}$ \\ \hline
				\parbox[c][1cm]{0.5cm}{$A_{3+}$} & $\left(\frac{2}{|\sigma|},\frac{\sqrt{4-\sigma^{2}}}{|\sigma|}\right)$ & N.H. always & $a(t)=(t-t_{*})^{\frac{3\sigma^{2}}{4}}, ~t\geq t_{*}$ \\ \hline
				\parbox[c][1cm]{0.5cm}{$A_{3-}$} & $\left(-\frac{2}{|\sigma|},\frac{\sqrt{4-\sigma^{2}}}{|\sigma|}\right)$ & N.H. always & $a(t)=(t_{*}-t)^{\frac{3\sigma^{2}}{4}}, ~t\leq t_{*}$ \\ \hline
				\parbox[c][1cm]{0.5cm}{$A_{4}$} & $\left(\frac{\sigma}{(1+\omega_{m})},\sqrt{\frac{1-\omega_{m}}{1+\omega_{m}}}\right)$ & N.H. always & $a(t)=(t-t_{*})^{\frac{2}{3(1+\omega_{m})}}, ~t\geq t_{*}$ \\ \hline
			\end{tabular}
		\end{center}
		\caption[Stability condition of physically viable fixed points given in Table \ref{tab:finite_fixed_pts_exp} along with their cosmological behavior.]{Stability condition of physically viable fixed points given in Table \ref{tab:finite_fixed_pts_exp} along with their cosmological behavior. Stability of $A_{2\pm},\,A_{3\pm}$ and $A_4$ are determined from the Jacobian eigenvalues whereas the stability of $A_{1\pm}$ are determined by examining the stability of the invariant submanifolds $y=0$ and $\Omega_m=0$ (see appendix \ref{app:stab_inv_sub}).}
		\label{tab:stab_finite_fixed_pts_exp}
	\end{table}
	Seven finite fixed points are present in the system \eqref{dynsys_exp}: $A_{1\pm},\,A_{2\pm},\,A_{3\pm}$, and $A_{4}$ (see the Table \ref{tab:finite_fixed_pts_exp}). Table \ref{tab:stab_finite_fixed_pts_exp} lists the stability and cosmic development associated with each fixed point. The fixed points $A_{1\pm}$ that exist only in the case of a canonical scalar field correspond to decelerated expanding and contracting phases, respectively, which is obtained for the power law potential as well. There is no equivalent finite fixed point for the power law potential among the remaining finite fixed points identified for the exponential potential. For a canonical scalar field, another pair of expanding and contracting solutions are noticed, $A_{3\pm}$, that can be accelerated or decelerated by $\sigma^2>\frac{4}{3}$ or $\sigma^2<\frac{4}{3}$, respectively. The values of $A_{3\pm}$ and $A_{1\pm}$ coincide in the limit $|\sigma|\rightarrow2$. In addition, we have an expanding power law solution $A_4$ for the canonical case, which is accelerated or decelerated based on $1+3\omega_m>0$ or $1+3\omega_m<0$, i.e., whether or not the matter component satisfies the SEC. Two solutions $A_{2\pm}$ were obtained for the situation of a phantom scalar field, which corresponds to finite time singularities in the past and future, respectively. The phase $A_{2+}$ is actually phantom-dominated and terminates in a big-rip singularity as expected.
 \paragraph{Fixed points at infinity}: To get a global picture of the 2-dimensional phase space the compact dynamical variables $\Bar{x},\,\Bar{y}$ are employed as in \eqref{3D_dynvar_comp_def}. For a constant $\sigma$ \eqref{dynsys_3d} reduces to
	\begin{subequations}\label{dynsys_2d}
 \footnotesize
		\begin{eqnarray}
			&& \frac{d\Bar{x}}{d\tau}=\frac{3}{2}(1-\Bar{x}^{2})^{\frac{3}{2}}\left[(\omega_{k}-\omega_{m})\left(\frac{\Bar{x}^{2}}{1-\Bar{x}^{2}}-\sgn(\beta)\right)+\left(1+\omega_{m}-\frac{\sigma\Bar{x}\sgn(\beta)}{\sqrt{1-\Bar{x}^{2}}}\right)\frac{\Bar{y}^{2}}{1-\Bar{y}^{2}}\right], \\
			&& \frac{d\Bar{y}}{d\tau}=\frac{3}{2}\Bar{y}(1-\Bar{y}^{2})\left[-\sigma +\frac{(\omega_{k}+1)\Bar{x}}{\sqrt{1-\Bar{x}^{2}}}-\frac{\sigma\Bar{y}^{2}}{1-\Bar{y}^{2}}\sgn(\beta)\right] \,.
		\end{eqnarray}
	\end{subequations}
	As in the previous case, to regularize the dynamical system  equation \eqref{dynsys_2d} the phase space-time variable redefined as
	\begin{equation}\label{time_redef_exp}
		d\tau \rightarrow d\Bar{\tau} = \frac{d\tau}{(1-\Bar{x}^{2})^{\frac{1}{2}}(1-\Bar{y}^{2})}.
	\end{equation}
	With respect to this redefined time variable, the dynamical system \eqref{dynsys_2d} can be rewritten as
	\begin{subequations}\label{dynsys_2d_new}
 \footnotesize
		\begin{eqnarray}
			\frac{d\Bar{x}}{d\Bar{\tau}} &=& \frac{3}{2}(\omega_{k}-\omega_{m})\left(\Bar{x}^{2}-(1-\Bar{x}^{2})\sgn(\beta)\right)(1-\Bar{x}^{2})(1-\Bar{y}^{2}) + \frac{3}{2}\bigg((1+\omega_{m})(1-\Bar{x}^{2})\nonumber\\
			&&  -\sigma\Bar{x}\sgn(\beta)\sqrt{1-\Bar{x}^{2}}\bigg)(1-\Bar{x}^{2})\Bar{y}^{2},\label{dynsys_2d_new_x}\\
			 \frac{d\Bar{y}}{d\Bar{\tau}} &=& \frac{3}{2}\Bar{y}(1-\Bar{y}^{2})\left[(-\sigma\sqrt{1-\Bar{x}^{2}} +(\omega_{k}+1)\Bar{x})(1-\Bar{y}^{2})-\sigma\Bar{y}^{2}\sqrt{1-\Bar{x}^{2}}\sgn(\beta)\right]. \label{dynsys_2d_new_y}
		\end{eqnarray}
	\end{subequations}
	\begin{table}[H]
		\begin{center}
			\begin{tabular}{|*{4}{c|}}
				\hline
				Point & Co-ordinate $(\Bar{x},\Bar{y})$ &  Existence &  \begin{tabular}{@{}c@{}}Physical viability \\ $(\Omega_{m}\geq0)$ \end{tabular}\\ \hline
				\parbox[c][0.6cm]{0.5cm}{$B_{1+}$} & $(1,1)$ & Always & Always \\ \hline
				\parbox[c][0.6cm]{0.5cm}{$B_{1-}$} & $(-1,1)$ & Always & Always \\ \hline
				\parbox[c][0.6cm]{0.5cm}{$B_{2+}$} & $(1,0)$ & Always & Always \\ \hline
				\parbox[c][0.6cm]{0.5cm}{$B_{2-}$} & $(-1,0)$ & Always & Always \\ \hline
				\parbox[c][1.2cm]{0.5cm}{$B_{3+}$} & $\left(\frac{1+\omega_{m}}{\sqrt{(1+\omega_{m})^{2}+\sigma^{2}}},1\right)$ & \begin{tabular}{@{}c@{}} $((1+\omega_{m})^{2}+\sigma^{2}>0)\wedge$\\ $\left((\beta<0)\wedge(\sigma\leq0)\right)\vee\left((\beta>0)\wedge(\sigma\geq0)\right)$ \end{tabular} & $\sigma=0$\\ \hline
				\parbox[c][1.2cm]{0.5cm}{$B_{3-}$} & $\left(-\frac{1+\omega_{m}}{\sqrt{(1+\omega_{m})^{2}+\sigma^{2}}},1\right)$ & \begin{tabular}{@{}c@{}}$((1+\omega_{m})^{2}+\sigma^{2}>0)\wedge$\\ $\left((\beta<0)\wedge(\sigma\geq0)\right)\vee\left((\beta>0)\wedge(\sigma\leq0)\right)$ \end{tabular} & $\sigma=0$ \\ \hline
			\end{tabular}
		\end{center}
		\caption[Existence and physical viability condition for fixed points at infinity for a scalar field with kinetic term $F(X)=\beta X$ and potential $V(\phi)=V_0 e^{-\lambda\phi/M_{Pl}}$.]{Existence and physical viability condition for fixed points at infinity for a scalar field with kinetic term $F(X)=\beta X$ and potential $V(\phi)=V_0 e^{-\lambda\phi/M_{Pl}}$, calculated from the system \eqref{dynsys_2d_new}. }
		\label{tab:infinite_fixed_pts_exp}
	\end{table}
	The constraints outlined in equations \eqref{physviab_can} and \eqref{physviab_noncan}, which define the physically viable region of the phase space for power-law potential, are also applicable to the exponential potential. This suggests that in order to achieve nonsingular bounces, we must solely concentrate on the phantom scalar field ($\beta<0$) in any subsequent analysis.
	
\begin{table}[H]
\begin{center}
			\begin{tabular}{|*{4}{c|}}
				\hline
				Point   & Co-ordinates $(\Bar{x},\Bar{y})$ & Stability & Cosmology \\ \hline
				\parbox[c][1.2cm]{0.5cm}{$B_{1+}$} & $(1,1)$ &  \begin{tabular}{@{}c@{}}stable for $\sigma=0$ \\or $~\{\sigma \neq 0, \sgn(\beta)\neq \sgn(\sigma)\}$ \\ \& saddle for \\ $\{\sigma \neq 0, \sgn(\beta)= \sgn(\sigma)\}$\end{tabular} & De Sitter\\ \hline
				\parbox[c][1.2cm]{0.5cm}{$B_{1-}$} & $(-1,1)$ & \begin{tabular}{@{}c@{}} unstable for $\sigma=0$ \\or $~\{\sigma \neq 0, \sgn(\beta)= \sgn(\sigma)\}$  \\ \& saddle for \\ $\{\sigma \neq 0, \sgn(\beta)\neq \sgn(\sigma)\}$\end{tabular} & De Sitter \\ \hline
				\parbox[c][0.8cm]{0.5cm}{$B_{2+}$}   & $(1,0)$ & saddle always & $a(t)=(t-t_{*})^{\frac{2}{3(1+\omega_{m})}}, ~t\geq t_{*}$\\ \hline
				\parbox[c][0.8cm]{0.5cm}{$B_{2-}$}   & $(-1,0)$ & saddle always & $a(t)=(t_{*}-t)^{\frac{2}{3(1+\omega_{m})}}, ~t\leq t_{*}$\\ \hline
			\end{tabular}
		\end{center}
		\caption[Stability condition of physically viable fixed points given in Table \ref{tab:infinite_fixed_pts_exp} along with their cosmological behavior.]{Stability condition of physically viable fixed points given in Table \ref{tab:infinite_fixed_pts_exp} along with their cosmological behavior. }
		\label{tab:stab_infinite_fixed_pts_exp}
	\end{table}
	
    The fixed points for the system \eqref{dynsys_2d_new} are presented in Table \ref{tab:infinite_fixed_pts_exp}, along with their corresponding stability conditions as shown in Table \ref{tab:stab_infinite_fixed_pts_exp}. Based on the information presented in Table \ref{tab:infinite_fixed_pts_exp}, it can be observed that there are three distinct pairs of fixed points at the infinity of the phase space. However, it is important to note that the pair denoted as $B_{3\pm}$ is only physically valid when $\sigma=0$, at which point it coincides with the pair denoted as $B_{1\pm}$. Hence, there exist only two distinct pairs of fixed points at infinity. The solutions denoted as $B_{1\pm}$ represent de-Sitter solutions, while $B_{2\pm}$ correspond to cosmological phases that are dominated by the hydrodynamic matter component. The fixed point $B_{2+}$ can be interpreted as the matter-dominated epoch when $\omega_m=0$. This fixed point is a saddle, which signifies an intermediate phase of evolution. It should be mentioned that at the limit $|\sigma|\rightarrow\infty$, $A_{2\pm}$ and $A_{3\pm}$ merge with $B_{1\pm}$.
    
    Compact 2-dimensional phase portraits are provided for various cases below. In all the plots the physically viable region as calculated by the constraints \eqref{physviab_can} or \eqref{physviab_noncan} is specified and the regions where the NEC and SEC are satisfied. The NEC and SEC are determined in terms of the ``effective'' EoS parameter defined in equation \eqref{eq:weff}.
	
	\begin{figure}[H]
		\centering
		\minipage{0.35\textwidth}
		\includegraphics[width=\textwidth]{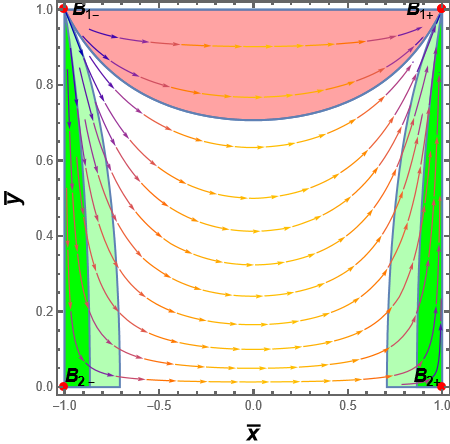}
		\endminipage
		\caption[Phase portrait for $F(X)=\beta X, V(\phi)=V_{0}e^{-\lambda \phi/M_{Pl}}$ with $\omega_{m}=0$, $\sigma=0$, and $\beta<0$.]{The phase space portrait for $\omega_{m}=0$ with $\sigma=0$, $\beta<0$ and $V_{0}>0$. Satisfaction of the NEC is shown in light green, the satisfaction of the SEC in dark green, and the red area denotes a non-physical part of the phase space.}\label{FIG.3}
	\end{figure}
 \begin{figure}[H]
     \centering
     \includegraphics[scale=0.6]{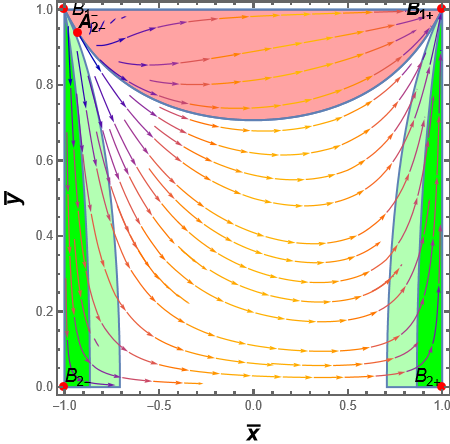}\hspace{1.5cm}
     \includegraphics[scale=0.6]{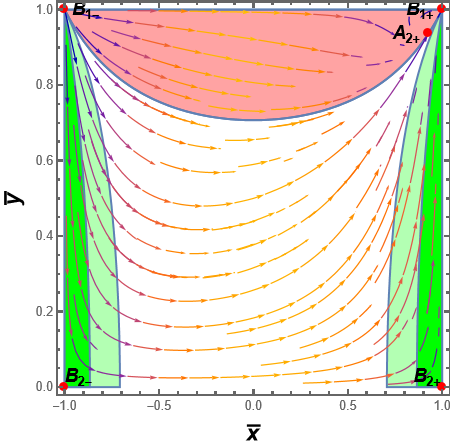}\vspace{0.8cm}
     \includegraphics[scale=0.6]{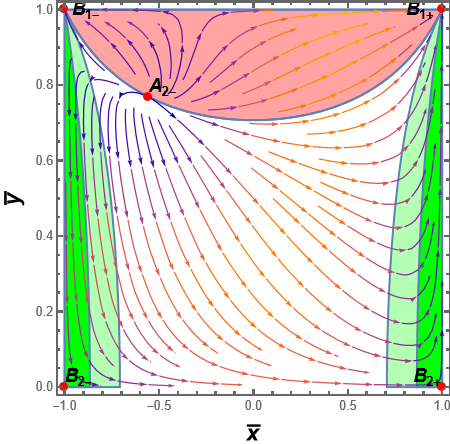}\hspace{1.5cm}
     \includegraphics[scale=0.6]{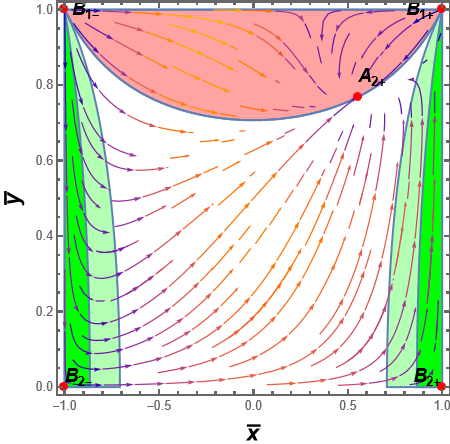}
     \caption[Phase portrait for $F(X)=\beta X, V(\phi)=V_{0}e^{-\lambda \phi/M_{Pl}}$ with $\omega_{m}=0$, $\beta<0$ moreover $\sigma=0.8$ (upper left panel) $\sigma=-0.8$ (upper right panel) $\sigma=3$ (lower left panel) $\sigma=-3$ (lower right panel).]{The phase space portrait for $\sigma=0.8$ (upper left panel) $\sigma=-0.8$ (upper right panel) $\sigma=3$ (lower left panel) $\sigma=-3$ (lower right panel) with $\omega_{m}=0$, $\beta<0$ and $V_{0}>0$. Satisfaction of the NEC is shown in light green, the satisfaction of the SEC in dark green, and the red area denotes a non-physical part of the phase space.}\label{FIG.4}
 \end{figure}
\begin{figure}[H]
		\centering
		\minipage{0.35\textwidth}
		\includegraphics[width=\textwidth]{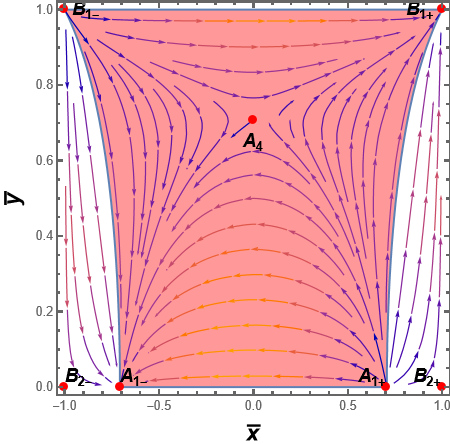}
		\endminipage
		\caption[Phase portrait for $F(X)=\beta X, V(\phi)=V_{0}e^{-\lambda \phi/M_{Pl}}$ with $\omega_{m}=0$, $\sigma=0$, and $\beta>0$.]{The phase space portrait for $\omega_{m}=0$ with $\sigma=0$, $\beta>0$ and $V_{0}>0$. The shaded region represents the non-physical viable region.}\label{FIG.5}
	\end{figure}

 It is observed from Figure \ref{FIG.3} and Figure \ref{FIG.4} that all phase trajectories inside the physically viable region exhibit a nonsingular bounce for a phantom scalar field ($\beta<0$). Nonsingular bounce is a generic feature in this case. In order to produce a bounce, the plot also confirms that the NEC and SEC have been violated. In Figure \ref{FIG.5}, the phase portrait for a specific example for $\beta>0$ demonstrates that all of the trajectories corresponding to nonsingular bounces fall within the physically non-viable region as predicted.     

    
\subsection{Specific case II: noncanonical scalar field \texorpdfstring{$F(X)=\beta X^{m}$ \quad ($m\neq1$)}{} }\label{subsec:general}

In this section, the scalar field Lagrangian is generalised to the noncanonical form
\begin{equation}
\mathcal{L}(\phi,X) = \beta X^m - V(\phi)\,,  \label{langragian_2}  
\end{equation}
 Since $\rho_{\phi}+p_{\phi}=2XF_X=2m\beta X^m$ and $X\geq0$, it corresponds to a non-phantom scalar field when $m\beta>0$ and to a phantom scalar field when $m\beta<0$. Again, one can notice from equation \eqref{eq:Raychaudhuri} that for $\dot{H}>0$ near the bounce one must necessarily need $m\beta<0$, i.e. a phantom scalar field. The case $m=1$ has been considered previously, we can concentrate on the case $m\neq1$ case.

 The analysis in this paper relies on the scalar field being a dynamical degree of freedom. The case $m=1/2$, i.e. $F(X)\propto\pm\sqrt{X}$ is a very special case for which the scalar field becomes non-dynamical in the homogeneous limit \cite{Afshordi:2006ad,Gomes:2017tzd}. In this case, the field is called Cuscuton. Since the field is nondynamical, violation of NEC during a bounce does not lead to a pathology \cite{Boruah:2018pvq}. This case is also left out of consideration in this paper. For more about Cuscustun bounce cosmology (See Ref. \cite{Afshordi:2007b,Bhattacharyya:2018,Iyonaga:2018}).
 
 One can calculate that, in this case,
\begin{equation}
  \rho_{k}=(2m-1)\beta X^{m}.  
\end{equation}
Using the same compact and non-compact dynamical variables as introduced earlier. The physical viability of the model requires the condition $\Omega_{m}\geq0$.  In terms of the non-compact dynamical variables $(x,y)$, it can be written as
 \begin{equation}
    \lbrace (x,y,\sigma) \in \mathbb{R}^3: x^2-y^2-\sgn((2m-1)\beta)\geq 0\rbrace\,.\label{viable:gen}
 \end{equation}
In terms of the compact dynamical variables $(\Bar{x},\Bar{y})$ defined in \eqref{3D_dynvar_comp_def}, one can write
    \begin{equation}
        \frac{\bar{x}^{2}}{1-\bar{x}^{2}}-\frac{\bar{y}^{2}}{1-\bar{y}^{2}}-\sgn{((2m-1)\beta)}=\Omega_{m}.
    \end{equation}
Note that the physical viability conditions for $m\neq1$ differ from that for $m=1$ only in the fact that $\sgn(\beta)$ is replaced by $\sgn((2m-1)\beta)$. When $\sgn((2m-1)\beta)=1$ one needs
\begin{equation}
        \frac{\bar{x}^{2}}{1-\bar{x}^{2}}-\frac{\bar{y}^{2}}{1-\bar{y}^{2}}-1=\Omega_{m}\geq 0.
\end{equation}
This leads to the necessary and sufficient conditions for the physical viability respectively as 
\begin{equation}
         \Bar{y}^{2}\leq 2-\frac{1}{\Bar{x}^{2}}, \qquad \Bar{x}^{2}\geq \frac{1}{2},   
\end{equation}
which are the same conditions obtained in equation \eqref{physviab_can}. When $\sgn((2m-1)\beta)=-1$ one needs
\begin{equation}
        \frac{\bar{x}^{2}}{1-\bar{x}^{2}}-\frac{\bar{y}^{2}}{1-\bar{y}^{2}}+1=\Omega_{m}\geq 0.
\end{equation}
This leads to the necessary and sufficient conditions as
\begin{equation}
         \Bar{y}^{2}\leq \frac{1}{2-\Bar{x}^{2}},  
\end{equation}
which is the same condition as obtained in equation \eqref{physviab_noncan}. Just like one could conclude for the case of $F(X)=\beta X$ that a physically viable nonsingular bounce requires $\beta<0$, one can conclude here that for the generic case $F(X)=\beta X^m$, a physically viable nonsingular bounce requires $(2m-1)\beta<0$. This implies the following parameter range
\begin{equation}
    \{(m>1/2)\wedge(\beta<0)\} \vee \{(m<1/2)\wedge(\beta>0)\}\,. 
\end{equation}
As we have already seen from \eqref{eq:Raychaudhuri} that achieving a nonsingular bounce necessarily requires a phantom scalar field, i.e. $m\beta<0$, together with the condition $(2m-1)\beta<0$ this slightly constrain the parameter range
\begin{equation}\label{param_space}
    \{(m>1/2)\wedge(\beta<0)\} \vee \{(m<0)\wedge(\beta>0)\}\,. 
\end{equation}
In particular, the parameter range $0\leq m\leq 1/2$ is not allowed if we want to achieve a nonsingular bounce.

\subsubsection{Power law potential \texorpdfstring{$V(\phi)=V_{0}\phi^{n}$}{}}\label{subsec:modelI_m}

 For the kinetic term $F(X)=\beta X^{m}$ and the potential $V(\phi)=V_{0}\phi^{n}$, we have
	\begin{equation}
		\Xi=m-1, \qquad \Gamma=1-\frac{1}{n}, \qquad \omega_{k}=\frac{1}{2m-1}, \qquad \rho_{k}=(2m-1)\beta X^{m}\,.
	\end{equation}

\paragraph{Finite fixed point analysis}:

In this case, the system \eqref{dynsys} reduces to
   \begin{subequations}\label{dynsys_pow_m}
   \footnotesize
		\begin{eqnarray}
		\frac{dx}{d\tau} &=& \frac{3}{2}x\left[\left(\frac{2m}{2m-1}\right)x - \sigma y^2 \sgn(V)\sgn[(2m-1)\beta]\right] - \frac{3}{2}\bigg[\left(\frac{1}{2m-1}-\omega_{m}\right)\sgn[(2m-1)\beta] \nonumber \\
            &&+ (1+\omega_{m})(x^{2}-y^2 \sgn(V))\bigg],\label{eq:ds_x_pow_m}  \\
		\frac{dy}{d\tau} &=& \frac{3}{2}y\left[-\sigma + \left(\frac{2m}{2m-1}\right)x - \sigma y^2 \sgn(V)\sgn[(2m-1)\beta]\right],\label{eq:ds_y_pow_m}\\
		\frac{d\sigma}{d\tau} &=& \frac{3}{n}\sigma^{2}+3\sigma\frac{(2m-3)m+1}{(4m-2)m}\left(\left(\frac{2m}{2m-1}\right)x-\sigma y^{2}\sgn(V)\right)\,.\label{eq:ds_s_pow_m}
		\end{eqnarray}
       \end{subequations}	
 The dynamical system is independent of the parameter $\lambda$. The above system reduces to the system \eqref{dynsys_pow} for $m=1$. As in the case of $m=1$, the system is symmetric under reflection around $y=0$, which happens to be an invariant submanifold. Therefore it suffices to consider only the part of the phase space given by
\begin{equation}
    \lbrace (x,y,\sigma) \in \mathbb{R}^3: y \geq 0, x^2-y^2-\sgn((2m-1)\beta)\geq 0\rbrace\,.\label{viable:gen_m}
 \end{equation}

\begin{table}[H]
		\begin{center}
			\begin{tabular}{|*{4}{c|}} 
				\hline
				Point & Co-ordinate $(x,y,\sigma)$ &  Existence &  \begin{tabular}{@{}c@{}}Physical viability \\ $(\Omega_{m}\geq0)$\end{tabular}\\ \hline
				\parbox[c][0.6cm]{0.5cm}{$A_{1+}$} & $(1,0,0)$ & $(2m-1)\beta>0\wedge m\neq0$ & Always \\ \hline
				\parbox[c][0.6cm]{0.5cm}{$A_{1-}$} & $(-1,0,0)$ & $(2m-1)\beta>0\wedge m\neq0$ & Always \\ \hline
			\end{tabular}
		\end{center}
		\caption[Existence and physical viability conditions for finite fixed points for a non-canonical scalar field with kinetic term $F(X)=\beta X^m$ ($m\neq 1$) and potential $V(\phi)=V_0 \phi^n$.]{Existence and physical viability conditions for finite fixed points for a non-canonical scalar field with kinetic term $F(X)=\beta X^m$ ($m\neq 1$) and potential $V(\phi)=V_0 \phi^n$, calculated from the system \eqref{dynsys_pow_m}.}
		\label{T:ex_pow_nc_m}
	\end{table} 
  The system \eqref{dynsys_pow_m} contains two finite fixed points $A_{1\pm}$, which exist only for non-phantom fields (see Table \ref{T:ex_pow_nc_m}). The stabilities and corresponding cosmologies are given in Table \ref{T:st_pow_nc_m}. Since these critical points exist only when $(2m-1)\beta>0$. When $(2m-1)\beta>0$, they cannot give rise to physically viable nonsingular bouncing trajectories.

 	\begin{table}[H]
		\begin{center}
			\begin{tabular}{|*{4}{c|}}
				\hline
				Point   & Co-ordinates $(x,y,\sigma)$ & Stability & Cosmology \\ \hline
				\parbox[c][1.6cm]{0.5cm}{$A_{1+}$} & $(1,0,0)$ &  \begin{tabular}{@{}c@{}} Stable for $0<m<\frac{1}{2}$ \&\\ unstable for $\frac{1}{2}<m\leq 1$\\ saddle otherwise\end{tabular} & $a(t)=(t-t_{*})^{\frac{2m-1}{3m}}, ~t\geq t_{*}$ \\ \hline
				\parbox[c][1.6cm]{0.5cm}{$A_{1-}$} & $(-1,0,0)$ & \begin{tabular}{@{}c@{}} Stable for $\frac{1}{2}<m<1$ \&\\ unstable for $0< m<\frac{1}{2}$\\ saddle otherwise\end{tabular} & $a(t)=(t_{*}-t)^{\frac{2m-1}{3m}}, ~t\leq t_{*}$ \\ \hline
			\end{tabular}
		\end{center}
		\caption[Stability condition of physically viable critical points given in Table \ref{T:ex_pow_nc_m} along with their cosmological behaviour.]{Stability condition of physically viable critical points given in Table \ref{T:ex_pow_nc_m} along with their cosmological behaviour.}
  \label{T:st_pow_nc_m}
	\end{table}
       
\begin{figure}[H]
		\centering
		\minipage{0.45\textwidth}
		\includegraphics[width=\textwidth]{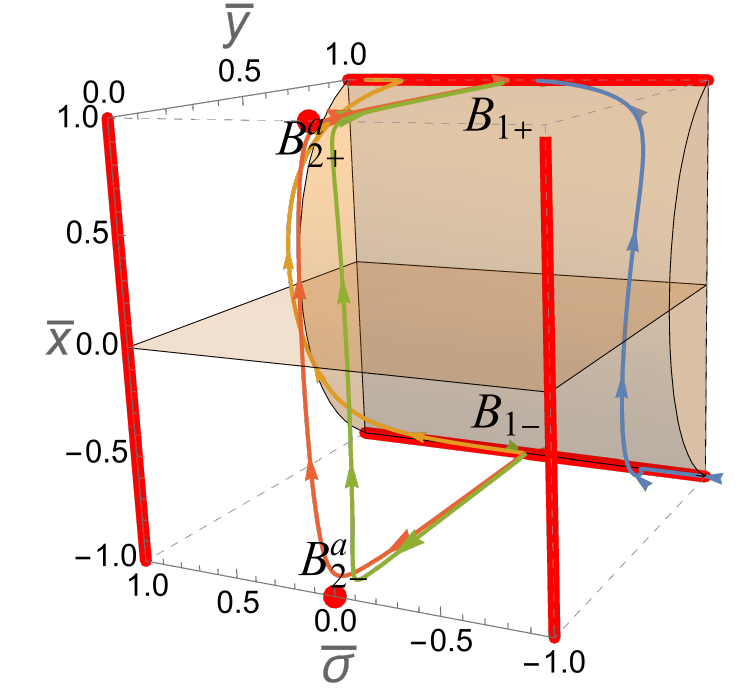}
		\endminipage
		\minipage{0.45\textwidth}
		\includegraphics[width=\textwidth]{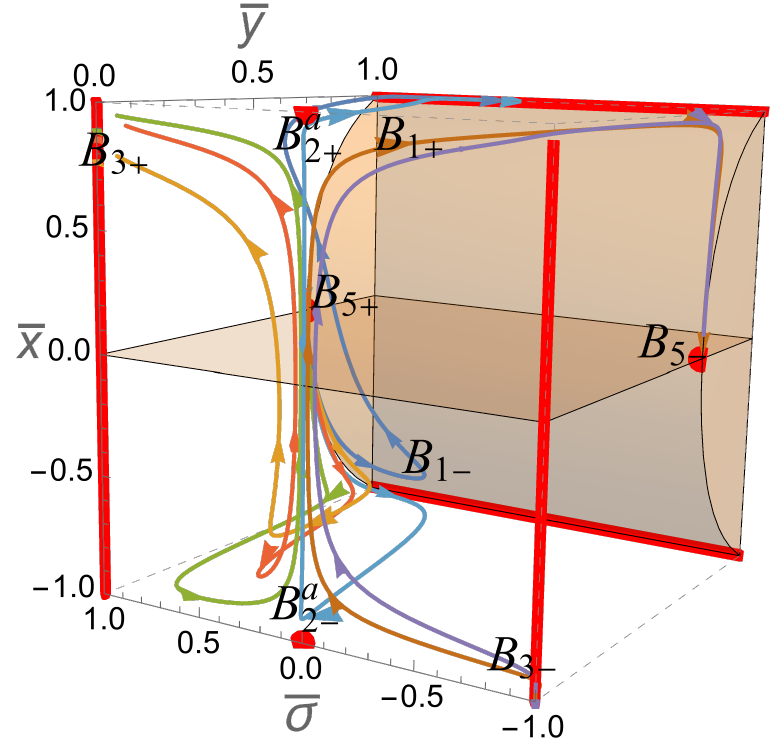}
		\endminipage
		\caption[Phase portrait for $F(X)=\beta X^{m},\, V(\phi)=V_0 \phi^n$ with $\omega_{m}=0$, the left plot is for $n=-5, m=\frac{2}{3}$ and the right plot is for $n=4, m=3$ moreover, $\sgn((2m-1)\beta)=-1$.]{Phase portrait of the system (\ref{dynsys_3d_pow_m}) with shaded region represents a non-physical region of the phase space. The left plot is for $n=-5, m=\frac{2}{3}$ and the right plot is for $n=4, m=3$ moreover, $\sgn((2m-1)\beta)=-1$.}
        \label{FIG.6}
	\end{figure}

\paragraph{Fixed points at infinity}:

In terms of the compact dynamical variables $(\Bar{x},\Bar{y},\Bar{\sigma})$ defined in equation \eqref{3D_dynvar_comp_def} and the redefined time variable defined in equation \eqref{time_redef_pow}, the dynamical system for $F(X)=\beta X^{m},\,V=V_0\phi^n$ can be rewritten in the following form 
\begin{subequations}\label{dynsys_3d_pow_m}
\footnotesize
\begin{eqnarray}
\frac{d\Bar{x}}{d\Bar{\tau}} &=& \frac{3}{2}\left(\frac{1}{2m-1}-\omega_{m}\right)\left(\Bar{x}^{2}-(1-\Bar{x}^{2})\sgn[(2m-1)\beta]\right) (1-\Bar{x}^{2})(1-\Bar{y}^{2})\sqrt{1-\Bar{\sigma}^2} \nonumber \\
&&+\frac{3}{2}\big((1+\omega_{m}) (1-\Bar{x}^{2}) \sqrt{1-\Bar{\sigma}^2}-\Bar{\sigma}\Bar{x}\sgn[(2m-1)\beta]\sqrt{1-\Bar{x}^{2}}\big)(1-\Bar{x}^{2})\Bar{y}^{2}, \label{dynsys_3d_pow_x_m}\\ 
\frac{d\Bar{y}}{d\Bar{\tau}}& =& \frac{3}{2}\Bar{y}(1-\Bar{y}^{2})\bigg[\left(-\Bar{\sigma}\sqrt{1-\Bar{x}^{2}}+\left(\frac{2m}{2m-1}\right)\Bar{x}\sqrt{1-\Bar{\sigma}^2}\right)(1-\Bar{y}^{2})\nonumber \\
&&-\Bar{\sigma}\Bar{y}^{2}\sqrt{1-\Bar{x}^{2}}\sgn[(2m-1)\beta]\bigg]\,, \label{dynsys_3d_pow_y_m}\\
\frac{d\Bar{\sigma}}{d\Bar{\tau}}&=&(1-\Bar{\sigma}^{2})\bigg[\frac{3}{n}\Bar{\sigma}^{2}\sqrt{1-\Bar{x}^{2}}(1-\Bar{y}^{2})+3\Bar{\sigma}\frac{(2m-3)m+1}{(4m-2)m}\bigg(\left(\frac{2m}{2m-1}\right)\Bar{x}\sqrt{1-\Bar{\sigma}^{2}}(1-\Bar{y}^{2})\nonumber \\
&&-\Bar{\sigma}\Bar{y}^{2}\sqrt{1-\Bar{x}^{2}}\bigg)\bigg]\,.\label{dynsys_3d_pow_s_m}
\end{eqnarray}
\end{subequations}	
The system \eqref{dynsys_3d_pow_m} presents six invariant submanifolds $\bar{x}=\pm1,\,\bar{y}=0,1$ and $\bar{\sigma}=\pm1$. Their stability is calculated in appendix \ref{app:stab_inv_sub_general}). The fixed points for the system \eqref{dynsys_3d_pow_m} are presented in Table \ref{T:ex_pow_com_m}. In the presence of pressureless dust, their stability conditions and corresponding cosmologies are presented in Table \ref{T:st_pow_com_m}. The lines of fixed points $B_{1\pm},\,B_{3\pm}$ and the isolated fixed points $B_{5\pm}$ are the same ones that were obtained earlier for the case $m=1$ (see Table \ref{tab:infinite_fixed_pts_pow}). However, unlike in the case of $m=1$, the entire lines $B_{2\pm}$ and $B_{4\pm}$ are not lines of fixed points when $m\neq1$. Instead, in this case, we only get two isolated fixed points $B^a_{2\pm}$ that lie on the line $B_{2\pm}$ respectively, and six isolated fixed points $B^{a,b,c}_{4\pm}$ that lie on the line $B_{4\pm}$ respectively. The points are listed in Table \ref{T:ex_pow_com_m}. Moreover, $B_{4\pm}^{a,b,c}$ are physically viable only for $\bar{\sigma}=0$, in which case they fall back into the lines of fixed points $B_{1\pm}$ respectively. Therefore at the infinity of the phase space there are only two lines of fixed points $B_{1\pm},\,B_{3\pm}$ and two pairs of isolated fixed points $B_{2\pm}^a$ and $B_{5\pm}$, whose nature of stability and the corresponding cosmology is listed in Table \ref{T:st_pow_com_m}. The cosmologies are the same as obtained for $m=1$ in the earlier case, as expected. Figure \ref{FIG.6} presents the 3D phase portrait of the system \eqref{dynsys_3d_pow_m} in the compact space for two cases, showing different types of physically viable nonsingular bouncing trajectories.
	\begin{table}[H]
		\begin{center}
			\begin{tabular}{|*{4}{c|}}
				\hline
				Point & Co-ordinate $(\Bar{x},\Bar{y},\Bar{\sigma})$ &  Existence &  \begin{tabular}{@{}c@{}}Physical viability \\$(\Omega_{m}\geq0)$\end{tabular}\\ \hline
				\parbox[c][0.6cm]{0.5cm}{$B_{1+}$} & $(1,1,\Bar{\sigma})$ & Always & Always \\ \hline
				\parbox[c][0.6cm]{0.5cm}{$B_{1-}$} & $(-1,1,\Bar{\sigma})$ & Always & Always \\ \hline         
                \parbox[c][0.6cm]{0.5cm}{$B_{2+}^{a}$} & $(1,0,0)$ & Always & Always \\ \hline 
                \parbox[c][0.6cm]{0.5cm}{$B_{2-}^{a}$} & $(-1,0,0)$ & Always & Always \\ \hline
				\parbox[c][1.2cm]{0.5cm}{$B_{3+}$} & $\left(\Bar{x},0,1\right)$ & Always & \begin{tabular}{@{}c@{}}$\frac{1}{2}\leq \Bar{x}^{2}\leq 1$\\ if $\sgn((2m-1)\beta)=1$ \\ $\Bar{x}^{2}\leq 1$\\ if $\sgn((2m-1)\beta)=-1$\end{tabular}\\ \hline
				\parbox[c][1.2cm]{0.5cm}{$B_{3-}$} & $\left(\Bar{x},0,-1\right)$ & Always & \begin{tabular}{@{}c@{}}$\frac{1}{2}\leq \Bar{x}^{2}\leq 1$\\ if $\sgn((2m-1)\beta)=1$ \\ $\Bar{x}^{2}\leq 1$\\ if $\sgn((2m-1)\beta)=-1$\end{tabular}\\ \hline
				\parbox[c][1.2cm]{0.5cm}{$B_{4+}^{a}$} & $\left(\frac{(1+\omega_{m})\sqrt{1-\Bar{\sigma}^2}}{\sqrt{(1+\omega_{{m}})^2-\Bar{\sigma}^2\omega_{m}(\omega_{m}+2)}},1,1\right)$ & \begin{tabular}{@{}c@{}}$\Bar{\sigma}^2\leq \frac{(1+\omega_{m})^2}{\omega_{m}(\omega_{m}+2)}$ \\ \end{tabular} &$\Bar{\sigma}=0$\\ \hline
                \parbox[c][1.2cm]{0.5cm}{$B_{4+}^{b}$} & $\left(\frac{(1+\omega_{m})\sqrt{1-\Bar{\sigma}^2}}{\sqrt{(1+\omega_{{m}})^2-\Bar{\sigma}^2\omega_{m}(\omega_{m}+2)}},1,-1\right)$ & \begin{tabular}{@{}c@{}}$\Bar{\sigma}^2\leq \frac{(1+\omega_{m})^2}{\omega_{m}(\omega_{m}+2)}$ \\ \end{tabular} &$\Bar{\sigma}=0$\\ \hline
				\parbox[c][1.2cm]{0.5cm}{$B_{4+}^{c}$} & $\left(\frac{(1+\omega_{m})\sqrt{1-\Bar{\sigma}^2}}{\sqrt{(1+\omega_{{m}})^2-\Bar{\sigma}^2\omega_{m}(\omega_{m}+2)}},1,0\right)$ & \begin{tabular}{@{}c@{}}$\Bar{\sigma}^2\leq \frac{(1+\omega_{m})^2}{\omega_{m}(\omega_{m}+2)}$ \\  \end{tabular} &$\Bar{\sigma}=0$\\ \hline
				\parbox[c][1.2cm]{0.5cm}{$B_{4-}^{a}$} & $\left(-\frac{(1+\omega_{m})\sqrt{1-\Bar{\sigma}^2}}{\sqrt{(1+\omega_{{m}})^2-\Bar{\sigma}^2\omega_{m}(\omega_{m}+2)}},1,1\right)$ & \begin{tabular}{@{}c@{}}$\Bar{\sigma}^2\leq \frac{(1+\omega_{m})^2}{\omega_{m}(\omega_{m}+2)}$ \\ \end{tabular} &$\Bar{\sigma}=0$\\ \hline
                \parbox[c][1.2cm]{0.5cm}{$B_{4-}^{b}$} & $\left(-\frac{(1+\omega_{m})\sqrt{1-\Bar{\sigma}^2}}{\sqrt{(1+\omega_{{m}})^2-\Bar{\sigma}^2\omega_{m}(\omega_{m}+2)}},1,-1\right)$ & \begin{tabular}{@{}c@{}}$\Bar{\sigma}^2\leq \frac{(1+\omega_{m})^2}{\omega_{m}(\omega_{m}+2)}$ \\ \end{tabular} &$\Bar{\sigma}=0$\\ \hline
				\parbox[c][1.2cm]{0.5cm}{$B_{4-}^{c}$} & $\left(-\frac{(1+\omega_{m})\sqrt{1-\Bar{\sigma}^2}}{\sqrt{(1+\omega_{{m}})^2-\Bar{\sigma}^2\omega_{m}(\omega_{m}+2)}},1,0\right)$ & \begin{tabular}{@{}c@{}}$\Bar{\sigma}^2\leq \frac{(1+\omega_{m})^2}{\omega_{m}(\omega_{m}+2)}$ \\  \end{tabular} &$\Bar{\sigma}=0$\\ \hline
				\parbox[c][1.2cm]{0.5cm}{$B_{5+}$} & $\left(0,\frac{1}{\sqrt{2}},1\right)$ & $(2m-1)\beta<0$ & Always \\ \hline
				\parbox[c][1.2cm]{0.5cm}{$B_{5-}$} &$\left(0,\frac{1}{\sqrt{2}},-1\right)$ & $(2m-1)\beta<0$ & Always \\ \hline
			\end{tabular}
		\end{center}
		\caption[Existence and physical viability condition for fixed points at infinity for a noncanonical scalar field with kinetic term $F(X)=\beta X^m$ ($m\neq 1$) and potential $V(\phi)=V_0 \phi^n$.]{Existence and physical viability condition for fixed points at infinity for a noncanonical scalar field with kinetic term $F(X)=\beta X^m$ ($m\neq 1$) and potential $V(\phi)=V_0 \phi^n$, calculated from the system (\ref{dynsys_3d_pow_m}). }
  \label{T:ex_pow_com_m}
	\end{table}
 \begin{table}[H]
		\begin{center}
			\begin{tabular}{|*{4}{c|}} 
				\hline
				Point   & Co-ordinates $(\Bar{x},\Bar{y},\Bar{\sigma})$ & Stability & Cosmology \\ \hline
				\parbox[c][1.6cm]{0.5cm}{$B_{1+}$} & $(1,1,\Bar{\sigma})$ & \begin{tabular}{@{}c@{}} Stable for \\
                $\left(m>\frac{1}{2}\right)\wedge\left(\sgn(\Bar{\sigma})\neq\sgn\beta\right)$\\ 
                or $\left(m>\frac{1}{2}\right)\wedge(\Bar{\sigma}=0)$ \\ saddle otherwise \end{tabular} & De Sitter\\ \hline
				\parbox[c][1.6cm]{0.5cm}{$B_{1-}$} &  $(-1,1,\Bar{\sigma})$& \begin{tabular}{@{}c@{}} Unstable for \\ $\left(m>\frac{1}{2}\right)\wedge\left(\sgn(\Bar{\sigma})=\sgn\beta\right)$ \\or $\left(m>\frac{1}{2}\right)\wedge(\Bar{\sigma}=0)$ \\ saddle otherwise \end{tabular} & De Sitter \\ \hline
                \parbox[c][1cm]{0.5cm}{$B_{2+}^{a}$} & $(1,0,0)$ & Saddle always & $a(t)=(t-t_{*})^{\frac{2}{3}}, ~t\geq t_{*}$ \\ \hline      \parbox[c][1cm]{0.5cm}{$B_{2-}^{a}$} & $(-1,0,0)$ & Saddle always & $a(t)=(t-t_{*})^{\frac{2}{3}}, ~t\leq t_{*}$ \\ \hline
				\parbox[c][1.2cm]{0.5cm}{$B_{3+}$} & $\left(\Bar{x},0,1\right)$ & \begin{tabular}{@{}c@{}} Stable for $n>0$ \\ saddle otherwise \end{tabular} & \begin{tabular}{@{}c@{}}Depending on $\Bar{x}$\\ and $\sgn(2m-1)\beta$ \end{tabular}\\ \hline
				\parbox[c][1.2cm]{0.5cm}{$B_{3-}$} & $\left(\Bar{x},0,-1\right)$ & \begin{tabular}{@{}c@{}} Unstable for $n>0$ \\ saddle otherwise \end{tabular} & \begin{tabular}{@{}c@{}}Depending on $\Bar{x}$\\ and $\sgn(2m-1)\beta$ \end{tabular}\\ \hline
				\parbox[c][1.2cm]{0.5cm}{$B_{5+}$} & $\left(0,\frac{1}{\sqrt{2}},1\right)$ & \begin{tabular}{@{}c@{}} Unstable for $\frac{3(mn-n-2m)}{2mn}\geq 0$ \\ saddle otherwise \end{tabular} & $a(t)=$ constant\\ \hline
				\parbox[c][1.2cm]{0.5cm}{$B_{5-}$} & $\left(0,\frac{1}{\sqrt{2}},-1\right)$ & \begin{tabular}{@{}c@{}} Stable for $\frac{3(mn-n-2m)}{2mn}\geq 0$ \\ saddle otherwise \end{tabular} & $a(t)=$ constant\\ \hline
			\end{tabular}
		\end{center}
		\caption[Stability condition and the cosmology of the fixed points given in Table \ref{T:ex_pow_com_m}.]{Stability condition and the cosmology of the fixed points given in Table \ref{T:ex_pow_com_m} in presence of pressureless dust $(\omega_m=0)$.} 
  \label{T:st_pow_com_m}
	\end{table}

The phase trajectories in the right panel of Figure \ref{FIG.6} shows nonsingular bouncing trajectories for the case $F(X)=\beta X^3\,(\beta<0),\,V(\phi)=V_{0}\phi^4$. Several different types of bouncing trajectories are possible, e.g. trajectories connecting a contracting de-Sitter phase to an expanding de-Sitter phase ($B_{1-}\rightarrow B_{1+}$), trajectories connecting a static and a de-Sitter phase ($B_{1-}\rightarrow B_{5-}$ and $B_{1+}\rightarrow B_{5+}$) and trajectories connecting $B_{3\pm}$ with $B_{1\pm}$ or $B_{5\pm}$. However, one cannot say this is the generic behavior of phase trajectories for $F(X)=\beta X^3\,(\beta<0),\,V(\phi)=V_{0}\phi^4$, as $B_{1-}$ and $B_{1+}$ are not global repellers and attractors. For $m=3,\,n=4$, there are two other attractor/repeller pairs $B_{3\pm}$ and $B_{5\pm}$. Therefore, there can exist heteroclinic trajectories connecting them, which may or may not correspond to a nonsingular bouncing cosmology, depending on how the trajectories evolve.

The phase trajectories in the left panel of Figure \ref{FIG.6}, which corresponds to $F(X)=\beta X^{2/3}\,(\beta<0),\,V(\phi)=V_{0}\phi^{-5}$, show nonsingular bouncing solutions connecting two de-Sitter phases ($B_{1-} \rightarrow B_{1+}$). For $m=\frac{2}{3},\,n=-5$, $B_{1+}$ and  $B_{1-}$ are the only attractor/repeller pair possible, i.e., they are global attractors and repellers. The trajectories connecting them must necessarily undergo a nonsingular bounce and it is shown explicitly in the figure four such characteristic trajectories. In this case, one can confidently say that a nonsingular bounce is a generic behavior of the phase trajectories.

 
 \subsubsection{Exponential potential \texorpdfstring{$V(\phi)=V_0 e^{-\lambda \phi/M_{Pl}}$}{}}\label{subsec:modelII_m}
 
 For the kinetic term $F(X)=\beta X^m$ and the potential $V(\phi)=V_0 e^{-\lambda \phi/M_{Pl}}$, we have
 \begin{equation}
 \Xi=m-1, \qquad \Gamma =1, \qquad \omega_{k}=\frac{1}{2m-1}, \qquad \rho_{k}=(2m-1)\beta X^{m}.    
 \end{equation}

\paragraph{Finite fixed point analysis}
 
In this case, the system \eqref{dynsys} reduces to
	 \begin{subequations}\label{dynsys_exp_m}
  \footnotesize
		\begin{eqnarray}
			 \frac{dx}{d\tau} &=& \frac{3x}{2}\left[\left(\frac{2m}{2m-1}\right)x - \sigma y^2\sgn[(2m-1)\beta]\right] - \frac{3}{2}\bigg[\left(\frac{1}{2m-1}-\omega_{m}\right)\sgn[(2m-1)\beta] \nonumber \\ 
            &&+ (1+\omega_{m})(x^{2}-y^2)\bigg],\label{eq:ds_x_exp_m}\\
		 \frac{dy}{d\tau} &=& \frac{3}{2}y\left[-\sigma + \left(\frac{2m}{2m-1}\right)x - \sigma y^2\sgn[(2m-1)\beta]\right],\label{eq:ds_y_exp_m}\\
		\frac{d\sigma}{d\tau} &=& 3\sigma\frac{(2m-3)m+1}{(4m-2)m}\left[\left(\frac{2m}{2m-1}\right)x-\sigma y^{2}\right]\,.\label{eq:ds_s_exp_m}
		\end{eqnarray}
       \end{subequations}
For $m=1$ the dynamical system for the exponential potential was 2D (\eqref{dynsys_exp}). For $m\neq1$, the dynamical system for the exponential potential becomes 3D. The dynamical system is independent of the parameter $\lambda$. The above system reduces to the system \eqref{dynsys_exp} for $m=1$. The system is symmetric under reflection around $y=0$, which happens to be an invariant submanifold. Therefore it suffices to consider only the part of the phase space given by
\begin{equation}
    \lbrace (x,y,\sigma) \in \mathbb{R}^3: y \geq 0, x^2-y^2-\sgn((2m-1)\beta)\geq 0\rbrace\,,
 \end{equation}
which corresponds to taking $V_0>0$.
	
	\begin{table}[H]
		\begin{center}
			\begin{tabular}{|*{4}{c|}}
				\hline
				Point & Co-ordinate $(x,y,\sigma)$ &  Existence &  Physical viability $(\Omega_{m}\geq0)$\\ \hline
				\parbox[c][0.6cm]{0.5cm}{$A_{1+}$} & $(1,0,0)$ & $(2m-1)\beta>0$ & Always \\ \hline
				\parbox[c][0.6cm]{0.5cm}{$A_{1-}$} & $(-1,0,0)$ & $(2m-1)\beta>0$ & Always \\ \hline
		\end{tabular}
		\end{center}
		\caption[Existence and the physical viability condition for finite fixed points for a noncanonical scalar field with kinetic term $F(X)=\beta X^m$ ($m\neq1$) and potential $V(\phi)=V_{0}e^{-\lambda\phi/M_{Pl}}$.]{Existence and the physical viability condition for finite fixed points for a noncanonical scalar field with kinetic term $F(X)=\beta X^m$ ($m\neq1$) and potential $V(\phi)=V_{0}e^{-\lambda\phi/M_{Pl}}$, calculated from the system (\ref{dynsys_exp_m}).}
  \label{T:ex_nc_m}
	\end{table}
	\begin{table}[H]
		\begin{center}
			\begin{tabular}{|*{4}{c|}} 
				\hline
				Point & Co-ordinate $(x,y,\sigma)$ &  Stability &  Cosmology \\ \hline
				\parbox[c][1cm]{0.5cm}{$A_{1+}$} & $(1,0,0)$ &\begin{tabular}{@{}c@{}} Unstable if $m> 1$\\ saddle otherwise \end{tabular}& $a(t)=(t-t_{*})^{\frac{2m-1}{3m}}, ~t\geq t_{*}$ \\ \hline
				\parbox[c][1cm]{0.5cm}{$A_{1-}$} & $(-1,0,0)$ &\begin{tabular}{@{}c@{}}Stable if $m>1$\\ saddle otherwise \end{tabular}& $a(t)=(t-t_{*})^{\frac{2m-1}{3m}}, ~t\leq t_{*}$ \\ \hline
				\end{tabular}
		\end{center}
		\caption[Stability and cosmological behavior of the physically viable fixed points given in Table \ref{T:ex_nc_m}.]{ Stability and cosmological behavior of the physically viable fixed points given in Table \ref{T:ex_nc_m}.}
    \label{T:st_exp_nc_m}
	\end{table}
 The system \eqref{dynsys_exp_m} contains two finite fixed points $A_{1\pm}$, which exist only for non-phantom fields (see Table \ref{T:ex_nc_m}). The stabilities and corresponding cosmologies are given in Table \ref{T:st_exp_nc_m}. Since these critical points exist only when $(2m-1)\beta>0$. When $(2m-1)\beta>0$, they cannot give rise to physically viable nonsingular bouncing trajectories. The finite fixed points $A_{2\pm},\,A_{3\pm}$ and $A_4$ that appeared in Table \ref{tab:finite_fixed_pts_exp}, are specific to the case $m=1$ and does not arise for $m\neq1$.

\paragraph{Fixed points at infinity}
In terms of the compact dynamical variables $(\Bar{x},\Bar{y},\Bar{\sigma})$ defined in equation \eqref{3D_dynvar_comp_def} and the redefined time variable defined in equation \eqref{time_redef_pow}, the dynamical system for $F(X)=\beta X^{m},\,V=V_0 e^{-\lambda\phi/M_{Pl}}$ can be rewritten in the following form 
\begin{subequations}\label{dynsys_3d_exp_m}
\footnotesize
\begin{eqnarray}
\frac{d\Bar{x}}{d\Bar{\tau}} &=& \frac{3}{2}\left(\frac{1}{2m-1}-\omega_{m}\right)\left(\Bar{x}^{2}-(1-\Bar{x}^{2})\sgn[(2m-1)\beta]\right) (1-\Bar{x}^{2})(1-\Bar{y}^{2})\sqrt{1-\Bar{\sigma}^2}\nonumber\\
&&+\frac{3}{2}\big((1+\omega_{m}) (1-\Bar{x}^{2}) \sqrt{1-\Bar{\sigma}^2}-\Bar{\sigma}\Bar{x}\sgn[(2m-1)\beta]\sqrt{1-\Bar{x}^{2}}\big)(1-\Bar{x}^{2})\Bar{y}^{2}, \label{dynsys_3d_exp_x_m}\\ 
\frac{d\Bar{y}}{d\Bar{\tau}}& =& \frac{3\Bar{y}(1-\Bar{y}^{2})}{2}\left[\left(-\Bar{\sigma}\sqrt{1-\Bar{x}^{2}}+\left(\frac{2m}{2m-1}\right)\Bar{x}\sqrt{1-\Bar{\sigma}^2}\right)(1-\Bar{y}^{2})-\Bar{\sigma}\Bar{y}^{2}\sqrt{1-\Bar{x}^{2}}\sgn[(2m-1)\beta]\right]\, \label{dynsys_3d_exp_y_m}\\ 
\frac{d\Bar{\sigma}}{d\Bar{\tau}}&=&3\Bar{\sigma}(1-\Bar{\sigma}^{2})
\left[\frac{(2m-3)m+1}{(4m-2)m}
\left(\left(\frac{2m}{2m-1}\right)\Bar{x}\sqrt{1-\Bar{\sigma}^{2}}(1-\Bar{y}^{2})-\Bar{\sigma}\Bar{y}^{2}\sqrt{1-\Bar{x}^{2}}\right)\right] \,.\label{dynsys_3d_exp_s_m}
\end{eqnarray}
\end{subequations}

\begin{figure}[H]
		\centering
		\minipage{0.45\textwidth}
		\includegraphics[width=\textwidth]{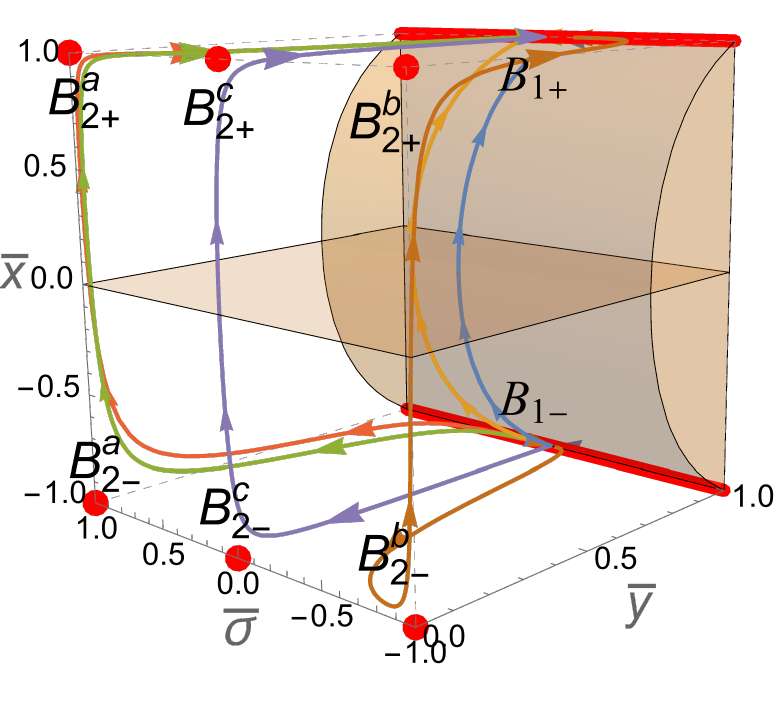}
		\endminipage
		\minipage{0.45\textwidth}
		\includegraphics[width=\textwidth]{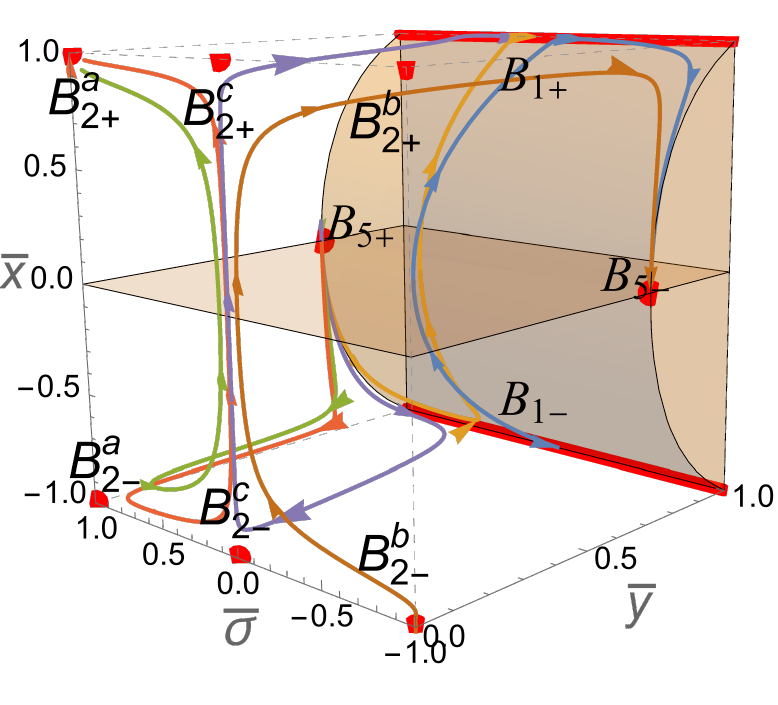}
		\endminipage
		\caption[Phase portrait for $F(X)=\beta X^{m},\,V(\phi)=V_0 e^{-\lambda\phi/M_{pl}}$ with $\omega_{m}=0$, for $m=2/3$ (left panel) and $m=3$ (right panel) moreover, $\sgn((2m-1)\beta)=-1$.]{Phase portrait of the system (\ref{dynsys_3d_exp_m}) with shaded region represents a non-physical region of the phase space. The plot is for $m=2/3$ (left panel) and $m=3$ (right panel) moreover, $\sgn((2m-1)\beta)=-1$.}
   \label{FIG. 7}
	\end{figure}

	\begin{table}[H]
		\begin{center}
			\begin{tabular}{|*{4}{c|}}
				\hline
				Point & Co-ordinate $(\Bar{x},\Bar{y},\Bar{\sigma})$ &  Existence &  \begin{tabular}{@{}c@{}}Physical viability \\$(\Omega_{m}\geq0)$\end{tabular}\\ \hline
				\parbox[c][0.5cm]{0.5cm}{$B_{1+}$} & $(1,1,\Bar{\sigma})$ & Always & Always \\ \hline
				\parbox[c][0.5cm]{0.5cm}{$B_{1-}$} & $(-1,1,\Bar{\sigma})$ & Always & Always \\ \hline
				\parbox[c][0.5cm]{0.5cm}{$B_{2+}^{a}$} & $(1,0,1)$ & Always & Always \\ \hline
                \parbox[c][0.5cm]{0.5cm}{$B_{2+}^{b}$} & $(1,0,-1)$ & Always & Always \\ \hline
				\parbox[c][0.5cm]{0.5cm}{$B_{2+}^{c}$} & $(1,0,0)$ & Always & Always \\ \hline
				\parbox[c][0.5cm]{0.5cm}{$B_{2-}^{a}$} & $(-1,0,1)$ & Always & Always \\ \hline
                \parbox[c][0.5cm]{0.5cm}{$B_{2-}^{b}$} & $(-1,0,-1)$ & Always & Always \\ \hline
				\parbox[c][0.5cm]{0.5cm}{$B_{2-}^{c}$} & $(-1,0,0)$ & Always & Always \\ \hline
				\parbox[c][1cm]{0.5cm}{$B_{3+}^{a}$} & $\left(\frac{(1+\omega_{m})\sqrt{1-\Bar{\sigma}^2}}{\sqrt{(1+\omega_{{m}})^2(1-\Bar{\sigma}^{2})+\Bar{\sigma}^2}},1,1\right)$ & \begin{tabular}{@{}c@{}} $(1+\omega_{{m}})^2(1-\Bar{\sigma}^{2})+\Bar{\sigma}^2\geq 0$\\ \end{tabular} & $\Bar{\sigma}=0$ \\ \hline
                \parbox[c][1cm]{0.5cm}{$B_{3+}^{b}$} & $\left(\frac{(1+\omega_{m})\sqrt{1-\Bar{\sigma}^2}}{\sqrt{(1+\omega_{{m}})^2(1-\Bar{\sigma}^{2})+\Bar{\sigma}^2}},1,-1\right)$ & \begin{tabular}{@{}c@{}} $(1+\omega_{{m}})^2(1-\Bar{\sigma}^{2})+\Bar{\sigma}^2\geq 0$ \\  \end{tabular} & $\Bar{\sigma}=0$ \\ \hline
				\parbox[c][1cm]{0.5cm}{$B_{3+}^{c}$} & $\left(\frac{(1+\omega_{m})\sqrt{1-\Bar{\sigma}^2}}{\sqrt{(1+\omega_{{m}})^2(1-\Bar{\sigma}^{2})+\Bar{\sigma}^2}},1,0\right)$ & \begin{tabular}{@{}c@{}} $(1+\omega_{{m}})^2(1-\Bar{\sigma}^{2})+\Bar{\sigma}^2\geq 0$\end{tabular} & $\Bar{\sigma}=0$ \\ \hline	\parbox[c][1cm]{0.5cm}{$B_{3-}^{a}$} & $\left(-\frac{(1+\omega_{m})\sqrt{1-\Bar{\sigma}^2}}{\sqrt{(1+\omega_{{m}})^2(1-\Bar{\sigma}^{2})+\Bar{\sigma}^2}},1,1\right)$ & \begin{tabular}{@{}c@{}} $(1+\omega_{{m}})^2(1-\Bar{\sigma}^{2})+\Bar{\sigma}^2\geq 0$ \\ \end{tabular} & $\Bar{\sigma}=0$ \\ \hline	
				\parbox[c][1cm]{0.5cm}{$B_{3-}^{b}$} & $\left(-\frac{(1+\omega_{m})\sqrt{1-\Bar{\sigma}^2}}{\sqrt{(1+\omega_{{m}})^2(1-\Bar{\sigma}^{2})+\Bar{\sigma}^2}},1,-1\right)$ & \begin{tabular}{@{}c@{}} $(1+\omega_{{m}})^2(1-\Bar{\sigma}^{2})+\Bar{\sigma}^2\geq 0$\\ \end{tabular} & $\Bar{\sigma}=0$ \\ \hline		
				\parbox[c][1cm]{0.5cm}{$B_{3-}^{c}$} & $\left(-\frac{(1+\omega_{m})\sqrt{1-\Bar{\sigma}^2}}{\sqrt{(1+\omega_{{m}})^2(1-\Bar{\sigma}^{2})+\Bar{\sigma}^2}},1,0\right)$ & \begin{tabular}{@{}c@{}} $(1+\omega_{{m}})^2(1-\Bar{\sigma}^{2})+\Bar{\sigma}^2\geq 0$\end{tabular} & $\Bar{\sigma}=0$ \\ \hline
                    \parbox[c][1.2cm]{0.5cm}{$B_{5+}$} & $\left(0,\frac{1}{\sqrt{2}},1\right)$ & $(2m-1)\beta<0$ & Always \\ \hline
				\parbox[c][1.2cm]{0.5cm}{$B_{5-}$} &$\left(0,\frac{1}{\sqrt{2}},-1\right)$ & $(2m-1)\beta<0$ & Always \\ \hline
				\end{tabular}
		\end{center}
		\caption[Existence and the physical viability condition for fixed points at infinity for a noncanonical scalar field with kinetic term $F(X)=\beta X^m$ ($m\neq1$) and $V(\phi)=V_{0}e^{-\lambda\phi/M_{Pl}}$.]{Existence and the physical viability condition for fixed points at infinity for a noncanonical scalar field with kinetic term $F(X)=\beta X^m$ ($m\neq1$) and $V(\phi)=V_{0}e^{-\lambda\phi/M_{Pl}}$, calculated from the system (\ref{dynsys_3d_exp_m}).}
  \label{T:ex_exp_com_m}
	\end{table}
The system \eqref{dynsys_3d_exp_m} presents six invariant submanifolds $\bar{x}=\pm1,\,\bar{y}=0,1$ and $\bar{\sigma}=\pm1$, same as in the power law case. Their stability is calculated in appendix \ref{app:stab_inv_sub_general}. The fixed points for the system \eqref{dynsys_3d_exp_m} are presented in Table \ref{T:ex_exp_com_m}. In the presence of pressureless dust, their stability conditions and corresponding cosmologies are presented in Table \ref{T:st_exp_com_m}. The six isolated fixed points $B_{3\pm}^{a,b,c}$ exist only for $\Bar{\sigma}=0$, for which they fall back on the lines of fixed points $B_{1\pm}$. The $\{x,y\}$ coordinates of the fixed points $B_{1\pm}$ and $B_{2\pm}^c$ are the same as the fixed points $B_{1\pm}$ and $B_{2\pm}$ of the 2D phase space for the case $m=1$ (see Table \ref{tab:infinite_fixed_pts_exp}). Since for the case $m\neq1$, $\sigma$ is a dynamical variable, the fixed points $B_{2\pm}^{a,b}$ and $B_{5\pm}$ obtained at the boundary $\sigma\rightarrow\pm\infty$. The corresponding point for the case $m=1$ does not arise because in that case, $\sigma$ was a parameter and is considered to be finite. In Figure \ref{FIG. 7} the 3D phase portrait of the system \eqref{dynsys_3d_exp_m} in the compact phase space for two cases is presented, showing different types of physically viable nonsingular bouncing trajectories. 
  
   	\begin{table}[H]
		\begin{center}
			\begin{tabular}{|*{4}{c|}} 
				\hline
				Point   & Co-ordinates $(\Bar{x},\Bar{y},\Bar{\sigma})$ & Stability & Cosmology \\ \hline
				\parbox[c][1.6cm]{0.5cm}{$B_{1+}$} & $(1,1,\Bar{\sigma})$ & \begin{tabular}{@{}c@{}} Stable for \\
                $\left(m>\frac{1}{2}\right)\wedge\left(\sgn(\Bar{\sigma})\neq\sgn\beta\right)$\\ 
                or $\left(m>\frac{1}{2}\right)\wedge(\Bar{\sigma}=0)$ \\ saddle otherwise \end{tabular}  & De Sitter\\ \hline
				\parbox[c][1.6cm]{0.5cm}{$B_{1-}$} &  $(-1,1,\Bar{\sigma})$ & \begin{tabular}{@{}c@{}} Unstable for \\ $\left(m>\frac{1}{2}\right)\wedge\left(\sgn(\Bar{\sigma})=\sgn\beta\right)$ \\or $\left(m>\frac{1}{2}\right)\wedge(\Bar{\sigma}=0)$ \\ saddle otherwise \end{tabular}  & De Sitter \\ \hline
		        \parbox[c][0.8cm]{0.5cm}{$B_{2+}^{a}$} & $(1,0,1)$ & Saddle always & $a(t)=(t-t_{*})^{\frac{2}{3}}, t\geq t_{*}$ \\ \hline
                \parbox[c][0.8cm]{0.5cm}{$B_{2+}^{b}$} & $(1,0,-1)$ & Saddle always & $a(t)=(t-t_{*})^{\frac{2}{3}}, t\geq t_{*}$ \\ \hline
				\parbox[c][0.8cm]{0.5cm}{$B_{2+}^{c}$} & $(1,0,0)$ & Saddle always & $a(t)=(t-t_{*})^{\frac{2}{3}}, t\geq t_{*}$ \\ \hline
				\parbox[c][0.8cm]{0.5cm}{$B_{2-}^{a}$} & $(-1,0,1)$ & Saddle always & $a(t)=(t-t_{*})^{\frac{2}{3}}, t\leq t_{*}$ \\ \hline
                \parbox[c][0.8cm]{0.5cm}{$B_{2-}^{b}$} & $(-1,0,-1)$ & Saddle always & $a(t)=(t-t_{*})^{\frac{2}{3}}, t\leq t_{*}$ \\ \hline
				\parbox[c][0.8cm]{0.5cm}{$B_{2-}^{c}$} & $(-1,0,0)$ & Saddle always & $a(t)=(t-t_{*})^{\frac{2}{3}}, t\leq t_{*}$ \\ \hline
                    \parbox[c][1.2cm]{0.5cm}{$B_{5+}$} & $\left(0,\frac{1}{\sqrt{2}},1\right)$ & \begin{tabular}{@{}c@{}} Unstable for $m<0$ or $m>1$ \\ saddle otherwise \end{tabular} & $a(t)=$ constant\\ \hline
				\parbox[c][1.2cm]{0.5cm}{$B_{5-}$} & $\left(0,\frac{1}{\sqrt{2}},-1\right)$ & \begin{tabular}{@{}c@{}} Stable for $m<0$ or $m>1$ \\ saddle otherwise \end{tabular} & $a(t)=$ constant\\ \hline
				\end{tabular}
		\end{center}
		\caption[Stability and the cosmological behavior of the physically viable fixed points defined in Table \ref{T:ex_exp_com_m}.]{Stability and the cosmological behavior of the physically viable fixed points defined in Table \ref{T:ex_exp_com_m}. Stability of the fixed points (or the line of fixed points) $B_{1\pm}$, $B_{2\pm}^{a,b,c}$ can be determined by investigating the stability of the invariant submanifolds $\Bar{x}=\pm1$ and $\Bar{y}=0,1$.}
  \label{T:st_exp_com_m}
	\end{table}
The phase trajectories in the right panel of Figure \ref{FIG. 7} shows nonsingular bouncing trajectories for the case $F(X)=\beta X^3\,(\beta<0),\,V(\phi)=V_0 e^{-\lambda\phi/M_{pl}}$. In this case, two types of bouncing trajectories are possible, e.g. trajectories connecting a contracting de-Sitter phase to an expanding de-Sitter phase ($B_{1-}\rightarrow B_{1+}$), trajectories connecting a static and a de-Sitter phase ($B_{1-}\rightarrow B_{5-}$ and $B_{5+}\rightarrow B_{1+}$). However, one cannot say this is the generic behavior of phase trajectories for $F(X)=\beta X^3\,(\beta<0),\,V(\phi)=V_0 e^{-\lambda\phi/M_{pl}}$, as $B_{1-}$ and $B_{1+}$ are not global repellers and attractors. For $m=3$, there are another attractor/repeller pair $B_{5\pm}$. Therefore, there can exist heteroclinic trajectories connecting them, which may or may not correspond to a nonsingular bouncing cosmology, depending on how the trajectories evolve.

The phase trajectories in the left panel of Figure \ref{FIG. 7}, which corresponds to $F(X)=\beta X^{2/3}\,(\beta<0),\,V(\phi)=V_0 e^{-\lambda\phi/M_{pl}}$, show nonsingular bouncing trajectories connecting two de-Sitter phases ($B_{1-} \rightarrow B_{1+}$). For $m=\frac{2}{3}$, $B_{1+}$ and  $B_{1-}$ are the only attractor/repeller pair possible, i.e., they are global attractors and repellers. The trajectories connecting them must necessarily undergo a nonsingular bounce and this is shown explicitly in figure five such characteristic trajectories. In this case, one can confidently say that a nonsingular bounce is a generic behavior of the phase trajectories.
	\section{Summary}\label{sec:summary}
	
	A careful analysis of the phase space structure of the models considered reveals some important physical aspects of the models. In this section, some of the main points of phase space analysis are summarized.
	\begin{itemize}
		\item \textbf{Matter dominated epoch:} At the fixed points $B_{2\pm}$, $(x,y)\rightarrow(\pm\infty,0)$ and one can write from equations \eqref{dynvar_def} and \eqref{constr}
		\begin{equation}
			\frac{\Omega_m}{x^2} = \frac{\rho_m}{3M_{Pl}^{2}H^2} \rightarrow 1.
		\end{equation}
		Therefore these two points are matter-dominated fixed points having a power law evolution (see Tables \ref{tab:stab_infinite_fixed_pts_pow}, \ref{tab:stab_infinite_fixed_pts_exp}, \ref{T:st_pow_com_m}, \ref{T:st_exp_com_m}). In all the cases considered above $B_{2\pm}$ are saddles. The saddle fixed point $B_{2+}$ can be interpreted as the intermediate matter-dominated epoch during the expanding phase of the universe.
  
		\item \textbf{Genericity of nonsingular bounces:} By genericity of nonsingular bouncing solutions, what it means is, no matter what initial state one may choose during the contracting phase (i.e. whatever phase space point one chooses in the region $\Bar{x}<0$), one always ends up with a nonsingular bounce. If there is a global repeller in the region $\Bar{x}<0$ and a global attractor in the region $\Bar{x}>0$, then all the heteroclinic trajectories must necessarily represent a nonsingular bounce and one can say that nonsingular bounce is generic. In this case, one can also say that the bouncing solutions are stable in both past and future directions. No perturbation of initial conditions alters the past and future asymptotic of the evolution.  
  
        Nonsingular bounces are only possible for phantom scalar fields within GR. When the kinetic term in the Lagrangian of the scalar field is $F(X)=\beta X$ ($\beta<0$), nonsingular bounce is generic for an exponential potential but not for a power law potential, as discussed in section \ref{subsec:modelI} and section \ref{subsec:modelII} and as is also clear from Figures \ref{fig:infinte_pow},\ref{FIG.3},\ref{FIG.4}. This is because the fixed points $B_{1\pm}$ are global attractors/repellers for an exponential potential but not for a power law potential. 
  
        When the kinetic term in the Lagrangian of the scalar field is $F(X)=\beta X^m$ with $\beta,\,m$ belonging to the range specified in \eqref{param_space}, nonsingular bounce is not generic for either power law or exponential potentials, as discussed in section \ref{subsec:modelI_m} and section \ref{subsec:modelII_m} and as is also clear from the examples presented in Figures \ref{FIG.6},\ref{FIG. 7}. This is because the fixed points $B_{1\pm}$ are not global attractors/repellers for either case. It is possible for heteroclinic trajectories to exist which do not undergo any bounce. However, there exist specific ranges for the model parameters for which nonsingular bounce is generic. For $F(X)=\beta X^m$ ($m\neq1$) and $V(\phi)=V_0 \phi^n$, when $\{m,\,n\}$ is within the range
        \begin{equation}
            \frac{1}{2} < m < 1\,, \qquad n < \frac{2m}{m-1}\,,
        \end{equation}
        the points $B_{3\pm}$ and $B_{5\pm}$ are saddles, and $B_{1\pm}$ becomes global attractors and repellers (see Table \ref{T:st_pow_com_m}). For $F(X)=\beta X^m$ ($m\neq1$) and $V(\phi)=V_0 e^{-\lambda\phi/M_Pl}$, when 
        \begin{equation}
            \frac{1}{2} < m < 1\,, 
        \end{equation}
        the points $B_{5\pm}$ are saddles, and $B_{1\pm}$ becomes global attractors and repellers (see Table \ref{T:st_exp_com_m}). In these cases, nonsingular bounce becomes generic, as can be seen in the examples presented in the left panels of Figures \ref{FIG.6} and \ref{FIG. 7}. 
        
		\item \textbf{Cosmic future when the bounce is generic:} It is interesting to investigate from the phase space the cosmic future of the nonsingular bouncing cosmologies when they are generic. 
        
        For $F(X)=\beta X$ ($\beta<0$) and an exponential potential, although mathematically the fixed points $A_{2\pm}$ are non-hyperbolic, from Figure \ref{FIG.4} one can see that they are actually global future and past attractors. The end state of a nonsingular bouncing cosmology is a future attractor which can be either a de-Sitter phase or a big-rip cosmology, depending on the choice of the model parameters. If $\sigma=\frac{\sqrt{{2}/{3}}\lambda}{\sqrt{-\beta}}\geq0$, the future attractor is a de-Sitter phase given by the point $B_{1+}$. In this case, the cosmology can be matched with the $\Lambda$CDM at the asymptotic future. If, however, $\sigma=\frac{\sqrt{{2}/{3}}\lambda}{\sqrt{-\beta}}<0$, then the future attractor is a big-rip singularity given by the point $A_{2+}$.

        On the contrary, when $F(X)=\beta X^m$ for either a power law or an exponential potential, and in cases when the bounce is generic, the only end state that is possible is the de-Sitter phases represented by the point $B_{1+}$. In these cases, the cosmology can be matched with $\Lambda$CDM at the asymptotic future.  
	\end{itemize}
	
	
	\section{Conclusion}\label{sec:concl}
	Nonsingular bouncing solutions are important candidates for early universe cosmology and it is necessary to investigate different aspects of them. This work deals with investigating them from the phase space point of view. For this analysis, nonsingular bouncing models in $F(X)-V(\phi)$ theory are considered, considering two simple choices for the potential, namely power law and exponential potential. The main motivation behind doing this exercise is to find how generic nonsingular bouncing solutions are. More precisely, even if for a theory a bouncing solution may exist, whether or not it arises from a large number of initial conditions. In the phase space picture, this question can be recast as to whether phase trajectories representing nonsingular bouncing solutions come from a large area of the phase space or only from some small patches. 
	
	For the purpose of dynamical system analysis, used the formulation of references \cite{De-Santiago:2012ibi, Panda:2015wya}. The formulation employs density-normalized dimensionless dynamical variables instead of the usual Hubble-normalized dynamical variables. This is because the Hubble normalized dynamical variables diverge at the bounce. The phase space analysis of \cite{De-Santiago:2012ibi} is extended by compactifying the phase space. Compactification of the phase space helps us visualize its global structure and answer questions regarding the past and future asymptotic of a cosmological model. In this case, the compact phase space analysis helps us investigate the genericity of solutions as well as answer questions about their past and future asymptotic dynamics.
	
	For both potentials, the existence of intermediate matter-dominated epochs which arise as saddle fixed points in the phase space is proved.  The result that for a nonsingular bounce to happen in an $F(X)$-$V(\phi)$ type scalar field Lagrangian in GR, the scalar field needs to be phantom is recovered.\footnote{If one goes beyond a scalar field with an $F(X)$-$V(\phi)$ type Lagrangians and GR, it is possible to achieve a nonsingular bounce even without non-phantom scalar fields e.g. ghost condensate or Galileon models}This result is also expected out of the fact that nonsingular bounce in spatially flat FLRW cosmology in GR requires violation of NEC \cite{Xue:2013iqy,Battefeld:2014uga}. In general, nonsingular bounce in these models is not generic due to the non-existence of global past or future attractors. However, the parameter range for which nonsingular bounce can be generic, for the kinetic term $F(X)=\beta X^m$ ($\beta<0$), these ranges are as follows:
    \begin{itemize}
        \item For a power law potential $V(\phi) = V_0 \phi^n$, the range is $\left\lbrace\frac{1}{2}<m<1,\,n<\frac{2m}{m-1}\right\rbrace$.
        \item For an exponential potential $V(\phi) = V_0 e^{-\lambda\phi/M_{Pl}}$, the range is $\left\lbrace\frac{1}{2}<m\leq1\right\rbrace$. 
    \end{itemize}
    When the model parameters lie outside this specific range, obtaining a bouncing solution is still feasible by carefully selecting a particular set of initial conditions and numerically evolving the system. However, in such cases, achieving a bounce may require fine-tuning of the initial conditions. Even a slight arbitrary alteration in the initial conditions may lead to a phase trajectory that does not exhibit a bounce.

Conversely, when the model parameters fall within the mentioned range, any arbitrary change in the initial conditions will still result in a bounce. In this scenario, there is no need for precise fine-tuning of the initial conditions to achieve the bouncing behavior.
    For the special case when $F(X) = \beta X$ ($\beta<0$) and $V(\phi) = V_0 e^{-\lambda\phi/M_{Pl}}$, the asymptotic future of the bouncing cosmology can be either de-Sitter or a big-rip. In all the other cases when the bounce is generic, the asymptotic future is definitely de-Sitter.

    Despite the challenges arising from inhomogeneous cosmological perturbations in the $F(X)$-$V(\phi)$ model, the model is chosen in a way to highlight the significance of exploring the genericity of bouncing solutions, specifically their stability concerning initial condition perturbations. The investigation employs the compact phase space formulation to address this question. It is asserted that, within the context of the theory being considered, cases where the bounce is generic hold greater interest than those where it is not. By examining the genericity of bounces, valuable insights into the stability and robustness of these solutions against variations in initial conditions are obtained. This knowledge is crucial in understanding bouncing cosmological models' overall behavior and viability. Furthermore, employing a compact phase space analysis enables us to explore the cosmic fate of nonsingular bouncing cosmologies when they are considered as generic solutions. The existing literature is replete with various nonsingular bouncing models, spanning both GR and modified gravity theories (for a comprehensive overview, refer to the reviews \cite{Battefeld:2014uga, Novello:2008ra}). Recent works \cite{Cai:2017dyi, Ilyas:2020qja, Vikman-2005} have also contributed to this intriguing area of research.
Given that a compact phase space analysis captures the system's global dynamics, it holds tremendous potential for studying theories that feature nonsingular bouncing solutions. It would be particularly captivating to explore the compact phase space for such theories, identifying cases where the bounce is generic and comprehending the cosmic future within these scenarios. These fascinating ideas remain as promising avenues for future projects, enriching our understanding of the broader implications of nonsingular bouncing cosmologies in the cosmos.

\chapter{Concluding Remarks and Future Perspectives} 

\label{Chapter7} 

\lhead{Chapter 7. \emph{Concluding Remarks and Future Perspectives}} 

\newpage
This chapter begins with an overview of the important findings that have been presented up to this point in the thesis, and then it moves on to discuss a wide variety of research.

The primary topic of this thesis centers around the phenomenon of the universe undergoing a bouncing scenario, which serves as an alternative to the standard cosmological model. The standard cosmological model has some problems with what happened in the early universe, and the bouncing cosmology is thought to be a possible solution. In the present thesis, the phenomenon of the universe undergoing a bouncing scenario has been extensively examined within several frameworks of modified gravity. These include modified curvature-based gravity models such as $f(R)$ and $f(R,T)$ gravity, as well as modified non-metricity-based gravity models such as $f(Q)$ and $f(Q,T)$ gravity. The possibility and the consequence of a bouncing scenario in modified gravity theories have been demonstrated in the preceding chapters.

In the preceding chapters, specifically in chapters 2, 3, and 5,  the occurrence of a bouncing scenario after a possible contraction phase of the universe has been studied within the frameworks of different geometrically modified gravity theories through the choice of various scale factors examined that exhibit bouncing behavior. Additionally, limitations are imposed on the scale factor parameters and model parameters to ensure the occurrence of a bounce. Bouncing cosmological models have been constructed and the dynamics of the universe are studied.  Different mathematical techniques have been employed to examine the stability of the models as well as the behavior of the dynamical parameters in the above-mentioned gravity theories. Based on the analysis of the cosmic Hubble radius, it has been observed that the matter bounce scenario is more effective compared to other bouncing scenarios. The matter bounce scenario is primarily studied because it demonstrates that the universe continues to evolve in a manner similar to a matter-dominated epoch, even in late times. It is possible to further develop this work in order to achieve a feasible bounce in the modified gravity theory. A different strategy was used in the context of extended symmetric teleparallel gravity in chapter 4, where a model was reconstructed to produce a scale factor that displayed a bouncing scenario. Dynamical system analysis has been used to study various states of the universe. The stability of the model has been investigated using the scalar perturbation methodology. The reconstructed model can be further studied to know more about the present universe.

A compact phase space analysis of a scalar field theory with a Lagrangian of the type $F(X)-V(\phi)$ is performed in Chapter 6. More specifically, a kinetic term of the type $F(X)=\beta X^{m}$ is considered with power law potential and the exponential potential of the scalar field. Nonsingular bouncing solutions are important options for early universe cosmology, and various aspects of them need to be studied. This work focuses on looking into them from a phase space perspective. The primary goal of this experiment was to determine the genericity of nonsingular bouncing solutions. It has been demonstrated that the occurrence of nonsingular bounce in these models is not a common phenomenon, mostly because there is a lack of global past or future attractors. Nevertheless, the range of parameters within which a nonsingular bounce can be considered a typical occurrence has been demonstrated. The ranges for the kinetic term $F(X)=\beta X^m$ ($\beta<0$) are as follows: The range of the power law potential $V(\phi) = V_0 \phi^n$ is given by the inequality $\left\lbrace\frac{1}{2}<m<1,\,n<\frac{2m}{m-1}\right\rbrace$. The range of the exponential potential $V(\phi) = V_0 e^{-\lambda\phi/M_{Pl}}$ is given by the interval $\left\lbrace\frac{1}{2}<m\leq1\right\rbrace$. 

The future prospects of the present work embodied in the thesis lie in the fact that one can take into consideration the current models with stable bouncing solutions to make the research interesting. The utilization of the Galileon field has been found to possess local stability. Since compact phase space analysis catches the global dynamic of the system, it would be more interesting to consider globally stable solutions. Such answers can be found outside of Horndeski's theory and in degenerate higher-order scalar-tensor (DHOST) theory in general. Furthermore, taking these models into account would support the application of the technique. The proposed method appears to be applicable to Lagrangians with involved DHOST terms. A case that we left out of consideration is the very special case of Cuscuton ($F(X)\propto\pm\sqrt{X}$). Phase space analysis of Cuscuton cosmology should be done with care as the Cuscuton field is nondynamical and provides an additional constraint. Since the field is nondynamical, it has been suggested that a bounce with a phantom Cuscuton does not lead to ghost instabilities \cite{Boruah:2018pvq}. It is therefore interesting to do the same exercise for for a Cuscuton bounce, something which we plan to address in a future work.




\appendix

\chapter[Appendix]{} 
\renewcommand{\theequation}{Appendix A.\arabic{equation}}
\setcounter{equation}{0}

  \section{Center manifold dynamics for point \texorpdfstring{$A_{1-}$}{} of model \texorpdfstring{$F(X)=\beta X, V(\phi)=V_0 \phi^n$}{}}\label{app:cmt}
		In this appendix, the center manifold theory to study the dynamics of the system \eqref{dynsys_pow} near a point $A_{1-}(-1,0,0)$. 
		
		Firstly, the point $(-1,0,0)$ is translated to $(0,0,0)$ via a transformation $x\rightarrow x-1$, $y\rightarrow y$, $\sigma \rightarrow \sigma$ under which
		the system \eqref{dynsys_pow} becomes
		\begin{subequations}\label{dynsys_pow_cmt}
  \footnotesize
			\begin{eqnarray}
				&& \frac{dx}{d\tau} = \frac{3}{2}(x-1)\left[2(x-1) - \sigma y^2 \right] - \frac{3}{2}\left[(1-\omega_{m}) + (1+\omega_{m})((x-1)^{2}-y^2 )\right],\label{eq:x_cmt}\\
				&& \frac{dy}{d\tau} = \frac{3}{2}y\left[-\sigma + 2(x-1) - \sigma y^2 \right],\label{eq:y_cmt}\\
				&& \frac{d\sigma}{d\tau} = \frac{3}{n}\sigma^{2}\,,\label{eq:s_cmt}
			\end{eqnarray}
		\end{subequations}
		which can be written as
		\[\left(\begin{array}{c}
			\frac{dx}{d\tau} \\
			\frac{dy}{d\tau}\\
			\frac{d\sigma}{d\tau} \end{array} \right)=\left(\begin{array}{ccc}
			-3(1-\omega_{m})  & 0 & 0  \\
			0  & -3& 0\\
			0  & 0 & 0   \end{array} \right) \left(\begin{array}{c}
			x\\
			y\\
			\sigma \end{array} \right)+\left(\begin{array}{c}
			g_1\\
			g_2\\
			f \end{array} \right)\,,\]
		where 
		\begin{eqnarray}
			g_1 &=& -\frac{3}{2}\,\sigma\,x{y}^{2}+\frac{3}{2}\,{x}^{2}+\frac{3}{2}\,\sigma\,{y}^{2}-\frac{3}{2}{x}^{2}\omega_{{m}}+\frac{3}{2}\,{y}^{2}\omega_{{m}}+\frac{3}{2}\,{y}^{2}\,,\nonumber\\
			g_2&=&-\frac{3}{2}\,\sigma\,{y}^{3}-\frac{3}{2}\sigma\,y+3\,xy\,,\nonumber\\
			f&=&\frac{3}{n}\sigma^{2}\,.\nonumber
		\end{eqnarray}
		
		The local center manifold is given by
		\begin{equation}
			\lbrace \mathbf{z}=\mathbf{h}(\sigma): \mathbf{h}(0)=\mathbf{0}, D\mathbf{h}(\mathbf{0})=\mathbf{0}\rbrace\,,
		\end{equation}
		where  
		\[\mathbf{h}=\left(\begin{array}{c}
			h_1\\
			h_2\ \end{array} \right)\,,\quad \mathbf{z}=\left(\begin{array}{c}
			x\\
			y\ \end{array} \right)\,.
		\] 
		The function $\mathbf{h}$  can be approximated   in terms of a power series as
		\begin{eqnarray}
			h_1(\sigma)&=&a_2 \sigma^2+a_3 \sigma^3+\mathcal{O}(\sigma^4),\\
			h_2(\sigma)&=&b_2 \sigma^2+b_3 \sigma^3+\mathcal{O}(\sigma^4)\,,
		\end{eqnarray}
		which is determined by the quasilinear partial differential equation \cite{Perko}
		\begin{equation}\label{quasi_C}
			D \mathbf{h}(\sigma)\left[A \sigma+\mathbf{F}(\sigma,\mathbf{h}(\sigma))\right]-B \mathbf{h}(\sigma)-\mathbf{g}(\sigma,\mathbf{h}(\sigma))=\mathbf{0} \,,
		\end{equation}
		with \[\mathbf{g}=\left(\begin{array}{c}
			g_1\\
			g_2 \end{array} \right),~~~~~ \mathbf{F}=f, ~~~~~B= \left(\begin{array}{cc}
			-3 (1-\omega_{m}) & 0 \\
			0 &-3 \end{array} \right),~~~~~ A=0. \]
		
		In order to solve the equation \eqref{quasi_C}, one can substitute $A$, $\textbf{h}$, $\mathbf{F}$, $B$, $\mathbf{g}$ in to it. On comparing like powers of $\sigma$ from equation \eqref{quasi_C} the constants can be obtained as $a_2=0$, $a_3=0$, $b_2=0$, $b_3=0$. Thus, the local center manifold is the $\sigma$-axis which coincides with the center subspace (a subspace generated by the eigenvectors corresponds to a vanishing eigenvalue of the corresponding Jacobian matrix).
		
		Finally, the dynamics in a local center manifold is determined by the equation
		\begin{equation}
			\frac{d\sigma}{d\tau} =A\,\sigma+\mathbf{F}(\sigma,\mathbf{h}(\sigma)),
		\end{equation}
		which is simply
		\begin{align}
			\frac{d\sigma}{d\tau} =\frac{3}{n}\sigma^2\,.
		\end{align}
		Hence, point $A_{1-}$ is always a saddle as an even-parity order term.
		
		\section{Stability at invariant sub-manifold}\label{app:stab_inv_sub}
		
		It is important to investigate the stability of invariant submanifolds because that helps us infer the nature of the stability of the fixed points that lie at the intersection of the invariant submanifolds. If a fixed point is located at the intersection of N-invariant submanifolds in an N-dimensional phase space, then the point is
		\begin{itemize}
			\item Stable if all the invariant submanifolds are of attracting nature.
			\item Unstable if all the invariant submanifolds are of repelling nature.
			\item Saddle if some of the invariant submanifolds are attracting and some are repelling.
		\end{itemize}
		For model II considered in section \ref{subsec:modelII}, the fixed points $A_{1\pm}$ are at the intersection of the invariant submanifolds $y=0,\,\Omega=0$. For both the models, the fixed points (or the lines of fixed points) $B_{1\pm}$ are at the intersection of the invariant submanifolds $\Bar{x}=\pm1,\,\Bar{y}=1$ and the fixed points $B_{2\pm}$ are at the intersection of the invariant submanifolds $\Bar{x}=\pm1,\,\Bar{y}=0$.
		
		Below is the stability of different invariant submanifolds have been investigated
		
		\subsection{\texorpdfstring{$\Bar{x}=\pm 1$}{}}
		
		Consider $\Bar{x}$ in the vicinity of $+1$ or $-1$, i.e. $0<\epsilon\equiv(1-\Bar{x}^{2})\ll1$. One can then rewrite equation \eqref{dynsys_2d_new_x} as
		\begin{equation}
			\frac{d\Bar{x}}{d\Bar{\tau}} = \frac{3}{2}(\omega_{k}-\omega_{m})\left(1-\epsilon-\epsilon \sgn(\beta)\right)\epsilon(1-\Bar{y}^{2}) + \frac{3}{2}\left((1+\omega_{m})\epsilon-\sigma\Bar{x}\sgn(\beta)\sqrt{\epsilon}\right)\epsilon\Bar{y}^{2}.   
		\end{equation}
		Since $\omega_{k}=1$, to the lowest order of $\epsilon$ one can write
		\begin{eqnarray}
			\frac{d\Bar{x}}{d\Bar{\tau}} \simeq \frac{3}{2}(1-\omega_{m})(1-\Bar{y}^{2})\epsilon.
		\end{eqnarray}
		Since all quantities on the right-hand side are positive. The flow will always be in positive $x$-direction in the vicinity of $\Bar{x}=\pm1$. Therefore the invariant submanifold $\Bar{x}=1$ is attracting while the invariant submanifold $\Bar{x}=-1$ is repelling. 
		
		
		\subsection{\texorpdfstring{$\Bar{y}=1$}{}}
		
		Consider $\Bar{y}$ in the vicinity of $1$, i.e. $0<\epsilon\equiv(1-\Bar{y}^{2})\ll1$. The cases $\sigma=0$ and $\sigma\neq0$ have been considered separately.
		
		
		\subsubsection{\texorpdfstring{$\sigma=0$}{}}
		
		One can rewrite equation \eqref{dynsys_2d_new_y} as
		\begin{equation}
			\frac{d\Bar{y}}{d\Bar{\tau}} = \frac{3}{2}\Bar{y}(1-\Bar{y}^{2})^{2}(\omega_{k}+1)\Bar{x} = \frac{3}{2}(\omega_{k}+1)\Bar{x}\epsilon^{2}\sqrt{1-\epsilon},   
		\end{equation}
		Since $\omega_k=1$, to the lowest order of $\epsilon$ one can write
		\begin{equation}
			\frac{d\Bar{y}}{d\Bar{\tau}} \simeq 3\Bar{x}\epsilon^{2}.
		\end{equation}
		From the above one can conclude that
		\begin{itemize}
			\item $\frac{d\Bar{y}}{d\Bar{\tau}}>0$ in the vicinity of $\Bar{y}=1$ i.e. the invariant submanifold $\Bar{y}=1$ is attracting for $\Bar{x}>0$.
			\item $\frac{d\Bar{y}}{d\Bar{\tau}}<0$ in the vicinity of $\Bar{y}=1$ i.e. the invariant submanifold $\Bar{y}=1$ is repelling for $\Bar{x}<0$.
		\end{itemize}
		
		
		\subsubsection{\texorpdfstring{$\sigma\neq0$}{}}
		
		One can rewrite equation \eqref{dynsys_2d_new_y} as
		\begin{eqnarray}
			\frac{d\Bar{y}}{d\Bar{\tau}} &=& \frac{3}{2}\Bar{y}(1-\Bar{y}^{2})\left[(-\sigma\sqrt{1-\Bar{x}^{2}} +(\omega_{k}+1)\Bar{x})(1-\Bar{y}^{2}) - \sigma\Bar{y}^{2}\sqrt{1-\Bar{x}^{2}}\sgn(\beta)\right], \nonumber\\
			&=& \frac{3}{2}\epsilon\sqrt{1-\epsilon}\left[(-\sigma\sqrt{1-\Bar{x}^{2}}+(\omega_{k}+1)\Bar{x})\epsilon - \sigma(1-\epsilon)\sqrt{1-\Bar{x}^{2}}\sgn(\beta)\right].\nonumber\\
            &&
		\end{eqnarray}
		Since $\omega_k=1$, to the lowest order of $\epsilon$ one can write
		\begin{equation}
			\frac{d\Bar{y}}{d\Bar{\tau}} \simeq -\frac{3}{2}\epsilon\sigma\sqrt{1-\Bar{x}^{2}}\sgn(\beta).   
		\end{equation}
		From the above one can conclude that
		\begin{itemize}
			\item $\frac{d\Bar{y}}{d\Bar{\tau}}>0$ in the vicinity of $\Bar{y}=1$ i.e. the invariant submanifold $\Bar{y}=1$ is attracting if $\sgn(\beta)\neq \sgn(\sigma)$.
			\item $\frac{d\Bar{y}}{d\Bar{\tau}}<0$ in the vicinity of $\Bar{y}=1$ i.e. the invariant submanifold $\Bar{y}=1$ is repelling if $\sgn(\beta)=\sgn(\sigma)$.
		\end{itemize}
		
		
		\subsection{\texorpdfstring{$\Bar{y}=0$}{}}
		
		Consider $\Bar{y}$ in the vicinity of $\Bar{y}=0$, i.e. $0<\Bar{y}\ll1$. The cases $\sigma=0$, $\sigma>0$ and $\sigma<0$ are considered separately.
		
		\subsubsection{\texorpdfstring{$\sigma=0$}{}}
		
		One can rewrite equation \eqref{dynsys_2d_new_y} as
		\begin{equation}
			\frac{d\Bar{y}}{d\Bar{\tau}} = \frac{3}{2}\Bar{y}(1-\Bar{y}^{2})^{2}(\omega_{k}+1)\Bar{x}.   
		\end{equation}
		Since $\omega_k=1$, to leading order of $\Bar{y}$ one can write
		\begin{equation}
			\frac{d\Bar{y}}{d\Bar{\tau}} \simeq 3\Bar{x}\Bar{y}.   
		\end{equation}
		From the above one can conclude that
		\begin{itemize}
			\item $\frac{d\Bar{y}}{d\Bar{\tau}}>0$ in the vicinity of $\Bar{y}=0$ i.e. the invariant submanifold $\Bar{y}=0$ is repelling for $\Bar{x}>0$.
			\item $\frac{d\Bar{y}}{d\Bar{\tau}}<0$ in the vicinity of $\Bar{y}=0$ i.e. the invariant submanifold $\Bar{y}=0$ is attracting for $\Bar{x}<0$.
		\end{itemize}
		
		
		\subsubsection{\texorpdfstring{$\sigma\neq0$}{}}
		
		One can rewrite equation \eqref{dynsys_2d_new_y} as
		\begin{equation}
			\frac{d\Bar{y}}{d\Bar{\tau}} = \frac{3}{2}\Bar{y}(1-\Bar{y}^{2})\left[(-\sigma\sqrt{1-\Bar{x}^{2}} +(\omega_{k}+1)\Bar{x})(1-\Bar{y}^{2}) - \sigma\Bar{y}^{2}\sqrt{1-\Bar{x}^{2}}\sgn(\beta)\right].    
		\end{equation}
		Since $\omega_k=1$, to the leading order of $\Bar{y}$ one can write
		\begin{equation}
			\frac{d\Bar{y}}{d\Bar{\tau}} \simeq \frac{3}{2}\Bar{y}\left[ -\sigma \sqrt{1-\Bar{x}^{2}}+2\Bar{x} \right].   
		\end{equation}
		From the above one can conclude that
		\begin{itemize}
			\item $\frac{d\Bar{y}}{d\Bar{\tau}}>0$ in the vicinity of $\Bar{y}=0$ i.e. the invariant submanifold $\Bar{y}=0$ is repelling for $\Bar{x}>\frac{\sigma}{\sqrt{4+\sigma^{2}}}$.
			\item $\frac{d\Bar{y}}{d\Bar{\tau}}<0$ in the vicinity of $\Bar{y}=0$ i.e. the invariant submanifold $\Bar{y}=0$ is attracting for $\Bar{x}<\frac{\sigma}{\sqrt{4+\sigma^{2}}}$. 
		\end{itemize}
		
		
		
		
		
		\subsection{\texorpdfstring{$\Omega_{m}=0$}{}}
		
		Apart from $\Bar{x}=\pm1$ and $\Bar{y}=0,1$, there is another invariant submanifold given by $\Omega_m=0$. The existence of this invariant submanifold is not apparent from the dynamical system \eqref{dynsys_2d_new}. One could have guessed the existence of this submanifold from the physical argument that, if cosmology is initially a vacuum, it remains so as there is no mechanism for matter creation in classical physics. That $\Omega_m=0$ is an invariant submanifold can be explicitly shown if one tries to write a dynamical equation for $\Omega_m$ in terms of $\Bar{\tau}$ using the dynamical equations \eqref{dynsys} and the definitions \eqref{constr_comp}, \eqref{time_redef_exp}.
		\begin{equation}
			\frac{d\Omega_{m}}{d\Bar{\tau}}=3\Omega_{m}[(1-\omega_{m})\Bar{x}(1-\Bar{y}^{2})-\sigma \Bar{y}^{2}\sqrt{1-\Bar{x}^{2}}\sgn(\beta)] .   
		\end{equation}
		From the above one can conclude that the invariant submanifold $\Omega_{m}=0$ is
		\begin{itemize}
			\item attracting for $(1-\omega_{m})\Bar{x}(1-\Bar{y}^{2})-\sigma \Bar{y}^{2}\sqrt{1-\Bar{x}^{2}}\sgn(\beta)<0$.
			\item repelling for $(1-\omega_{m})\Bar{x}(1-\Bar{y}^{2})-\sigma \Bar{y}^{2}\sqrt{1-\Bar{x}^{2}}\sgn(\beta)>0$.
		\end{itemize}

	\section{Stability at invariant sub-manifold of \texorpdfstring{$F(X)=\beta X^{m}$}{}}\label{app:stab_inv_sub_general}	
 
 As we have seen the stability through the invariant submanifold, the fixed points (or the lines of fixed points) $B_{1\pm}$ are at the intersection of the invariant submanifolds $\Bar{x}=\pm1,\,\Bar{y}=1$, the fixed points $B_{2\pm}^{a},\, B_{2\pm}^{b},\, B_{2\pm}^{c}$ are at the intersection of the invariant submanifolds $\Bar{x}=\pm1,\,\Bar{y}=0$ and the fixed points $B_{3\pm}$ are at the intersection of the invariant submanifolds $\Bar{y}=0,\,\Bar{\sigma}=\pm 1$.
\subsection{\texorpdfstring{$\Bar{x}=\pm 1$}{}}
Consider $\Bar{x}$ in the vicinity of $+1$ or $-1$, i.e. $0<\epsilon\equiv (1-\Bar{x}^{2})\ll 1$.
\begin{equation}
 \frac{d\Bar{x}}{d\Bar{\tau}}=\frac{3}{2}\epsilon\left(\frac{1}{2m-1}-\omega_{m}\right)(1-\Bar{y}^{2})\sqrt{1-\Bar{\sigma}^{2}}   
\end{equation}
Since all the quantities $\epsilon,(1-\Bar{y}^{2}), \sqrt{1-\Bar{\sigma}^{2}}$ are positive in the right hand side. If $\omega_{m}=0$, the quantity $\frac{1}{2m-1}$ will decide the attracting or repelling behaviour of the $\Bar{x}=\pm 1$.

\begin{itemize}
\item $\frac{d\Bar{x}}{d\Bar{\tau}}>0$ if $m>\frac{1}{2}$ the invariant submanifold $\Bar{x}=1$ is attracting and the invariant submanifold $\Bar{x}=-1$ is repelling.

\item $\frac{d\Bar{x}}{d\Bar{\tau}}<0$ if $m<\frac{1}{2}$ the invariant submanifold $\Bar{x}=1$ is repelling and the invariant submanifold $\Bar{x}=-1$ is attracting.

\end{itemize}
\subsection{\texorpdfstring{$\Bar{y}= 1$}{}}
\subsubsection{\texorpdfstring{$\Bar{\sigma}\neq 0$}{}}
Consider $\Bar{y}$ in the vicinity of 1, i.e. $0<\epsilon\equiv (1-\Bar{y}^{2})\ll 1$.
\begin{equation}
\frac{d\Bar{y}}{d\Bar{\tau}}=-\frac{3}{2}\epsilon \Bar{\sigma}\sqrt{1-\Bar{x}^{2}}\sgn((2m-1)\beta)    
\end{equation}
\begin{itemize}
    \item $\frac{d\Bar{y}}{d\Bar{\tau}}>0$ in the vicinity of $\Bar{y}=1$ i.e. the invariant submanifold $\Bar{y}=1$ is attracting if $\sgn((2m-1)\beta)\neq \sgn(\Bar{\sigma})$.
    \item $\frac{d\Bar{y}}{d\Bar{\tau}}<0$ in the vicinity of $\Bar{y}=1$ i.e. the invariant submanifold $\Bar{y}=1$ is repelling if $\sgn((2m-1)\beta)= \sgn(\Bar{\sigma})$.
\end{itemize}
\subsubsection{\texorpdfstring{$\Bar{\sigma}= 0$}{}}
\begin{equation}
 \frac{d\Bar{y}}{d\Bar{\tau}}= \left(\frac{3m}{2m-1}\right)\bar{x}\epsilon^{2} 
\end{equation}
\begin{itemize}
    \item $\frac{d\Bar{y}}{d\Bar{\tau}}>0$ in the vicinity of $\Bar{y}=1$ i.e. the invariant submanifold $\Bar{y}=1$ is attracting if $(\Bar{x}>0)\wedge\left((m<0)\vee(m>\frac{1}{2})\right)$ or $(\Bar{x}<0)\wedge(0<m<\frac{1}{2})$.
    \item $\frac{d\Bar{y}}{d\Bar{\tau}}<0$ in the vicinity of $\Bar{y}=1$ i.e. the invariant submanifold $\Bar{y}=1$ is repelling if $(\Bar{x}<0)\wedge\left((m<0)\vee(m>\frac{1}{2})\right)$ or $(\Bar{x}>0)\wedge(0<m<\frac{1}{2})$.
\end{itemize}

\subsection{\texorpdfstring{$\Bar{y}=0$}{}}
\subsubsection{\texorpdfstring{$\Bar{\sigma}\neq 0$}{}}
Consider $\Bar{y}$ in the vicinity of 0, i.e. $0<\bar{y}\ll 1$.
\begin{equation}
 \frac{d\Bar{y}}{d\Bar{\tau}}= \frac{3}{2}\Bar{y}\left(-\Bar{\sigma}\sqrt{1-\bar{x}^{2}}+\left(\frac{2m}{2m-1}\right)\bar{x}\sqrt{1-\bar{\sigma}^{2}}\right)  
\end{equation}
\begin{itemize}
\item $\frac{d\Bar{y}}{d\Bar{\tau}}>0$ in the vicinity of $\Bar{y}=0$ i.e. the invariant submanifold $\Bar{y}=0$ is repelling if $\Bar{\sigma}=-1$.

\item $\frac{d\Bar{y}}{d\Bar{\tau}}<0$ in the vicinity of $\Bar{y}=0$ i.e. the invariant submanifold $\Bar{y}=0$ is attracting if $\Bar{\sigma}=1$.

\item $\frac{d\Bar{y}}{d\Bar{\tau}}>0$ in the vicinity of $\Bar{y}=0$ i.e. the invariant submanifold $\Bar{y}=0$ is repelling if $(\Bar{x}=1)\wedge\left((m<0)\vee (m>\frac{1}{2})\right)$ or $(\Bar{x}=-1)\wedge(0<m<\frac{1}{2})$.

\item $\frac{d\Bar{y}}{d\Bar{\tau}}<0$ in the vicinity of $\Bar{y}=0$ i.e. the invariant submanifold $\Bar{y}=0$ is attracting if $(\Bar{x}=-1)\wedge\left((m<0)\vee (m>\frac{1}{2})\right)$ or $(\Bar{x}=1)\wedge(0<m<\frac{1}{2})$.
\end{itemize}
\subsubsection{\texorpdfstring{$\Bar{\sigma}= 0$}{}}
\begin{equation}
 \frac{d\Bar{y}}{d\Bar{\tau}}= \left(\frac{3m}{2m-1}\right)\bar{x}\Bar{y} 
\end{equation}
\begin{itemize}
    \item $\frac{d\Bar{y}}{d\Bar{\tau}}>0$ in the vicinity of $\Bar{y}=0$ i.e. the invariant submanifold $\Bar{y}=0$ is repelling if $(\Bar{x}>0)\wedge\left((m<0)\vee(m>\frac{1}{2})\right)$ or $(\Bar{x}<0)\wedge(0<m<\frac{1}{2})$.
    \item $\frac{d\Bar{y}}{d\Bar{\tau}}<0$ in the vicinity of $\Bar{y}=0$ i.e. the invariant submanifold $\Bar{y}=0$ is attracting if $(\Bar{x}<0)\wedge\left((m<0)\vee(m>\frac{1}{2})\right)$ or $(\Bar{x}>0)\wedge(0<m<\frac{1}{2})$.
\end{itemize}

\subsection{\texorpdfstring{$\Bar{\sigma}=\pm 1$}{}}
\begin{equation*}
\frac{d\Bar{\sigma}}{d\Bar{\tau}}=-3\epsilon\sqrt{1-\Bar{x}^{2}}\left((1-\Bar{y}^{2})(\Gamma -1)+\frac{(2m -3)m+1}{(4m-2)m}\Bar{y}^{2}\right)    
\end{equation*} 
 \begin{equation}
\frac{d\Bar{\sigma}}{d\Bar{\tau}}=\begin{cases}
			-3\epsilon\sqrt{1-\Bar{x}^{2}}\left(-\frac{(1-\Bar{y}^{2})}{n}+\frac{(2m -3)m+1}{(4m-2)m}\Bar{y}^{2}\right), & \text{For power law potential}\\
                -\frac{3((2m -3)m+1)}{(4m-2)m}\epsilon\sqrt{1-\Bar{x}^{2}}\Bar{y}^{2}, & \text{For exponential law potential}
		 \end{cases}    
\end{equation}
For the power law potential
\subsubsection{\texorpdfstring{$\Bar{y}\neq 0$}{}}
\begin{itemize}
\item $\frac{d\Bar{\sigma}}{d\bar{\tau}}>0$ if $-\frac{1}{n}+\left(\frac{(2m -3)m+1}{(4m-2)m}+\frac{1}{n}\right)\Bar{y}^{2}<0$ the invariant submanifold $\Bar{\sigma}=-1$ is repelling and $\Bar{\sigma}=1$ is attracting. 

\item  $\frac{d\Bar{\sigma}}{d\bar{\tau}}<0$ if $-\frac{1}{n}+\left(\frac{(2m -3)m+1}{(4m-2)m}+\frac{1}{n}\right)\Bar{y}^{2}>0$ the invariant submanifold $\Bar{\sigma}=-1$ is attracting and $\Bar{\sigma}=1$ is repelling.
\end{itemize}
\subsubsection{\texorpdfstring{$\Bar{y}= 0$}{}}
\begin{itemize}
\item $\frac{d\Bar{\sigma}}{d\bar{\tau}}>0$ if $\frac{1}{n}>0$ the invariant submanifold $\Bar{\sigma}=-1$ is repelling and $\Bar{\sigma}=1$ is attracting.

\item  $\frac{d\Bar{\sigma}}{d\bar{\tau}}<0$ if $\frac{1}{n}<0$ the invariant submanifold $\Bar{\sigma}=-1$ is attracting and $\Bar{\sigma}=1$ is repelling.
\end{itemize}

For the exponential law potential 
\begin{itemize}
\item $\frac{d\Bar{\sigma}}{d\bar{\tau}}>0$ if $\frac{(2m -3)m+1}{(4m-2)m}<0$ i.e. $(0<m<1)$ the invariant submanifold $\Bar{\sigma}=-1$ is repeling and $\Bar{\sigma}=1$ is attracting.

\item $\frac{d\Bar{\sigma}}{d\bar{\tau}}<0$ if $\frac{(2m -3)m+1}{(4m-2)m}>0$ i.e. $(m<0)\vee(m>1)$ the invariant submanifold $\Bar{\sigma}=-1$ is attracting and $\Bar{\sigma}=1$ is repelling. 
\end{itemize}

\addtocontents{toc}{\vspace{1em}} 

\backmatter


\label{References}
\lhead{\emph{References}}
\bibliographystyle{IEEEtranN}
\bibliography{References.bib}
\cleardoublepage
\pagestyle{fancy}
\label{Publications}
\chapter{List of Publications}
\lhead{\emph{List of Publications}}

\paragraph{List of publications included in thesis}
\begin{enumerate}
    \item \textbf{A.S. Agrawal}, L. Pati, S.K. Tripathy, and B. Mishra, ``Matter bounce scenario and the dynamical aspects in $f(Q,T)$ gravity", \textit{Physics of the Dark universe}, \textbf{33}, 100863 (2021).

    \item \textbf{A.S. Agrawal}, F. Tello-Ortiz, B. Mishra and S.K. Tripathy, ``Bouncing cosmology in extended gravity and its reconstruction as dark energy model", \textit{Fortschritte der Physik}, \textbf{70}, 202100065 (2022).

    \item \textbf{A.S. Agrawal}, B. Mishra and P.K. Agrawal, ``Matter bounce scenario in extended symmetric teleparallel gravity", \textit{European Physical Journal C}, \textbf{83}, 113 (2023).

    \item \textbf{A.S. Agrawal}, S. Mishra, B. Mishra and S.K. Tripathy ``Bouncing cosmological models in the functional form of $f(R)$ gravity", \textit{Gravitation and Cosmology}, \textbf{29}, 293-303 (2023). 

    \item  \textbf{A.S. Agrawal}, S. Chakraborty, B. Mishra, J. Dutta, and W. Khyllep, ``Global phase space analysis for a class of single scalar field bouncing solutions in general relativity”, \textit{European Physical Journal C}, \textbf{84}, 56 (2024).
    
\end{enumerate}

\paragraph{List of other publications}

\begin{enumerate}
\item \textbf{A.S. Agrawal}, S.K. Tripathy and B. Mishra, ``Gravitational baryogenesis models comparison in $f(R)$ gravity", \textit{Chinese Journal of Physics}, \textbf{71} 333-340, (2021).

\item B. Mishra, \textbf{A.S. Agrawal}, S.K. Tripathy and S. Ray, ``Wormhole solutions in $f(R)$ gravity", \textit{International Journal of Modern Physics D}, \textbf{30} 08, 2150061 (2021).

\item \textbf{A.S. Agrawal}, S.K. Tripathy, S. Pal, and B. Mishra, ``Role of extended gravity theory in matter bounce dynamics", \textit{Physica Scripta}, \textbf{97}, 2, 025002 {\bf}(2022).

\item B. Mishra, \textbf{A.S. Agrawal}, S.K. Tripathy, and S. Ray, ``Traversable wormhole models in $f(R)$ gravity", \textit{International Journal of Modern Physics A}, \textbf{37}, 05, 2250010 (2022).

\item \textbf{A.S. Agrawal}, B. Mishra, F. Tello-Ortiz, and A. Alvarez, ``$f(R)$ wormholes embedded in a Pseudo–Euclidean space $E^{5}$", \textit{Fortschritte der Physik}, \textbf{70}, 2100177 (2022).

\item \textbf{A.S. Agrawal}, B. Mishra and P. Moraes, ``Unimodular gravity traversable wormholes", \textit{European Physical Journal Plus}, \textbf{138}, 1-9 (2023).

\item \textbf{A.S. Agrawal}, B. Mishra, and S. Tripathy ``Observationally constrained accelerating cosmological model with higher power of non-metricity and squared trace", \textit{Journal of High Energy Astrophysics}, \textbf{38}, 41-48 (2023).

\end{enumerate}
\newpage
\paragraph{Workshop}
\begin{enumerate}
    \item Workshop on General Relativity and Cosmology, GLA University Mathura. (Oct 11-13, 2019).
    \item TEQIP-III Workshop on Relativity, Cosmology and Astrophysics. (Jan 27-30, 2020).
    \item TEQIP-III Sponsored International Webinar on Trends of Current Research in Physical Science, (Oct 17, 2020).
    \item “Teachers Enrichment Workshop” (Linear Algebra, Differential equations and its applications) at BITS-Pilani Hyderabad Campus (Jan 09-14, 2023).
\end{enumerate}

\paragraph{Conferences}  
\begin{enumerate}
     \item	TEQIP-III Sponsored International Webinar Recent Advances in Science and Technology. (Online) delivered an invited talk on “Wormhole solution in $f(R)$ Gravity” in the international webinar on Recent Advances in Sciences and Technology (RAST-2020). Held at Indira Gandhi Institute of Technology Sarang, Odisha, India (Nov 06-08, 2020).

    \item	$26^{th}$ International Conference of International Academy of Physical Sciences (CONIAPS-XXVI). On Advances in Relativistic Astrophysics and Cosmology (ARAC-2020). (Online) presented a research paper entitled “Gravitational Baryogenesis Models Comparisons in $f(R)$ Gravity”. Held at Department of Mathematics Sant Longowal Institute of Engineering and Technology, Punjab (Dec 18-20, 2020).
    
    \item	International webinar on Recent Developments in Modified Gravity and Cosmology (RDCM-2021). (Online) presented a research paper entitled “Gravitational Baryogenesis Models Comparisons in $f(R)$ Gravity” Held at the Department of Mathematics, Birla Institute of Technology and Science – Pilani, Hyderabad Campus, Hyderabad (March 9-11, 2021).
    
    \item	International Conference on Physical Interpretations of Relativity Theory-2021 (PIRT-2021). (Online) presented the research paper entitled “Matter Bounce Scenario in an Extended Gravity”. Held at Bauman Moscow State Technical University, Russia (July 5-9, 2021).
    
    \item	International webinar Recent Advances in Science and Technology (RAST-2021). (Online) delivered an invited talk on “Matter Bounce Cosmology in FLRW Metric with $f(R)$ Model”. Held at Indira Gandhi Institute of Technology Sarang, Odisha, India (July 28, 2021).
    
    \item	International Conference on Gravitation (Gravitex-2021). (Online) presented the research paper entitled “Bouncing Cosmology in Extended Gravity and its Reconstruction as Dark Energy Model.” Held at Durban, KwaZulu-Natal, South Africa (Aug 9-12, 2021).
    
    \item	$27^{th}$ International Conference of International Academy of Physical Sciences (CONIAPS XXVII). On Advances in Relativity and Cosmology (PARC-2021). (Online) presented the research paper entitled “Matter Bounce Scenario in the Functional Form of $F(R)$ Gravity”. Held at Department of Mathematics, Birla Institute of Technology and Science – Pilani, Hyderabad Campus, Hyderabad. (Oct 26-28, 2021).
    
    \item	(Online) flash talk delivered in Cosmology from Home-2021 entitled “Bouncing Cosmology in Extended Gravity and its Reconstruction as Dark Energy Model” (July 05-16, 2021).
    
    \item	Prof. P. C. Vaidya National Conference on Mathematical Sciences. (Online) presented a paper entitled “Bouncing Cosmology in Extended Gravity and its Reconstruction as Dark Energy Model.” Held at Department of Mathematics, Sardar Patel University, Vallabh Vidyanagar, Gujarat, India. (March 15-16, 2022).
    
    \item	Metric-Affine Frameworks for Gravity 2022. (Online) presented a paper entitled “Matter Bounce Scenario in the Functional Form of $F(R)$ Gravity”. Organized by the Laboratory of Theoretical Physics, Institute of Physics. The University of Tartu, Estonia (June 27- July 01, 2022).
    
    \item	(Online) talk delivered in Cosmology from Home-2022 Conference Entitled “Dynamical Stability of Bouncing Cosmology in Extended Gravity” (July 04-15-2022).
    
    \item	International Conference on Mathematical Sciences and its Applications-2022 (ICMSA-2022), presented a paper entitled “Dynamical Stability of Bouncing Cosmology in Extended Gravity”. Organized by the School of Mathematical Sciences, Swami Ramanand Tirth Marathwada University, Nanded (July 28-30, 2022).

    \item  The $16^{th}$ workshop on the Dark Side of the universe (DSU-2022). Delivered a talk on “Matter Bounce Scenario in Extended Symmetric Teleparallel Gravity” at the University of New South Wales, Sydney, Australia (December 05-09, 2022).
    
    \item	(Online) talk delivered in Cosmology from Home-2023 entitled “Matter Bounce Scenario in Extended Symmetric Teleparallel Gravity” (July 03-14, 2023)
    
    \item	International Conference on Physical Interpretations of Relativity Theory-2023 (PIRT-2023). (Online) presented the research paper entitled “Non-singular Bouncing Solution of the universe in Extended Symmetric Teleparallel Gravity”. Held at Bauman Moscow State Technical University, Russia  (July 03-06, 2023).

\end{enumerate}
\cleardoublepage

\pagestyle{fancy}
\lhead{\emph{Biography}}
\chapter{Biography}
\textbf{Brief Biography of Candidate}\\
\textbf{Mr. Agrawal Amarkumar Shyamsunder} completed his M.Sc. from N.E.S. Science College Nanded, Maharashtra, in 2016. He got a UGC NET-JRF all-India rank of 38 in June 2017. He published several research papers in reputed national and international journals. He has presented research papers at several national and international conferences. He was given an international travel grant by DST-SERB through the International Travel Scheme (ITS) to attend a conference at the University of New South Wales in Sydney, Australia, on December 5–9, 2022. The title of the talk was ``Matter Bounce Scenario in Extended Symmetric Teleparallel Gravity".

\textbf{Brief Biography of Supervisor}\\
\textbf{Prof. Bivudutta Mishra} is presently working as a Professor of Mathematics at Birla Institute of Technology and Science Pilani, Hyderabad Campus. He received his Ph.D. degree from Sambalpur University, Odisha, India in 2003. His main area of research is Cosmology and Relativity and currently working on the theoretical aspects of Dark Energy and Modified Theory of Gravity. He has already published around 132 research papers in these areas of research in journals of national and international repute. He has presented several research papers and delivered invited talks in these areas in national and international conferences held in India and abroad. He has conducted several academic and scientific events in the department. He has become a member of a scientific advisory committee of national and international academic events. He has successfully completed two sponsored projects: one each from the University Grants Commission (UGC), New Delhi, and the Science and Engineering Research Board, Department of Science and Technology, (SERB-DST), New Delhi. Presently he is handling one project from Council of Scientific and Industrial Research (CSIR), New Delhi. He is also an awardee of DAAD-RISE, 2019. He has also reviewed several research papers in highly reputed journals, Ph.D. examiner, and BoS member of several universities. He has also visited countries like Canada, UAE, Germany, the Republic of China, Malaysia, Italy, Czech Republic, Russia, Australia, Switzerland, Japan, Poland, UK and presented his research work at different scientific events. He is also a Visiting Associate of the Inter-University Centre for Astronomy and Astrophysics (IUCAA), Pune. At the same, he is also active in academic administration like Head of the Department, of Mathematics from 2012-2016,  Associate Dean of International Programmes and Collaborations since July 2018. He is also an awardee of DAAD-RISE, 2019. Fellow of Royal Astronomical Society, UK. Fellow of Institute of Mathematics and Applications, UK. Foreign member of the Russian Gravitational Society, Moscow.

\end{document}